\makeatletter\let\latex@xfloat\@xfloat\makeatother

\documentclass[12pt,a4paper]{report}

\usepackage{thesis}
\usepackage{lmodern}
\usepackage[semicolon]{natbib}
   \setcitestyle{aysep={}} 
\usepackage{har2nat}
\usepackage[hidelinks, colorlinks=true, linkcolor=black, citecolor=blue]{hyperref}
\usepackage[left=2.5cm,top=2.5cm,right=2.5cm,bottom=2.5cm]{geometry}

\usepackage{etoolbox}
\makeatletter
\let\@xfloat\latex@xfloat
\apptocmd{\@xfloat}{\linespread{1}\normalsize}{}{}
\makeatother

\usepackage{color}

\usepackage{physics}
\usepackage{amsmath}
\usepackage{graphicx}
\usepackage{multirow}
\usepackage{multicol}
\usepackage{amsfonts}
\usepackage{lscape}
\usepackage{caption}
\DeclareCaptionLabelFormat{adja-page}{\hrulefill\\#1 #2 (p.85-86)}
\usepackage{afterpage}
\newcommand\myemptypage{
    \null
    \thispagestyle{empty}
    \addtocounter{page}{-1}
    \newpage
    }
\usepackage{tocloft}

\newcommand{\thesistitle}{System Identification near a Hopf Bifurcation via the Noise-Induced Dynamics in the Fixed-Point Regime}
\newcommand{\thesisauthor}{Minwoo LEE}
\newcommand{\programname}{Mechanical Engineering}
\newcommand{\departmentname}{Department of Mechanical and Aerospace Engineering}
\newcommand{\thesisdate}{May 2020}
\newcommand{\signdate}{15 May 2020}
\newcommand{\thesisdegree}{PhD}
\newcommand{\dedicate}{\textit{To Aram, Yerin, and Sid Meier, \\ \vspace{5mm} who encouraged me to go on the next journey.}}
\newcommand{\supervisorinfo}{Prof. Larry K.B. LI, Thesis Supervisor}
\newcommand{\depheadinfo}{Prof. Huihe QIU, Head of Department}

\begin{document}

\pagenumbering{roman}
\pagestyle{plain}
\setcounter{page}{1}
\addcontentsline{toc}{chapter}{Title Page}
\thispagestyle{empty}
\null\vskip0.5in
\begin{center}
  \begin{LARGE}
    \thesistitle
  \end{LARGE}
  \vfill
  \vspace{20mm}

  by

  \vspace{4mm}

  \thesisauthor \\
  \vfill
  \vspace{20mm}

  A Thesis Submitted to\\
  The Hong Kong University of Science and Technology \\
  in Partial Fulfillment of the Requirements for\\
  the Degree of Doctor of Philosophy \\
  in \programname \\
  \vfill \vfill
  \thesisdate, Hong Kong
  \vfill
\end{center}

\vfill

\myemptypage{}

\newpage
\addcontentsline{toc}{chapter}{Authorization Page}
\null\skip0.2in
\begin{center}
{\bf \Large \underline{Authorization}}
\end{center}
\vspace{12mm}

I hereby declare that I am the sole author of the thesis.

\vspace{10mm}

I authorize the Hong Kong University of Science and Technology to lend this
thesis to other institutions or individuals for the purpose of scholarly research.

\vspace{10mm}

I further authorize the Hong Kong University of Science and Technology to
reproduce the thesis by photocopying or by other means, in total or in part, at the
request of other institutions or individuals for the purpose of scholarly research.

\vspace{30mm}

\begin{center}
\underline{~~~~~~~~~~~~~~~~~~~~~~~~~~~~~~~~~~~~~~~~~~~~~~~~~~~~~~~~~~~~~~~~~~~~~~}\\
~~~~\thesisauthor \\
~~~~\signdate

\end{center}

\myemptypage{}

\newpage
\addcontentsline{toc}{chapter}{Signature Page}
\begin{center}
{\Large \thesistitle}\\
\vspace{5mm}
by\\
\vspace{3mm}
\thesisauthor\\
\vspace{5mm}
This is to certify that I have examined the above PhD thesis\\
and have found that it is complete and satisfactory in all respects,\\
and that any and all revisions required by\\
the thesis examination committee have been made.
\end{center}

\vspace{15mm}

\begin{center}
\underline{~~~~~~~~~~~~~~~~~~~~~~~~~~~~~~~~~~~~~~~~~~~~~~~~~~~~~~~~~~~~~~~~~~~~~~~~~~~ }\\
\supervisorinfo
\end{center}


\vspace{15mm}
\begin{center}
\underline{~~~~~~~~~~~~~~~~~~~~~~~~~~~~~~~~~~~~~~~~~~~~~~~~~~~~~~~~~~~~~~~~~~~~~~~~~~~ }\\
\depheadinfo
\end{center}

\vspace{5mm}
\begin{center}
\departmentname\\
\vspace{5mm}
\signdate
\end{center}

\myemptypage{}

\newpage
\thispagestyle{empty}
\null\vskip0.5in
\begin{center}

  \vspace{20mm}

  \begin{LARGE}
    \dedicate 
  \end{LARGE}

  \vspace{4mm}

\end{center}

\vfill

\myemptypage{}

\newpage
\addcontentsline{toc}{chapter}{Acknowledgments}
\centerline{{\bf \Large Acknowledgments}} \vspace{15mm} \noindent

First and foremost, I would like to express my deepest appreciation to my supervisor, Prof. Larry K.B. Li, for his invaluable guidance and support. I have been very fortunate to have an advisor who patiently taught me how to develop ideas and formulate arguments in a scientific manner. His encouragement has been the greatest contributor to the completion of this thesis. 

\vspace{2mm}

I would also like to extend my sincere gratitude to Prof. Vikrant Gupta at SUSTech for his constructive advice throughout my study. His insightful comments steered me in the right direction. I am also grateful to Prof. Kyutae Kim and his team at KAIST for their kind help and cooperation at the final stage of my research. 

\vspace{2mm}

I have great pleasure in acknowledging Mr. Yuanhang Zhu at Brown University and Dr. Yu Guan at HKUST who carried out all the experiments in Chapters 2 and 3, respectively. This thesis would not have been possible without their help.

\vspace{2mm}

Most importantly, I am thankful to my family for their incredible support and understanding. I am grateful to my two sets of parents, who have given me wise counsel. I thank my daughter, Yerin, for her ability to make me smile. Above all, I heartily thank my wife Aram for her encouragement and inspiration throughout my study.

\vspace{2mm}

Last but not least, I would like to thank the Research Grants Council of Hong Kong and LG CNS for their financial and academic support.

\myemptypage{}

\newpage
\addcontentsline{toc}{chapter}{Table of Contents}
\tableofcontents

\newpage
\addcontentsline{toc}{chapter}{List of Figures}
\listoffigures

\newpage
\addcontentsline{toc}{chapter}{List of Tables}
\listoftables

\newpage
\addcontentsline{toc}{chapter}{Abstract}
\begin{center}
{\Large \thesistitle}\\
\vspace{20mm}
by \thesisauthor\\
\departmentname\\
The Hong Kong University of Science and Technology
\end{center}
\vspace{8mm}
\begin{center}
Abstract
\end{center}
\par
\noindent

A Hopf bifurcation, where a fixed-point solution loses stability and a limit cycle is born, is prevalent in many nonlinear dynamical systems. When a system prior to a Hopf bifurcation is exposed to a sufficient level of noise, its noise-induced dynamics can provide valuable information about the impending bifurcation and the post-bifurcation dynamics. In this thesis, we present a system identification (SI) framework that exploits the noise-induced dynamics prior to a supercritical or subcritical Hopf bifurcation. The framework is novel in that it is capable of predicting the bifurcation point and the post-bifurcation (limit-cycle) dynamics using only pre-bifurcation data. Specifically, we present two different versions of the framework: input-output and output-only. For the input-output version, the system is forced with additive noise generated by an external actuator, and its response is measured. For the output-only version, the intrinsic noise of the system acts as the noise source, so no external actuator is required, and only the output signal is measured. In both versions, the Fokker--Planck equations, which describe the probability density function of the fluctuation amplitude, are derived from self-excited oscillator models. Then, the coefficients of these models are extracted from the experimental probability density functions characterizing the noise-induced response in the fixed-point regime, prior to the Hopf point itself. These two versions of the SI framework are tested on three different experimental systems: a hydrodynamic system (a low-density jet), a laminar thermoacoustic system (a flame-driven Rijke tube), and a turbulent thermoacoustic system (a gas-turbine combustor). For these systems, we demonstrate that the proposed framework can identify the super/subcritical nature of the Hopf bifurcation and the system's order of nonlinearity. Moreover, by extrapolating the identified model coefficients, we are able to forecast the locations of the bifurcation points and the limit-cycle features after those points. To the best of our knowledge, this is the first time that SI has been performed using data from only the pre-bifurcation (fixed-point) regime, without the need for \textit{a priori} knowledge of the location of the bifurcation point. Given that such noise-induced dynamics are universal near a Hopf bifurcation, the proposed SI framework should be applicable to a variety of nonlinear dynamical systems in nature and engineering.


\myemptypage{}

\newpage
\pagenumbering{arabic}
\pagestyle{plain}
\setcounter{page}{1}
\chapter{Introduction} \label{chap:intro}

\begin{quote}
   \textit{``Like the mythical perpetual motion machine, self-oscillation succeeds in driving itself, but does so in a way that is compatible with the known laws of physics.''}\\
   \vspace{-2mm}
   \hfill--- \textcolor{blue}{A.} \citet{jenkins2013}
\end{quote}

\vspace{5mm}

A self-excited oscillation\footnote{It is also known as `self-oscillation', `self-sustained oscillation' or `autonomous oscillation'.} is a special type of oscillation---sustained by a balance between an energy source and some dissipation mechanism---that is only observable in nonlinear systems \citep{pikovsky2003}. Self-excited oscillations are characterized by three key features:  (i) they neither grow nor decay in time, (ii) they are driven by an internal energy source, not by external rhythmic forcing, and (iii) the amplitude, frequency and shape of the oscillations are determined by the system itself, rather than by the initial conditions \citep{balanov2008}. Such self-excited oscillations can be found in many natural phenomena, such as the heartbeat, firefly flashes, and stellar pulsations \citep{pikovsky2003, jenkins2013}.


In engineered systems, however, self-excited oscillations are often detrimental, as they can excite unwanted acoustic or structural resonances. For example, self-excited thermoacoustic oscillations, which manifest as high-amplitude pressure fluctuations (i.e. thermoacoustic instabilities), may induce cyclic fatigue loading on the hardware of gas turbines and rocket engines, potentially leading to catastrophic mechanical failure \citep{culick1992, lieuwen2012unsteady}. It is therefore crucial to be able to predict self-excited oscillations in a dynamical system before they actually occur.

Accordingly, numerous methods for predicting the onset of self-excited oscillations have been studied. In particular, researchers have found that when a system is about to exhibit self-excited oscillations, certain parameters of the system gradually change. By setting a threshold on these parameters, researchers have been able to use them as early-warning indicators of an impending self-excited instability \citep{chisholm2009, Gopalakrishnan2016}. However, to determine the instability threshold, one needs to operate the system in the self-excited oscillatory state, which is often unsafe. Moreover, although existing early-warning indicators can predict the location where the self-excited oscillations are born, they cannot reliably predict the characteristics of the resultant limit cycle. In this thesis, we develop a predictive framework that can overcome these limitations, by exploiting the noise-induced dynamics arising in the fixed-point regime, before the bifurcation itself.


Unlike the conventional perspective that noise is contamination to a system signal, recent studies have shown that the response of a system to noise can provide valuable information about the system dynamics \citep{Horsthemke1984, Neiman1997, Gammaitoni1998}. Specifically, when the system is exposed to an optimal level of noise, its dynamics exhibits a peak in coherence \citep{Wiesenfeld1985, Neiman1997}. This phenomenon---known as coherence resonance---will be used in this research to develop a forecasting strategy for the onset of self-excited oscillations. In particular, we will show that it is possible to predict not only the onset of self-excited oscillations but also the dynamics of their nonlinearly saturated dynamics without having to collect data from the potentially dangerous oscillatory state.

\section{Hopf bifurcation} \label{sec:hopf}


In dynamical systems, a bifurcation occurs when small variations in its parameter---the bifurcation parameter---cause a qualitative change in the system dynamics. For example, for a dynamical system with bifurcation parameter $\epsilon$: 

\begin{equation}\label{bif}
    \frac{\mathrm{d}{X}}{\mathrm{d}t}=F(X,\epsilon),
\end{equation}
a bifurcation exists at a critical value of $\epsilon = \epsilon_c$, if a change in the stability and/or the number of system equilibria occurs at this point \citep{verhulst1990}. There are many different types of bifurcation, such as Hopf, saddle-node (fold), pitchfork, flip, and transcritical bifurcations, to name just a few \citep{strogatz2000}. 




In this thesis, we focus on the Hopf bifurcation, which is a common route through which self-excited oscillations arise \citep{marsden1976}. In equation~\ref{bif}, let $X^{*}$ be a steady solution $F(X^{*},\epsilon)=0$, while $\epsilon$ is varied. When linearized near $X^{*}$, the Jacobian matrix $J$ of the system becomes:
\begin{equation}\label{jaco}
    J=\pdv{{F(X,\epsilon)}}{X}\bigg\rvert_{X=X^{*}}.
\end{equation}

Suppose there exists a pair of nonzero complex conjugate eigenvalues of $J$. When this pair of eigenvalues crosses the imaginary axis as a result of variations in $\epsilon$, the fixed-point solution loses stability, giving rise to a limit cycle at the Hopf point $\epsilon=\epsilon_c$ \citep{marsden1976, strogatz2000}. The normal-form equation for a Hopf bifurcation in terms of variable $a$ is\footnote{In fluid mechanics, this equation is known as the Stuart--Landau equation, which \citet{landau1944problem} originally proposed and \citet{stuart1960non} and \citet{watson1960non} later formally derived from the hydrodynamic equations using an energy balance.}:
\begin{equation}\label{norm_Hopf}
    \frac{\mathrm{d}{a}}{\mathrm{d}t} = a((\epsilon+i)+\alpha_1{|a|}^2+\cdots),
\end{equation}
where $\alpha_1=\alpha_{R1}+\alpha_{I1} i$. Both $\alpha_{R1}$ and $\alpha_{I1}$ are real, and $\alpha_{R1}$ is referred to as the first Lyapunov constant. The amplitude of the limit cycle is determined by $\epsilon$ and $\alpha_{R1}$, while its angular frequency is determined by $\alpha_{I1}$ \citep{karaaslanli2012}. 

\begin{figure}
    \centering
    \includegraphics[width=0.48\textwidth,trim={0 0 0 0},clip]{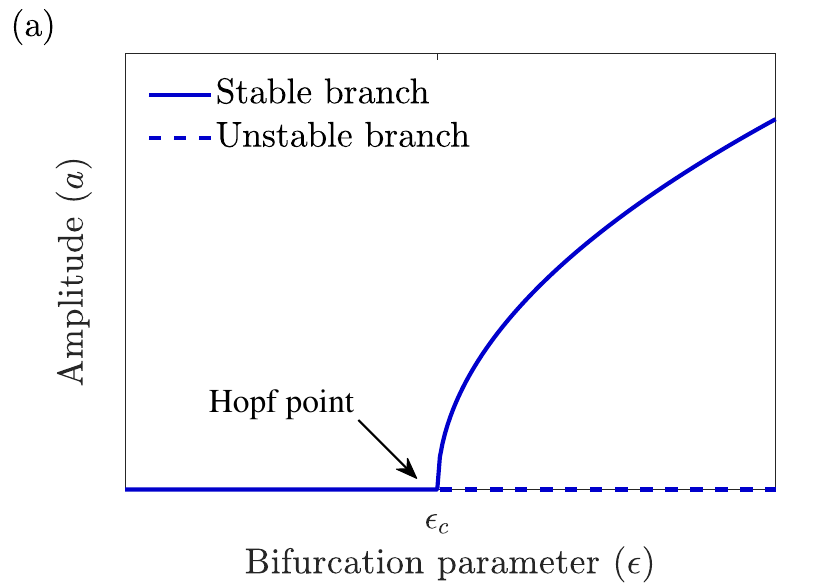}%
    \includegraphics[width=0.48\textwidth,trim={0 0 0 0},clip]{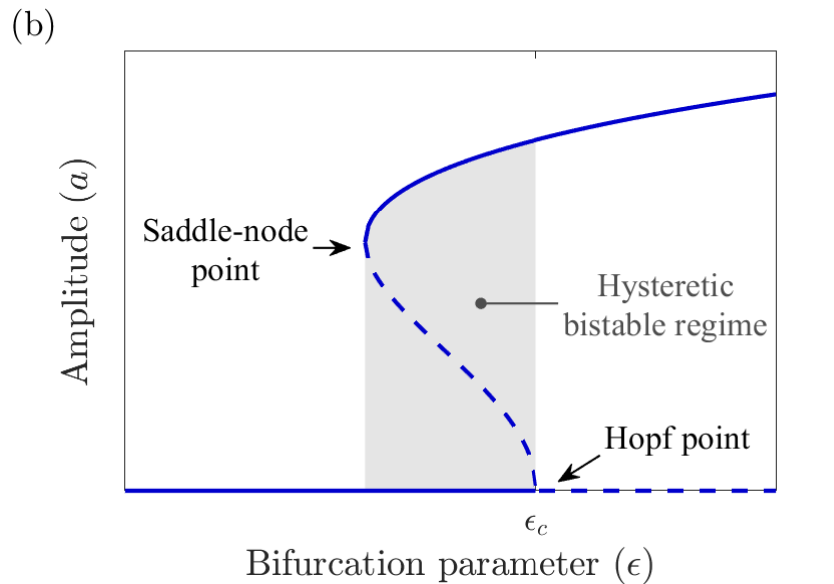}%
    \caption{Two classic types of Hopf bifurcation: (a) supercritical and (b) subcritical. Only in the subcritical case is there a hysteretic region of bistability.}\label{fig:Hopf}
\end{figure}

A Hopf bifurcation can be classified into two types. The first type arises when $\alpha_{R1}$ is negative: in this case, a \textit{supercritical} Hopf bifurcation occurs where a limit cycle is observed only after the critical point ($\epsilon>\epsilon_c$) and its amplitude increases gradually as the bifurcation parameter increases. The second type arises when $\alpha_{R1}$ is positive: in this case, a \textit{subcritical} Hopf bifurcation occurs where a limit cycle can also be found in the hysteretic bistable regime ($\epsilon<\epsilon_c$) between the Hopf and saddle-node points. In a subcritical Hopf bifurcation, an abrupt jump in the oscillation amplitude occurs as the bifurcation parameter increases past the Hopf point (see figure \ref{fig:Hopf}b) \citep{marsden1976}. 

Hopf bifurcations occur in many oscillatory systems in nature and engineering, such as chemical systems \citep{Kopell1973,vanag2000}, biological systems \citep{fussmann2000,liu2011}, electrical systems \citep{tomim2005, Divshali2009}, financial systems \citep{gao2009}, fluid systems \citep{mathis1984,jackson_1987,Provansal1987,monk1990self,Raghu1991,zhu2017onset,juniper2009forcing,balusamy2017extracting,murugesan2017intermittency,ren2018global,ren2018spatiotemporal} and fluid-structural systems \citep{ghadami2016,ghadami2017,he2017ground,ghadami2018,he2018non,he2018ground,he2019stability,he2019non}. The resultant limit-cycle oscillations are desirable in some devices, such as acoustic instruments \citep{abel2009} and pulsed combustors \citep{putnam1986}, but they are detrimental in others, such as aeroelastic systems \citep{blevins1977}, gas turbines and rocket engines \citep{Lieuwen2005,jegal2019mutual,moon2020mutual} and fatigue-prone structures \citep{schijve2009fatigue}. Regardless of the exact situation, it is important to be able to forecast the location and type of a Hopf bifurcation as well as its limit-cycle dynamics.

\section{Early warning indicators of a bifurcation} \label{sec:hopf_pr}

Given the importance of forecasting a bifurcation, it is no surprise that much effort has been devoted over the past several decades to developing early warning indicators of an impending instability. Below we review some of the main outcomes of such efforts.

\subsection{Critical slowing down}

\begin{figure}
    \centering
    \includegraphics[width=0.7\textwidth,trim={0 0 0 0},clip]{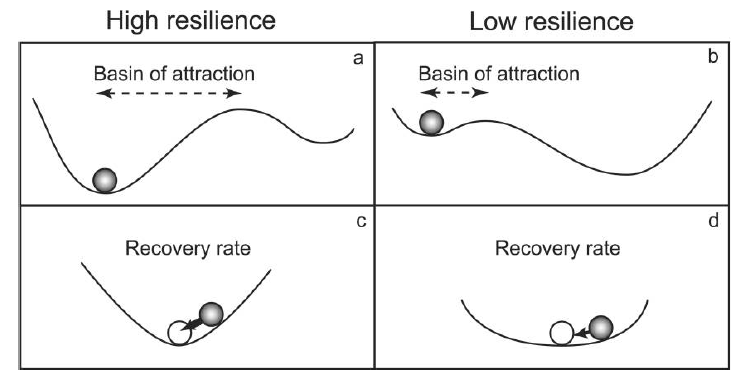}%
    \caption{Illustration of critical slowing down (CSD). (a,c) Far from a bifurcation, the system recovers quickly from a perturbation. (b,d) Close to a bifurcation, the recovery rate decreases. Reproduced from \citet{vannes2007}.}\label{fig:resil}
\end{figure}

When subjected to a small perturbation, a stable system tends to return to its original equilibrium state. If the system is close to the `critical point' (i.e. a bifurcation), the rate at which it recovers from the perturbation decreases (see figure \ref{fig:resil} for an illustration). Such `slowing down' near a bifurcation is called critical slowing down (CSD). The phenomenon of CSD has long been used to develop early warning indicators of  bifurcations \citep{ma1976,wissel1984}. Analysis of various systems has shown that CSD starts well before the bifurcation point itself, with the recovery rate decreasing monotonically as the bifurcation is approached \citep{vannes2007}. CSD can be inferred from numerous metrics, including an increasing return time \citep{kramer1985, scholz1987, tredicce2004, dai2012, veraart2012}, a higher lag-1 autocorrelation \citep{dakos2008, hines2011, dai2012, veraart2012, Gopalakrishnan2016b}, and an increasing variance \citep{dai2012, Gopalakrishnan2016b}.

CSD has been observed for various types of bifurcations (see \citealp{scheffer2012} for a review), including Hopf bifurcations \citep{chisholm2009, lim2011, Gopalakrishnan2016b, ghadami2016, ghadami2017, ghadami2018}. For example, during investigations of a prototypical thermoacoustic system, \citet{Gopalakrishnan2016b} showed that CSD can be used to predict the onset of a Hopf bifurcation. In particular, these researchers found an increase in the lag-1 autocorrelation and variance well before the onset of instability. In another study, \citet{lim2011} proposed a methodology for forecasting the post-bifurcation dynamics, based on an analysis of the CSD features under large-amplitude perturbations. In this method, with prior knowledge of the location of the bifurcation point, the type of the Hopf bifurcation (i.e. whether it is supercritical or subcritical) and the resultant limit-cycle amplitude can be predicted from only pre-bifurcation data. Using this technique, \citet{ghadami2016, ghadami2017} were able to predict the post-bifurcation dynamics of a nonlinear aeroelastic model exposed to gust perturbations. Although the post-bifurcation limit-cycle amplitude was well predicted with only pre-bifurcation data, this method requires \textit{a priori} information about how close the system is to the bifurcation point.

\subsection{Asymmetric distribution}
\begin{figure}[t]
    \centering
    \includegraphics[width=0.8\textwidth,trim={0 0 0 0},clip]{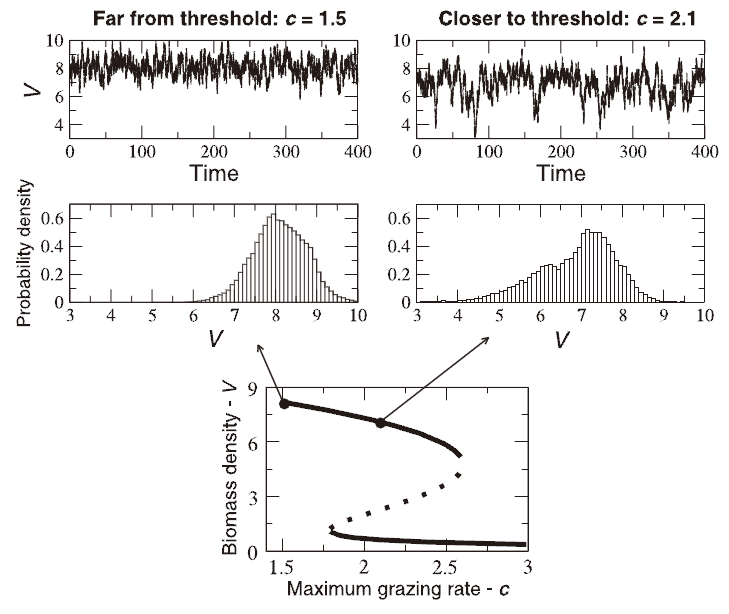}%
    \caption{Illustration showing the skewness of the biomass density $V$ as the bifurcation parameter $c$ is varied. The asymmetry intensifies as the system approaches the bifurcation point. Reproduced from \citet{guttal2008}.}\label{fig:skew}
\end{figure}

Another early warning indicator of an impending instability is the asymmetry present in the distribution of a time series \citep{guttal2008, carpenter2011, guttal2013}. Specifically, the probability density function (PDF) of a time trace is nearly symmetric with respect to the median, but its skewness tends to increase as the system approaches the bifurcation point (see figure \ref{fig:skew}) \citep{guttal2008}. This `skewness' of the signal occurs on both sides of the bifurcation, which allows it to serve as a useful precursor \citep{guttal2013}. However, this precursor lacks the ability to forecast the exact location of the bifurcation point. 

\subsection{Noise-driven precursors}
Early warning indicators of an impending instability can also be extracted under a highly stochastic (noisy) environment. Such precursors have been developed for systems exhibiting strong intrinsic noise, such as the turbulent combustors used in gas-turbine engines. Such precursors are different from those that rely on the noise-induced dynamics under extrinsic forcing (e.g. coherence resonance: to be discussed in \S\ref{subsec:CR}), because they make use of only intrinsic noise. Such precursors, therefore, have a practical advantage in that no external forcing, and thus no actuation device, is required.

\subsubsection{Damping coefficient}
If a system is in the fixed-point regime of a Hopf bifurcation and is subjected to strong intrinsic noise, its instantaneous state is governed by a competition between its inherent damping and the stochastic driving \citep{Lieuwen2005a}. The overall level of damping decreases as the system approaches the bifurcation point, which implies that the damping coefficient can be used as an early warning indicator. A celebrated example of this was reported by \citet{Lieuwen2005a}, who proposed a strategy for forecasting the stability margins (i.e. how far a system is from the bifurcation) of a gas-turbine combustor using the damping coefficient, as computed from the autocorrelation function. \citet{yi2008} later extended this approach to the frequency domain, obtaining noise-driven precursors for multi-mode oscillations. Compared with other existing precursors, however, the damping coefficient is limited in that the system must be relatively close to the bifurcation in order for the forecast to be accurate. Furthermore, if a transition between two states (e.g. flickering) occurs, it can be difficult to compute the damping coefficient.

\subsubsection{Flickering} \label{flick}
In highly stochastic systems, the transition from one state to another can occur even well before the bifurcation point \citep{scheffer2009}. This is because when an alternative basin of attraction emerges, noise can cause the system to flip to that state \citep{scheffer2012}. The phenomenon in which a system rapidly switches back and forth between two states is referred to as flickering\footnote{Rapid alternations involving chaos are separately classified as intermittency (see \S\ref{intermit})} \citep{vannes2007}. Flickering can serve as an early warning indicator of an impending instability because it occurs increasingly frequently as a system approaches the bifurcation point \citep{scheffer2012, dakos2013}. For example, \citet{wang2012} studied an ecological system and found flickering in the form of eutrophication events and algal blooms. These researchers showed that such flickering presages a bifurcation from the oligotrophic state to the eutrophic state. However, like most other precursors, flickering only provides a qualitative index of an impending bifurcation.

\subsection{Intermittency} \label{intermit}
Intermittency is a phenomenon in which a system repeatedly alternates between two qualitatively different states \citep{schuster2006}. By analyzing low-order dissipative dynamical systems, \citet{pomeau1980} identified three different types of intermittency, each corresponding to a different bifurcation: (i) type-I, which occurs near a saddle-node bifurcation, (ii) type-II, which occurs near a subcritical Hopf bifurcation and (iii) type-III, which occurs near an inverse period-doubling bifurcation. Later, \citet{platt1993} identified another type of intermittency involving aperiodic temporal fluctuations in the bifurcation parameter. This `on-off' intermittency causes a system to alternate abruptly between a quiescent (off) state and a bursting (on) state \citep{platt1993}. At first glance, intermittency appears similar to flickering, but the former typically involves a transition to or from chaos.

For example, recognizing that combustion noise is made up of high-dimensional chaotic fluctuations \citep{nair2013}, \citet{nair2014a} showed that intermittency presages the transition from (chaotic) combustion noise to self-excited periodic thermoacoustic oscillations. Because such intermittency lasts longer in time as the system approaches the bifurcation point, these researchers suggested that the onset of thermoacoustic instability could be forecasted by quantifying the loss of chaos in the intermittent state \citep{nair2014a}. Intermittency has also been identified as a route to self-excited chaotic thermoacoustic oscillations \citep{guan2020intermittency}. Furthermore, \citet{venkatramani2016, venkatramani2018} showed that the onset of flutter in an aeroelastic system is presaged by on-off intermittency, and can be forecasted with statistical metrics computed via recurrence quantification analysis. Given that the instability threshold is defined \textit{a priori}, such precursors can forecast an impending instability further in advance than can other precursors, such as the damping coefficient \citep{nair2014a}. However, this class of precursors lacks the ability to forecast the exact location of the bifurcation point and the post-bifurcation dynamics.

\subsection{Multi-fractality}
When a system exhibits chaos, its signal contains a self-similar structure across different time scales. In other words, a chaotic time signal has fractal patterns, whose individual sub-sections resemble the whole signal. The complexity of a fractal signal can be quantified with the fractal dimension $D$ \citep{falconer2004}. A fractal time signal $x(t)$ has the relationship $x(ct)=x(t)/c^H$ for some scaling $c$ \citep{falconer2004}, where $H$ is the Hurst exponent characterizing the fractal and is related to $D$ via $D=2-H$ \citep{bassingthwaighte1994}.

In practical systems, however, a signal is never perfectly self-similar, and a single fractal dimension is never enough to describe the dynamics perfectly. Such systems feature an interwoven subset of different fractal dimensions, and are often called multi-fractal systems \citep{harte2001}. In such systems, a single Hurst exponent is not enough to describe the whole system. Instead, the singularity spectrum ($f(\alpha)$, also called the multi-fractal spectrum) is used to characterize the system \citep{paladin1987}. Specifically, the width of the singularity spectrum ($\Delta\alpha$) provides a measure of the multi-fractality of a signal. Over the last several decades, multi-fractality has been found in various systems, such as fluid systems \citep{Sreenivasan1986, Sreenivasan1991}, combustion systems \citep{gotoda2012, nair2014b}, mechanical systems \citep{lin2013} and electromagnetic systems \citep{hikihara1997}, among others.

\begin{figure}
    \centering
    \includegraphics[width=0.85\textwidth,trim={0 0 0 0},clip]{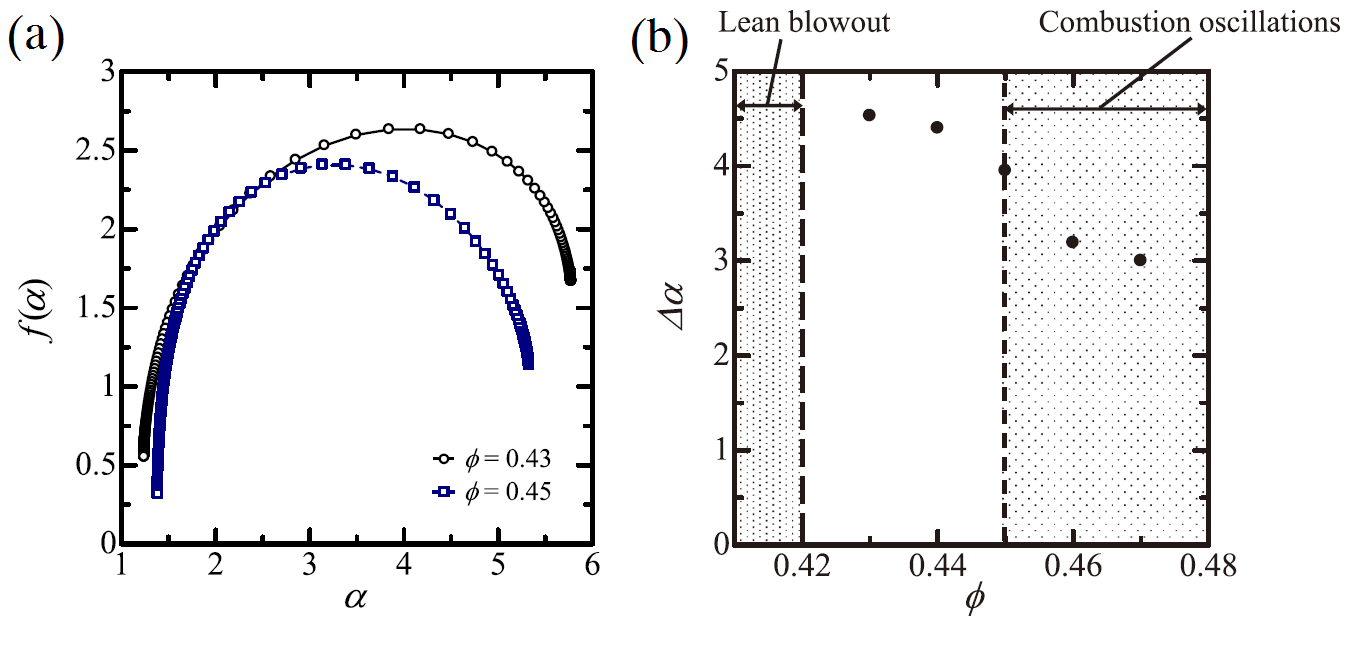}%
    \caption{(a) Singularity spectra and (b) its widths for varying equivalence ratios ($\phi$), which are used as indicators of multi-fractality \citep{gotoda2012}.}\label{fig:multif}
\end{figure}

Recently, \citet{gotoda2012} found that the multi-fractality of a gas-turbine system weakens before the onset of self-excited thermoacoustic oscillations. In particular, these authors showed that the width of the singularity spectrum $f(\alpha)$ starts to decrease before the system reaches a self-excited state (see figure \ref{fig:multif}). \citet{nair2013} and \citet{nair2014b} further showed that a multi-fractal signature also exists in combustion noise, and can serve as a precursor to thermoacoustic instability. In another study, \citet{venkatramani2017} analyzed the multi-fractal characteristics of an aeroelastic system and showed that precursors such as the generalized Hurst exponent are capable of predicting a Hopf bifurcation, prior to the onset of flutter. However, as noted by \citet{nair2014b}, an ad-hoc instability threshold for the multi-fractal measure is required in order to be able to accurately track the proximity to the bifurcation point.

\subsection{Permutation entropy}
Another measure of the complexity of a time series is the permutation entropy \citep{bandt2002}, which encapsulates the relationship between different segments of a time series by extracting the probability distribution of the ordinal patterns \citep{henry2019}. The normalized permutation entropy ($\Bar{h}$) takes on a value between 0 and 1. If $\Bar{h}$ is close to 1, the system dynamics is complex, and thus the nature of the time series is stochastic. By contrast, if $\Bar{h}$ is close to 0, the system is more deterministic. For a signal contaminated with noise, the permutation entropy is easier to compute than other complexity measures such as the fractal dimension \citep{bandt2002}. Also, owing to its algorithmic simplicity, the permutation entropy is computationally inexpensive.

In a study on quantifying system complexity, \citet{lamberti2004} proposed the statistical complexity measure (SCM), which defines a quantifier of Jensen--Shannon statistical complexity ($C_{JS}$). Specifically, $C_{JS}$ is obtained from the product of $\Bar{h}$ and the discrepancy between $P$ and $P_e$ \citep{lamberti2004}. Here, $P$ is the probability distribution of the input time series, and $P_e$ is that of the uniform distribution. $C_{JS}$ provides significant additional information about the complexity of a system, especially regarding its underlying probability distribution \citep{dong2018}. Later, \citet{rosso2007} suggested that combining $\Bar{h}$ and $C_{JS}$, specifically by creating a two-dimensional plane with $\Bar{h}$ on the $x$-axis and $C_{JS}$ on the $y$-axis, provides a powerful tool for examining the complexity in time-series data. This plane is now known as the complexity-entropy causality plane (CECP) and is often used to distinguish between the dynamical states of nonlinear systems \citep{ribeiro2012, dong2018, hachijo2019}.

Besides their ability to identify the dynamical states of nonlinear systems, the permutation entropy and CECP can be used to predict bifurcations. For example, by analyzing brain signals (electroencephalogram) from genetic absence epilepsy rats, \citet{li2007} found that the permutation entropy decreases well before the onset of seizure. From this observation, the researchers postulated that seizures can be predicted by setting a predefined threshold for the permutation entropy. \citet{gotoda2012} analyzed a combustion system and found a decrease in the permutation entropy at the onset of thermoacoustic instability. Furthermore, \citet{hachijo2019} projected to the CECP the transition from small-amplitude aperiodic oscillations to large-amplitude self-excited periodic oscillations. Aided by machine learning, these researchers showed that the transition to self-excited thermoacoustic oscillations can be predicted well before they emerge. This was achieved by assigning zones to the CECP (i.e. feature space) via k-means clustering so as to delineate the different dynamical states.

\subsection{Complex networks}
Recently, complex networks have been increasingly used to model the dynamics of complex systems \citep{boccaletti2006}. In \emph{network analysis} (also known as \emph{graph analysis} in mathematics), a system is modeled as a set of items (nodes) with connections between them (edges) \citep{newman2003}. For a system with many co-interacting subsystems (i.e. a complex system), the network properties can provide valuable information about the system dynamics. Converting time-series data into a complex network is an effective way of uncovering hidden patterns in complex data \citep{small2013}. Over the last two decades, complex networks have been used extensively to analyze various systems (see \citealt{strogatz2001} and \citealt{newman2003} for comprehensive reviews), such as the world-wide-web \citep{albert1999}, socio-biological networks \citep{girvan2002}, power grids \citep{pagani2013}, turbulent jet flow \citep{charakopoulos2014}, low-density jets \citep{murugesan2016recurrence,murugesan2019complex}, and thermoacoustic systems \citep{murugesan2016}. 

\begin{figure}
    \centering
    \includegraphics[width=0.85\textwidth,trim={0 0 0 0},clip]{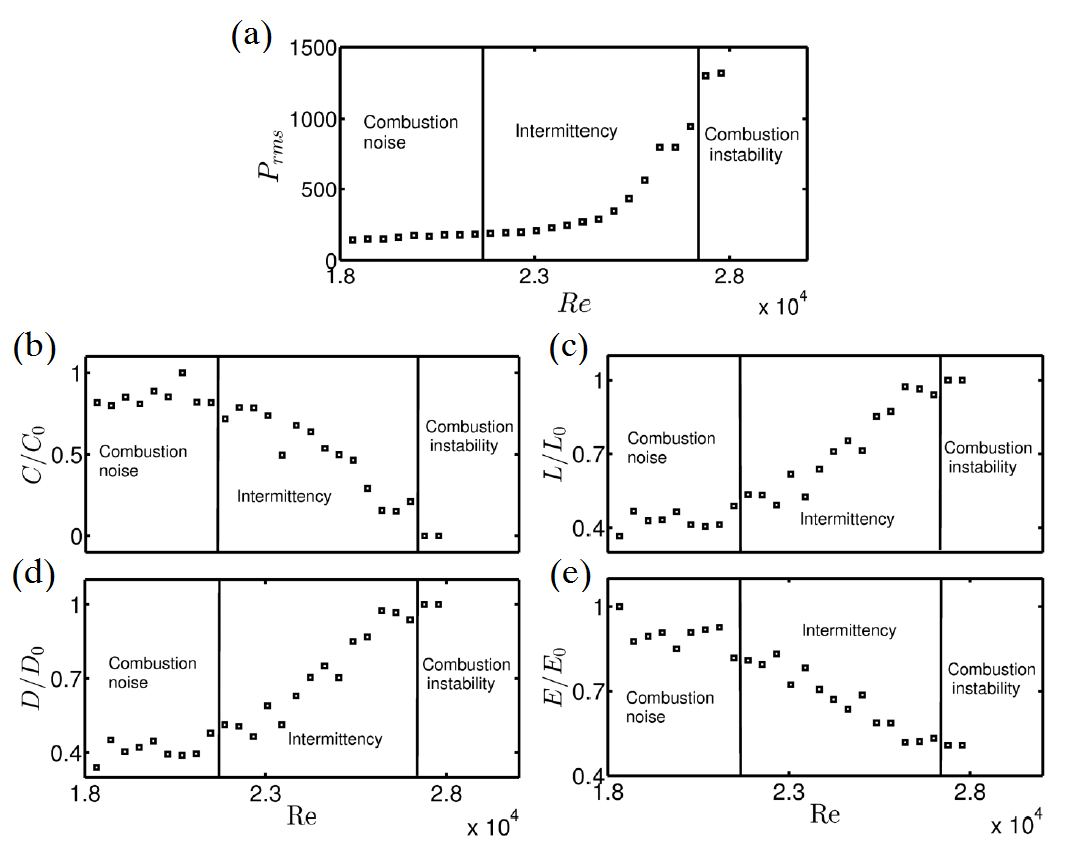}%
    \caption{Variation of (a) rms unsteady pressure and (b-e) network properties in a combustor \citep{murugesan2016}. The network properties shown are (b) the normalized clustering coefficient, (c) the normalized characteristic path length, (d) the normalized network diameter, and (d) the normalized global efficiency.}\label{fig:netw}
\end{figure}

Close to a bifurcation, changes in the dynamics of a system can be detected in its network properties. For example, \citet{peng2019} showed that the network properties--- such as the entropy of transition networks and the mean edge betweenness of visibility graphs--- decrease as the system approaches a pitchfork bifurcation. By analyzing a mathematical model of lake eutrophication, these researchers showed that these network-based precursors can forecast real-world bifurcations. For a thermoacoustic system, \citet{murugesan2016} analyzed four network properties, namely the clustering coefficient, the characteristic path length, the network diameter, and the global efficiency. As the system approaches the onset of thermoacoustic instability (i.e. self-excited limit-cycle oscillations), these properties are found to vary monotonically (see figure \ref{fig:netw}), suggesting that they can serve as early warning indicators of an impending bifurcation. \citet{kobayashi2019} proposed a method for early detection of thermoacoustic instability using complex networks that combines the use of transition patterns in the ordinal partition transition network and machine learning. As with \citet{hachijo2019}, thresholds in the feature space were defined using support vector machines.

\vspace{13mm}
We have reviewed several precursors of impending bifurcations. Although they can all forecast the onset of instability, they suffer from two key limitations. First, to accurately predict the location of a bifurcation, it is necessary to define ad-hoc instability thresholds. Such thresholds generally require experimental or numerical data from both the pre- and post-bifurcation regime; the latter regime can be dangerous to enter because the oscillations within it are usually large in amplitude and self-excited. Second, without \textit{a priori} information about the bifurcation point, existing early warning indicators tend to have difficulty predicting the limit-cycle dynamics after the Hopf point. In this thesis, we present a framework that can overcome both of these limitations, while remaining generic enough to be applicable to a variety of nonlinear dynamical systems in nature and engineering. To achieve this goal, we analyze the noise-induced dynamics of the system, which is discussed in \S\ref{subsec:NID}.

\section{Noise-induced dynamics} \label{subsec:NID}
Practical systems are usually contaminated by noise, which may arise as perturbations to the system parameters (parametric noise), as perturbations that depend on the system state (multiplicative noise), or as perturbations independent of the system parameters or its state (additive noise). It is therefore intuitive to view noise as a source of contamination on the dynamics of a system \citep{Horsthemke1984}. However, studies have shown that noise can induce counterintutive phenomena, providing information about the system that might otherwise be overlooked \citep{Horsthemke1984, Neiman1997, Gammaitoni1998}.

For this reason, the response of nonlinear dynamical systems to noise has attracted much attention from mathematicians, scientists and engineers over the last several decades. Studies have shown that noise-induced dynamics can arise in a variety of mechanical, biological, and chemical processes, ranging from micro-optical transport \citep{Bhaban2016} to plasma fluctuations \citep{Nurujjaman2008} to bursting neurons \citep{Lang2010} to glacial climate changes \citep{Ganopolski2002}. Examples of phenomena arising from the influence of noise include stochastic resonance \citep{Benzi1981, Gammaitoni1998, Benzi2010}, coherence resonance \citep{Neiman1997, pikovsky1997coherence, Kabiraj2015, Gupta2017, zhu2017onset}, noise-induced transition \citep{Horsthemke1984, Doering1986, Landa2000}, noise-induced pattern formation \citep{Parrondo1996}. Below, we review the salient features of some of these phenomena.

\subsection{Stochastic resonance}
Let us consider a dynamical system initially perturbed by weak periodic forcing. When an optimal level of white noise is subsequently added to the system, the original periodic signal perturbing the system becomes amplified. This phenomenon, in which noise `resonates' with a periodic signal, is known as stochastic resonance (SR) \citep{Gammaitoni1998}. Benzi and coworkers \citep{Benzi1981, Benzi1982, Benzi1983} were the first to observe SR and did so in periodically recurrent ice ages. They found that the modulation of Earth's orbital eccentricity (weak periodic forcing), which by itself is too weak to induce a global climate shift, is amplified by short-term climate fluctuations (white noise), inducing the periodically recurrent ice age. The first experimental verification of SR was reported by \citet{Fauve1983}, who applied white noise to a Schmidt trigger and observed an increase in the signal-to-noise ratio (SNR) of the voltage output. Since then, SR has been used to amplify weak periodic signals in fields as diverse as physics \citep{mcnamara1988}, chemistry \citep{leonard1994}, biology \citep{mcdonnell2009}, and engineering \citep{jerome2014}. A comprehensive review of SR and its applications has been conducted by \citet{Gammaitoni1998} and \citet{wellens2003}. 

\subsection{Coherence resonance} \label{subsec:CR}
Studies on SR have uncovered insight into the coherent behavior of various nonlinear systems under external periodic forcing. The behavior of some systems, however, is governed by their intrinsic dynamics, rather than by external periodic forcing \citep{gang1993}. A natural question to ask is whether applied noise can induce coherence in the absence of an external periodic signal.

\citet{gang1993} found that even in the absence of external periodic forcing, noise can stimulate coherent motion in a system, producing a peak in the power spectrum. They also found that the position and height of this spectral peak are strongly dependent on the noise amplitude. This phenomenon was interpreted as `stochastic resonance without external periodic force' \citep{gang1993} or `autonomous stochastic resonance' \citep{longtin1997}. Later, in excitable systems, \citet{pikovsky1997coherence} found that the coherence in the noise-induced dynamics first increases, reaches a maximum, and then decreases as the noise amplitude increases. They termed this phenomenon coherence resonance (CR). \citet{ushakov2005coherence} later formally defined CR in terms of the coherence factor $\beta=H \omega_p / \Delta_\omega$, where $H$ is height of the spectral peak, $\omega_p$ is the peak frequency and $\Delta_\omega$ is the peak width at half maximum. In their study, they showed that CR is a universal feature of Hopf bifurcations, and can be used to distinguish between supercritical and subcritical Hopf bifurcations. This application of CR was later put on firmer footing by \citet{Gupta2017}, who used CR to identify the specific type of Hopf bifurcation generated in low-order models of thermoacoustic systems.

\begin{figure}
    \centering
    \includegraphics[width=0.98\textwidth]{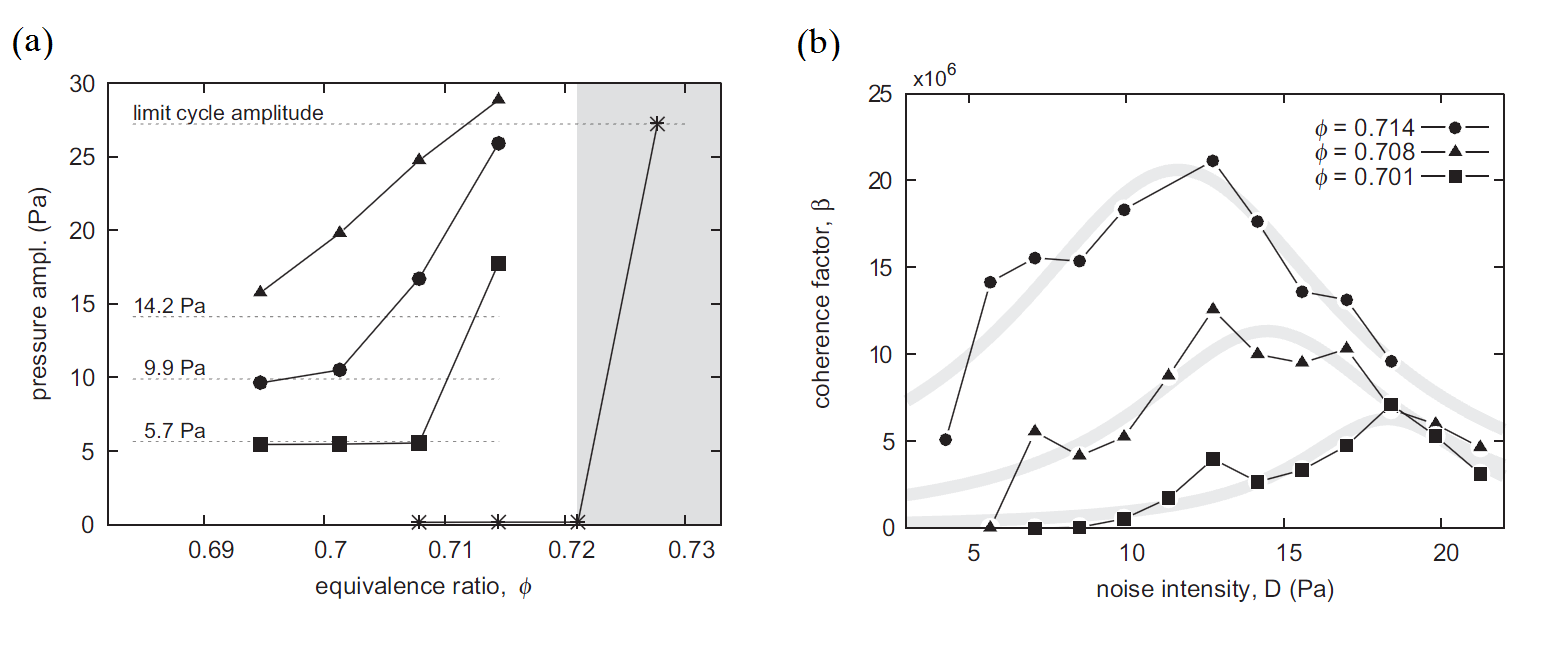}
    \caption{(a) Pressure fluctuation amplitude under varying noise amplitudes in the unconditionally stable regime. (b) Coherence factor ($\beta$) as a function of the noise amplitude for different equivalence ratios. $\beta$ reaches a peak at an intermediate noise amplitude \citep{Kabiraj2015}.}
    \label{fig:CR_K}
\end{figure}

For the purpose of forecasting, it has been found that CR can be used to predict the onset of a Hopf bifurcation. Before the concept of CR was fully established, \citet{Wiesenfeld1985} showed in pioneering work that the spectrum of a noise-perturbed system contains precursors capable of forecasting the onset of impending oscillations. In particular, it was found that the coherence of a system's response to noise increases as the bifurcation is approached. This observation was later revisited by \citet{Neiman1997}, who used CR as a `noisy precursor' to impending periodic oscillations. More recently, \citet{Kabiraj2015} experimentally observed CR in the unconditionally stable regime (i.e. the subthreshold regime) prior to a subcritical Hopf bifurcation (see figure \ref{fig:CR_K}), and suggested that CR can be used to forecast the onset of thermoacoustic instability. Along similar lines, \citet{zhu2019} reported CR in a prototypical hydrodynamic system (an axisymmetric low-density jet) and used CR to forecast the onset of global instability. However, CR-based forecasting techniques share the same limitation as the other precursors discussed in \S\ref{sec:hopf_pr}: to obtain quantitative predictions of the bifurcation point, it is necessary to define an ad-hoc instability threshold.

\subsection{Mathematical modeling of noise-perturbed systems}

We now turn our attention to mathematical models that can describe the dynamics of noise-perturbed systems. When a system is perturbed by noise, its dynamics are often characterized by a random process \citep{risken1984, parzen1999}. In a random process, also known as a stochastic process, the variables evolve in time by some random mechanism \citep{lax2006}. If the probability distribution of a random variable is known at all possible times, a random process is considered completely described \citep{lax2006}. 

The formal mathematical study of a random process was pioneered by \citet{einstein1905}, who solved a partial differential equation describing the time evolution of a Brownian particle. Three years later, \citet{langevin1908} applied Newton's second law to a Brownian particle, yielding a more versatile description of a random process. This description is now known as the Langevin equation:
\begin{equation} \label{eq:lang}
    m\frac{\mathrm{d}v}{\mathrm{d}t} = -\gamma v + \eta (t),
\end{equation}
where $m$ is the mass of the particle, $v$ is its velocity, $\gamma$ is the damping coefficient, and $\eta(t)$ is a delta-correlated Gaussian noise term, which has the form:
\begin{equation}
    \langle \eta(t) \eta(t') \rangle = 2d \delta(t-t'),
\end{equation}
where $d$ is the strength of the noise and $\delta$ is a Dirac delta function.

Because equation \ref{eq:lang} contains a random component, its solution, $v(t)$, cannot be expressed deterministically. However, if a \emph{set} of particles is considered, it then becomes possible to predict how many particles in this set will have velocity $v$ at time $t$ via the Fokker--Planck equation \citep{fokker1914, planck1917} that corresponds to equation \ref{eq:lang}:
\begin{subequations} \label{fp_gen}
\begin{align}
    \pdv{P(v,t)}{t} = -\pdv{}{v} \big[D^{(1)} &P(v,t) \big] + \pdv[2]{}{v} \big[ D^{(2)} P(v,t) \big], \\
    D^{(1)} &= -\gamma v P, \\
    D^{(2)} &= \frac{d}{2},
\end{align}
\end{subequations}
where $P(v,t)$ is the transitional PDF of $v$ at time $t$. $D^{(1)}$ is the \emph{drift} term, which represents the deterministic components, while $D^{(2)}$ is the \emph{diffusion} term, which represents the stochastic components \citep{gitterman2013}. 

Equation \ref{fp_gen}a is called the standard Fokker--Planck equation, or the Fokker--Planck--Kolmogorov equation. In general, the Fokker--Planck equation is very difficult to solve analytically \citep{risken1984}. However, if $P(v,t)$ does not change with time, an analytical solution is given by the \emph{stationary} Fokker--Planck equation, which can be obtained by integrating the standard Fokker--Planck equation over time.

Although \citet{fokker1914} and \citet{planck1917} only focused on the case of Brownian particle motion, subsequent studies have shown that the Fokker--Planck equation can be used to describe the dynamics of various noise-perturbed systems, such as the intensity fluctuations of a laser \citep{risken1965, hempstead1967}, the molecular concentration in a chemical reaction \citep{lotstedt2006}, the pressure fluctuations in a combustor \citep{culick1992}, the population growth in a region \citep{sikdar1982}, and the wealth distribution in a simple market economy \citep{cordier2005}, to name just a few. Various examples of the Fokker--Planck equation and the methods to solve them have been discussed by \citet{risken1984}.

From a practical perspective, the Fokker--Planck equation can also be used to extract the unknown parameters of a system (i.e. for system identification). A discussion of the methodology and examples behind this application are reserved for the next section (\S\ref{sec:SI}).

\section{System identification} \label{sec:SI}

\begin{figure}
    \centering
    \includegraphics[width=0.6\textwidth]{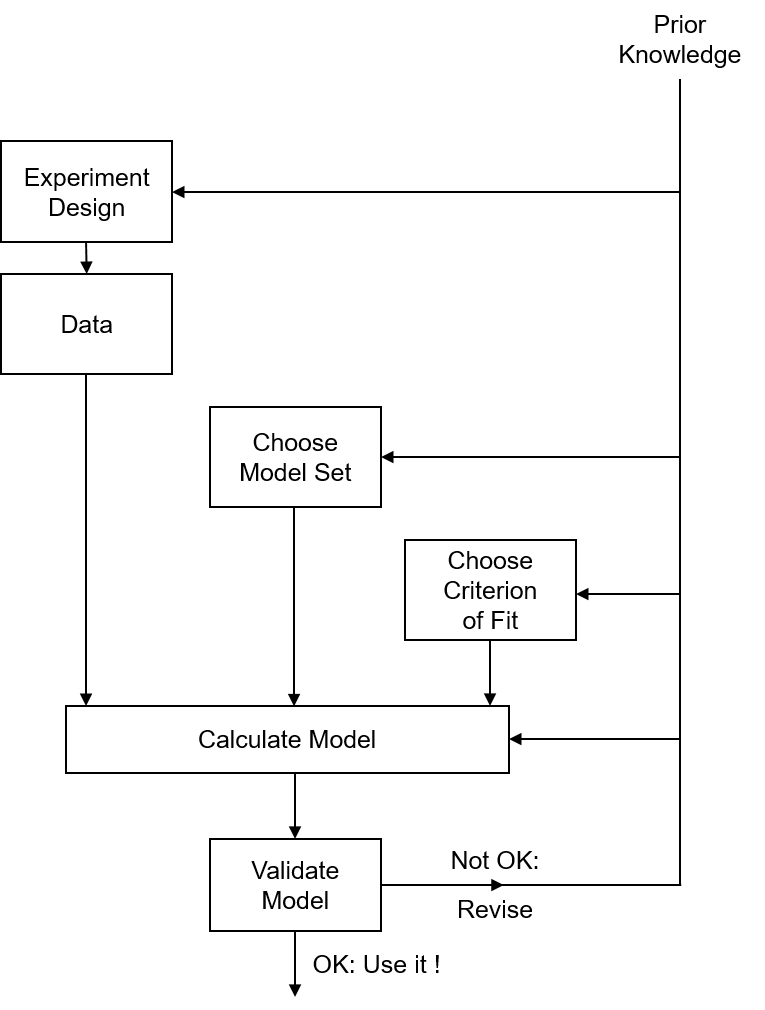}
    \caption{The system identification loop \citep{Ljung1999}.}
    \label{fig:SIloop}
\end{figure}

System identification (SI), also known as model identification, is a statistical method of building a mathematical model of a system from input and/or output data \citep{Fu2013}. In conventional SI, three basic elements are required to build such a model: data, a set of candidate models, and a rule to evaluate the model \citep{Ljung1999}. These elements, along with the validation process that follows, form an SI loop, which is illustrated in figure \ref{fig:SIloop}. Conventional mathematical methods for SI have been reviewed by \citet{Cuenod1968}, \citet{Astrom1971} and \citet{Kalaba1982}.

There are two main purposes of SI. In control problems, the goal is to design a control strategy, but in other problems, the goal is to analyze the system's dynamics \citep{Astrom1971}. In the latter case, the final goal is to find the values of the system parameters \citep{Eykhoff1968}, and to determine how those parameters relate to the dynamics of the system \citep{Ljung1999}.  For this purpose, SI has proven to be useful in a variety of fields \citep{Astrom1971}.

\begin{figure}[t]
    \centering
    \includegraphics[width=0.95\textwidth]{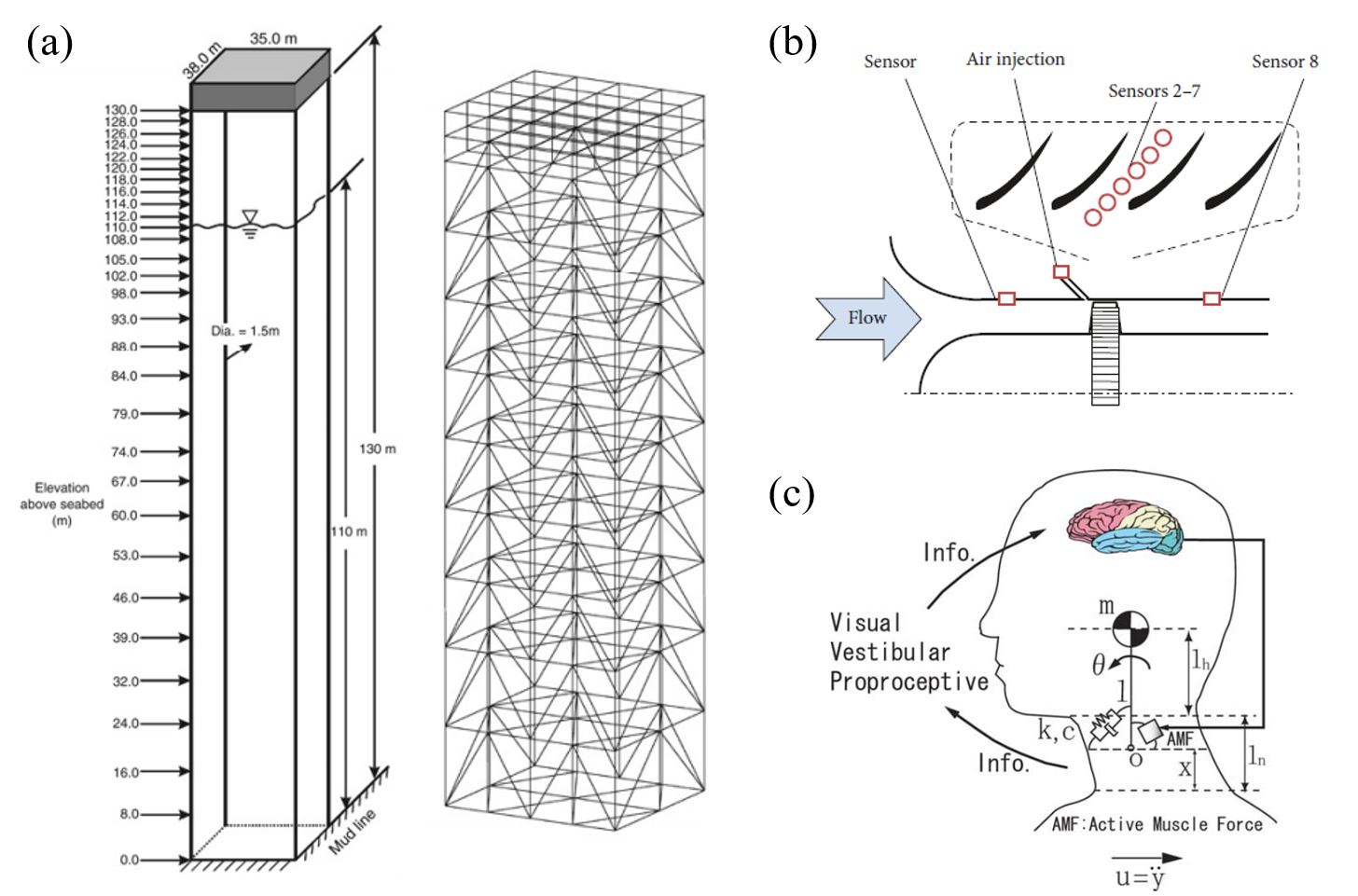}
    \caption{Examples of SI application. (a) Offshore structure \citep{NAJAFIAN2007b} (b) axial compressor \citep{schoen2017} (c) human head-neck complex \citep{Fard2004}}
    \label{fig:SIexample}
\end{figure}

In mechanical and aerospace engineering, for example, SI is often performed on structural or thermofluids systems, so as to find the system parameters and to estimate the system response. For example, \citet{NAJAFIAN2007a,NAJAFIAN2007b} evaluated the dynamic response of an offshore structure by identifying the system parameters such as the nonlinear drag component and the linear inertial component (see figure \ref{fig:SIexample}a). \citet{schoen2017} performed SI on an axial compressor (see figure \ref{fig:SIexample}b), with the aim of investigating the way tip air injection and throttle activation affect the overall compressor dynamics. NASA Langley Research Center \citep{morelli2005} applied SI to flight test data in order to build an analytical model of aircraft aerodynamics. This model has been used for aircraft simulation, the design of control systems, and dynamic analysis of general transport aircraft, fighter aircraft, space shuttle and other test aircraft.

In biology, an increasing number of researchers are turning to SI to better understand and model physiological systems \citep{BEKEY1978}. \citet{Peslin1975} described the frequency response of a respiratory system with a fourth-order mechanical model, and identified the governing parameters such as tissue compliance, alveolar gas compressibility and airway resistance.  \citet{Fard2004} proposed a method of identifying vibration in human head-neck complex in a seated human body, using a simple spring-mass-damper model with a single degree of freedom (see figure \ref{fig:SIexample}c).

SI has also been performed in econometrics.  For example, \citet{Granger1986} suggested that a stochastic model, based on both time-domain and frequency-domain approaches, can be used to analyze and forecast a financial time series. Furthermore, \citet{LOS2006} performed SI of stock markets in several countries, with the aim of separating the systematic signal from noise. Other fields in which SI has been applied include geophysics \citep{robinson2000, mendel2013}, environmental science \citep{Beck1983}, and electromagnetism \citep{DUDLEY1983}.

\begin{figure}[t]
    \centering
    \includegraphics[width=0.98\textwidth]{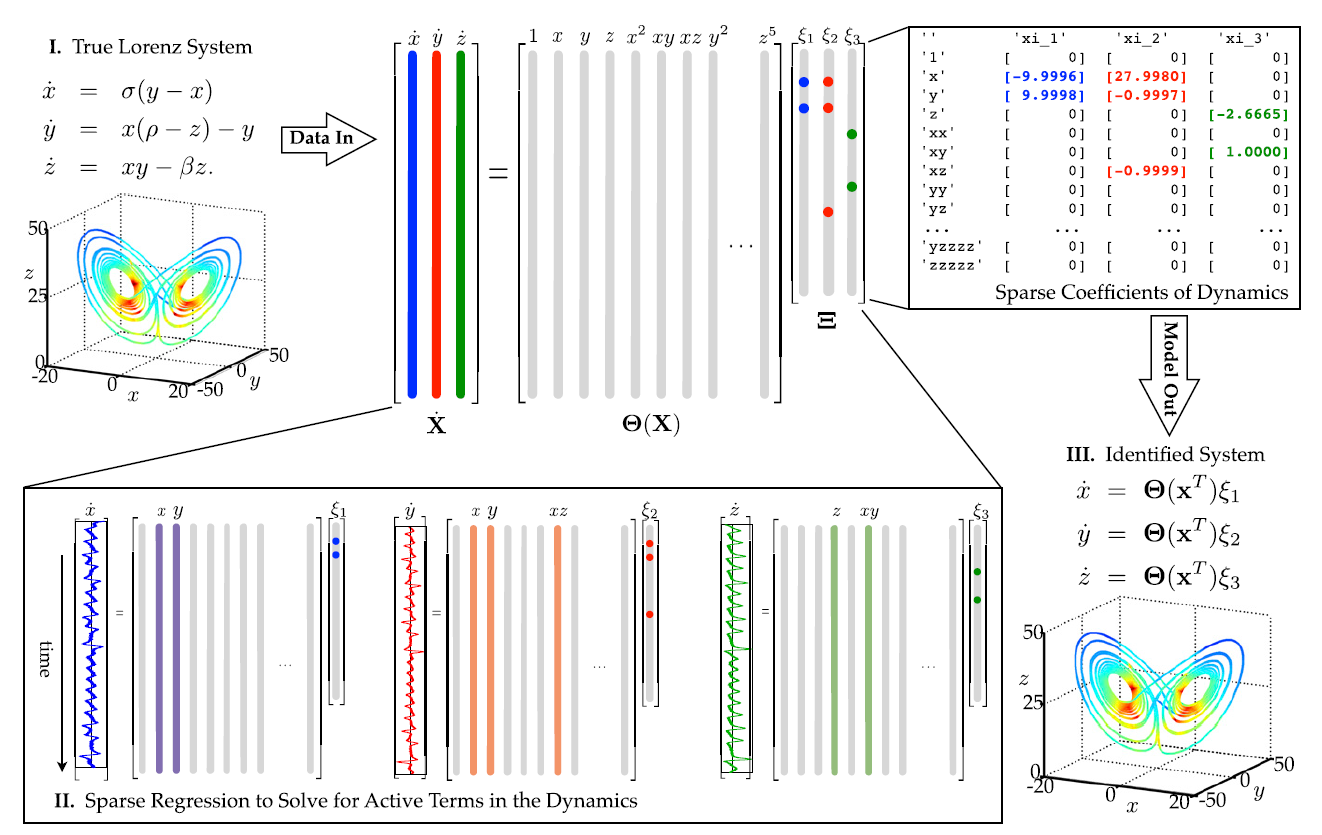}
    \caption{Schematic of Sparse Identification of Nonlinear Dynamics (SINDy) algorithm \citep{Brunton2016}. The active terms are identified from iterative sparse regression, and the original dynamics of the Lorentz system is captured in the reconstructed trajectory.}
    \label{fig:SINDY}
\end{figure}

Recent studies have shown that SI can be performed without the use of candidate models \citep{Schmidt2009,Brunton2016}. These methods are purely data-driven, requiring no \textit{a priori} knowledge of the system. For example, \citet{Schmidt2009} exploited symbolic regression to identify the governing equations of oscillatory systems. \citet{Brunton2016} developed an SI framework that exploits sparsity promotion and machine learning to identify low-dimensional models of physical systems (see figure \ref{fig:SINDY}). However, although such data-driven frameworks can successfully identify a system model, their applicability is limited to cases where abundant data are available. In most engineering systems, collecting abundant data is difficult or expensive, so conventional model-based SI methods are still widely adopted.

When a system is in a noisy environment, an effective SI strategy is to make use of its noise-induced response \citep{sura2002, jafari2003, bottcher2006, van2006, NOIRAY2013152, boujo2017robust, Pau2017, Boujo2020}. This approach works on the principle that the PDF of a stochastic time series is determined by only two factors: (i) the deterministic dynamics of the system and (ii) the dynamic noise that affects (i) \citep{siegert1998}. Therefore, by modeling the system dynamics with one or more stochastic differential equations, it is possible to extract the deterministic component from the corresponding Fokker--Planck equation. 

Specifically, \citet{friedrich1997} and \citet{siegert1998} proposed that the drift and diffusion terms of the Fokker--Planck equation---which correspond to the deterministic and the random parts of the system, respectively---can be extracted from the PDF of noisy time series data. Exploiting this feature, \citet{friedrich2000} proposed a method for identifying model equations from noisy experimental data, specifically by fitting analytical functions to the numerically determined drift and diffusion terms. Subsequently, SI via the Fokker--Planck equation has been widely carried out to identify the dynamics of stochastic systems \citep{sura2002, jafari2003, bottcher2006, van2006, NOIRAY2013152, bonciolini2017output}.

Notably, the idea that the noise-induced response of a system can be used to extract deterministic components has been used extensively in the field of combustion engineering, where the governing processes occur in a turbulent environment. The statistical features of pressure fluctuations have long been studied for SI of gas turbines and rocket engines, so as to predict their stability margins \citep{seywert2001, Kabiraj2020}. In a pioneering study, \citet{culick1992} numerically analyzed the effect of noise on a combustor, by using a set of stochastic differential equations to model multiple acoustic modes. The statistical features, such as the log-normal distribution of the pressure fluctuation, were analytically found from the corresponding Fokker--Planck equation and were shown to match well with the numerical simulations. \citet{seywert2001} later used a similar approach, with the system represented by four stochastic differential equations. By curve-fitting the power spectrum, the author was able to identify the frequencies and linear growth rates from the time series perturbed by multiplicative noise. Yet, this method was not successful in the limit-cycle regime where nonlinear mechanisms are dominant.

In a more recent study, \citet{NOIRAY2013152} conducted SI of a gas turbine combustor by assuming that only a single acoustic mode is dominant and that it can be modeled with a stochastic van der Pol (VDP) equation. In this work, four different approaches for extracting deterministic quantities, namely the linear growth rate and the nonlinear coefficient, were proposed. In the first method, which neglects the nonlinear term, an analytical equation for the power spectral density (PSD) of the acoustic pressure is derived from the stochastic VDP oscillator perturbed by additive noise. The linear growth rate and the noise amplitude are then found by curve-fitting the experimental PSD. Like the study by \citet{seywert2001}, this method neglects the nonlinearity of the system, and therefore cannot be used in the limit-cycle regime. 

\begin{figure}
    \centering
    \includegraphics[width=0.55\textwidth]{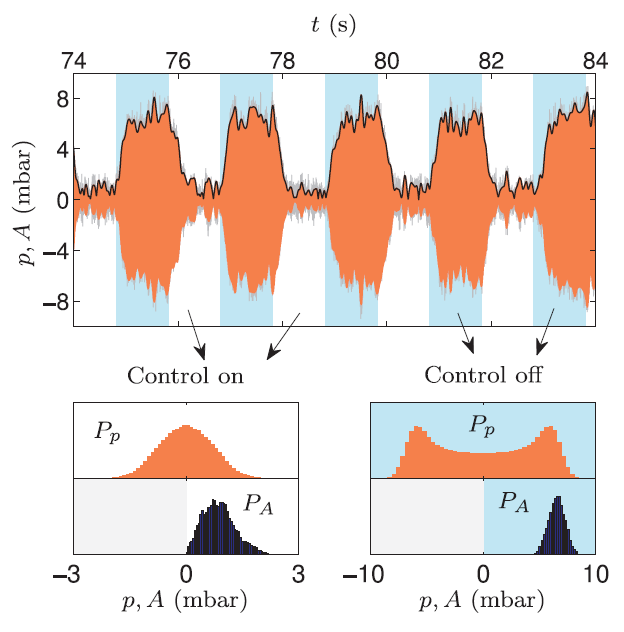}
    \caption{Periodic feedback control of limit-cycling combustor using the output-only SI results \citep{noiray2017method}. Orange and black lines show the pressure signal and its amplitude, respectively.}
    \label{fig:control}
\end{figure}

The other three methods proposed by \citet{NOIRAY2013152}, by contrast, can be used in the limit-cycle regime, after the Hopf bifurcation. In the second method, the authors consider a weakly perturbed limit cycle, where the amplitude of the applied noise is small. The authors assume that the fluctuation of the limit-cycle amplitude is much smaller than its deterministic amplitude. Accordingly, a stochastic differential equation that describes the time evolution of the limit-cycle amplitude is linearized about the stationary solution (i.e. deterministic limit-cycle amplitude). Consequently, the linear growth rate and the noise amplitude could be identified by fitting the PSD. In the third method, the authors make use of the stationary Fokker--Planck equation. In this approach, the linear and nonlinear coefficients are found by fitting the experimental PDF of the fluctuation amplitude to analytical solution of the stationary Fokker--Planck equation. Unlike the second method, this method does not require the system to be weakly perturbed. In the fourth method, the drift and diffusion terms of the Fokker--Planck equation are used, and the SI method proposed by \citet{siegert1998} is applied. Specifically, by analyzing the transitional PDF of the oscillation amplitude, the drift and diffusion terms of the Fokker--Planck equation are obtained. The deterministic quantities, such as the linear and nonlinear coefficients, are then calculated from these terms. The fourth approach has been further validated by \citet{noiray2017method}, who used the results obtained from SI for feedback control of a lab-scale combustion chamber (see figure \ref{fig:control}). In all four of the methods proposed by \citet{NOIRAY2013152}, the intrinsic noise of the system, namely turbulence, was used as the perturbation source, and only the output signal was measured. In other words, output-only SI was conducted.

In a follow-up study, \citet{bonciolini2017output} showed that regardless of the color of noise, output-only SI based on the Fokker--Planck equation can be successfully performed. Specifically, the authors showed that colored noise could be approximated by white noise if an appropriate band-pass filter around the oscillator eigenfrequency is applied. The fact that SI can be performed successfully regardless of the spectral composition of the noise implies that it is not necessary to know what color the noise is for SI to work.  This is very important from a practical perspective as it is often difficult or impossible to know the noise characteristics \textit{a priori}. However, the authors found that a band-pass filter, especially when its bandwidth is narrow, alters the original signal and adversely affects the SI results.

This limitation was overcome by \citet{boujo2017robust}, who further developed the fourth output-only SI strategy proposed by \citet{NOIRAY2013152}. In particular, the authors adopted a coefficient optimization algorithm that makes use of the adjoint Fokker--Planck equation. In this algorithm, the original coefficients of the oscillator equation, which are obtained from the aforementioned output-only SI methods \citep{NOIRAY2013152, noiray2017method}, are optimized by minimizing the discrepancy between the experimental and mathematical (adjoint-based) drift and diffusion terms of the Fokker--Planck equation. The authors showed that the adverse effect of a band-pass filter, namely the finite-time effect, can be mitigated by using this optimization algorithm. This method was later revisited by \citet{Boujo2020}, who applied output-only SI and adjoint-based optimization to an aeroacoustic system.

\vspace{8mm}

As discussed above, various versions of SI have been successfully applied to nonlinear dynamical systems. However, in all the established SI methods, at least some data from the limit-cycle regime is required in order to identify the bifurcation point and the post-bifurcation dynamics. An SI method capable of predicting an impending bifurcation using only pre-bifurcation data has yet to be developed, but is desperately needed if one is to be able to forecast the properties of a Hopf bifurcation without having to enter the potentially dangerous limit-cycle regime after the Hopf point.

\section{Motivation and aim}

As the foregoing review has shown, Hopf bifurcations are ubiquitous to nonlinear dynamical systems, and it is of interest to be able to predict their properties before the emergence of limit-cycle oscillations. However, as noted in \S\ref{sec:hopf_pr}, existing prediction methods suffer from two key limitations: (i) they require ad-hoc instability thresholds to be defined, and (ii) they cannot predict the post-bifurcation dynamics.

In this thesis, we develop a mathematical framework capable of identifying the properties of a Hopf bifurcation---such as its location, its degree of nonlinearity and the post-bifurcation behavior---without \textit{a priori} knowledge of the system. In particular, we exploit the phenomenon of coherence resonance, which enables information about the system dynamics to be readily extracted even in the fixed-point regime (see \S\ref{subsec:NID}).

However, as noted in \S\ref{sec:SI}, for existing SI methods to predict the location and type of a Hopf bifurcation and the dynamics beyond requires data in the limit-cycle regime, after the bifurcation has already occurred. In practical systems, collecting data in such a high-amplitude regime can often be difficult or even dangerous. Therefore, a unique advantage of our SI framework is that it requires data from only the fixed-point regime, before the Hopf point itself, where the oscillation amplitudes, and by extension the risk of catastrophic damage, are low.

The primary goal of this research is to develop a robust SI framework that relies on the noise-induced dynamics in the fixed-point regime. This framework can (i) provide precursors to a Hopf bifurcation, (ii) identify the degree of nonlinearity of a Hopf bifurcation (i.e. predict whether it is supercritical or subcritical), and (iii) predict the limit-cycle dynamics after the bifurcation. 

We present two versions of this SI framework: input-output and output-only. In the input-output version, which is particularly suitable for systems with weak intrinsic noise, we apply external noise to the system. By contrast, for systems with strong intrinsic noise, we use an output-only version of our SI framework. In both versions, a self-excited oscillator model (VDP equation) is used to model the system oscillations, and the experimental PDF is compared with the analytical solutions obtained from the Fokker--Planck equation corresponding to the system model.

We demonstrate our SI framework on three different experimental systems: a laminar hydrodynamic system (\S\ref{chap:low}: a low-density jet), a laminar thermoacoustic system (\S\ref{chap:Rij}: a flame-driven Rijke tube), and a turbulent thermoacoustic system (\S\ref{chap:gas}: a gas turbine combustor). Input-output SI is demonstrated on the first two systems, while output-only SI is demonstrated on the third system. The details of the SI methodology for each system will be described in the respective chapters. Furthermore, in \S\ref{chap:io_oo}, we investigate just how much noise is required for SI to work.

A key limitation of our SI framework is that it assumes the presence of only a single oscillatory mode near a Hopf bifurcation. To overcome this limitation, we consider in \S\ref{chap:coupled} an alternative system model consisting of two coupled oscillators. We then derive its corresponding Fokker--Planck equation and analyze the effect of three factors: (i) noise amplitude, (ii) coupling strength, and (iii) coupling type. Furthermore, we model in \S\ref{chap:ntau} the dynamics of a Rijke tube using the momentum and energy equations with a stochastic forcing term. We derive the corresponding Fokker--Planck equation and investigate the effect of noise and heater power, before concluding in \S\ref{chap:concl}.


\chapter{System identification of a low-density jet via its noise-induced dynamics} \label{chap:low}

Published in \textit{Journal of Fluid Mechanics}, vol. 862, pp. 200-215 (2019).

\section*{Abstract}
Low-density jets are central to many natural and industrial processes.  Under certain conditions, they can develop global oscillations at a limit cycle, behaving as a prototypical example of a self-excited hydrodynamic oscillator.  In this study, we perform system identification of a low-density jet using measurements of its noise-induced dynamics in the unconditionally stable regime, prior to both the Hopf and saddle-node points.  We show that this approach can enable prediction of (i) the order of nonlinearity, (ii) the locations and types of the bifurcation points (and hence the stability boundaries), and (iii) the resulting limit-cycle oscillations.  The only assumption made about the system is that it obeys a Stuart--Landau equation in the vicinity of the Hopf point, thus making the method applicable to a variety of hydrodynamic systems.  This study constitutes the first experimental demonstration of system identification using the noise-induced dynamics in only the unconditionally stable regime, i.e. away from the regimes where limit-cycle oscillations may occur.  This opens up new possibilities for the prediction and analysis of the stability and nonlinear behaviour of hydrodynamic systems.

\section{Introduction}\label{sec:intro}
Low-density jets have attracted considerable attention over the last few decades as a result of their role in industrial processes such as fuel injection and plasma spraying.  Under certain conditions, such jets can develop global hydrodynamic instability, leading to self-excited oscillations at a limit cycle \citep{Sreenivasan1989,huerre1990local,monk1990self}.  On the one hand, such oscillations can be beneficial in situations where mixing is desired.  On the other hand, they can be detrimental in situations where they excite unwanted acoustic or structural resonances.  Therefore, it is important to be able to predict the onset of global hydrodynamic instability as well as the frequency and amplitude of the resulting limit-cycle oscillations (LCOs).

\subsection{Bifurcation of a low-density jet}
\citet{Raghu1991} have shown that a low-density jet becomes globally unstable via a Hopf bifurcation: after a critical point (the Hopf point), the jet becomes unstable to infinitesimal perturbations and transitions to a self-excited state characterised by LCOs.  Near the Hopf point, the growth rate is small, implying that the oscillation amplitude ($a$) evolves much more slowly than the oscillation frequency ($\omega$) \citep{Raghu1991}.  \citet{landau1944problem} proposed an equation to model the amplitude evolution in this specific regime, which \citet{stuart1960non} later formulated for plane Poiseuille flow using an energy balance.  This has become known as the Stuart--Landau equation:
\begin{equation}\label{eq:landau0}
    \frac{\mathrm{d}{a}}{\mathrm{d}t}=k_1{a}+k_2{a}^3+\cdots,
\end{equation}
where $t$ is time, $k_1$ is a linear driving/damping parameter, and $k_2$ is a nonlinear parameter.  The Hopf point is at $k_1 = 0$, after which ($k_1 > 0$) the system becomes linearly unstable.

The Hopf bifurcation in low-density jets is usually considered to be supercritical \citep{monk1990self,Raghu1991}, i.e. LCOs cannot occur before the Hopf point ($k_1 < 0$).  Therefore, the linear parameter ($k_1$) alone determines the stability boundaries of the system.  However, \citet{Sreenivasan1989} observed a hysteretic regime in which LCOs can occur even when $k_1 < 0$.  This led \citet{kyle_sreenivasan_1993} to suggest that the Hopf bifurcation in low-density jets can also be subcritical, which \citet{zhu2016subcritical,zhu2017onset} later formally established.  In a system with a subcritical Hopf bifurcation, a finite-amplitude perturbation can trigger the system to LCOs via contributions from the nonlinear terms (such as $k_2 a^3$) even when $k_1 < 0$ \citep{zhu2018noise}.  This regime, where LCOs can occur despite the system being linearly stable, is called the bistable regime.

An important implication from the existence of a subcritical bifurcation in a system is that the nonlinear terms need to be calculated before the stability boundaries can be determined. The challenge, however, is that existing methods applied to jets \citep{Raghu1991} and wakes \citep{Provansal1987,Dusek1994,Sipp2007} can only calculate the nonlinear terms from measurements of the system dynamics after the emergence of LCOs.  In other words, such methods can describe the system behaviour via postprocessing, which is itself useful, but they lack predictive capabilities, particularly for nonlinearities.

\subsection{Bifurcation analysis and system identification of fluid dynamical systems}
In most fluid dynamical systems, it is important to know where the bifurcation points are, as they determine the stability boundaries.  The most direct way of finding the bifurcation points is to solve the time-dependent Navier--Stokes equations and determine the parameter value (e.g. the Reynolds number, $Re$) at which the flow undergoes a qualitative change in behaviour.  Alternatively, one can obtain steady solutions of the system at a lower computational cost, and then solve for the eigenvalues of its Jacobian matrix \citep{jackson_1987,dijkstra_2014}.  However, if applied to systems with complex geometries or boundaries, such direct methods can be expensive and unreliable, as it is often difficult to define the boundary conditions with sufficient accuracy to produce meaningful numerical solutions \citep{kim1985application, thompson1997general}.  In such cases, one needs to first identify the system using the data available and then determine its bifurcation points.  System identification (SI) methods for this purpose can be divided into two classes: (i) purely data-driven methods and (ii) hybrid methods.

In purely data-driven methods, the governing equations of a physical system are found exclusively from experimental data, without the need to assume a system model \textit{a priori}.  For example, \citet{Schmidt2009} used symbolic regression to identify the nonlinear differential equations governing a variety of physical systems, ranging from simple harmonic oscillators to chaotic double pendula.  In that procedure, experimental data are fitted to simple mathematical building blocks based on Hamiltonians and Lagrangians.  New equations are then added to these via genetic programming.  Although useful for simple systems, symbolic regression becomes impractical for systems containing a large number of degrees of freedom.  To overcome this problem, \citet{Brunton2016} recognised that the key dynamics of most physical systems are usually simple enough to be described by just a few leading terms.  This makes it possible to use sparsity-promoting tools and machine learning to identify low-dimensional models of physical systems at a reduced computational cost.  Recently, \citet{Shimizu2018} also used machine learning to determine the low-dimensional equations governing low-$Re$ turbulence in plane Couette flow, enabling the entire bifurcation cascade to be reproduced and studied.

Purely data-driven methods for SI are useful for their role in explaining many naturally occurring phenomena for which there is an abundance of experimental data but nearly no knowledge of the governing equations.  In engineering situations, however, collecting experimental data is usually expensive, but there is often some knowledge of the underlying system dynamics.  Therefore, for such situations, a hybrid method may be more suitable.  In hybrid methods, an appropriate low-dimensional model is assumed for the system, and then experimental data is used to determine the parameter values of the model and their variations with the physical parameters of the system \citep{PRICE1990419,thothadri2005nonlinear}.  For example, equation~(\ref{eq:landau0}) can be assumed to be a low-dimensional model of a jet or wake in the vicinity of a Hopf bifurcation.  Variations in $k_1$ and $k_2$ with $Re$ can then be extracted from experimental data, as demonstrated by \citet{Provansal1987} and \citet{Raghu1991}.  These conventional methods, however, are limited to nearly noise-free measurements and to systems with a supercritical Hopf bifurcation.

Recently, \citet{NOIRAY2013152} and \citet{boujo2017robust} have extended the aforementioned SI methods to exploit the influence of noise, which, in their experiments, came from background turbulence in the flow field of a thermoacoustic system.  They replaced the Stuart--Landau equation with its corresponding Fokker--Planck equation, yielding expressions for the probability density function, which is equivalent to the long-time average of the noise-affected measurements.  \citet{Bonciolini172078} further extended this method to enable SI of a laboratory-scale combustor undergoing a subcritical Hopf bifurcation.  However, to be able to determine the nonlinear terms, all of these SI methods require at least some data from the LCO regime.  Consequently, these methods cannot predict the nature of a bifurcation or the resulting LCO dynamics.  In fact, in most of these methods, the nonlinear terms are ignored in the regime before the Hopf point ($k_1 < 0$) \citep{Provansal1987}.  By contrast, \citet[][Chapter 3]{zhu2017mphil} has shown from the noise-induced dynamics of a low-density jet that the nonlinear terms are active even before the stability boundaries are reached, i.e. in the unconditionally stable regime, where the system is stable to infinitesimal as well as finite-amplitude perturbations.

\subsection{Noise-induced dynamics: Coherence resonance}
In pioneering work, \citet{Wiesenfeld1985} explored the effect of noise on oscillatory systems and found that the spectra of the noise-induced dynamics contain precursors capable of forecasting impending nonlinear instabilities.  In particular, it was found that the system response to noise becomes more coherent (or less noisy) on approach to the Hopf point.  Later, \citet{pikovsky1997coherence} found for the FitzHugh--Nagumo system that the coherence in the noise-induced dynamics first increases, reaches a maximum, and then decreases as the noise amplitude increases.  They termed this phenomenon coherence resonance.  \citet{ushakov2005coherence} formally defined coherence resonance in terms of the coherence factor and showed that systems with Hopf bifurcations generally exhibit some degree of coherence resonance.

Recently, \citet{Kabiraj2015} and \citet{zhu2017mphil,zhu2019} reported coherence resonance in two different fluid dynamical systems: a thermoacoustic oscillator and a low-density jet, respectively.  \citet{Gupta2017} phenomenologically modelled coherence resonance in a thermoacoustic system, enabling the noise-induced dynamics arising from supercritical and subcritical Hopf bifurcations to be explored in detail.  Moreover, \citet{zhu2017mphil} experimentally demonstrated the use of coherence resonance to identify the different types of Hopf bifurcation in a low-density jet via the noise-induced dynamics in only the unconditionally stable regime.  However, information obtained in this specific regime has yet to be exploited for SI of any experimental system -- fluid dynamical or otherwise.

\subsection{Contributions of the present study}
In this paper, we develop an SI framework that uses data from \emph{only} the unconditionally stable regime to predict the nonlinear behaviour of a low-density jet in the vicinity of its Hopf bifurcation.  Specifically, we aim to predict (i) the order of nonlinearity, (ii) the locations and types of the bifurcation points (and hence the stability boundaries), and (iii) the resulting LCO dynamics -- without having to operate the system in the potentially dangerous linearly unstable or bistable regimes.

Below, we present the experimental data and SI methodology in \S\ref{exp} and \S\ref{num}, respectively.  We then show the results in \S\ref{res} in terms of the order of nonlinearity, dynamic and stochastic bifurcations, and the LCO dynamics beyond the bifurcation points, before concluding in \S\ref{conc}.

\section{Experimental data}\label{exp}

\begin{figure}
    \centering
    \includegraphics[width=0.95\textwidth]{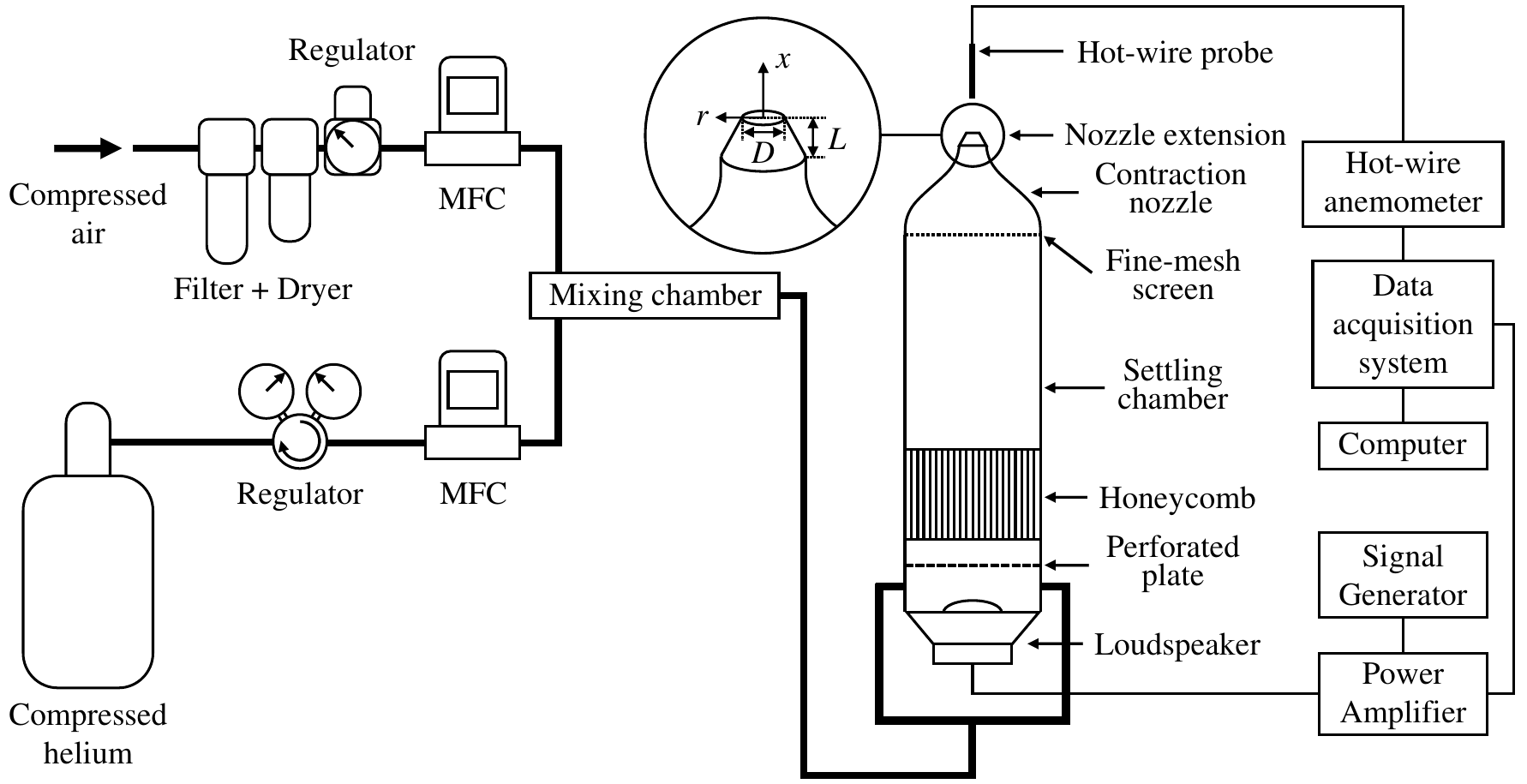}
    \caption{A schematic of the experimental setup used to produce a low-density jet perturbed by external noise \citep{zhu2017mphil}.  MFC: mass flow controller.}
\label{fig:setup}
\end{figure}

\begin{figure}
  \includegraphics[width=0.48\textwidth,trim={0 0 0 0},clip]{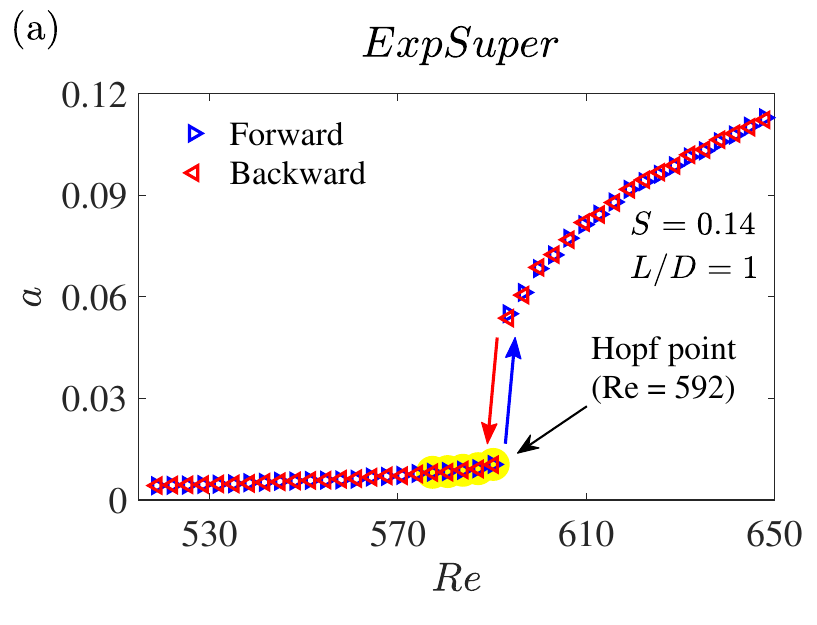}
  \includegraphics[width=0.48\textwidth,trim={0 0 0 0},clip]{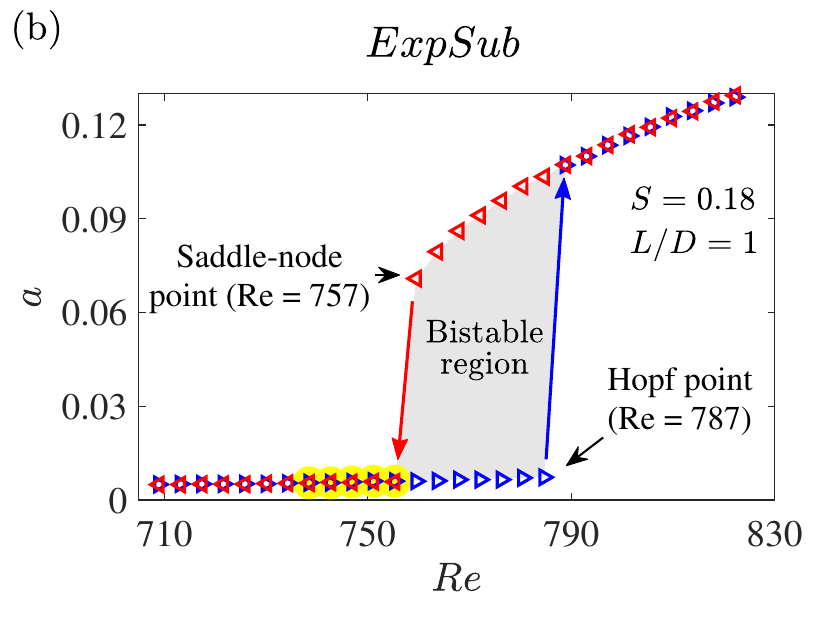}%
  \caption{Bifurcation diagrams for two experimental cases: (a) $ExpSuper$ and (b) $ExpSub$.  In the legend, the terms `forward' and `backward' refer to data collected by increasing and decreasing $Re$, respectively.  The data used for SI are collected exclusively in the unconditionally stable regime, as highlighted in yellow.}
  \label{fig:bifexp}
\end{figure}

We use the experimental data from \citet{zhu2017mphil}.  Figure~\ref{fig:setup} shows the setup used to collect the data, which consists of an axisymmetric nozzle assembly, an acoustic forcing system, gas supply lines, and a hot-wire anemometer.  In this setup, a laminar helium--air jet discharging into quiescent ambient air is perturbed by external noise.  There are three main independent control parameters governing the stability boundaries of the jet and its LCO dynamics.  These are (i) the jet-to-ambient density ratio, $S \equiv \rho_j/\rho_\infty$; (ii) the aspect ratio of the nozzle tip, $L/D$, which controls the thickness of the initial shear layer; and (iii) the jet Reynolds number, $Re \equiv \rho_j U_j D/ \mu_j$, where $U_j$ is the jet centreline velocity, $D$ is the nozzle exit diameter, and $\mu_j$ is the dynamic viscosity of the jet fluid.  In this paper, we keep the first two parameters fixed and vary only $Re$.

The acoustic forcing system consists of three components: (i) a signal generator (Keysight 33512B), (ii) a power amplifier (Alesis RA150), and (iii) a loudspeaker (FaitalPRO 3FE25).  The signal generator produces Gaussian noise with a bandwidth of $0$--$20$~MHz.  The upper frequency limit of the noise (20~MHz) is four orders of magnitude higher than the natural global frequency of the jet.  Therefore, the noise felt by the jet is essentially white.  The noise amplitude is controlled by regulating the input voltage into the loudspeaker ($V$) with the power amplifier.  The noise-induced dynamics of the jet is measured in terms of the local streamwise velocity in the potential core, using a hot-wire probe positioned on the jet centreline, 1.5$D$ downstream from the jet exit.  The output voltage from the hot-wire probe is digitised at a frequency of 32768~Hz.  Further details on these measurements can be found in \citet{zhu2017mphil}.

We consider two representative flow conditions, whose bifurcation diagrams are shown in figure~\ref{fig:bifexp}.  In figure~\ref{fig:bifexp}(a), where $S = 0.14$ and $L/D = 1$, the Hopf point is at $Re = 592$, below which LCOs are not observed.  Thus, this condition is experimentally determined to be supercritical and is called $ExpSuper$ here.  In figure~\ref{fig:bifexp}(b), where $S = 0.18$ and $L/D = 1$, the Hopf point is at $Re = 787$, below which LCOs are observed down to $Re = 757$, which is a saddle-node point.  Thus, this condition is experimentally determined to be subcritical and is called $ExpSub$ here.  In this study, the data used for SI are collected exclusively in the unconditionally stable regime, which is highlighted in yellow in figure~\ref{fig:bifexp}.

\section{Methodology for system identification}\label{num}

\subsection{System model}\label{s_model}
\begin{figure}
    \centering
    \includegraphics[width=0.48\textwidth,trim={0 0 0 0},clip]{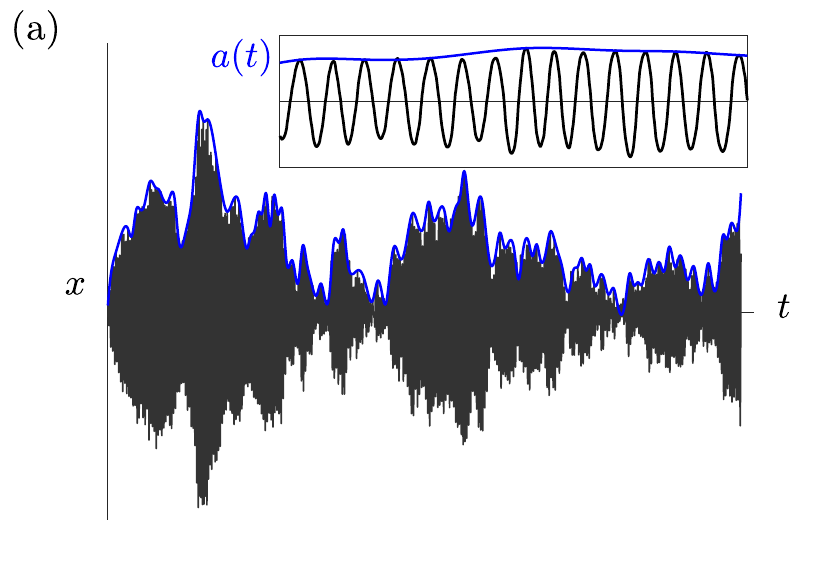}%
    \includegraphics[width=0.48\textwidth,trim={0 0 0 0},clip]{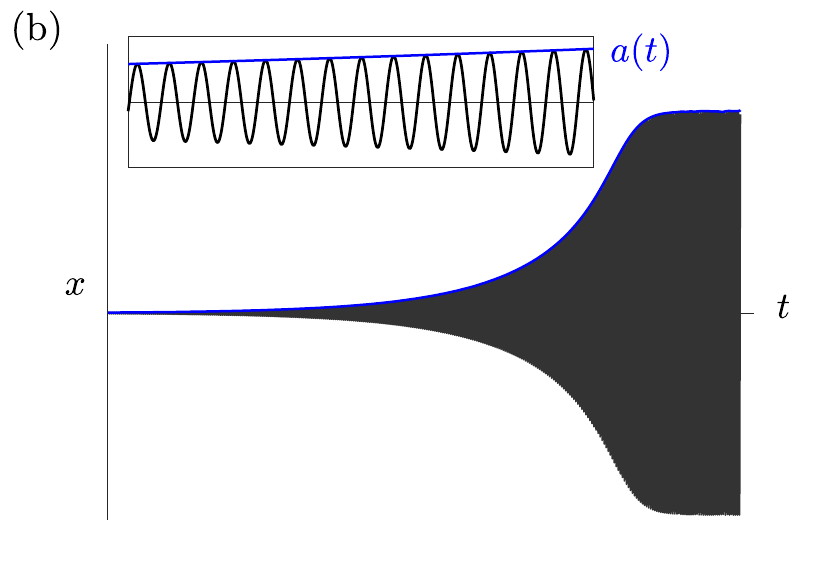}%
    \caption{Evolution of $a(t)$ and $x(t)$ for (a) noise-induced dynamics in the marginally unconditionally stable regime and (b) noise-free limit-cycle development in the marginally linearly unstable regime.  In both cases, the evolution of $a(t)$ is slower than that of $x(t)$.}\label{figXA}
\end{figure}

Figure~\ref{figXA} shows cartoon drawings relating the oscillation amplitude, $a(t)$, to the instantaneous state of the system, $x(t)$, e.g. velocity measurements from a hot-wire probe, for (a) a marginally unconditionally stable regime and (b) a marginally linearly unstable regime.  For both regimes, the evolution of $a(t)$, which can be approximated by a Stuart--Landau equation, is at a much slower rate than that of $x(t)$.  The effect of noise on the system is felt via $x(t)$, for which we assume the following governing equation:
\begin{equation}\label{fracdp}
    \ddot{x}-(\epsilon+\alpha_1x^2+\alpha_2x^4+\alpha_3x^6+\alpha_4x^8+\cdots)\dot{x}+x+{\beta}x^3=\sqrt{2d}\eta(t),
\end{equation}
where $\eta(t)$ is a unit additive white Gaussian noise term representing the effect of the loudspeaker, $d$ is its amplitude, $\epsilon$ is the linear growth/damping term, $\alpha_1, \alpha_2, \alpha_3, \alpha_4$, $\dots$ are the nonlinear system parameters, and $\beta$ is the anisochronicity factor, which controls the shift in oscillation frequency with amplitude. 
%
Equation~(\ref{fracdp}) is non-dimensionalised such that (i) the natural frequency is fixed at 1 for all $Re$ and (ii) $x \equiv u^{\prime}/\overline{u}$, where $u^{\prime}$ is the measured velocity fluctuation and $\overline{u}$ is its time average.  Here $\alpha_1$ is a counterpart to $k_2$ in equation~(\ref{eq:landau0}) and determines the nature of the Hopf bifurcation.  

In order to derive the probabilistic solution of equation~(\ref{fracdp}), we first perform a variation of parameters \citep{nayfeh1979nonlinear, nayfeh1981introduction}, where we transform the instantaneous state of the system ($x$) in terms of its amplitude ($a$) and phase ($\phi$) as: \\
\begin{equation}\label{transform}
    x(t) = a(t)\cos(t+\phi(t)).
\end{equation}
Here, we have two equations ((\ref{fracdp}) and (\ref{transform})) for three unknowns: $x(t)$, $a(t)$ and $\phi(t)$. Thus, we can impose a third condition that is independent of equations (\ref{fracdp}) and (\ref{transform}). Following \citet{nayfeh1981introduction}, we assume a convenient condition given as: 
\begin{equation}\label{condition}
    \dot{x}(t) = -a(t)\sin(t+\phi(t)).
\end{equation}
It should be noted that so far we have made no assumptions about $a$ and $\phi$ being slow variables. This transformation simply allows us to derive two first-order differential equations from one second-order differential equation and is popular in analysing noisy nonlinear oscillators~\citep{roberts1986stochastic, ZHU1987421, xu2011stochastic, yamapi2012effective}. Its effectiveness, particularly when $a$ and $\phi$ are slow variables, will be soon clear. By (i) differentiating equation (\ref{transform}) and subtracting equation (\ref{condition}) from it and (ii) by differentiating equation (\ref{condition}), we get the following two equations, respectively, as:
\begin{subequations}\label{dat}
\begin{align}
	0             &= \dot{a}(t)\cos(t+\phi(t)) - a(t)\dot{\phi}(t)\sin(t+\phi(t)), \label{dat_a}\\
	\ddot{x}(t) &= -\dot{a}(t)\sin(t+\phi(t)) -a(t)\cos(t+\phi(t)) - a(t)\dot{\phi}(t)\cos(t+\phi(t)). \label{dat_b}
\end{align}
\end{subequations}
We substitute equations~(\ref{transform}),~(\ref{condition}), and~(\ref{dat}) into equation~(\ref{fracdp}) and further use the trigonometric identities (such as $\sin^2(\theta) = \frac{1}{2} - \frac{1}{2}\cos(2\theta)$), to obtain the transformed first-order equations in $a$ and $\phi$ as:
\begin{subequations}\label{derived}
\begin{align}
    \dot{a}&=\underbrace{\Big(\frac{\epsilon}{2}a+\frac{\alpha_1}{8}a^3+\frac{\alpha_2}{16}a^5+\frac{5\alpha_3}{128}a^7+\frac{7\alpha_4}{256}a^9+\cdots\Big)+P(\Phi)}_{f_1}-\underbrace{\vphantom{\frac{1}{1}}\big(\sqrt{2d}\sin{\Phi}\big)}_{g_1}\eta_1\label{derived11},\\
    \dot{\phi}&=\underbrace{\frac{3\beta}{8}a^2+Q(\Phi)}_{f_2}-\underbrace{\Big(\frac{\sqrt{2d}}{a}\cos{\Phi}\Big)}_{g_2}\eta_2\label{derived12},
\end{align}
\end{subequations}
where $\eta_1$ and $\eta_2$ are independent white noise terms and $\Phi(t) = t + \phi(t)$.  Lastly, $P(\Phi)$ and $Q(\Phi)$ are the sum of all the terms with first-order cosine components (i.e. in the form of $a^{n_1}\cos{n_2\Phi}$, where $n_1$ and $n_2$ are integers). The equations are exact until now, however, in subsequent averaging procedures we will assume $a$ and $\phi$ to be slow variables,  which leads to the terms of the form $\int_0^{2\pi} a^{n_1}\cos{n_2\Phi}$ equal to zero. Consequently, on time-averaging and for $d=0$ (no noise), equation~(\ref{derived11}) gives the same form as the Stuart--Landau equation, which justifies our choice of the governing equation for $x(t)$. Equation~(\ref{derived}) contains deterministic parts ($f_1$, $f_2$) and stochastic parts ($g_1$, $g_2$).  When stochastically averaged as per \citet{stratonovich1963,stratonovich1967}, equation~(\ref{derived}) transforms into a stochastic differential equation for $a$, which can be written in It{\^o} sense as:
\begin{subequations}
\begin{gather}
\mathrm{d}a=\boldsymbol{m}\mathrm{d}t+\boldsymbol{\sigma}\mathrm{d}W,\\
\boldsymbol{m}=T^{av}\big\{{f_1}\big\}+T^{av}\bigg\{\int_{-\infty}^{0}\Big(\pdv{g_1(s)}{a}g_1(s+\tau)+\pdv{g_1(s)}{\phi}g_2(s+\tau)\Big)\langle\eta(s)\eta(s+\tau)\rangle\mathrm{d}\tau\bigg\}\notag\\
=\Big(\frac{{\epsilon}}{2}a+\frac{\alpha_1}{8}a^3+\frac{\alpha_2}{16}a^5+\frac{5\alpha_3}{128}a^7+\frac{7\alpha_4}{256}a^9+\cdots\Big)+\frac{d}{2a},\\
\boldsymbol{\sigma}^2=T^{av}\bigg\{\int_{-\infty}^{\infty}g_1(s)g_1(s+\tau)\langle\eta(s)\eta(s+\tau)\rangle\mathrm{d}\tau\bigg\}=d,
\end{gather}
\end{subequations}
where $\mathrm{d}W$ is a unit Wiener process, $T^{av}$ denotes the time average of the functions, and $\boldsymbol{m}$ and $\boldsymbol{\sigma}$ represent the drift and diffusion terms of $a$, respectively.  Finally, the equation for $a$ is transformed into equations for the probability density function of $a$, yielding the Fokker--Planck equation:
\begin{equation}
    \pdv{}{t}P(a,t)=-\pdv{}{a}\Big[\boldsymbol{m}(a,t)P(a,t)\Big]+\pdv[2]{}{a}\Big[\frac{\boldsymbol{\sigma}^2(a,t)}{2}P(a,t)\Big],
\end{equation}
%
%
\begin{equation}\label{stationary}
    P(a)=Ca\exp[\frac{a^2}{d}\Big(\frac{\epsilon}{2}+\frac{\alpha_1}{16}a^2+\frac{\alpha_2}{48}a^4+\frac{5\alpha_3}{512}a^6+\frac{7\alpha_4}{1280}a^8+\cdots\Big)].
\end{equation}
Here $P(a,t)$ denotes the probability that the oscillation amplitude has a value of $a$ at a given time $t$, $P(a)$ is its stationary solution, and $C$ is a normalisation constant.  These equations are independent of the anisochronicity factor $\beta$.


\subsection{Actuator model}\label{act}
\begin{figure}
    \centering
    \includegraphics[width=0.48\textwidth,trim={0 0 0 0},clip]{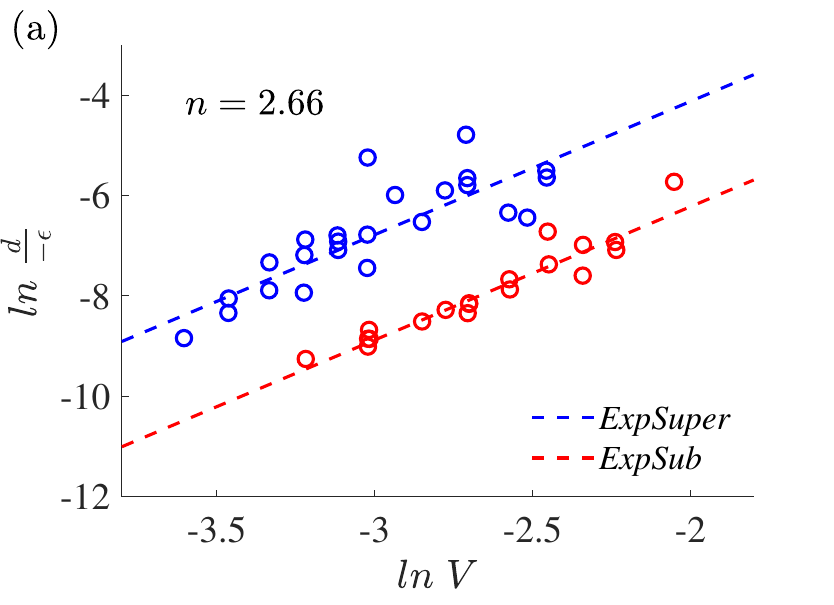}
    \includegraphics[width=0.48\textwidth,trim={0 0 0 0},clip]{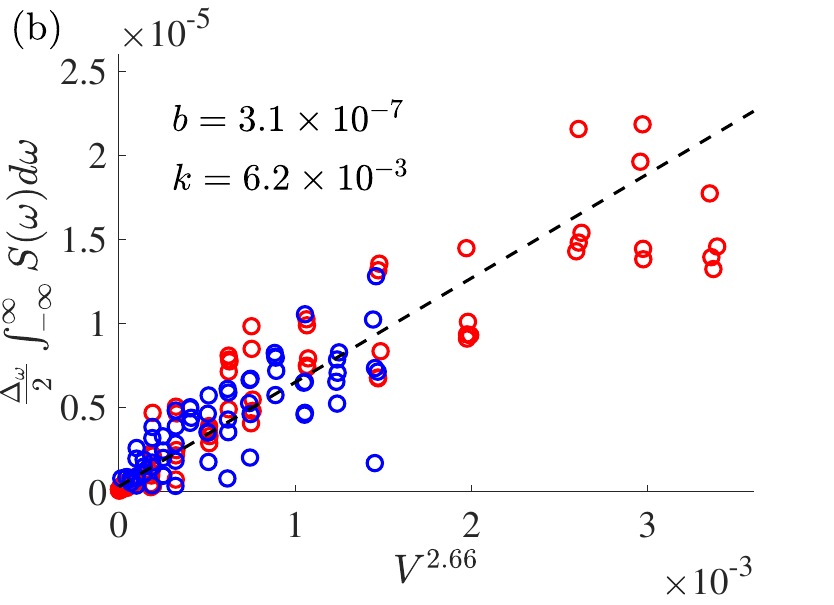}
    \caption{Modelling of the actuator to determine (a) $n$ and (b) $b$ and $k$.  The markers are experimental data, and the dotted lines are linear fits.}
    \label{fig:noise}
\end{figure}

One of the key challenges in SI is modelling the effect of an actuator on an experimental system.  This is because the way in which an actuator input, e.g. the loudspeaker voltage ($V$), is fed into a system, via the noise amplitude ($d$), is unique to that particular system.  This difficulty can be circumvented by turning to output-only SI, in which the actuator input is not modelled \citep{NOIRAY2013152,boujo2017robust}.  We will discuss this further in \S\ref{conc}.  Here, we derive a relationship between $V$ and $d$ based on experiments, with only two assumptions: (i) a power-law relationship exists between $V$ and $d$, such that $d=b+kV^n$, where $b$ is the inherent amplitude of background noise, $k$ is the proportionality constant, and $n$ is the exponent; and (ii) $b \ll d$.  Thus, we can write:
\begin{equation}\label{noise2}
    \ln{\Big(\frac{d}{-\epsilon}\Big)} \approx n \ln{V} + \ln{\Big(\frac{k}{-\epsilon}\Big)}.
\end{equation}
The logarithm of equation~(\ref{stationary}) gives the ratio between $d$ and one of the system parameters ($\epsilon$, $\alpha_1$, $\cdots$) at each value of $V$ (see the matrix in equation~(\ref{mat})).  We choose $\epsilon$ based on its smallest variance in multiple experimental replications, and plot $\ln{(\frac{d}{-\epsilon})}$ against $\ln{V}$ in figure \ref{fig:noise}(a).  The data for both $ExpSuper$ and $ExpSub$ fit well with a common slope of $n = 2.66$.  The $y$-intercept directly gives $\ln{(\frac{k}{-\epsilon})}$, but neither $k$ nor $\epsilon$ is known at this stage.

To find $k$ and $b$, we use information in the spectral domain.  Following \citet{ushakov2005coherence}, we derive an equation for the jet spectrum ($S_u$):
\begin{equation}\label{noise3}
    \int_{-\infty}^{\infty}S_u(\omega)d\omega=\frac{2d}{\Delta_\omega}=\frac{2(b+kV^n)}{\Delta_\omega},
\end{equation}
where $\Delta_\omega$ is the half-width at half-maximum when a Lorentzian curve is fitted to $S_u$.  The coefficients $k$ and $b$ are then extracted from the $y$-intercept and gradient of the data in figure~\ref{fig:noise}(b), respectively.  Thus, the relationship between the input loudspeaker voltage and the noise amplitude is $d=(3.1\times10^{-7})+(6.2\times10^{-3})V^{2.66}$.  The fact that data for both $ExpSuper$ and $ExpSub$ fit this power-law model and that $b$ is indeed very small justifies our modelling assumptions for the actuator.

\subsection{System identification}\label{SI}
\begin{figure}
    \centering
    \includegraphics[width=0.48\textwidth,clip]{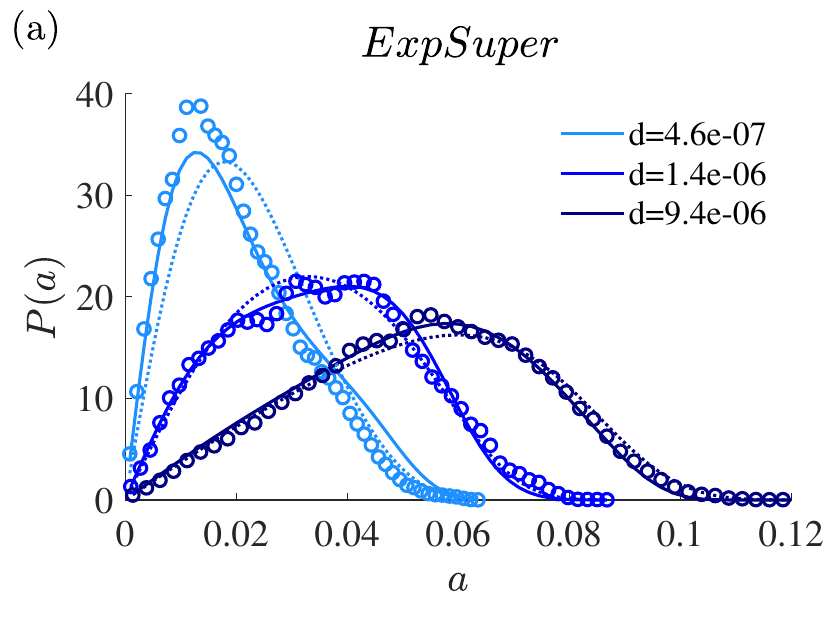}%
    \includegraphics[width=0.48\textwidth,clip]{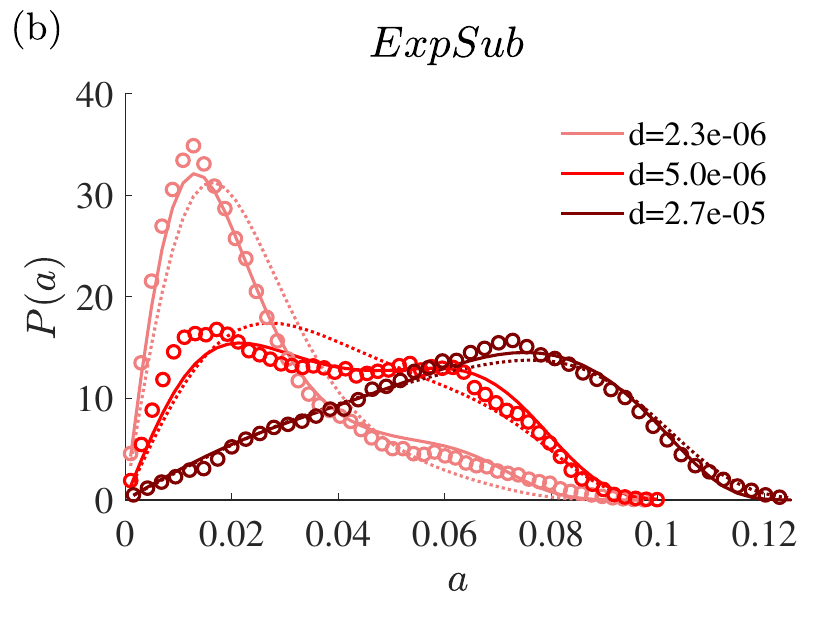}
    \caption{Probability density function $P(a)$ at different noise amplitudes for (a) $ExpSuper$ ($Re=584$) and (b) $ExpSub$ ($Re=755$).  The markers are experimental data, and the dashed and solid lines are numerical estimates from the N05 model (up to $\alpha_2$, quintic order) and the N09 model (up to $\alpha_4$, nonic order), respectively.}
    \label{fig:model}
\end{figure}

Figure~\ref{fig:model} shows the probability density function of the velocity fluctuation amplitude, $P(a)$, in the jet for (a) $ExpSuper$ and (b) $ExpSub$ under increasing noise amplitudes.  The model coefficients ($\epsilon$ and $\alpha_{1,\cdots}$) are found by fitting polynomials to equation~(\ref{stationary}) with the measured $P(a)$.  More specifically, the model coefficients are found via a linear least-squares fitting solution of the following matrix problem:
\begin{equation}
\renewcommand \arraystretch{1.2}
\begin{bmatrix}
\ln {P(a_{b1})} - \ln{a_{b1}} \\
\ln {P(a_{b2})} - \ln{a_{b2}} \\
\multirow{2}{*}{\vdots} \\
\\
\ln {P(a_{bN})} - \ln{a_{bN}}
\end{bmatrix}
=
\begin{bmatrix}
1 && a_{b1}^2 && a_{b1}^4 && a_{b1}^6 && \cdots \\
1 && a_{b2}^2 && a_{b2}^4 && a_{b2}^6 && \cdots \\
\multirow{2}{*}{\vdots} && \multirow{2}{*}{\vdots} && \multirow{2}{*}{\vdots} && \multirow{2}{*}{\vdots} && \\
\\
1 && a_{bN}^2 && a_{bN}^4 && a_{bN}^6 && \cdots \\
\end{bmatrix}
\begin{bmatrix}
\ln {C} \\
\frac{\epsilon}{2d} \\
\frac{\alpha_{1}}{16d} \\
\frac{\alpha_{2}}{48d} \\
\multirow{2}{*}{\vdots} \\
\\
\end{bmatrix},
\label{mat}
\end{equation}
where $a_{b1}$, $a_{b2}$, $\cdots$, $a_{bN}$ are uniformly distributed bins of $a$ (i.e. the $x$ axis of figure~\ref{fig:model}).  At each $Re$, there are 20 and 19 different levels of $d$ in $ExpSuper$ and $ExpSub$, respectively.  For each value of $d$, there are five sets of data. The final values of the model coefficients are determined by averaging across all levels and sets of $d$ at each $Re$.  In this averaging procedure, we exclude outliers by discarding the data points within 20$\%$ of the extrema. Figure~\ref{fig:model} shows that the ability of the model to reproduce $P(a)$ improves as the number of nonlinear terms in the model increases.

\section{Results and discussion}\label{res}

\subsection{Determination of the order of nonlinearity}\label{nl}
The order of nonlinearity in the system is determined based on the number of nonlinear terms required to reproduce the measured $P(a)$. This is achieved by successively adding higher-order nonlinear terms to equation~(\ref{fracdp}) until the rank of the matrix in equation~(\ref{mat}) becomes deficient. Figure~\ref{fig:model}(b) shows that, at an intermediate noise amplitude, two peaks appear in $P(a)$.  This behaviour is called bimodality and is observed in both $ExpSuper$ and $ExpSub$.  We derive a condition for the amplitude ($a_m$) at which extrema of $P(a)$ occur:
\begin{equation}\label{max_sup}
    d+\epsilon a_m^2 + \frac{\alpha_1}{4} a_m^4 + \frac{\alpha_2}{8} a_m^6 + \frac{5\alpha_3}{64} a_m^8 + \frac{7\alpha_4}{128} a_m^{10} +\cdots = 0.
\end{equation}
For bimodality to exist, there must be two positive solutions of $a_m^2$ at some values of $d$.  For this to occur, the model must have nonlinear terms up to at least quintic order ($\alpha_2$).

To reproduce $P(a)$, we show three different models.  The first model, called N05, has up to 5th-order (quintic) nonlinearity, requiring up to $\alpha_2$, which is the minimum for bimodality.  The second and third models have up to 9th-order (called N09; up to $\alpha_4$) and 13th-order (called N13; up to $\alpha_6$) nonlinearity, respectively.
Table~\ref{tab:my_label2} lists the coefficients for the three models.  It can be seen that when going from N05 to N09, the coefficients change significantly -- by an order of magnitude in many cases.  However, when going from N09 to N13, only a small change in the coefficients is observed, with $\alpha_5$ and $\alpha_6$ being negligible.  In figure~\ref{fig:model}, we observe that the N09 model (solid lines) reproduces $P(a)$ satisfactorily.  Thus, we conclude that the nonlinearity in this system is up to 9th order.

\begin{landscape}
\begin{table}[]
\small
    \centering
    \begin{tabular}{c c | c c c c c | c c c c c}
    \hline
    \vspace{2mm}
      \multirow{2}{*}{Coef.} & \multirow{2}{*}{Model} & \multicolumn{5}{c}{$Re$ for $ExpSuper$} & \multicolumn{5}{c}{$Re$ for $ExpSub$}\\
      \vspace{2mm}
       & & 577.4 & 580.5 & 583.8 & 587.0 & 590.3 & 738.5 & 742.7 & 746.9 & 751.1 & 755.3 \\

      \multirow{3}{*}{$\epsilon$} & N05 & -2.4e-3 & -2.2e-3 & -1.5e-3 & -1.0e-3 & -7.4e-4 & -1.7e-2 & -1.4e-2 & -1.3e-2 & -1.1e-2 & -9.1e-3\\
      & N09 & -4.0e-3 & -3.6e-3 & -3.3e-3 & -2.7e-3 & -2.1e-3 & -2.2e-2 & -2.1e-2 & -1.8e-2 & -1.7e-2 & -1.5e-2\\

      \vspace{2mm}
      & N13 & -4.0e-3 & -3.6e-3 & -3.0e-3 & -2.8e-3 & -2.2e-3 & -2.2e-2 & -2.0e-2 & -1.8e-2 & -1.6e-2 & -1.5e-2\\

      \multirow{3}{*}{$\alpha_1$} & N05 & 8.1e-1 & 1.8e+0 & 1.5e+0 & 1.8e+0 & 2.1e+0 & 1.4e+1 & 1.4e+1 & 1.1e+1 & 1.4e+1 & 1.3e+1\\
      & N09 & 1.6e+1 & 1.6e+1 & 1.5e+1 & 1.5e+1 & 1.5e+1 & 5.0e+1 & 4.9e+1 & 4.7e+1 & 4.5e+1 & 4.3e+1 \\
      \vspace{2mm}
      & N13 & 1.6e+1 & 1.6e+1 & 1.3e+1 & 1.5e+1 & 1.5e+1 & 4.9e+1 & 4.5e+1 & 4.5e+1 & 4.1e+1 & 4.4e+1\\

      \multirow{3}{*}{$\alpha_2$} & N05 & -1.5e+3 & -1.8e+3 & -1.5e+3 & -1.5e+3 & -1.5e+3 & -5.9e+3 & -6.1e+3 & -4.2e+3 & -4.5e+3 & -3.5e+3\\
      & N09 & -1.8e+4 & -1.6e+4 & -1.4e+4 & -1.3e+4 & -1.2e+4 & -4.0e+4 & -3.2e+4 & -3.1e+4 & -2.6e+4 & -2.3e+4 \\
      \vspace{2mm}
      & N13 & -1.8e+4 & -1.6e+4 & -1.4e+4 & -1.3e+4 & -1.2e+4 & -3.9e+4 & -3.0e+4 & -3.0e+4 & -2.5e+4 & -2.3e+4\\

      \multirow{2}{*}{$\alpha_3$} & N09 & 5.3e+6 & 4.2e+6 & 3.5e+6 & 3.0e+6 & 2.4e+6 & 1.0e+7 & 6.2e+6 & 6.1e+6 & 4.6e+6 & 3.9e+6\\
      \vspace{2mm}
      & N13 & 5.1e+6 & 4.1e+6 & 3.5e+6 & 3.2e+6 & 2.4e+6 & 9.7e+6 & 6.2e+6 & 6.1e+6 & 4.6e+6 & 3.4e+6\\

      \multirow{2}{*}{$\alpha_4$} & N09 & -5.1e+8 & -3.8e+8 & -2.9e+8 & -2.2e+8 & -1.6e+8 & -8.4e+8 & -4.5e+8 & -4.1e+8 & -3.0e+8 & -2.4e+8 \\
      \vspace{2mm}
      & N13 & -4.9e+8 & -3.6e+8 & -3.1e+8 & -2.4e+8 & -1.6e+8 & -8.0e+8 & -4.7e+8 & -4.1e+8 & -2.9e+8 & -1.8e+8\\

      \vspace{2mm}
      $\alpha_5$/$\alpha_6$ & N13 & 0.0e+0 & 0.0e+0 & 0.0e+0 & 0.0e+0 & 0.0e+0 & 0.0e+0 & 0.0e+0 & 0.0e+0 & 0.0e+0 & 0.0e+0\\

      \hline
    \end{tabular}
    \caption{Model coefficients for $ExpSub$ and $ExpSuper$.  The models N05, N09 and N13 have up to $\alpha_2$, $\alpha_4$ and $\alpha_6$ terms, respectively.  Increasing the order of nonlinearity above the nonic term ($\alpha_4$) does not further improve the agreement with the experimental data.}
    \label{tab:my_label2}
\end{table}
\end{landscape}

\subsection{Prediction of dynamic and stochastic bifurcations} \label{sec:extrapol}
The main motivation for SI is to be able to predict dynamic bifurcations (i.e. the Hopf and saddle-node points) and, hence, the stability boundaries.  In addition to this, we also predict stochastic P-bifurcations, i.e. when the system switches from unimodal to bimodal behaviour \citep{Zakh2010}.  Stochastic P-bifurcations are important for determining the dynamic bifurcations of noisy systems \citep{Zakh2010}.

\begin{figure}
    \centering
    \includegraphics[width=0.48\textwidth,clip]{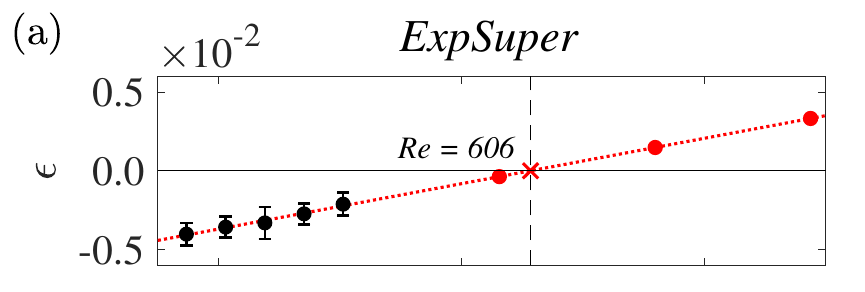}
    \includegraphics[width=0.48\textwidth,clip]{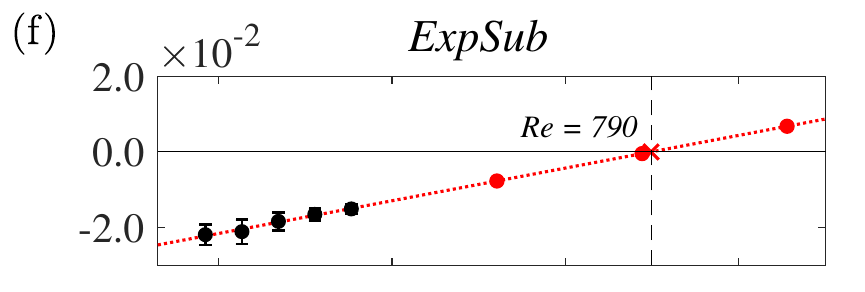}\\
    \includegraphics[width=0.48\textwidth,clip]{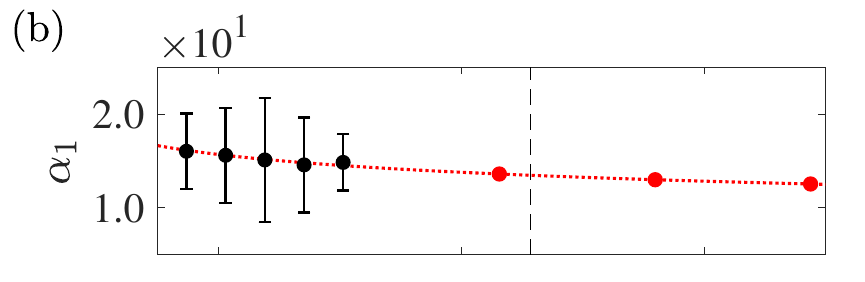}
    \includegraphics[width=0.48\textwidth,clip]{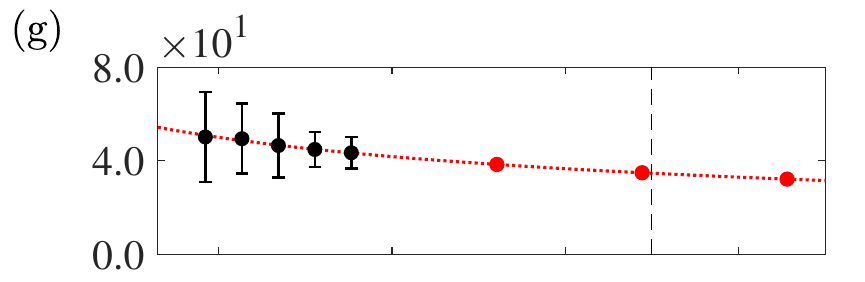}\\
    \includegraphics[width=0.48\textwidth,clip]{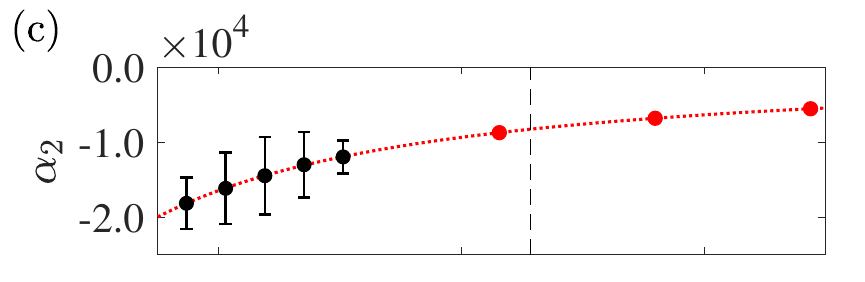}
    \includegraphics[width=0.48\textwidth,clip]{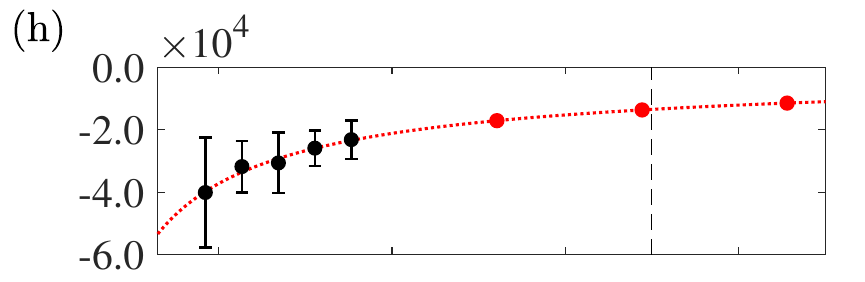}\\
    \includegraphics[width=0.48\textwidth,clip]{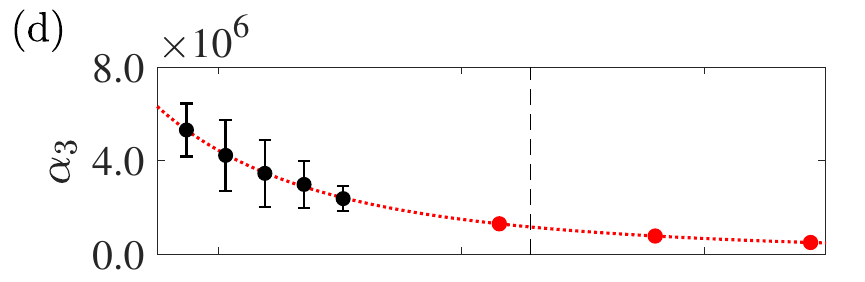}
    \includegraphics[width=0.48\textwidth,clip]{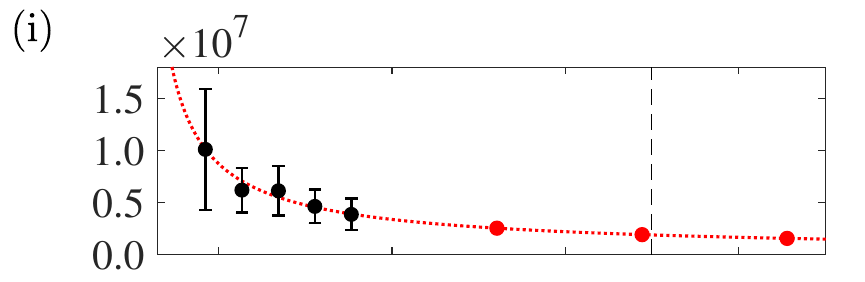}\\
    \includegraphics[width=0.48\textwidth,clip]{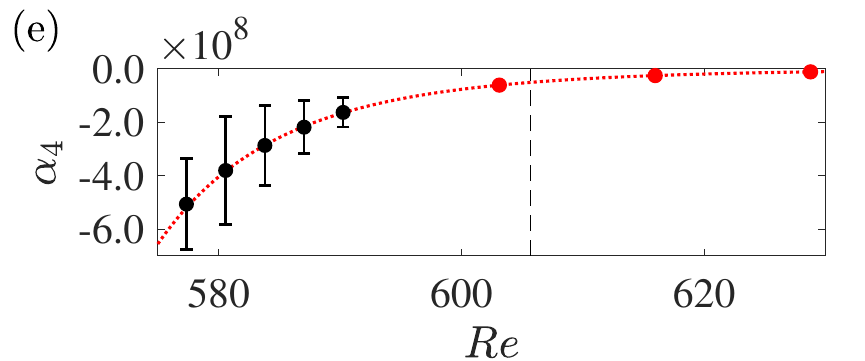}
    \includegraphics[width=0.48\textwidth,clip]{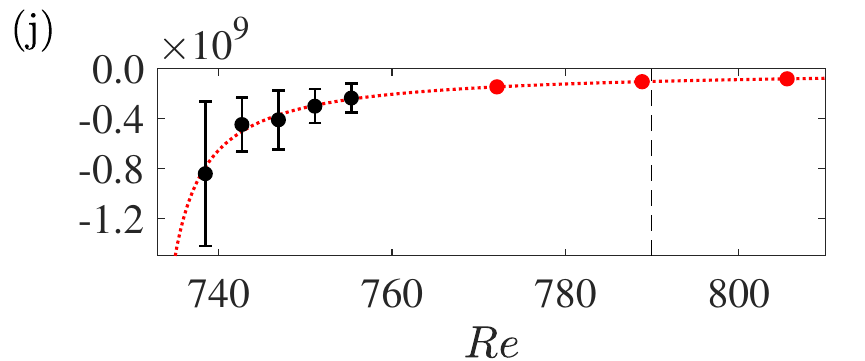}
  \caption{Model coefficients with respect to $Re$ for (a--e) $ExpSuper$ and (f--j) $ExpSub$.  The black markers with error bars are the experimental data. The dotted red lines are (a,f) linear fits and (b--e,g--j) power-law fits.  The red markers are the extrapolated points for $ExpSuper$ ($Re=603$, $616$, $629$) and for $ExpSub$ ($Re=772$, $789$, $806$).}\label{fig:coeff}
\end{figure}

\begin{figure}
    \centering
    \includegraphics[width=0.48\textwidth,trim={0 0 0 0},clip]{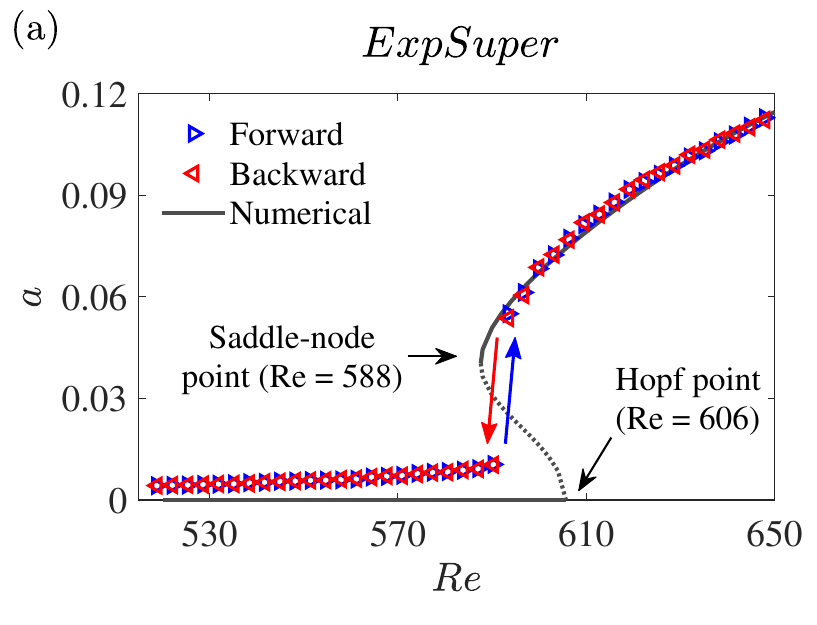}
    \includegraphics[width=0.48\textwidth,trim={0 0 0 0},clip]{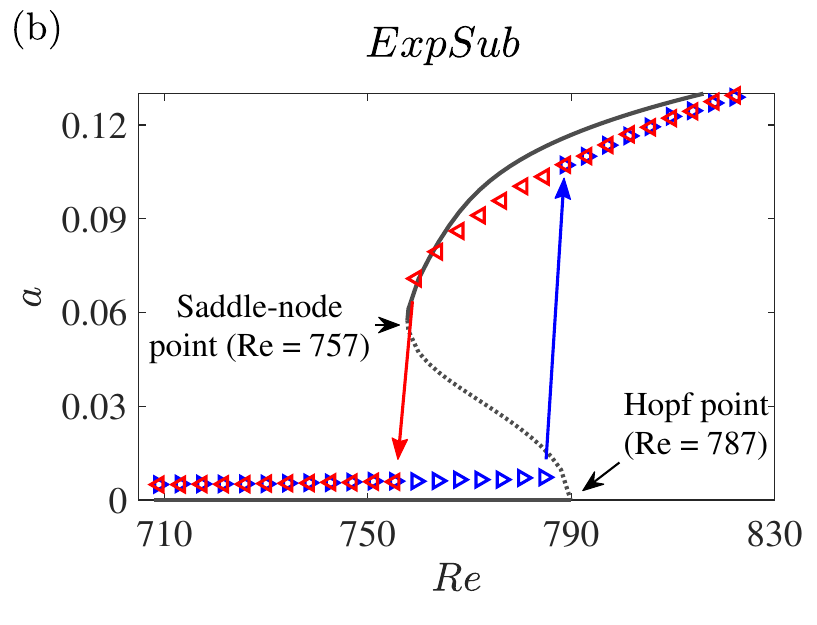}\\
    \includegraphics[width=0.48\textwidth,trim={0 0 0 0},clip]{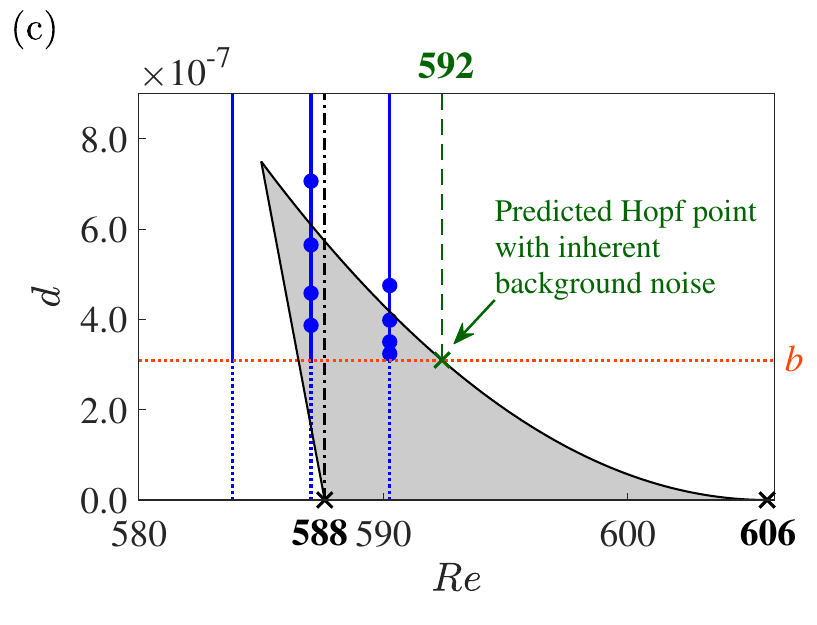}%
    \includegraphics[width=0.48\textwidth,trim={0 0 0 0},clip]{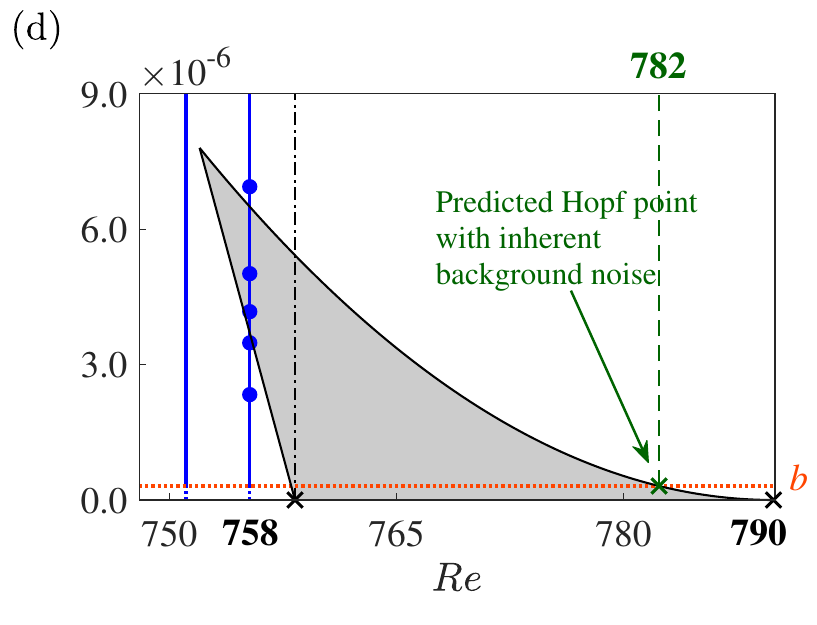}%
    \caption{Dynamic and P-bifurcation plots for (a,c) $ExpSuper$ and (b,d) $ExpSub$. In (a,b), the solid and dotted lines denote the stable and unstable solutions, respectively, as calculated from the model.  In (c,d), the grey areas denote the bimodal regime calculated from the model, the blue vertical lines denote where the experiments were conducted, and the blue circular markers denote where bimodality is observed experimentally.  The orange horizontal lines denote the inherent amplitude of the background noise.}
    \label{fig:bif}
\end{figure}

To predict the bifurcation points, we extrapolate the model coefficients calculated in \S\ref{nl} to higher $Re$, as shown in figure~\ref{fig:coeff}.  We use a linear regression for $\epsilon$, much as \citet{Provansal1987} did in their experiments on a cylinder wake.  For the higher-order coefficients, we use a power-law fit: $\alpha_n \propto (Re-{m_1})^{-m_2}$, where $m_1$ and $m_2$ are positive constants obtained from least-squares fitting of the experimental data. From the extrapolated coefficients, we generate dynamic and stochastic bifurcation plots for $ExpSuper$ and $ExpSub$, and compare them in figure~\ref{fig:bif} with our experimental data. The dynamic bifurcation plots are generated by solving equation~(\ref{derived11}) without the effect of noise, whereas the stochastic bifurcation plots are generated by finding solutions of equation~(\ref{max_sup}) that have two positive $a_m^2$.

Figure~\ref{fig:bif}(a,b) shows that, without noise ($d=0$), the numerically predicted Hopf and saddle-node points are, respectively, at $Re=606$ and $588$ for $ExpSuper$, and at $Re=790$ and $758$ for $ExpSub$.  The model correctly identifies $ExpSub$ to be subcritical but, curiously, it identifies $ExpSuper$ to be subcritical as well, which might seem to contradict the experiments.  However, a careful examination of the experimental data (figure~\ref{fig:bif}a) shows a marked jump in the oscillation amplitude at the bifurcation point.  We speculate that this jump occurs because the Hopf and saddle-node points have either collided or moved so close to each other as to be indistinguishable within the limits of experimental uncertainty.  This interpretation of supercritical-like behaviour can also explain previous observations of a similar amplitude jump in the low-density jet experiments of \citet{hallberg2006universality} and \citet{zhu2017onset}.  Moreover, the presence of background noise shrinks the hysteretic bistable region by triggering LCOs.  Next, we examine the effect of background noise on dynamic bifurcations using P-bifurcation plots.

Bimodality is usually associated with subcritical Hopf bifurcations \citep{Zakh2010}.  As shown in figure~\ref{fig:bif}(c,d), bimodality (grey areas) exists between the Hopf and saddle-node points, even for infinitesimally weak noise.  Bimodality represents the tendency of a system to switch between the zero-amplitude state and the LCO state in the bistable regime.  In the presence of finite-amplitude noise, this tendency can be observed even before the system reaches the saddle-node point.  This is seen in our experiments (figure~\ref{fig:bif}c,d: blue markers) and is well predicted by our model.  Background noise, however, can shrink the bimodal region by triggering LCOs.  In figure~\ref{fig:bif}(c,d), this shrinkage can be seen as a tapering of the bimodal region (grey area) above the inherent amplitude of the background noise, $b$ (orange horizontal line).  Therefore, for an accurate comparison between the predicted and experimentally observed bifurcation points, we must account for the effect of noise.  We do this by locating the points (green crosses) at which $b$ (orange horizontal line) intersects the bimodal region (grey area).  For $ExpSuper$ (figure~\ref{fig:bif}c), this gives a predicted Hopf point of $Re = 592$, which matches exactly with the experimentally observed value at $Re = 592$ (figure~\ref{fig:bifexp}a).  As mentioned earlier, the absence of a bistable region can be understood because it is exceedingly small.  For $ExpSub$ (figure~\ref{fig:bif}d), the intersection of $b$ and the bimodal region gives predicted Hopf and saddle-node points of $Re = 782$ and $758$, respectively, which match well with the experimentally observed values at $Re = 787$ and $757$ (figure~\ref{fig:bifexp}b).

\subsection{Prediction of the system dynamics beyond the bifurcation points}
\begin{figure}
    \centering
    \includegraphics[width=0.4\textwidth,clip]{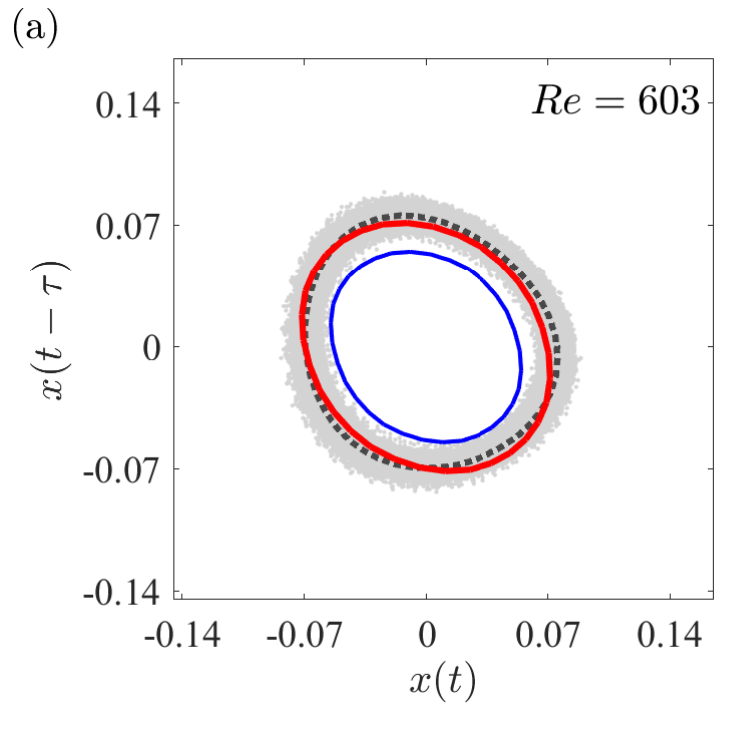}%
    \includegraphics[width=0.4\textwidth,clip]{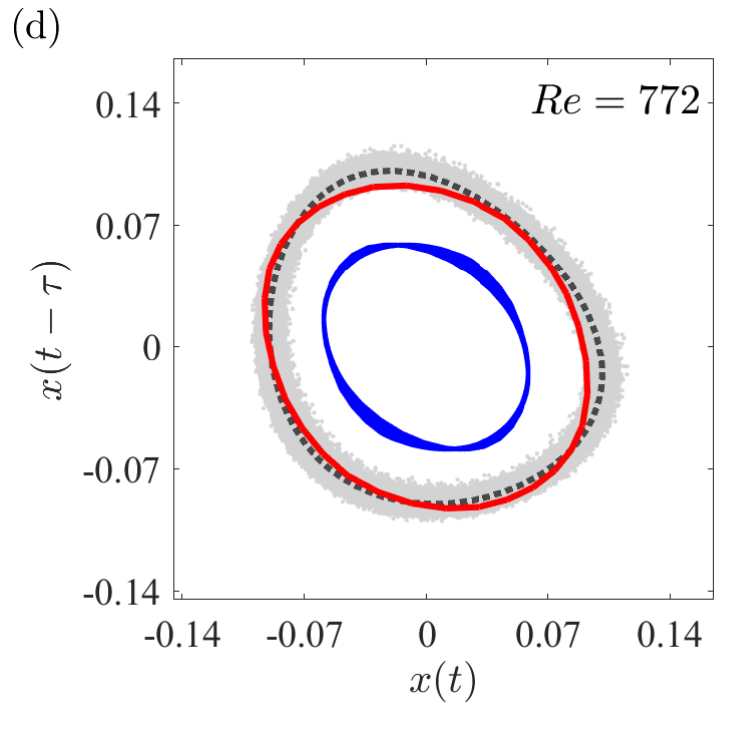}\\%
    \includegraphics[width=0.4\textwidth,clip]{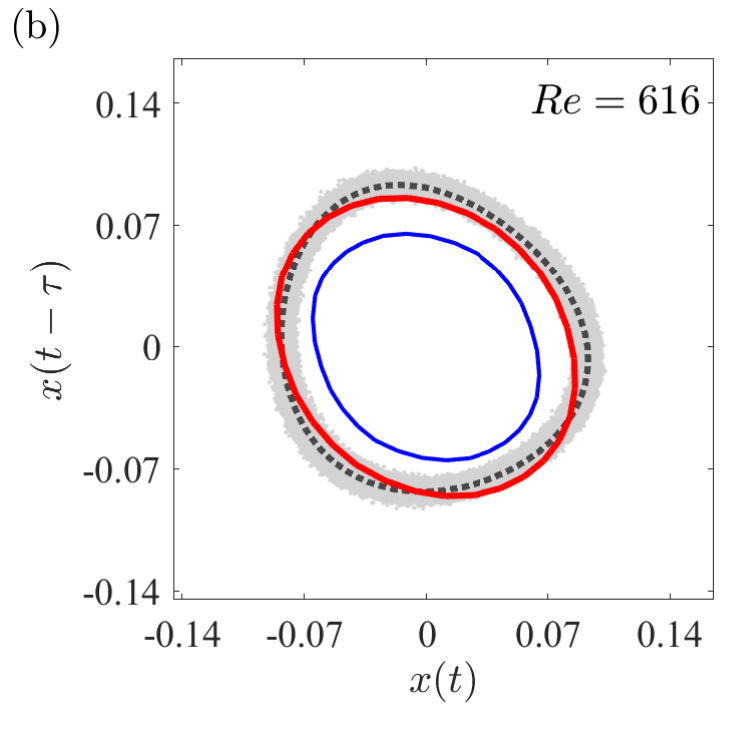}
    \includegraphics[width=0.4\textwidth,clip]{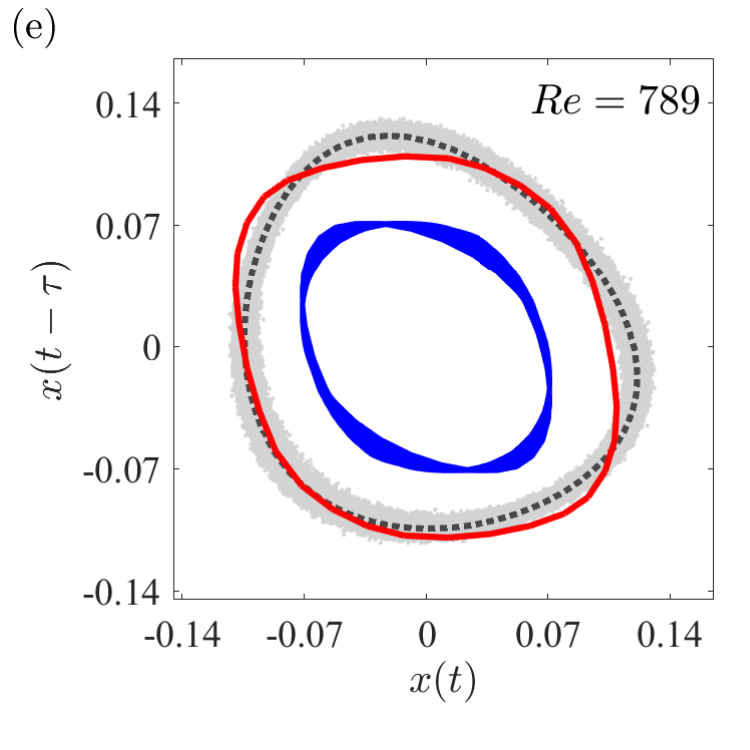}\\
    \includegraphics[width=0.4\textwidth,clip]{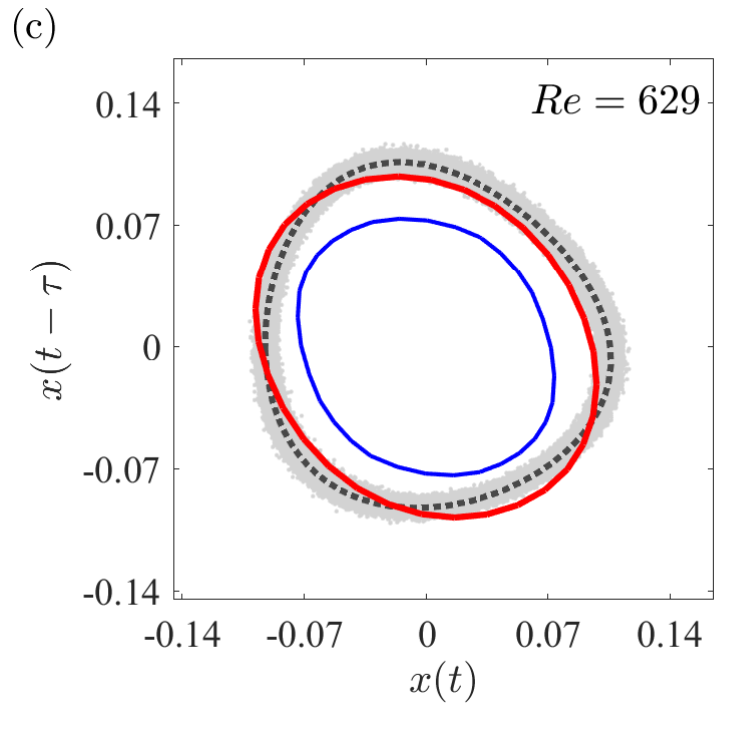}
    \includegraphics[width=0.4\textwidth,clip]{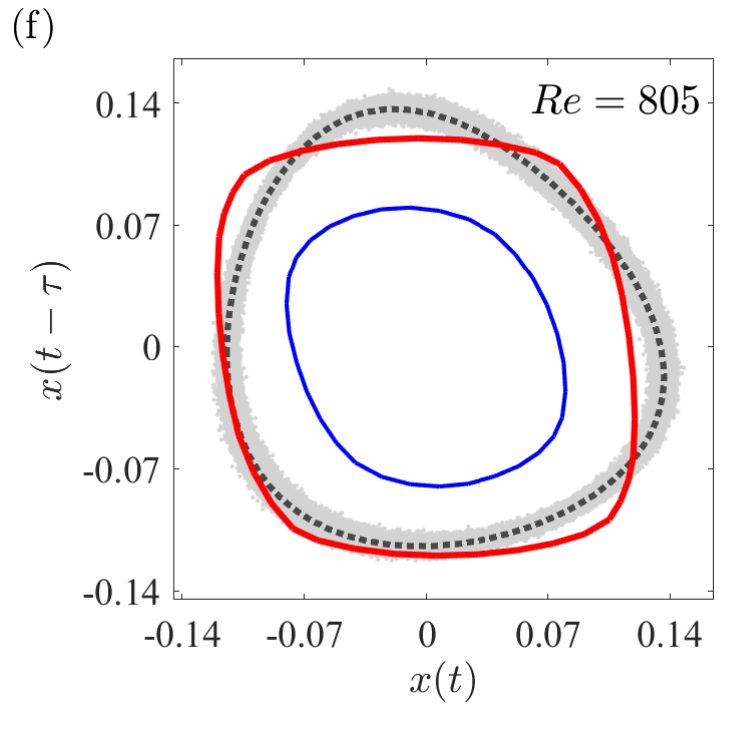}
    \caption{Phase portraits of the LCOs for (a--c) $ExpSuper$ and (d--f) $ExpSub$.  The experimental LCO orbits (grey bands) are shown alongside their mean orbits (black dotted lines).  These can be compared with the numerically obtained LCO orbits from the N05 model (blue line) and the N09 model (red line).}
    \label{fig:lc}
\end{figure}

We now turn to predicting the system dynamics away from the bifurcation points. Figure~\ref{fig:lc} shows phase portraits of the LCOs for (a--c) $ExpSuper$ at $Re=603, 616$ and $629$, and (d--f) $ExpSub$ at $Re=772, 789$ and $806$.  These are compared with the corresponding LCOs from the experiments.  In seminal work, \citet{takens1981detecting} showed that the dynamical properties of a system containing many degrees of freedom can be represented by just a single scalar time series with an appropriately chosen time delay ($\tau$).  Here, we show the phase portrait in two dimensions with $\tau$ calculated using the average mutual information method of \citet{fraser1986}.  The comparison between the experimental and numerical LCOs shows that the N09 model can accurately predict both the amplitude and shape of the LCO orbits.  This further highlights the important role that the higher-order nonlinear terms have in determining the system dynamics.

\section{Conclusions}\label{conc}
We perform SI of a low-density jet from its noise-induced dynamics, using a low-order oscillator model and its corresponding Fokker--Planck equation.  To the best of our knowledge, this is the first time that SI has been achieved on an experimental system using the noise-induced dynamics in only the unconditionally stable regime, i.e. without having to operate in the regimes where LCOs may occur.  We show that our estimated numerical model can accurately predict three key system properties: (i) the order of nonlinearity, (ii) the locations and types of the bifurcation points (and hence the stability boundaries), and (iii) the limit-cycle dynamics beyond the bifurcation points.

There are two main implications of this work that go beyond low-density jets. First, the SI methodology proposed here should be applicable to many other dynamical systems, as the only inherent assumption made about the system is that it obeys the Stuart--Landau equation.  This assumption is, in fact, valid in the vicinity of the Hopf point for many dynamical systems -- hydrodynamic or otherwise.  Consequently, the Stuart--Landau equation has been used in a number of other SI methodologies in the literature (see \S\ref{sec:intro}).  However, in all of those studies, it has been assumed that the nonlinear terms can only be obtained from data collected during the occurrence of LCOs, in the unstable or bistable regime.  With our SI methodology, however, we show that data from the noise-induced dynamics in the unconditionally stable regime is itself enough to determine the bifurcation points and to predict the LCO dynamics beyond those points.  Thus, our SI methodology opens up new pathways for the development of early-warning indicators and active-control strategies against unwanted oscillations in systems operating near a Hopf point.  This is particularly useful for the design of systems prone to exhibiting dangerously energetic LCOs, such as thermoacoustic oscillations in gas turbines and rocket engines.

Second, the prediction of system nonlinearity -- in particular, the order and signs of the nonlinear terms -- can provide physical insight into the system.  For plane Poiseuille flow, \citet{stuart1960non} was able to explain that the physical meaning of a positive $k_2$ term is that the distortion of the fundamental instability mode is dominant over the combination of the distortion of the mean motion and the generation of harmonics.  It is beyond the scope of this paper to perform an equivalent analysis for the low-density jet and extend it to the higher-order terms.  \citet{stuart1960non}, however, did not attempt to calculate the nonlinear terms, which we have done here.

As for improvements to this SI methodology, we should be able to relax the assumption that the background noise amplitude is small.  In many natural and engineered systems, background noise can be significant, making the development of an actuator model difficult.  An instinctive solution is to turn to output-only SI methods, but these are usually only reliable when the input data size is large \citep{MEVEL2006531}.  This problem can be alleviated through the use of adjoint equations, as demonstrated by \citet{boujo2017robust}.  Furthermore, in the simple axisymmetric jet studied here, we have used information collected at only one spatial location.  This keeps the system size small without adversely affecting the quality of the predictions for the bifurcation points and LCO dynamics.  However, there could be other, more complicated, flows for which it may be useful to include information about the spatial structure of the global instability mode.  In such cases, we may need to incorporate the use of sparsity-promoting tools and machine learning in this SI framework to deal with the larger data matrices.

This work was funded by the Research Grants Council of Hong Kong (Project Nos. 16235716 and 26202815).

\chapter{Input-output system identification of a thermoacoustic oscillator near a Hopf bifurcation using only fixed-point data} \label{chap:Rij}

Published in \textit{Physical Review E}, vol. 101(1), pp. 013102 (2020).


\section*{Abstract}
We present a framework for performing input-output system identification near a Hopf bifurcation using data from only the fixed-point branch, prior to the Hopf point itself.  The framework models the system with a van der Pol-type equation perturbed by additive noise, and identifies the system parameters via the corresponding Fokker--Planck equation.  We demonstrate the framework on a prototypical thermoacoustic oscillator (a flame-driven Rijke tube) undergoing a supercritical Hopf bifurcation.  We find that the framework can accurately predict the properties of the Hopf bifurcation and the limit cycle beyond it.  This study constitutes the first experimental demonstration of system identification on a reacting flow using only pre-bifurcation data, opening up new pathways to the development of early warning indicators for nonlinear dynamical systems near a Hopf bifurcation.

\section{Introduction} \label{sec:4intro}
Many natural and engineered systems are nonlinear and can develop self-sustained oscillations \citep{strogatz2000}.  Such oscillations are desirable in some systems (e.g. musical instruments \citep{abel2009}, pendulum clocks \citep{huygens1986} and pulsed combustors \citep{putnam1986}) but they are undesirable in other systems (e.g. bridge structures \citep{blevins1977}, predator-prey systems \citep{yoshida2003} and gas turbines \citep{Lieuwen2005}.  A canonical way for self-sustained oscillations to arise is via a Hopf bifurcation \citep{strogatz2000}, in which a fixed point loses stability and a complex conjugate pair of eigenvalues crosses the imaginary axis in response to changes in a control parameter \citep{Thompson2002}.  The result is a transition from a fixed point to a limit cycle \citep{Thompson2002}.  If the limit cycle arises only after the Hopf point and its amplitude increases gradually with changes in the control parameter, then the Hopf bifurcation is supercritical.  If the limit cycle arises in a hysteric bistable regime, between the Hopf and saddle-node points, and its amplitude increases abruptly, then the Hopf bifurcation is subcritical.  Whether a Hopf bifurcation is supercritical or subcritical depends on the specific form of nonlinearity in the system \citep{Thompson2002}.  In many practical systems, it is advantageous to be able to predict the type and location of the Hopf bifurcation, because this can enable users to avoid destructive acoustic or structural resonances.  Thus, there is a need for robust methods capable of identifying the nonlinear properties of dynamical systems using only pre-bifurcation data.  In this paper, we demonstrate a framework for this that uses the noise-perturbed data on the fixed-point branch, prior to the Hopf point itself.

\subsection{Thermoacoustic instability via a Hopf bifurcation}
Despite significant research, thermoacoustic instability continues to hamper the development of combustion devices such as gas turbines and rocket engines \citep{culick2006,Poinsot2017}.  The underlying cause of this instability is the positive feedback between the heat-release-rate (HRR) oscillations of an unsteady flame and the pressure oscillations of its surrounding combustor \citep{Candel2002}.  If the HRR oscillations are sufficiently in phase with the pressure oscillations, the former can transfer energy to the latter via the Rayleigh mechanism \citep{Rayleigh1878}, leading to self-sustained flow oscillations at one or more of the natural acoustic modes of the system \citep{lieuwen2012unsteady,guan2019chaos}.  If severe, such thermoacoustic oscillations can exacerbate vibration, mechanical fatigue and thermal loading, reducing the reliability of the overall system.  This problem is especially concerning in modern gas turbines because the conditions under which such devices must operate to achieve low pollutant emissions are also those that provoke thermoacoustic instability \citep{Lieuwen2005}.

Like other self-sustained oscillations, thermoacoustic oscillations often arise via a Hopf bifurcation, making them amenable to a weakly nonlinear analysis near the Hopf point \citep{juniper2018}.  Such an analysis can be performed with the normal-form equation for a Hopf bifurcation, which, in fluid mechanics, is known as the Stuart--Landau equation \citep{drazin2004hydrodynamic}:
\begin{equation}\label{eq:landau40}
\frac{\mathrm{d}{a}}{\mathrm{d}t}=k_1{a}+k_2{a}^3+\cdots,
\end{equation}
where $a$ is the complex mode amplitude, $k_1$ is the linear driving/damping coefficient, $k_2$ is a nonlinear coefficient, and $t$ is time.  A Hopf bifurcation occurs at $k_1=0$.  The Stuart--Landau equation can capture the amplitude evolution of a system near the Hopf point, where the growth rate, which controls the amplitude evolution, is still much smaller than the oscillation frequency.  Weakly nonlinear analyses based on the Stuart--Landau equation have been used before to study hydrodynamic systems \citep{Sipp2007,Li13a,Li13b,zhu2017onset,zhu2019} and thermoacoustic systems \citep{subramanian2013,Etikyala2017}.  For example, \citet{orchini2016} recently carried out a weakly nonlinear analysis of a Rijke tube and showed that such an approach can reduce the computational cost of investigating oscillatory phenomena near a Hopf bifurcation.

\subsection{Noise-induced dynamics of thermoacoustic systems}  \label{sec:NIDthermo}
Thermoacoustic systems often exhibit combustion noise, which can arise from direct sources, such as the HRR fluctuations of an unsteady flame, and indirect sources, such as the acceleration of entropy or vortical inhomogeneities through a nozzle \citep{dowling2015combustion}.  Previous studies on the noise-induced dynamics of thermoacoustic systems have focused primarily on two objectives: (i) to investigate the dynamical effect of noise, such as how it shifts the stability boundaries \citep{lieuwen2005b,gopalakrishnan2015} and how it triggers limit-cycle oscillations in the bistable regime \citep{burnley2000, jegadeesan2013}; and (ii) to gather information about the system from its noise-induced dynamics.  Such information can then be used to predict the onset of instability \citep{nair2014b,Kabiraj2015,Gopalakrishnan2016b,murayama2018characterization,hashimoto2019spatiotemporal,kobayashi2019,li2019coherence}, to distinguish between supercritical and subcritical bifurcations \citep{Gupta2017}, and to extract deterministic quantities \citep{NOIRAY2013152,boujo2017robust,noiray2017linear,noiray2017method}.

The noise-induced dynamics of a system can be determined by measuring its response to extrinsic or intrinsic perturbations \citep{risken1984}.  In the early years of rocket development, extrinsic perturbations in the form of bomb detonations were used in combustors to determine their stochastic properties and stability boundaries \citep{harrje1972}.  In recent years, such noise-induced dynamics has been used to forecast the onset of thermoacoustic instability.  For example, \citet{Kabiraj2015} applied extrinsic perturbations to a thermoacoustic system and found that its degree of coherence peaks at an intermediate noise amplitude -- a phenomenon called coherence resonance.  These researchers noted that such dynamics could be used as a precursor to a Hopf bifurcation.  Other metrics capable of forecasting the onset of thermoacoustic instability include the Hurst exponent \citep{nair2014b}, ordinal partition transition networks \citep{kobayashi2019}, the phase parameter \citep{hashimoto2019spatiotemporal}, sequential horizontal-visibility-graph motifs \citep{murayama2018characterization}, and the autocorrelation function and variance \citep{Gopalakrishnan2016b}.  In the present study, we build on these contributions by showing that it is possible to predict the properties of a Hopf bifurcation and the resultant limit cycle, using only pre-bifurcation data and without the need to set ad-hoc instability thresholds.

\subsection{System identification} \label{sec:SIintro}
System identification (SI) refers to the use of statistical methods to construct mathematical models of dynamical systems from input and/or output data.  There are two main ways in which SI can be performed: data-driven SI and model-based SI.  In data-driven SI, \textit{a priori} knowledge of the system physics is not required.  Instead, a model of the system is found solely from data using techniques such as symbolic regression \citep{Schmidt2009} and machine learning \citep{Brunton2016}.  Data-driven SI is useful when abundant data are available, either from experiments or simulations.  However, in practical systems, it is often difficult and costly to acquire sufficient data.  In such cases, it may be more efficient to use model-based SI, in which a low-dimensional model is assumed or developed for a system using information about its physics, and then the coefficients of the model are determined from data \citep{jaensch2014grey}.

In thermoacoustics, most studies relying on the noise-induced dynamics for SI have used a model-based approach \citep{culick2006, polifke2010system}.  For example, Noiray's group used a self-sustained oscillator equation perturbed by additive noise to model the dynamics of a gas-turbine combustor perturbed by its own turbulence \citep{NOIRAY2013152,noiray2017linear,boujo2017robust,noiray2017method}. Specifically, \citet{NOIRAY2013152} used stochastic differential equations, based on the Fokker--Planck formalism, to extract deterministic quantities from noise-perturbed data.  Recently, \citet{boujo2017robust} improved the accuracy of this SI method by incorporating adjoint-based optimization.  In these studies \citep{NOIRAY2013152,noiray2017linear,boujo2017robust,noiray2017method}, an intrinsic noise source, namely turbulence, was used to extract the system coefficients.  This output-only approach is convenient in that like most SI methods, it requires at least some data from the limit-cycle branch.  By contrast, \citet{lee2018system,lee2019systemt,lee2019exploiting,lee_2019} recently proposed an input-output SI framework in which extrinsic noise is fed into the system to enable prediction of its bifurcation properties and limit-cycle amplitudes, using data from only the fixed-point branch, before the Hopf point itself.  To date, however, the SI framework of \citet{lee_2019} has only been demonstrated on a simple hydrodynamic system, a low-density jet, which has none of the complexities of a thermoacoustic system such as nonlinear coupling between HRR oscillations and sound waves.

\subsection{Contributions of the present study}
In this study, we apply the SI framework of \citet{lee_2019} to a prototypical thermoacoustic system, a flame-driven Rijke tube, undergoing a supercritical Hopf bifurcation.  We show that this framework can enable accurate prediction of the properties of the Hopf bifurcation and the limit cycle beyond it, using nothing more than the noise-perturbed data on the fixed-point branch, prior to the Hopf point itself.  Crucially, we show that, unlike most other forecasting methods, ours does not require empirical instability thresholds to be set ad hoc, implying that our method can give objective predictions for a variety of nonlinear dynamical systems.  Below we present the experimental setup (\S\ref{sec:4exp}), review the SI framework of \citet{lee_2019} (\S\ref{sec:4method}), and apply that framework to a thermoacoustic system (\S\ref{sec:4result}), before concluding with the key implications and limitations of this study (\S\ref{sec:4conc}).

\section{Experimental setup} \label{sec:4exp}
\begin{figure}
\centering
\includegraphics[width=0.6\textwidth]{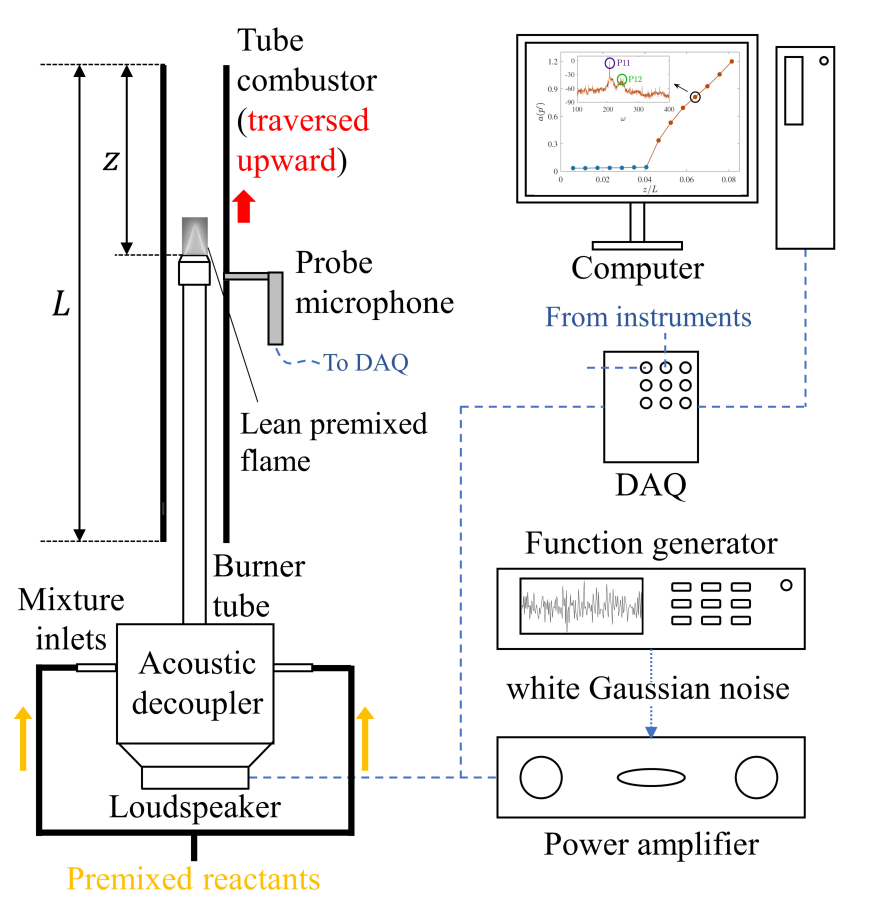}
\caption{Schematic of the experimental setup consisting of a prototypical thermoacoustic system (a flame-driven Rijke tube) perturbed by extrinsic noise from a loudspeaker.  DAQ: data acquisition system.}
\label{fig:Rijke}
\end{figure}

The thermoacoustic system under study consists of a vertical tube combustor containing a laminar conical premixed flame.  This system, which is also known as a flame-driven Rijke tube, can exhibit a variety of nonlinear states and bifurcations, making it an ideal platform for studying thermoacoustic phenomena \citep{guan2018strange,guan2019open,guan2019QP,guan2019control}. Shown in figure~\ref{fig:Rijke}, the system features a stainless-steel tube burner (inner diameter, ID~=~16.8~mm; length~=~800~mm), a double-open-ended quartz tube combustor (ID~=~44~mm; length, $L=$~860~mm) and an acoustic decoupler (ID~=~180~mm; length~=~200~mm).  The flame is stabilized on a copper extension tip (ID: $D=12$~mm; length~=~30~mm) mounted at the burner outlet.  Extrinsic perturbations are applied to the system via a loudspeaker (FaitalPRO 6FE100) mounted in the acoustic decoupler.  The loudspeaker is driven by a white Gaussian noise signal from a function generator (Keysight 33512B) via a power amplifier (Alesis RA150). The fuel used for the flame is liquefied petroleum gas (70\% butane, 30\% propane).  The fuel flow rate is controlled with a rotameter ($\pm2.5\%$), and the air flow rate is controlled with a mass flow controller (Alicat MCR: $\pm0.2\%$).  In this study, the system is operated at an equivalence ratio of 0.62 ($\pm3.2$\%), a bulk reactant velocity of $\bar{u}=$~1.6~m/s ($\pm0.2$\%), and a Reynolds number of $Re = 1300$ ($\pm1.7$\%) based on $\bar{u}$ and $D$.

To induce a Hopf bifurcation, we traverse the combustor upward relative to the stationary burner.  The non-dimensional flame position ($z/L$) is defined as the distance from the top of the combustor to the burner extension tip ($z$) normalized by the combustor length ($L$).  To determine the state of the system, we use the acoustic pressure fluctuation ($p^\prime$), which is measured with a probe microphone (GRAS 40SA, $\pm2.5\times10^{-5}$ Pa sensitivity) mounted 387~mm from the bottom of the combustor.  For each test run, we collect 8~s long time traces of $p^\prime$ at a sampling frequency of 32768~Hz, which is more than 150 times the frequency of the incipient limit cycle.


\section{System-identification framework} \label{sec:4method}

\subsection{System model} \label{sec:4syst}
To model the thermoacoustic system, we use a high-order Duffing--van der Pol (DVDP) oscillator perturbed by additive white Gaussian noise \citep{lee_2019}:
\begin{equation}\label{eq:4vdp}
    \ddot{x}-(\epsilon+\alpha_1x^2+\alpha_2x^4+\alpha_3x^6+\cdots)\dot{x}+x+{\beta}x^3 =\sqrt{2d}\eta(t),
\end{equation}
where $x$ represents the pressure fluctuation in the combustor ($p^{\prime}$ in units of Pa), $\eta(t)$ is a unit noise term, $d$ is the noise amplitude, $\epsilon$ is the linear growth (positive) or damping (negative) term, $\alpha_1, \alpha_2, \alpha_3, \cdots$ are the nonlinear terms, and $\beta$ is the anisochronicity factor, which determines the frequency shift as a function of amplitude.  The point at which $\epsilon$ crosses zero is the Hopf point, with the sign of $\alpha_1$ determining whether the Hopf bifurcation is supercritical (negative) or subcritical (positive).

The probabilistic solution to Eq.~\ref{eq:4vdp} can be found via the method of variation of parameters \citep{nayfeh1981introduction}.  On substitution of $x$ and $\dot{x}$ as $x(t)=a(t)\cos{(t+\phi(t))}$ and $\dot{x}(t)=-a(t)\sin{(t+\phi(t))}$, we obtain two first-order equations for the amplitude ($a$) and the phase ($\phi$):
\begin{subequations}\label{4derived}
\begin{align}
   \dot{a}&=\Big(\frac{\epsilon}{2}a+\frac{\alpha_1}{8}a^3+\frac{\alpha_2}{16}a^5+\frac{5\alpha_3}{128}a^7+\cdots\Big)+Q_{1}(a,\Phi)-\big(\sqrt{2d}\sin{\Phi}\big)\eta\label{4derived11},\\
    \dot{\phi}&=\frac{3\beta}{8}a^2+Q_{2}(a,\Phi)-\Big(\frac{\sqrt{2d}}{a}\cos{\Phi}\Big)\eta\label{4derived12},
\end{align}
\end{subequations}
where $\eta$ is a unit white Gaussian noise term, $\Phi(t)=t+\phi(t)$, and $Q_1(a,\Phi)$ and $Q_2(a,\Phi)$ are the sum of all the terms with first-order sine and cosine terms.  Assuming that $a$ and $\phi$ vary much more slowly than $x$ itself, we can justifiably neglect $Q_1(a,\Phi)$ and $Q_2(a,\Phi)$ via time averaging \citep{nayfeh1981introduction}.  Thus, for zero noise ($d=0$), Eq.~\ref{4derived11} can be rewritten as:
\begin{equation}\label{4st_ld}
    \frac{\mathrm{d}a}{\mathrm{d}t}=\frac{\epsilon}{2}a+\frac{\alpha_1}{8}a^3+\frac{\alpha_2}{16}a^5+\frac{5\alpha_3}{128}a^7+\cdots.
\end{equation}

Equation~\ref{4st_ld} takes the form of a Stuart--Landau equation, which is often used to model fluid-mechanical systems near a Hopf bifurcation \citep{chomaz2005,Sipp2007,Li13,NOIRAY2013152,boujo2017robust,zhu2017onset,lee_2019,zhu2019}. This equation will later be used to calculate the noise-free bifurcation diagram.  However, if the noise amplitude is finite ($d > 0$), stochastic averaging can be applied to Eq.~\ref{4derived}, yielding the following stochastic differential equation, expressed here in It{\^o} sense \citep{stratonovich1963,risken1984}:
\begin{equation}\label{4ito}
    \mathrm{d}a=\underbrace{\Big(\frac{d}{2a}+\frac{{\epsilon}}{2}a+\frac{\alpha_1}{8}a^3+\frac{\alpha_2}{16}a^5+\frac{5\alpha_3}{128}a^7+\cdots\Big)}_{\mathrm{m}(a,t)}\mathrm{d}t+\underbrace{\Big(\sqrt{d}\Big)}_{\mathrm{\sigma}(a,t)}\mathrm{d}W,
\end{equation}
where $\mathrm{d}W$ is a unit Wiener process, and $\mathrm{m}(a,t)$ and $\mathrm{\sigma}(a,t)$ appear in the drift and diffusion terms of $a$, respectively.  These two terms can be used to derive the classic Fokker--Planck equation:

\begin{equation}\label{4FPK}
    \pdv{}{t}P(a,t)=-\pdv{}{a}\Big[\boldsymbol{m}(a,t)P(a,t)\Big]+\pdv[2]{}{a}\Big[\frac{\boldsymbol{\sigma}^2(a,t)}{2}P(a,t)\Big],
\end{equation}
where $P(a,t)$ is the probability density function at time $t$.  The stationary probability density function, $P(a)$, is found by integrating Eq.~\ref{4FPK}:
\begin{equation}\label{4stationary}
    P(a)=Ca\exp\Big[\frac{a^2}{d}\Big(\frac{\epsilon}{2}+\frac{\alpha_1}{16}a^2+\frac{\alpha_2}{48}a^4+\frac{5\alpha_3}{512}a^6+\cdots\Big)\Big],
\end{equation}
where $C$ is a normalization constant, and $P(a)$ is independent of the anisochronicity factor $\beta$ \citep{Zakh2010}.

\subsection{System identification} \label{sec:4SI}

\begin{figure}[t]
\centering
\includegraphics[width=0.7\textwidth]{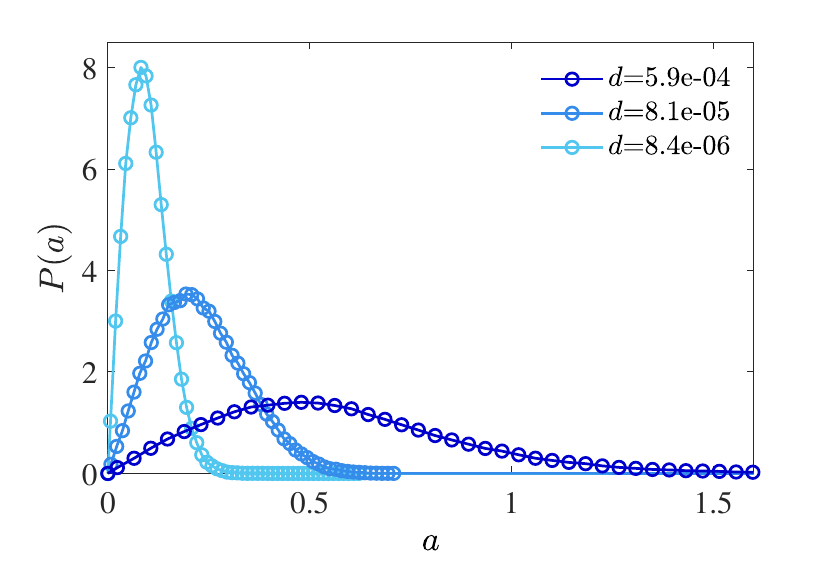}
\caption{Probability density function of the pressure fluctuation amplitude on the fixed-point branch ($z/L=0.256$) for three different noise amplitudes ($d$).}
\label{fig:4_2}
\end{figure}

We perform SI using probability density functions of the pressure fluctuation amplitude on the fixed-point branch, a subset of which is shown in figure~\ref{fig:4_2}.  By taking the logarithm of Eq.~\ref{4stationary}, we obtain:
\begin{equation}\label{4log}
\ln{P(a)}-\ln{a}=\ln{C}+\frac{\epsilon}{2d}a^{2}+\frac{\alpha_1}{16d}a^{4}+\frac{\alpha_2}{48d}a^{6}+\cdots.
\end{equation}
The left-hand side of Eq.~\ref{4log} is measured experimentally for a given value of $a$.  Thus, the ratio of $d$ to the unknown coefficients on the right-hand side ($\epsilon/2d$, $\alpha_1/16d$, $\cdots$) can be extracted via polynomial regression.  The number of terms on the right-hand side, which defines the order of nonlinearity, is determined by incrementally adding higher-order terms until the rank of the polynomial regression becomes deficient.

At each flame position ($z/L$), 16 different noise amplitudes ($d$) are applied, with three replications performed at each $d$.  The system coefficients at each $z/L$ are then found by averaging the results across all values of $d$.

\subsection{Actuator model} \label{sec:4actu}

\begin{figure}
\centering
\includegraphics[width=0.7\textwidth]{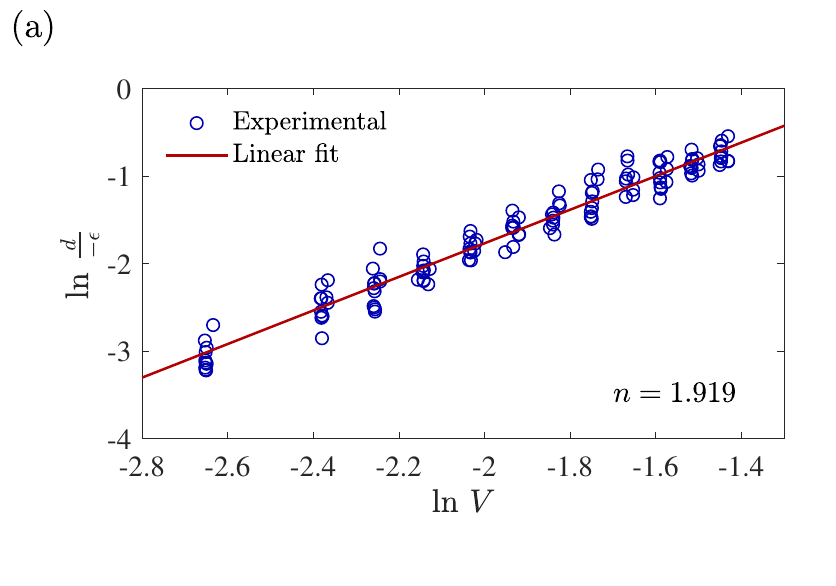}
\includegraphics[width=0.7\textwidth]{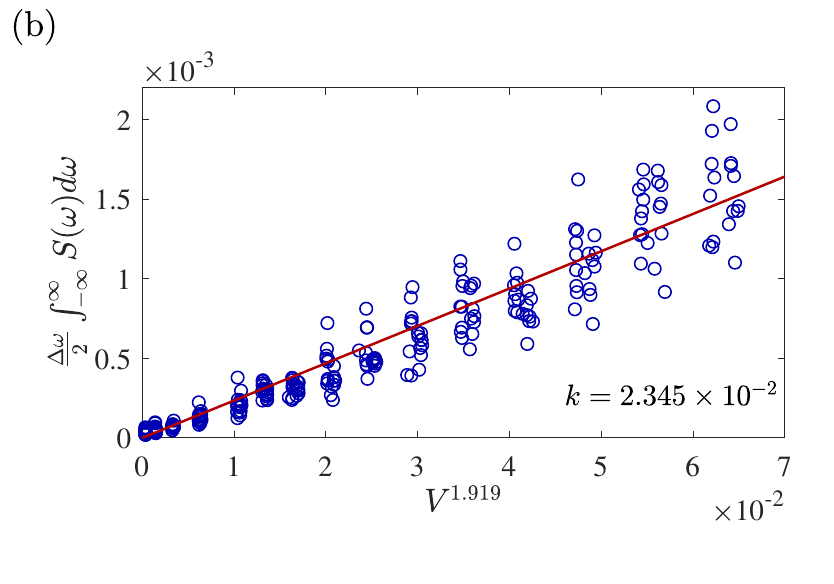}
\caption{Identification of the actuator model coefficients, where $n$ is the gradient of subfigure (a) and $k$ is the gradient of subfigure (b).  The vertical intercept of subfigure (b) is the background noise amplitude ($b$), which is negligible and thus consistent with our modeling assumptions.}
\label{fig:4_3}
\end{figure}

\begin{figure}
\centering
\includegraphics[width=0.7\textwidth]{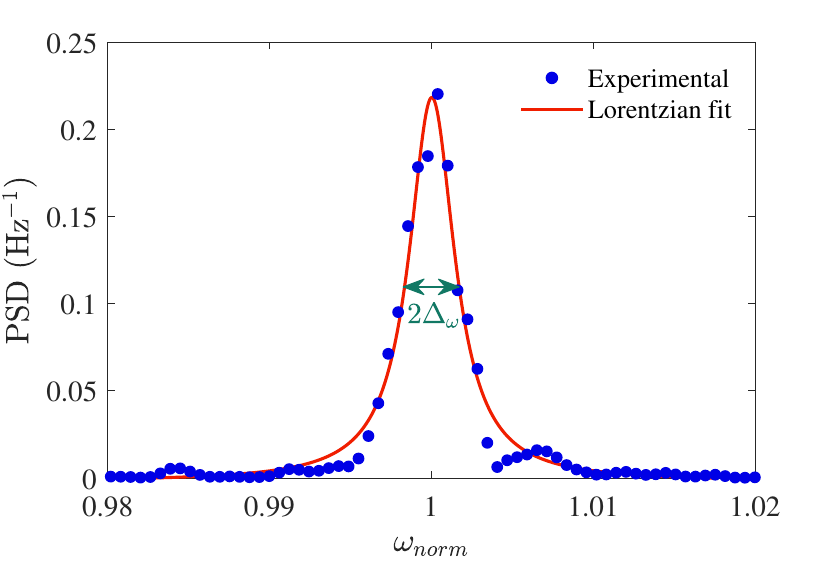}
\caption{Power spectral density showing a noise-induced peak and its Lorentzian fit on the fixed-point branch ($z/L=0.267$) for $d=4.0$e$-04$.  The horizontal axis is the normalized frequency ($\hat{\omega}$), with $2\Delta \omega$ denoting the width at half-maximum of the Lorentzian fit.}
\label{fig:4_4}
\end{figure}

A key component of input-output SI is the actuator model, which here is a function that transforms the loudspeaker voltage ($V$) into input noise ($d$).  We use a power-law relationship for the actuator model, $d=b+kV^{n}$, where $b$ is the background noise amplitude and $k$ and $n$ are constants.  Assuming $b \ll d$, we take the logarithm of this power-law equation, yielding:
\begin{equation}\label{4noise2}
    \ln{\Big(\frac{d}{c}\Big)} \approx n \ln{V} + \ln{\Big(\frac{k}{c}\Big)},
\end{equation}
where $c$ is an arbitrary constant.  From \S\ref{sec:4SI}, a ratio between $d$ and one of the system parameters ($\epsilon$, $\alpha_1$, $\alpha_2$, $\cdots$) can be found before applying the actuator model.  Thus, by replacing $c$ with one of those parameters, $n$ can be found from linear regression (figure~\ref{fig:4_3}a).  In this study, $\lvert\epsilon\rvert$ on the fixed-point branch is chosen because its values, sampled over multiple experimental runs, are the most consistent among all the DVDP coefficients.  To find the remaining constants ($k$ and $b$), we use information in the spectral domain, as per \citet{ushakov2005coherence}:
\begin{equation}\label{4noise3}
    d=b+kV^{n}=\frac{\Delta \omega}{2}\int_{-\infty}^{\infty}S_u(\omega)d\omega,
\end{equation}
where $S_u$ is the spectrum and $\Delta \omega$ is the half-width at half-maximum of a Lorentzian fit to the noise-induced peak in the power spectral density (PSD), as shown in figure~\ref{fig:4_4}.  Finally, $k$ and $b$ are determined by linear regression, as per figure~\ref{fig:4_3}(b).  In this way, we determine the relationship between $d$ and $V$ to be $d=(2.345\times10^{-2})V^{1.919}$.  In our experiments, the background noise amplitude ($b$) is negligible, as evidenced by the zero vertical intercept of the data shown in figure~\ref{fig:4_3}(b).

\section{Results and discussion} \label{sec:4result}
\begin{figure}
\centering
\includegraphics[width=0.7\textwidth]{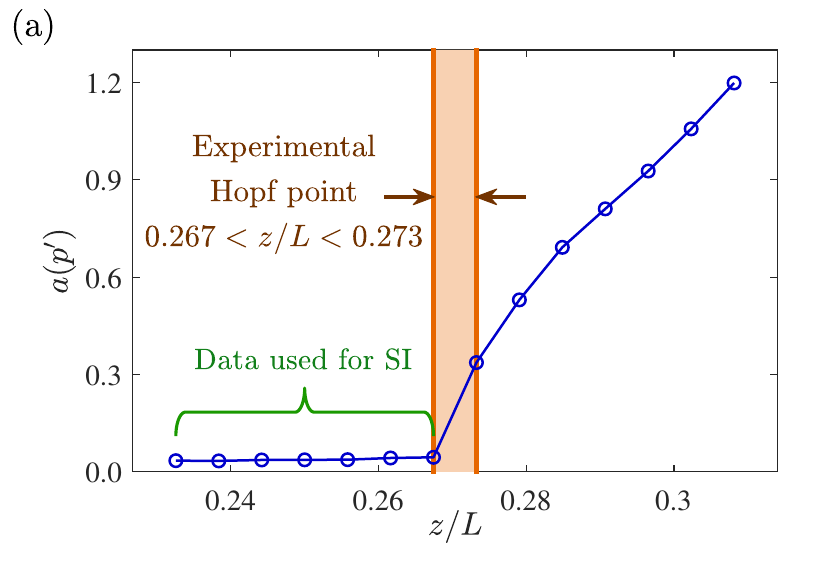}
\includegraphics[width=0.7\textwidth]{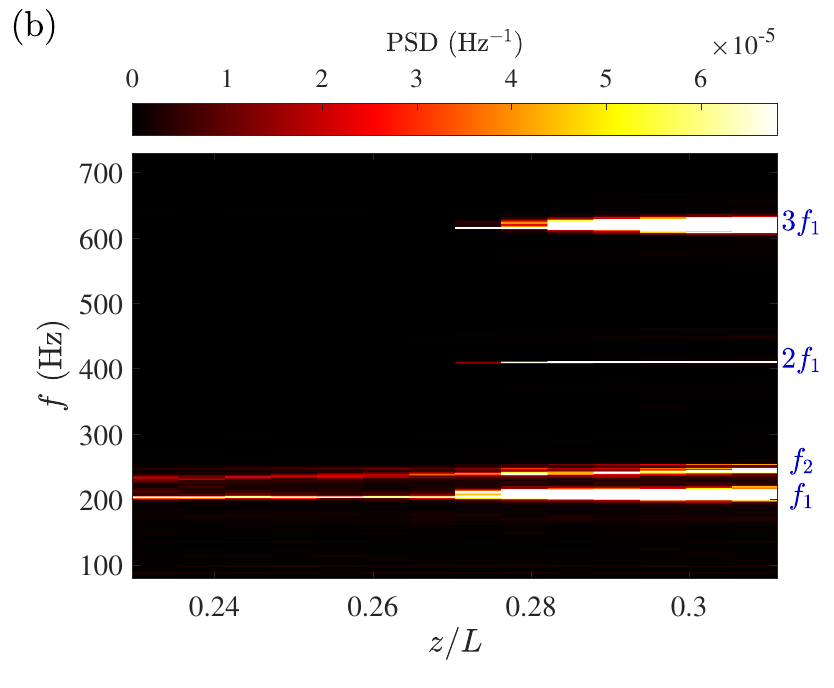}
\caption{(a) Experimental bifurcation diagram of the system, where the horizontal axis is the normalized flame position ($z/L$) measured from the top of the combustor.  Also shown is (b) the PSD of the pressure fluctuations as a function of $z/L$.}
\label{fig:4_5}
\end{figure}

\begin{figure}
\centering
\includegraphics[width=0.7\textwidth]{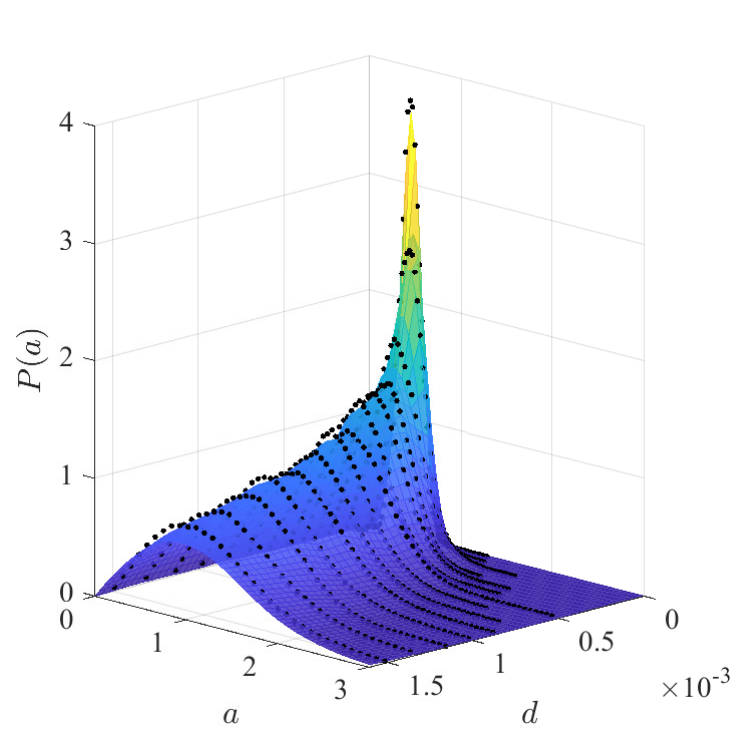}
\caption{Experimental probability density function (black dots) and its surface interpolation on the fixed-point branch ($z/L=0.267$), just before the Hopf point. For all the noise amplitudes tested, $P(a)$ is unimodal, confirming the supercritical nature of the Hopf bifurcation.}
\label{fig:4_6}
\end{figure}

Figure~\ref{fig:4_5}(a) shows a bifurcation diagram of the system.  When the flame reaches a position of $0.267 < z/L < 0.273$, the system transitions from a fixed point to a limit cycle via a supercritical Hopf bifurcation.  The supercritical nature of this bifurcation can be confirmed by examining the probability density function $P(a)$.  If $P(a)$ shows two local maxima with respect to $a$ at intermediate noise amplitudes $d$ (a feature called bimodality \citep{Zakh2010}), then the system is undergoing a subcritical Hopf bifurcation.  However, if $P(a)$ is unimodal, exhibiting only one peak at every value of $d$, then the system is undergoing a supercritical Hopf bifurcation.  Figure~\ref{fig:4_6} shows $P(a)$ and its surface interpolation on the fixed-point branch ($z/L=0.267$), just before the Hopf point.  There is only one peak for every value of $d$, confirming that the Hopf bifurcation is indeed supercritical.

Figure~\ref{fig:4_5}(b) shows the PSD as a function of the flame position ($z/L$).  Before the Hopf point ($z/L < 0.267$), the PSD contains mostly broadband noise, with slight increases around 200--250~Hz due to incipient modes.  Just after the Hopf point ($0.273 < z/L < 0.285$), the PSD is dominated by sharp peaks at $f_1=208$~Hz and its higher harmonics, indicating a limit cycle.  Accompanying this primary mode is a weaker secondary mode at $f_2=243$~Hz.  This secondary mode, however, is more than 100 times weaker than the primary mode, so the system dynamics is still dominated by the limit cycle at $f_1$.  Further from the Hopf point ($z/L > 0.285$), the secondary mode ($f_2$) remains relatively unchanged, but the primary mode ($f_1$) and its higher harmonics ($2f_1$ and $3f_1$) continue to grow.  This is particularly true for the third harmonic ($3f_1$), which grows to nearly the same amplitude as the fundamental itself ($f_1$).  As we will see later, the growth of these higher harmonics has a significant influence on the limit-cycle amplitude.

\begin{figure}
\centering
\includegraphics[width=0.6\textwidth]{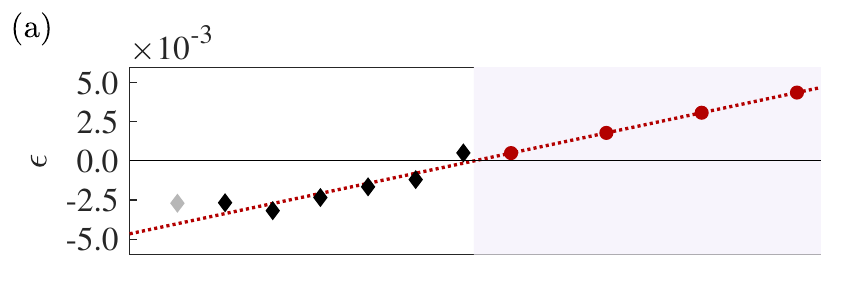}
\includegraphics[width=0.6\textwidth]{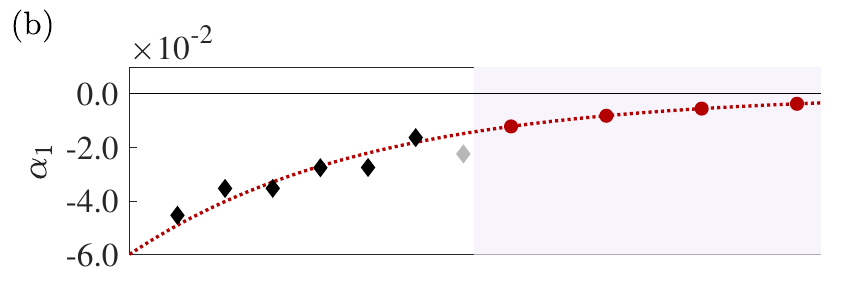}
\includegraphics[width=0.6\textwidth]{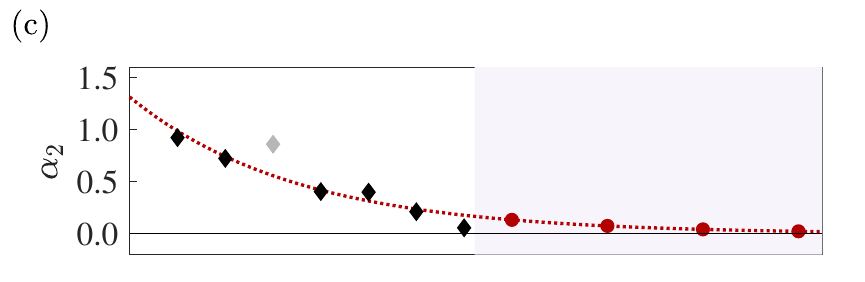}
\includegraphics[width=0.6\textwidth]{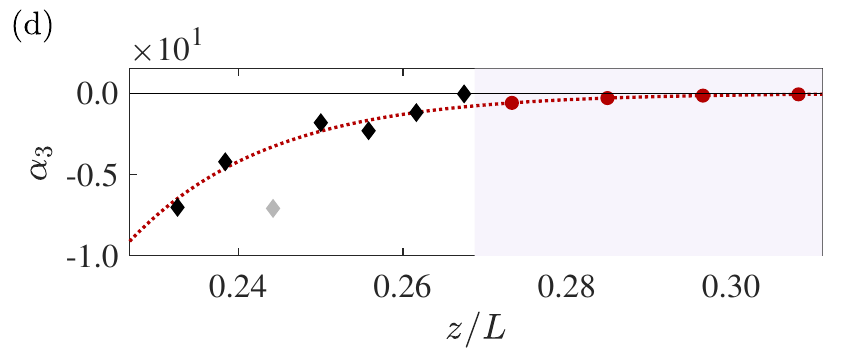}
\caption{Determining the DVDP coefficients via SI.  Extrapolation is performed using data from only the fixed-point branch (black diamonds), after the removal of outliers (gray diamonds), which are defined here as being outside three standard deviations.  The extrapolation is performed with a linear model for the linear coefficient ($\epsilon$) and with a power-law model for the nonlinear coefficients ($\alpha_1$, $\alpha_2$, $\alpha_3$).  The predicted data (red circles) are on the limit-cycle branch, whose features are examined in figure~\ref{fig:4_10}.}
\label{fig:4_7}
\end{figure}

\begin{figure}
\centering
\includegraphics[width=0.7\textwidth]{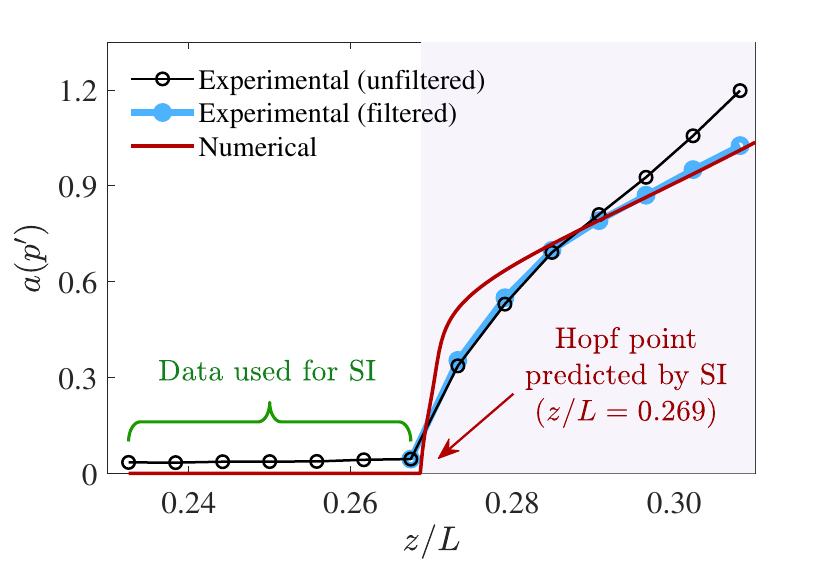}
\caption{Comparison of bifurcation diagrams between the experimental system and the numerical model found via SI.  The Hopf point predicted by the model is at $z/L=0.269$, which is within the experimentally observed range: $0.267 < z/L < 0.273$.  The blue line represents the experimental data bandpass filtered around the limit-cycle frequency ($f_1 \pm 10$~Hz).}
\label{fig:4_8}
\end{figure}

The DVDP coefficients found via SI are shown in figure~\ref{fig:4_7}.  The highest nonlinear term of Eq.~\ref{eq:4vdp} is $(\alpha_3x^6)\dot{x}$, and the signs of the nonlinear coefficients ($\alpha_1$, $\alpha_2$, $\alpha_3$) remain unchanged across the entire range of $z/L$.

To predict the Hopf point and the resultant limit cycle using data from only the fixed-point branch, we build a mathematical relationship between $z/L$ and the system coefficients.  In a Hopf bifurcation, the linear coefficient $\epsilon$ is known to be linearly proportional to the control parameter \citep{Provansal1987}.  We therefore linearly extrapolate $\epsilon$ from within the fixed-point branch (i.e. from the smallest $z/L$ to the largest $z/L$ with negative $\epsilon$) to the limit-cycle branch.  For the nonlinear coefficients ($\alpha_1$, $\alpha_2$, $\alpha_3$), we repeat this process but with a power law, $\alpha_{n} \propto (z/L-m_1)^{m_2}$, as per our previous study \citep{lee_2019}.  Figure~\ref{fig:4_7} shows that the absolute values of the nonlinear coefficients ($|\alpha_1|$, $|\alpha_2|$, $|\alpha_3|$) decrease with increasing $z/L$.  In particular, the higher the order of the nonlinear coefficients, the faster they decay, confirming that the system is indeed weakly nonlinear near the Hopf point.

Next we reconstruct the bifurcation diagram by solving the Stuart--Landau equation (Eq.~\ref{4st_ld}) with the extrapolated coefficients, as shown in figure~\ref{fig:4_8}.  The numerical model found via SI predicts that a supercritical Hopf bifurcation occurs at $z/L=0.269$, which agrees well with the experimentally observed Hopf point at $0.267 < z/L < 0.273$.  After the Hopf point, however, the numerical predictions agree less well with the experimental data.  As alluded to earlier, we speculate that this is due to the growth of the higher harmonics ($2f_1$ and $3f_1$) with increasing $z/L$.  To test this, we bandpass filter the experimental limit-cycle data using different filter widths.  We find improved agreement only when the higher harmonics ($2f_1$ and $3f_1$) are removed (figure~\ref{fig:4_8}); no significant difference is found when only the secondary mode ($f_2$) is removed (not shown here for brevity).  The improved agreement occurs far from the Hopf point ($z/L > 0.285$), which is consistent with where the harmonics are strongest.  The agreement close to the Hopf point ($z/L < 0.285$), however, remains relatively unaffected by the filtering, with the numerical model over-predicting the experimental data (both unfiltered and filtered; figure~\ref{fig:4_8}).  This over-prediction could be due to nonlinear interactions between the harmonics, which the Stuart--Landau equation (Eq.~\ref{4st_ld}) cannot capture because it was derived on the basis of weak nonlinearity (\S\ref{sec:4syst}).  In the experiments, there is substantial energy transfer from the fundamental mode ($f_1$) to its higher harmonics ($2f_1$ and $3f_1$).  The absence of such energy transfer in the model may explain why it over-predicts the experimental data in this regime ($0.273 < z/L < 0.285$).  Overall these findings show that although the presence of strong harmonics affects the limit-cycle predictions, it does not affect the Hopf-point predictions.

\begin{figure}
\centering
\includegraphics[width=0.7\textwidth]{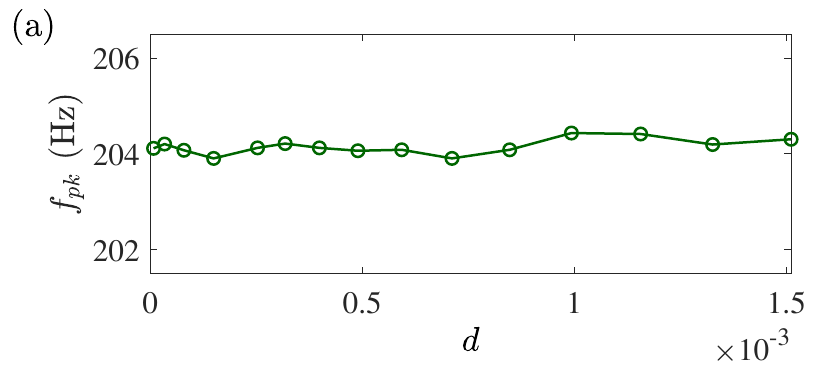}
\includegraphics[width=0.7\textwidth]{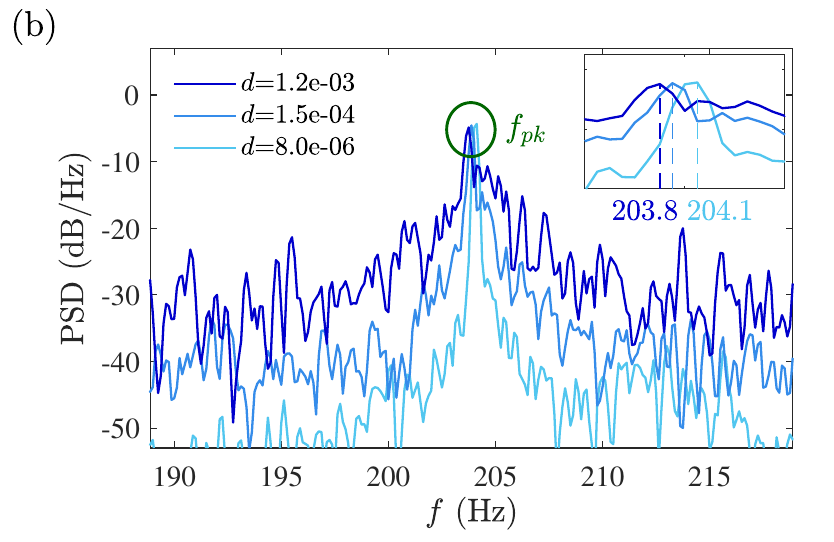}
\caption{Anisochronicity of the experimental system: (a) peak frequency $f_{pk}$ as a function of the noise amplitude $d$ and (b) power spectral density at different values of $d$ on the fixed-point branch ($z/L=0.267$), just before the Hopf point.  The frequency shift is observed to be less than $0.3\%$.}
\label{fig:4_9}
\end{figure}

To determine $\beta$ in Eq.~\ref{eq:4vdp}, we analyze the anisochronicity of the experimental system in the frequency domain.  Figure~\ref{fig:4_9} shows that the dominant frequency ($f_{pk}$) shifts by less than $0.3\%$ as $d$ increases, supporting our original assumption of a negligible frequency shift.

\begin{figure}[t!]
\centering
\includegraphics[width=0.48\textwidth]{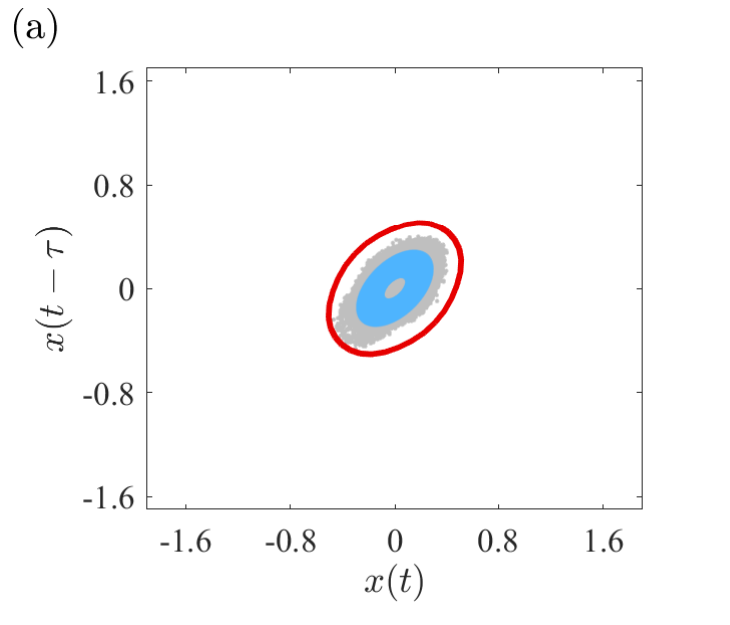}
\includegraphics[width=0.48\textwidth]{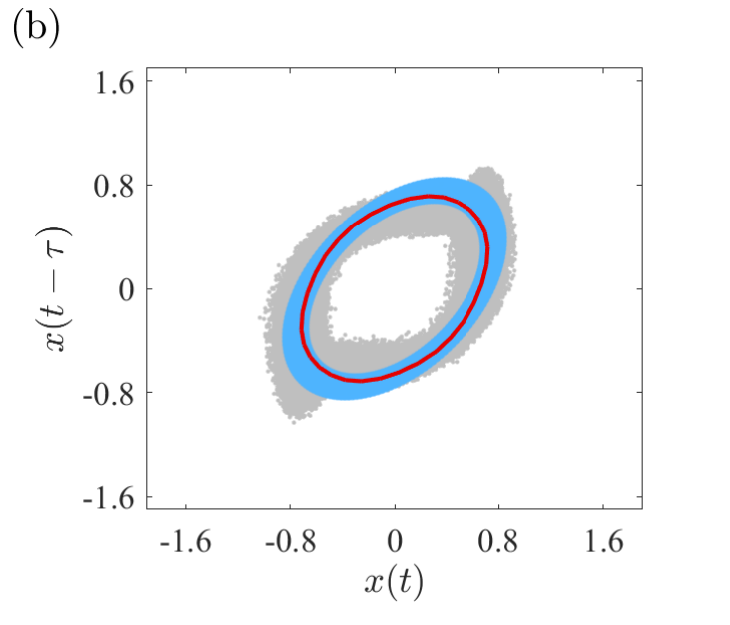}
\includegraphics[width=0.48\textwidth]{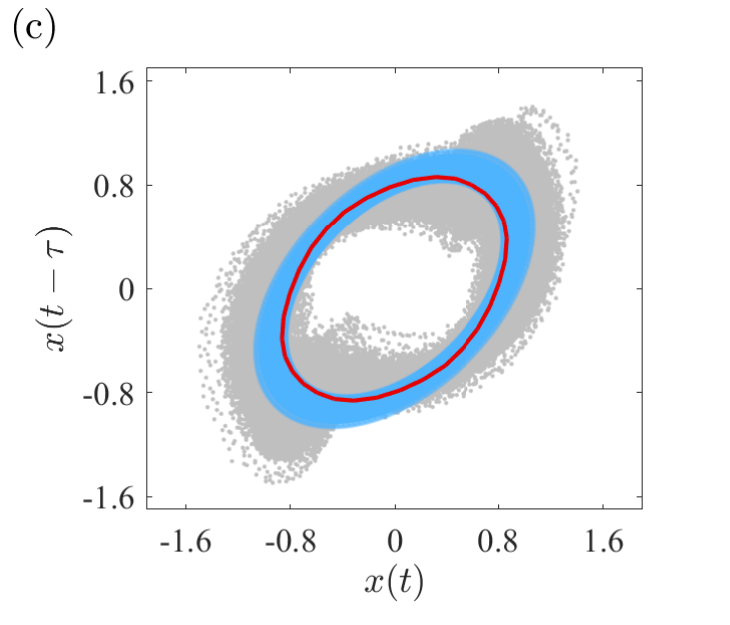}
\includegraphics[width=0.48\textwidth]{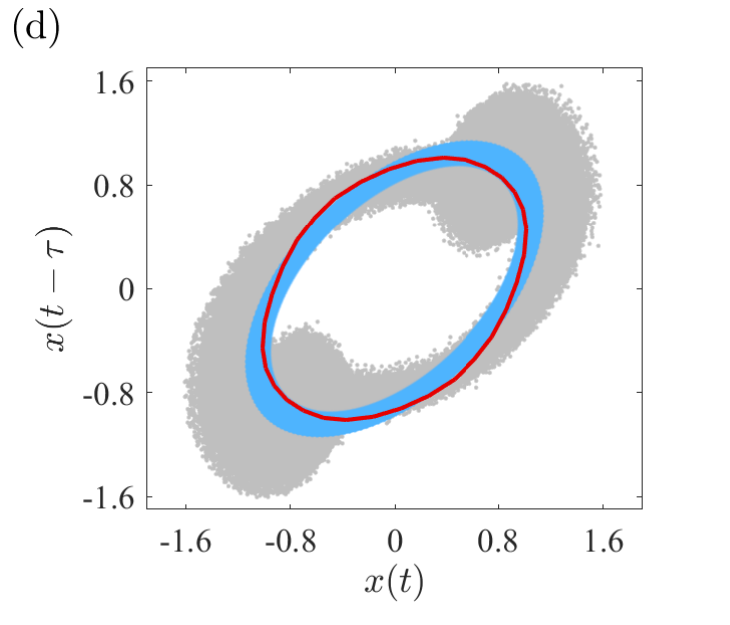}
\caption{Comparison of phase portraits between the experimental system and the numerical model found via SI at four different positions on the limit-cycle branch: $z/L=$ (a) $0.273$, (b) $0.285$, (c) $0.297$, and (d) $0.308$.  The experimental data are shown both in unfiltered form (gray) and in bandpass-filtered form (blue: $f_1 \pm 10$~Hz), while the numerical data are shown in unfiltered form only (red).}
\label{fig:4_10}
\end{figure}

Finally, we examine the limit-cycle features of the numerical model using the time-delay embedding technique of \citet{takens1981detecting}.  This technique, which has seen widespread use in thermoacoustics \citep{balusamy2015nonlinear,lee2016nonlinear,guan2018nonlinear,kashinath2018forced}, enables an attractor to be reconstructed in phase space using just a single scalar time series shifted by an appropriate time delay ($\tau$).  A typical choice of $\tau$ is the first minimum of the average mutual information function \citep{fraser1986}.  Figure~\ref{fig:4_10} compares the phase portraits of the experimental system (both unfiltered and bandpass-filtered signals) with those of the numerical model found via SI.  Owing to the presence of higher harmonics ($2f_1$ and $3f_1$), the unfiltered experimental data are seen to develop `circular swelling' as $z/L$ increases (figure~\ref{fig:4_10}).  Our SI framework, however, cannot predict this feature because it assumes weak nonlinearity and hence weak harmonics.  Nevertheless, if the primary mode is isolated via bandpass filtering around its fundamental frequency ($f_1 \pm 10$~Hz), the agreement between the experimental and numerical data improves far from the Hopf point (figure~\ref{fig:4_10}b--d: $z/L \geq 0.285$), although it remains relatively unchanged close to the Hopf point (figure~\ref{fig:4_10}a: $z/L = 0.273$).  These trends are consistent with our discussion of figure~\ref{fig:4_8}.

\section{Conclusions} \label{sec:4conc}

We have presented a framework for performing input-output SI near a Hopf bifurcation using data from only the fixed-point branch, prior to the Hopf point itself.  The framework models the system with a DVDP-type equation perturbed by additive noise, and identifies the system parameters via the corresponding Fokker--Planck equation.  We have demonstrated the framework on a prototypical thermoacoustic oscillator (a flame-driven Rijke tube) undergoing a supercritical Hopf bifurcation.  We find that the properties of the Hopf bifurcation -- such as its location and its super/subcritical nature -- can be accurately predicted even before the onset of limit-cycle oscillations.  We believe that this marks the first time that input-output SI has been successfully performed on a reacting flow using only pre-bifurcation data, paving the way for the development of new early warning indicators of thermoacoustic instability in combustion devices.

Compared with existing early warning indicators used in thermoacoustics, the SI framework presented here has two advantages: (i) it can predict the properties of a Hopf bifurcation without the need to set ad-hoc instability thresholds, and (ii) it can predict post-bifurcation behavior such as limit-cycle amplitudes.  Although demonstrated here on a thermoacoustic system, this SI framework should be applicable to other nonlinear dynamical systems as well, provided that they obey the normal-form equation for a Hopf bifurcation (i.e. the Stuart--Landau equation).  Examples of such systems include open shear flows \citep{Provansal1987}, chemical reactions \citep{kuramoto2003chemical}, and optical lasers \citep{ludge2012nonlinear} -- among many other systems in nature and engineering.

This SI framework has two notable limitations.  First, it assumes that the background noise amplitude is low.  This assumption, however, may not be valid in turbulent systems, which could complicate the development of an accurate actuator model.  Output-only SI methods can offer a way out of this, but they typically require large datasets, which could be difficult to acquire in practical systems \citep{MEVEL2006531}.  Nevertheless, this problem can be circumvented with the use of adjoint equations, as \citet{boujo2017robust} have shown.  Second, our SI framework makes use of time-series data collected at a single location.  This works well for the thermoacoustic system studied here because its temporal dynamics is globally synchronized at every location in the flow domain.  Such localized sampling keeps the matrix sizes manageable without compromising the accuracy of the numerical predictions.  Other systems, however, may show elaborate spatial variations in their dynamics, requiring data to be sampled at multiple locations.  In such a scenario, it may be necessary to use sparsity-promoting techniques and machine learning to process the larger data matrices \citep{Brunton2016}.

This work was supported by the Research Grants Council of Hong Kong (Project Nos. 16235716, 26202815 and 16210418). 

\chapter{System identification and early warning detection of thermoacoustic oscillations in a turbulent combustor using its noise-induced dynamics} \label{chap:gas}

Under review by the \textit{Proceedings of the Combustion Institute}.


\section*{Abstract}
Despite significant research, self-excited thermoacoustic oscillations continue to hinder the development of lean-premixed gas turbines, making the early detection of such oscillations a key priority. We perform output-only system identification of a turbulent lean-premixed combustor near a Hopf bifurcation using the noise-induced dynamics generated by inherent turbulence in the fixed-point regime, prior to the Hopf point itself. We model the pressure fluctuations in the combustor with a van der Pol-type equation and its corresponding Stuart–Landau equation. We extract the drift and diffusion terms of the equivalent Fokker–Planck equation via the transitional probability density function of the pressure amplitude. We then optimize the extracted terms with the adjoint Fokker–Planck equation. Through comparisons with experimental data, we show that this approach can enable prediction of (i) the location of the Hopf point and (ii) the limit-cycle amplitude after the Hopf point. This study shows that output-only system identification can be performed on a turbulent combustor using only pre-bifurcation data, opening up new pathways to the development of early warning indicators of thermoacoustic instability in practical combustion systems.

\section{Introduction}
Despite decades of research, thermoacoustic instability remains a key challenge in the development and operation of lean-premixed gas turbines \citep{Poinsot2017}.  This problem arises from the coupling between the heat-release-rate (HRR) oscillations of an unsteady flame and the pressure oscillations of the combustor \citep{Candel2002}. If these two types of oscillations are sufficiently in phase, thermal energy can be transferred from the flame to the pressure field via the \citet{Rayleigh1878} mechanism, leading to self-excited flow oscillations at the natural acoustic modes of the system.  At high amplitudes, such thermoacoustic oscillations can impair flame stability and increase thermal stresses \citep{Lieuwen2005}.  It is thus important to be able to predict the onset of such thermoacoustic oscillations so that preventative measures can be taken \citep{juniper2018}.

\subsection{Thermoacoustic oscillations via a Hopf bifurcation}
Thermoacoustic oscillations often arise via a Hopf bifurcation \citep{lieuwen2002,Lieuwen2005,juniper2018}.  Here, when a control parameter reaches the Hopf point, a fixed-point solution becomes unstable as a complex conjugate pair of eigenvalues crosses the imaginary axis, inducing a transition from a fixed point to a limit cycle \citep{strogatz2000}. If the limit cycle appears only after the Hopf point, the Hopf bifurcation is supercritical.  If the limit cycle appears in a hysteretic bistable regime, between the Hopf and saddle-node points, the Hopf bifurcation is subcritical.  The super/subcritical nature of a Hopf bifurcation is determined by the degree of nonlinearity present in the system \citep{strogatz2000}.  To avoid the detrimental effects of thermoacoustic instability, it is important to be able to predict not just the location of the Hopf point but also the degree of nonlinearity. In this paper, we present a technique for doing this that exploits the noise-induced dynamics generated by inherent turbulence in the fixed-point regime, prior to the Hopf point itself.

\subsection{Precursors of thermoacoustic instability} \label{sec:NIcomb}
The development of early warning indicators of thermoacoustic instability has long been an active area of research. Early on, extrinsic perturbations were applied to rockets \citep{harrje1972} and gas turbines \citep{johnson2000} to quantify their stability margins.  \citet{Lieuwen2005a} then showed that the intrinsic dynamics of combustion noise contains sufficient information to form an instability precursor, namely the damping rate extracted from the autocorrelation function.  \citet{yi2008} later extended this approach to the frequency domain, establishing precursors for multi-mode oscillations.  Additional precursors of thermoacoustic instability have since been proposed by Sujith's group (e.g. the Hurst exponent \citep{nair2014b} and recurrence quantification metrics \citep{nair2014a}) and Gotoda's group (e.g. the permutation entropy \citep{gotoda2012} and ordinal partition transition networks \citep{kobayashi2019}).  Most precursors, however, require \emph{ad-hoc} instability thresholds to be set, typically using experimental data collected in both the fixed-point and limit-cycle regimes. Moreover, although such precursors can predict the location of a Hopf point, they usually cannot predict the limit-cycle dynamics after it. In this study, we build on these seminal contributions by exploiting a classic feature known as coherence resonance, which is universal to all nonlinear dynamical systems near a Hopf bifurcation \citep{ushakov2005coherence} and which causes the noise-induced response to become more coherent as the Hopf point is approached \citep{Kabiraj2015,li2019coherence,zhu2019}.  Here we use this feature to predict the properties of a Hopf bifurcation and the resultant limit cycle, without the need for \emph{ad-hoc} instability thresholds.

\subsection{System identification via noise-induced dynamics}
System identification (SI) involves using statistical tools to build mathematical models of dynamical systems based on observed data \citep{soderstrom1988}.  In thermoacoustics, SI has been used to identify the flame transfer function \citep{merk2019}, the flame describing function \citep{krediet2012}, and the parameters of a low-order oscillator model representing the pressure fluctuations in a turbulent combustor \citep{seywert2001,NOIRAY2013152,boujo2017robust}.  Here we focus on the last case, in which the pressure fluctuations are modeled with one or more harmonic oscillators. In pioneering work, \citet{paparizos1989} and \citet{culick1992} used stochastic differential equations containing noise terms to model multiple acoustic modes in a combustor.  Through a similar approach, \citet{seywert2001} was able to identify the frequencies and linear growth rates by curve fitting the pressure spectrum.  \citet{NOIRAY2013152} later extended this approach to show that it can predict not only the linear parameters but also the nonlinear parameters of a turbulent combustor. These researchers modeled the pressure oscillations as a single acoustic mode, represented by a van der Pol (VDP) equation perturbed by additive noise.  Thus, they were able to extract the linear and nonlinear coefficients by analyzing the probability density function (PDF) given by the Fokker--Planck equation. More recently, \citet{boujo2017robust} improved the accuracy of this output-only SI technique by incorporating adjoint-based optimization.  However, most existing SI techniques require at least some data from the limit-cycle regime to be able to predict the properties of a Hopf bifurcation and the resultant limit cycle.


Recognizing this, we \citep{lee_2019,lee_tbr} recently adapted the SI framework proposed by \citet{NOIRAY2013152} such that it works with only pre-bifurcation data. This involved determining the VDP coefficients by measuring the stochastic response of a system (e.g. a low-density jet \citealp{lee_2019} or a laminar flame-driven Rijke tube \citealp{lee_tbr}) to extrinsic noise of a known amplitude. By analyzing the relationship between the identified coefficients and the bifurcation parameters, we were able to predict, using only pre-bifurcation data, the locations and types of the bifurcation points as well as the limit-cycle dynamics \citep{lee_2019,lee_tbr}. This input-output SI framework, however, works only when the background noise is weak and when extrinsic perturbations can be readily applied, making the framework unsuitable for practical combustors perturbed by turbulence. An open research question is whether and how the bifurcation point and the post-bifurcation dynamics of a turbulent combustor can be predicted from only pre-bifurcation data.

\subsection{Contributions of the present study}
We perform output-only SI on a turbulent lean-premixed combustor operating near a supercritical Hopf bifurcation, using the noise-induced dynamics generated by inherent turbulence in the fixed-point regime, prior to the Hopf point itself.  Our aim is not only to demonstrate a viable way of predicting the properties of a Hopf bifurcation, but also to open up new pathways to the development of early warning indicators of thermoacoustic instability in turbulent combustors such as gas turbines.  Below we present the experimental setup and data (\S\ref{sec:5_2}), describe the SI framework (\S\ref{sec:5_3}) and apply it to a turbulent combustor (\S\ref{sec:5_4}), before concluding with the key implications and limitations of this study (\S\ref{sec:5_5}).

\section{Experimental setup and data} \label{sec:5_2}
\begin{figure}
    \centering
    \includegraphics[clip,width=\textwidth]{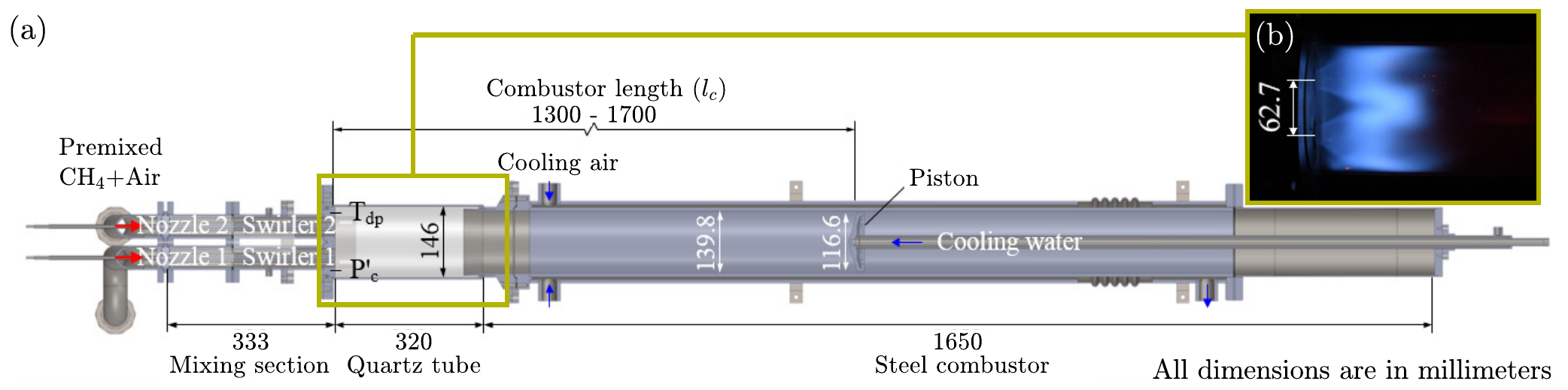}\\
    \includegraphics[clip,width=\textwidth]{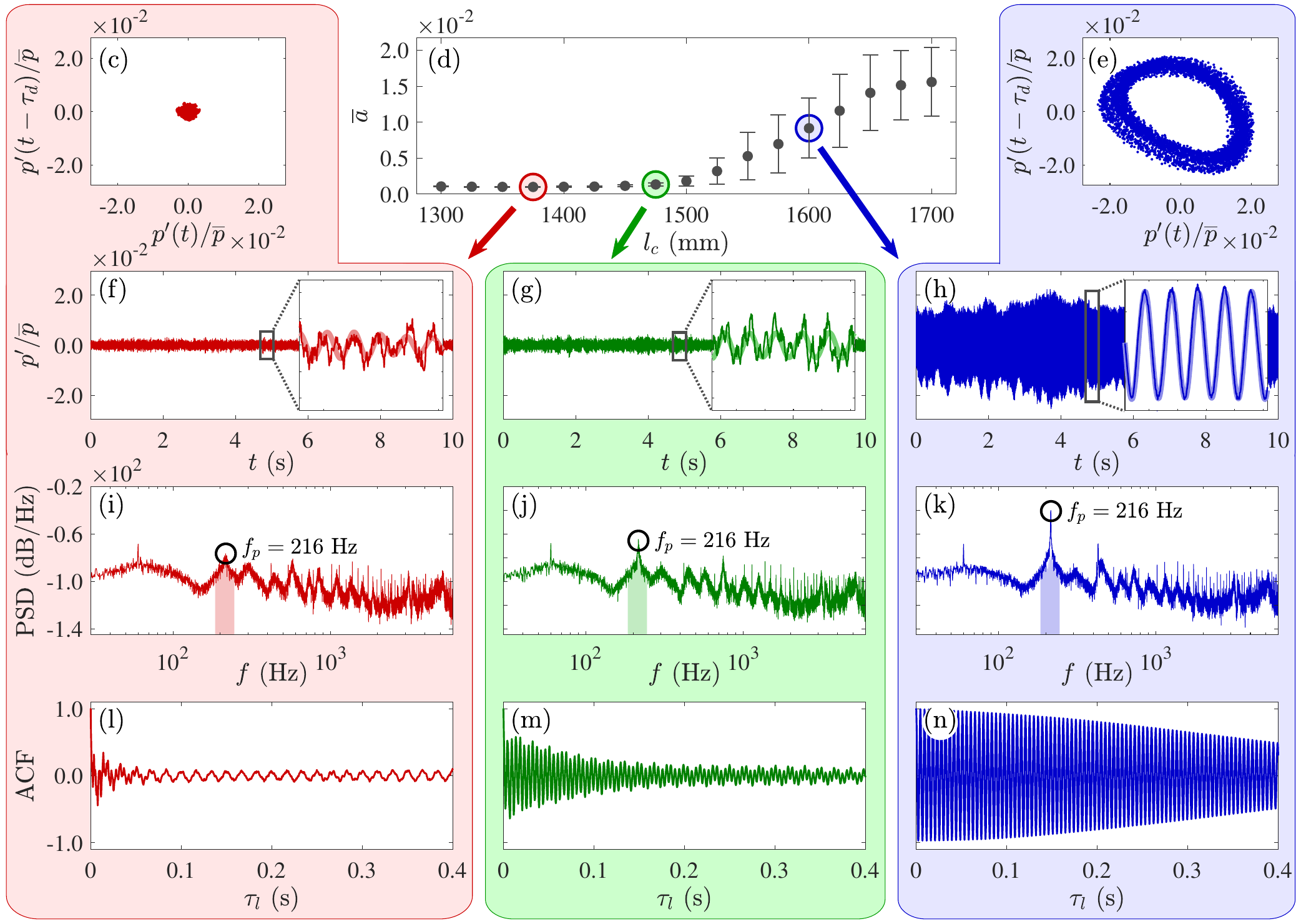}\\
    \caption{Experimental setup and data: (a) schematic of the variable-length combustor, (b) broadband chemiluminescence snapshot of the premixed flame, and (d) bifurcation diagram showing the normalized time-averaged amplitude of $p^{\prime}$ ($\overline{a}$) as a function of the combustor length ($l_c$), with error bars denoting the standard deviation. Also shown are the (c,e) phase portrait, (f--h) time series, (i--k) PSD, and (l--n) normalized ACF for three values of $l_c$: (left) $1375$~mm, (center) $1475$~mm, and (right) $1600$~mm.  In subfigures (c,e), the average mutual information function is used to determine the embedding time delay $\tau_d$ for phase-space reconstruction. In subfigures (f--h), the bandpass filtered time traces are shown as thick lines in the insets. In subfigures (i--k), the shaded regions indicate the width of the bandpass filter.}
\label{fig:fig5_1}
\end{figure}

Figure~\ref{fig:fig5_1}(a) shows a cross-sectional view of the variable-length combustor used in this study.  The combustor is similar to that used by \citet{LEE20195137}, so only a brief overview is given here.  The combustor is equipped with two identical swirling nozzles from which fully-premixed reactants (CH$_{4}$--air) are injected via a mixing section. The mixing section has an annular geometry defined by a mixing tube (inner diameter: 38.1~mm) and a centerbody (outer diameter: 19.1~mm), both 333~mm long. In each of the two nozzles, a six-vane axial swirler (swirl number: 0.65) is mounted 76.2~mm upstream of the dump plane. The centerline-to-centerline distance between the two nozzles is 62.7~mm. The combustor consists of two sections: an optically accessible quartz tube (length: 320~mm; diameter: 146~mm) and a steel tube.  The combustor length ($l_c$) is varied from 1300 to 1700~mm (in 25~mm steps) via an internal piston.

The air and CH$_{4}$ flow rates are metered with thermal mass flow controllers (air: Sierra FlatTrak 780S, 200~SCFM, $\pm$0.5\%; CH$_{4}$: Teledyne HFM-D-301, 500~SLM, $\pm$0.6\%).  The premixed CH$_{4}$--air mixture has an equivalence ratio of 0.65, an inlet temperature of 200$^{\circ}$C, and a bulk velocity of 40~m/s, for a Reynolds number of 44,000.  The pressure fluctuations in the combustor ($p^{\prime}$) is measured at the dump plane with a piezoelectric transducer (PCB 112A22, 14.5~mV/kPa) sampled at 12,000~Hz for 10~s. For reliable statistics, each test condition is replicated ten times.  Figure~\ref{fig:fig5_1}(b) shows a broadband chemiluminescence snapshot of the flame.

Figure~\ref{fig:fig5_1}(d) shows the bifurcation diagram, where the normalized time-averaged amplitude of $p^{\prime}$ ($a \equiv |p^{\prime}|/\overline{p}$, where $|p^{\prime}|$ is computed with the Hilbert transform) is plotted as a function of $l_c$. As $l_c$ increases, $\overline{a}$ increases smoothly, indicating that the system is transitioning from a fixed point to a limit cycle via a supercritical Hopf bifurcation. The states before and after the Hopf point, which sits at $l_c \approx 1500$--$1550$~mm, are examined in figure~\ref{fig:fig5_1}(c,e--n).  Well before the Hopf point ($l_c=1375$~mm), the time series of $p^{\prime}/\overline{p}$ shows low-amplitude aperiodic fluctuations, with no little of coherent motion (figure~\ref{fig:fig5_1}f), while the phase portrait shows a single blob (figure~\ref{fig:fig5_1}c). These features are characteristic of a stable fixed-point attractor in phase space \citep{strogatz2000}. However, an inspection of the power spectral density (PSD) reveals initial signs of an impending instability at $f_p \approx 216$~Hz (figure~\ref{fig:fig5_1}i).  Closer to the Hopf point ($l_c=1475$~mm), the time series remains aperiodic and low in amplitude (figure~\ref{fig:fig5_1}g), but the PSD shows a strengthening of the incipient self-excited mode at $f_p$ (figure~\ref{fig:fig5_1}j). After the Hopf point ($l_c=1600$~mm), the system evolves on a stable limit-cycle attractor, as evidenced by a closed trajectory in the phase portrait (figure~\ref{fig:fig5_1}e), high-amplitude coherent oscillations in the time series (figure~\ref{fig:fig5_1}h), and a sharp discrete peak at $f_p$ in the PSD (figure~\ref{fig:fig5_1}k).  This transition from a fixed point to a limit cycle is also corroborated by the normalized auto-correlation function (ACF), whose amplitude envelope is seen to decay increasingly slowly as $l_c$ increases, indicating that the system becomes more coherent as it approaches---and then crosses---the Hopf point (figure~\ref{fig:fig5_1}l--n).  For SI (\S\ref{sec:5_3}), we isolate the dominant thermoacoustic mode by bandpass filtering the pressure signal around $f_p$ with a bandwidth of $\Delta f = 0.3 f_p$ (see the insets in figure~\ref{fig:fig5_1}f--h, and the shaded regions in figure~\ref{fig:fig5_1}i--k).

\section{System-identification technique} \label{sec:5_3}

Following \citet{NOIRAY2013152}, we model the combustor pressure fluctuations with a VDP-type oscillator perturbed by additive white Gaussian noise:
\begin{subequations}\label{fracdp5}
\begin{align}
    \dv[2]{x}{t}-\Big(\epsilon+\alpha_1x^2+&\alpha_2x^4+\alpha_kx^{2k}+\cdots\Big)\dv{x}{t}+\omega^2  x=\sqrt{2d}\eta,\\
    x=a&\cos(\omega t+\phi),
\end{align}
\end{subequations}
where $x$ represents $p^{\prime}/\overline{p}$ (with amplitude $a$ and phase $\phi$), $\eta$ is a unit white Gaussian noise term, $d$ is the noise amplitude, $\omega$ is the angular frequency, $\epsilon$ is the linear growth/decay coefficient (with the Hopf point at $\epsilon=0$), $\alpha_1$ is the cubic coefficient that determines whether the system is supercritical (negative) or subcritical (positive), and $\alpha_2$ and $\alpha_k$ are quintic and higher-order coefficients, respectively. We use this high-order VDP model (equation \ref{fracdp5}) rather than the classical VDP model, so as to enable the degree of nonlinearity to be identified by the SI framework itself (see below). In other words, we do not limit ourselves to only supercritical bifurcations (it just so happens that the present combustor is supercritical) but can admit subcritical bifurcations as well, as our recent work on low-density jets have shown \citep{lee_2019}.

By applying the method of variation of parameters to equation~\ref{fracdp5} (see \citealt{lee_2019} for details), we obtain one equation for the amplitude and another for the phase:
\begin{subequations}\label{derived5}
\begin{align}
    \dot{a}&=\Big(\frac{\epsilon}{2}a+\frac{\alpha_1}{8}a^3+\cdots\Big)+Q_1(\Phi)-\big(\frac{\sqrt{2d}}{\omega}\sin{\Phi}\big)\eta\label{derived51},\\
    \dot{\phi}&=Q_2(\Phi)-\Big(\frac{\sqrt{2d}}{\omega a}\cos{\Phi}\Big)\eta\label{derived52},
\end{align}
\end{subequations}
where $\Phi \equiv t+\phi$, and $Q_1(\Phi)$ and $Q_2(\Phi)$ are the sum of all the terms with first-order cosine components, which become zero when time averaged under the assumption that $a$ and $\phi$ vary much more slowly than the oscillations themselves. This assumption is generally valid near a Hopf bifurcation \citep{strogatz2000}. Thus, in the noise-free limit ($d=0$), equation~\ref{derived51} can be reduced to:
\begin{equation}\label{st_ld5}
    \frac{\mathrm{d}a}{\mathrm{d}t}=\frac{\epsilon}{2}a+\frac{\alpha_1}{8}a^3+\cdots,
\end{equation}
which is the normal-form equation for a Hopf bifurcation, justifying our choice of the system model in equation~\ref{fracdp5}. Equation \ref{st_ld5} is known as the Stuart--Landau equation and has been used to analyze various thermoacoustic systems \citep{subramanian2013, orchini2016, murthy2019analysis}. The fact that equation \ref{st_ld5} is in normal form implies that it is universal, applicable to any nonlinear dynamical system near a Hopf bifurcation–--regardless of its super/subcritical nature and regardless of its exact physical form \citep{strogatz2000}. Thus, this SI framework is not limited to just the present combustor, but is applicable to a variety of nonlinear dynamical systems.

For finite noise ($d > 0$), equation~\ref{derived5} can be stochastically averaged, yielding the Fokker--Planck equation:
\begin{subequations} \label{fpk5}
\begin{gather}
    \pdv{}{t}P(a,t)=-\pdv{}{a}\Big[D^{(1)}P(a,t)\Big]+\pdv[2]{}{a}\Big[D^{(2)}P(a,t)\Big],\\
    D^{(1)}=\Big(\frac{{\epsilon}}{2}a+\frac{\alpha_1}{8}a^3+\cdots\Big)+ \frac{d}{2 \omega^2 a}, \qquad D^{(2)}=\frac{d}{2 \omega^2}, \label{drdf5}
\end{gather}
\end{subequations}
where $P(a,t)$ is the transitional PDF of $a$ at time $t$, and $D^{(1)}$ and $D^{(2)}$ represent the drift and diffusion terms, respectively. These terms can be estimated from the time correlation of the output signal \citep{siegert1998}:
\begin{subequations}\label{transient}
\begin{align}
    D^{(n)}&=\lim_{\tau\to0}D_{\tau}^{(n)}, \label{Dr1}\\
    D_{\tau}^{(n)}&=\frac{1}{n!\tau}\int_{0}^{\infty}(A-a)^n P(A,t+\tau|a,t)\mathrm{d}A, \label{Df2}
\end{align}
\end{subequations}
where $P(A,t+\tau|a,t)$ is the conditional probability of the amplitude being $A$ at time $t+\tau$ when the amplitude is $a$ at time $t$, which can be found from the amplitude envelope $a(t)$ \citep{siegert1998}. Following \citet{friedrich2000} and \citet{NOIRAY2013152}, we determine $D_{\tau}^{(n)}$ by numerically integrating equation \ref{transient}b over $A$. Finally, from equation \ref{transform}a, we determine the drift and diffusion terms ($D^{(n)}$) by extrapolating $D_{\tau}^{(n)}$ to $\tau=0$ using an exponential function.

\begin{figure}
\centering
\includegraphics[width=0.6\textwidth]{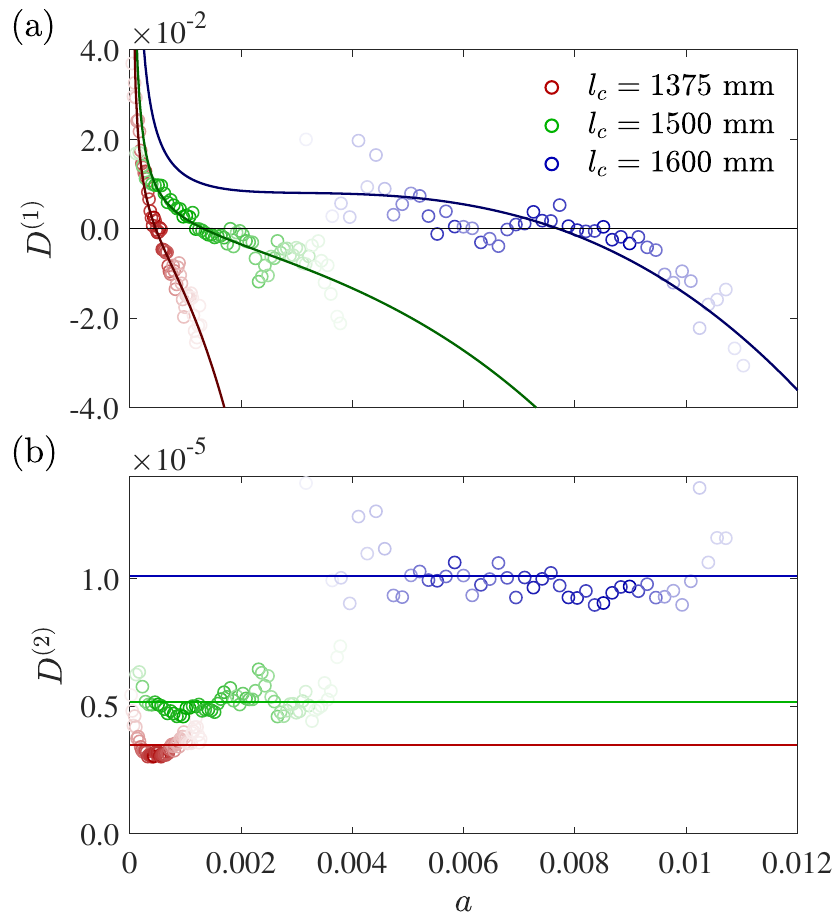}
\caption{(a) Drift and (b) diffusion terms measured experimentally (colored markers) and their regression (solid lines) based on equation~\ref{drdf5}.  The intensity of the marker colors indicates the magnitude of $P(a)$.}
\label{fig:fig5_2}
\end{figure}

The advantage of calculating the drift and diffusion terms from equation~\ref{transient} is that the mathematical form of $D_\tau^{(n)}$ need not be pre-specified. Thus, the system's degree of nonlinearity, which determines the number of nonlinear terms required to reproduce the system dynamics, can be found by combining equations \ref{fpk5}b and \ref{transient}. Specifically, we add successively higher terms (containing $\alpha_1$, $\cdots$, $\alpha_k$) to equation \ref{fpk5}b and perform polynomial regression (figure \ref{fig:fig5_2}). If the addition of term $\alpha_k$ leads to a rank-deficient fit, this implies that the system can be represented by terms up to only $\alpha_{k-1}$ In this way, we enable the degree of nonlinearity to be identified by the SI framework itself \citep{lee_2019}, rather than letting it take on a predefined value.

In experiments, noise is never perfectly white.  This causes finite-time effects, giving rise to discrepancies between the actual $D^{(n)}$ and the experimentally measured values. Nevertheless, \citet{boujo2017robust} have shown that this effect can be reduced by optimizing the VDP coefficients through a minimization of the difference between the coefficient-based approximate solution and the experimental estimation of the drift and diffusion terms in the Fokker–Planck equation. Following \citet{lade2009} and \citet{boujo2017robust}, we find an approximate solution to equation \ref{fpk5} by solving the adjoint Fokker--Planck equation:
\begin{equation}\label{AFP}
    \pdv{}{t}P^{\dagger}(A,t)=D^{(1)}\pdv{}{A}P^{\dagger}(A,t)+D^{(2)}\pdv[2]{}{A}P^{\dagger}(A,t).
\end{equation}
Given some initial conditions, the solution of equation~\ref{AFP} at $A=a$ and $t=\tau$ can be related to equation~\ref{transient} \citep{lade2009}:
\begin{subequations}\label{AFPsol}
\begin{align}
    P^{\dagger}(A,0)=(A-a)^n, \label{D1}\\
    P^{\dagger}(a,\tau)=n!\tau D_\tau^{(n)}(a).
\end{align}
\end{subequations}
Thus, equation~\ref{AFP} can be numerically solved using the initial coefficients, and $D_\tau^{n}$ can be computed from equation~\ref{AFPsol}b. The PDF-weighted error ($E$) between the experimental estimate and the numerical (coefficient-based) $D_\tau^{n}$ can then be found by comparing equations~\ref{transient} and \ref{AFPsol} \citep{boujo2017robust}:
\begin{equation}\label{err}
    E=\frac{1}{2N_{\tau}N_a}\sum_{n=1}^{2}\sum_{i=1}^{N_a}\sum_{j=1}^{N_{\tau}} P_{ij}(\hat{D}_{\tau_j}^{(n)}(a_i)-D_{\tau_j}^{(n)}(a_i;[\epsilon,\alpha,d]))^2,
\end{equation}
where $N_a$ and $N_\tau$ are the number of amplitude and time-shift values used in the optimization, and $P_{ij}$ is the experimentally measured PDF at $a=i$ and $\tau=j$.  The term $\hat{D}_{\tau j}^{(n)}$ is an experimental estimate of the drift $(n=1)$ or diffusion $(n=2)$ term, and $D_{\tau j}^{(n)}(a_i;[\epsilon,\alpha,d]))$ is computed numerically from equations~\ref{AFP} and \ref{AFPsol} using finite difference algorithm\footnote{We used a MATLAB partial differential equation solver (\emph{pdepe} solver), which numerically solves initial-boundary value problems of one-dimensional parabolic and elliptic partial differential equations. This solver gives an accurate approximation to the analytical solution even with a large mesh size \citep{yudianto2010comparison}}. Here the values of $\tau_j$ are distributed uniformly between the lower bound ($\tau_1 = \Delta f^{-1}$) and the upper bound ($\tau_2$), where $\rm{ACF}_{\tau_2}=0.25$.

\begin{figure}
\centering
\includegraphics[width=0.6\textwidth]{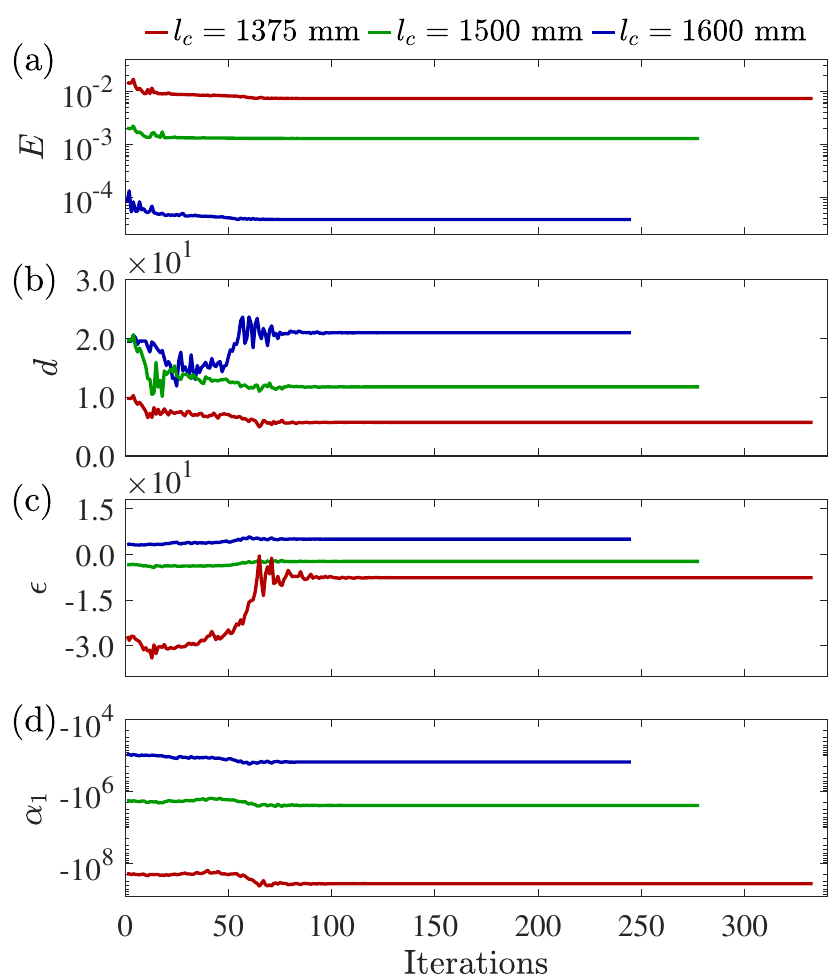}
\caption{Convergence history of (a) $E$, (b) $d$, (c) $\epsilon$, and (d) $\alpha_1$. Initial values at the first iteration are determined from equations~\ref{drdf5} and \ref{transient} (see figure~\ref{fig:fig5_2}).  Convergence is achieved at the 333$^{\rm{rd}}$ ($l_c=1375$~mm), 278$^{\rm{th}}$ ($l_c=1500$~mm), and 245$^{\rm{th}}$ ($l_c=1600$~mm) iteration.}
\label{fig:fig5_3}
\end{figure}

In the optimization, we minimize $E$ by updating [$\epsilon$, $\alpha_1$, $d$] via the Nelder--Mead simplex algorithm, as per \citet{boujo2017robust}. This algorithm does not require a function derivative and is often used for nonlinear optimization \citep{lagarias1998}. Here we update [$\epsilon$, $\alpha_1$, $d$] until [$E$, $\epsilon$, $\alpha_1$, $d$] have converged, i.e. until every quantity varies by less than $10^{-4}$. From the convergence history shown in figure~\ref{fig:fig5_3}, we can see that the optimization reduces the error between the experimental and numerical data.

\section{Results and discussion} \label{sec:5_4}
\begin{figure}
\centering
\includegraphics[width=0.65\textwidth]{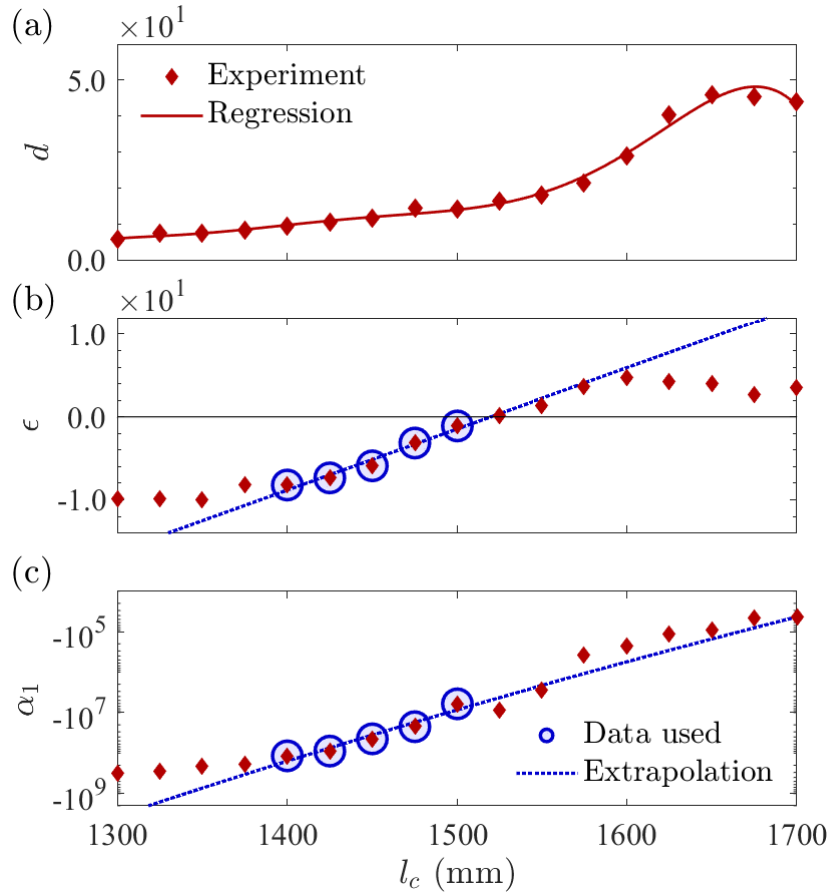}
\caption{(a) Noise amplitude, (b) linear coefficient, and (c) cubic coefficient computed via SI.  The blue dotted lines denote extrapolation from the fixed-point regime to the limit-cycle regime using (b) a linear model and (c) a power-law model. The vertical axis in (c) is on a logarithmic scale.}
\label{fig:fig5_4}
\end{figure}

Figure~\ref{fig:fig5_4} shows the noise amplitude ($d$) and the VDP coefficients ($\epsilon$, $\alpha_1$), with $\alpha_1$ being the highest identifiable nonlinear term (cubic). To perform SI with only pre-bifurcation data, we extrapolate the identified coefficients into the limit-cycle regime.  Here we recall that our system model (equations~\ref{fracdp5} and \ref{st_ld5}) is valid only near the Hopf point.  We thus limit the $l_c$ range over which the extrapolation is performed.  Specifically, recognizing that the linear coefficient $\epsilon$ varies linearly only within a finite range of $l_c$ near the Hopf point, we perform SI within this specific range only.  We find $\epsilon_l$ such that $\epsilon_l<\epsilon_{l+1}<\epsilon_{l+2}<\epsilon_{l+3}$, and use $\epsilon_l$, $\epsilon_{l+1}$, $\cdots$, $\epsilon_{l+k}$ and $\alpha_l$, $\alpha_{l+1}$, $\cdots$, $\alpha_{l+k}$ where $\epsilon_{l+k}<0$.  In other words, we initiate the prediction when four consecutive increases in $\epsilon$ have occurred.  Using fixed-point data at $1400 \le l_c \le 1500$~mm, we extrapolate $\epsilon$ with a linear model and $\alpha_1$ with a power-law model, [$\alpha_1 \propto (l_c - m_1)^{m_2}$] as per \citet{lee_2019,lee_tbr}.  In this way, we identify the Hopf point---across which $\epsilon$ changes sign---to be at $l_c = 1523$~mm.  Crucially, this is done without the use of \textit{ad-hoc} instability thresholds, in contrast to most other early warning indicators of thermoacoustic instability (\S\ref{sec:NIcomb}).

\begin{figure}
\centering
\includegraphics[clip,width=0.65\textwidth]{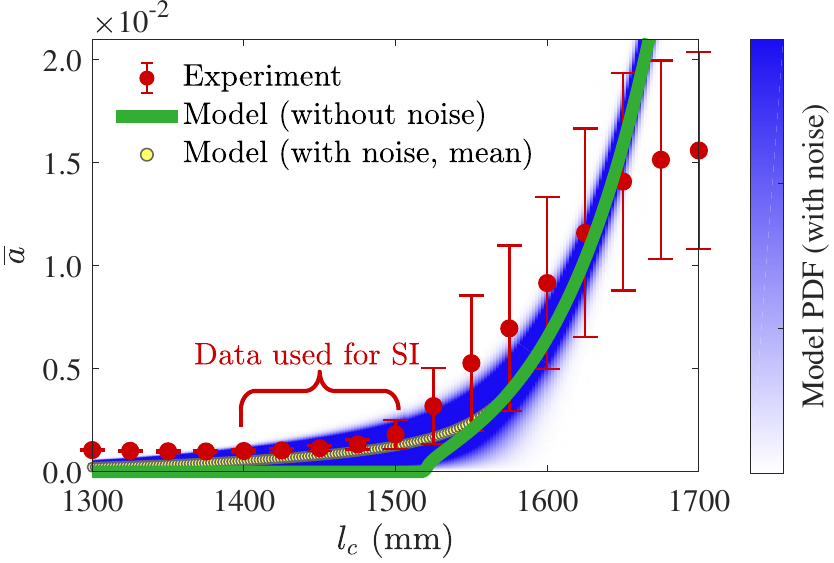}
\caption{Comparison of bifurcation diagrams between the experiments (red markers, with error bars denoting the standard deviation), the noise-free model (green line), and the noise-perturbed model (yellow markers).  The blue shading in the background represents the model PDF computed after adding the identified noise profile, whose mean values are plotted as yellow markers.}
\label{fig:fig5_5}
\end{figure}

Next we use the extrapolated coefficients to solve the Stuart--Landau equation (equation~\ref{st_ld5}), thereby reconstructing the noise-free bifurcation diagram. Here we use the VDP coefficients extracted and optimized for a specific value of $d$, on the assumption that these coefficients do not change with $d$. Figure~\ref{fig:fig5_5} compares the SI predictions with the experimental data.  Despite the linear assumption of $\epsilon$ and the fact that the highest identifiable nonlinearity is only cubic ($\alpha_1$), the agreement between the noise-free model and the experiments is reasonable.  However, a closer inspection of figure~\ref{fig:fig5_5} reveals that the experimental amplitude begins to grow before the SI-predicted Hopf point. Hypothesizing that this discrepancy might be due to the high level of intrinsic noise arising from turbulence in the combustor, we extract the noise profile using SI (see figure~\ref{fig:fig5_4}a). Instead of solving the noise-free Stuart--Landau equation (equation~\ref{st_ld5}), we solve the stationary Fokker--Planck equation:
\begin{equation}
    P(a)=Ca\exp\Big[\frac{\omega^2 a^2}{d} \Big(\frac{\epsilon}{2}+\frac{\alpha_1}{16}a^2\Big)\Big],
\end{equation}
where $P(a)$ is the stationary PDF of $a$, and $C$ is a normalization constant.  In this way, we build a PDF-based bifurcation diagram that matches the experimental data more closely than the noise-free predictions (figure~\ref{fig:fig5_5}).  The limit-cycle amplitude after the Hopf point is also well predicted with the noise-perturbed model, highlighting the important role that stochastic processes play in determining the bifurcation diagram.

\section{Conclusions} \label{sec:5_5}

Using the noise-induced dynamics generated by inherent turbulence in the fixed-point regime, we have performed output-only SI of a turbulent lean-premixed combustor near a Hopf bifurcation. We modeled the pressure fluctuations in the combustor with a VDP-type equation and its corresponding Stuart--Landau equation. We determined the model coefficients with the Fokker--Planck equation, and optimized them with an adjoint-based algorithm.  We accounted for the effects of intrinsic noise due to turbulence by solving the stationary Fokker--Planck equation with an identified noise profile. Through comparisons with experimental data, we showed that this approach can enable prediction of (i) the location of the Hopf point and (ii) the limit-cycle amplitude after the Hopf point, using only data before the Hopf point itself.

It is worth recalling that while the method used to extract and optimize the VDP coefficients is based on the work of \citet{NOIRAY2013152} and \citet{boujo2017robust}, we go further in three distinct ways. First, as mentioned in \S\ref{sec:5_3}, instead of using the classic cubic VDP model, we use a high-order VDP model (equation \ref{fracdp5}), thus enabling the degree of nonlinearity to be identified by the SI framework itself. Although this ability to identify high-order nonlinearities could not be fully demonstrated here owing to the supercritical (cubic) nature of the present combustor, it should nevertheless prove to be demonstrable in systems with subcritical Hopf bifurcations, provided that the noise amplitude is sufficiently high \citep{lee_2019}. Second, we show that output-only SI can be performed using data from only the fixed-point regime, before the Hopf point itself, without ever needing to enter the limit-cycle regime. This is an important distinction because it shows that the SI framework is capable of early warning detection of thermoacoustic instability, which could be useful in practical combustion systems. Third, unlike most existing early warning indicators, the present framework requires no \textit{ad-hoc} instability thresholds, implying that it is applicable to a wide variety of turbulent thermoacoustic systems.

This SI framework, however, is not without its limitations. First, it works only when the system is close to a Hopf bifurcation and only when the control parameter (in this case $l_c$) is adjusted in small enough increments to allow for sufficient data collection. Second, owing to noise in the combustor, there is some discrepancy in the bifurcation diagrams of the experiments and the noise-free model.  Although we show that this discrepancy can be reduced near the Hopf point by incorporating the noise profile from SI, the noise profile itself cannot be identified without first observing the system in its limit-cycle regime. Thus, to be able to reconstruct a bifurcation diagram that accounts for the probabilistic effects of noise with only fixed-point data, one would have to develop a more advanced framework for identifying the noise profile.


\chapter{Minimum noise level for system identification using the noise-induced dynamics} \label{chap:io_oo}

\section{Introduction}

In chapters \S\ref{chap:low}, \S\ref{chap:Rij} and \S\ref{chap:gas}, we demonstrated two versions of the system identification (SI) framework: input-output and output-only (see figure \ref{fig:block}). Input-output SI is applied to systems that have a low level of background noise, and is performed by perturbing the system with an external source of noise (input) and measuring the system's response (output). By contrast, output-only SI is applied to systems that have a high level of intrinsic noise, so no external forcing is required, and only the output signal is measured. Because both versions of the SI framework utilize the effect of noise on a dynamical system, two questions naturally arise: (i) how strongly do we have to perturb the system for input-output SI to work?, and (ii) how noisy does the system have to be for output-only SI to work?

To answer these questions, we determine the minimum level of noise that input-output and output-only SI require to function reliably. We define these minimum thresholds using a new measure of the noise level, and show from numerical simulations that these minimum thresholds are independent of the system parameters. Finally, using these criteria, we verify whether the choice of input-output or output-only SI was appropriate in the previous chapters (\S\ref{chap:low} - \S\ref{chap:gas}).



\begin{figure}
\captionsetup{labelformat=empty}
    \centering
    \includegraphics[width=\textwidth,trim={2cm 0cm 0cm 0cm},clip]{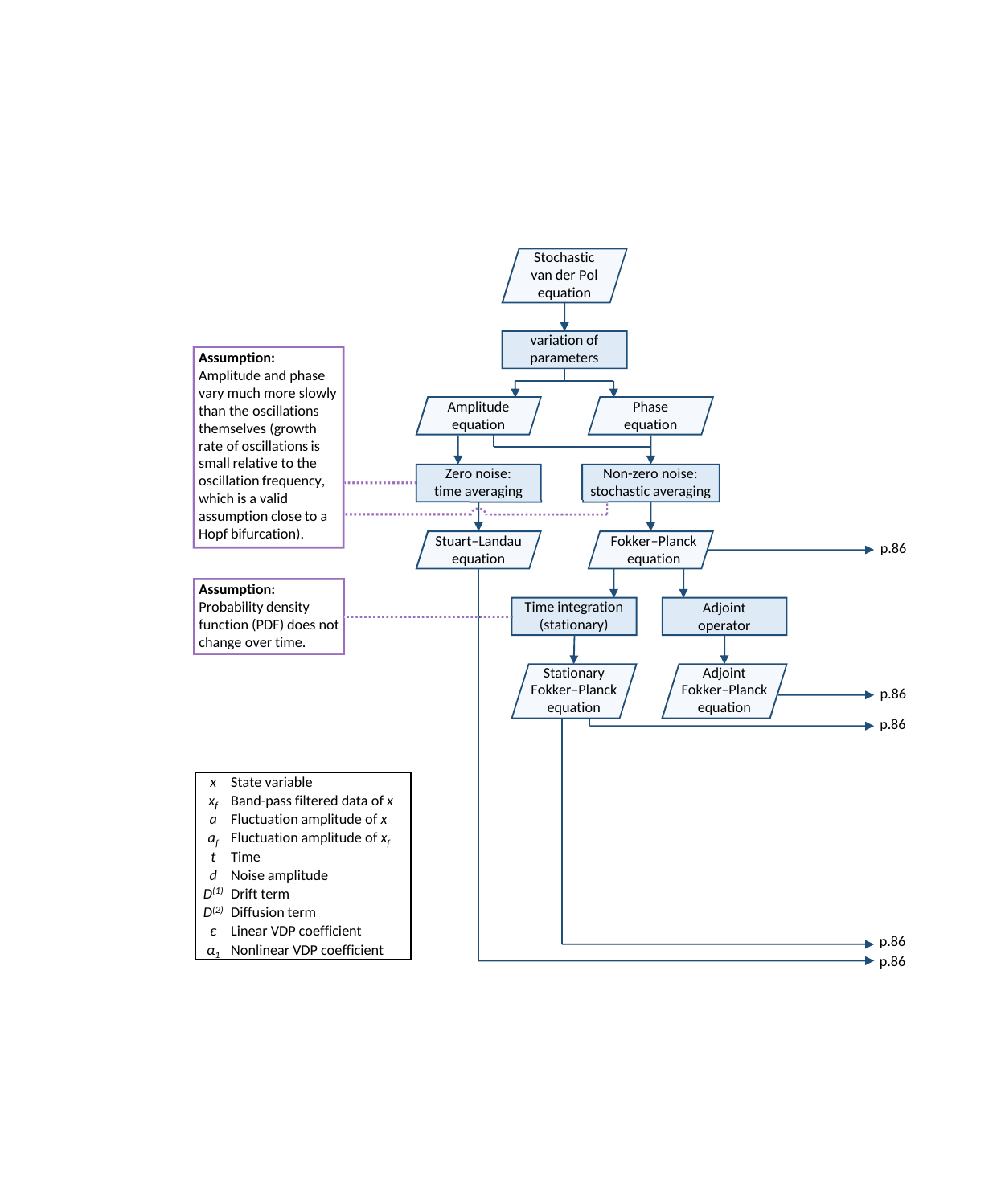}%
    \caption[]{Figure continued on the next page. For caption see p.87.}
\end{figure}
\clearpage
\begin{figure}
\captionsetup{labelformat=empty}
    \centering
    \includegraphics[width=\textwidth,trim={0cm 0cm 2cm 0cm},clip]{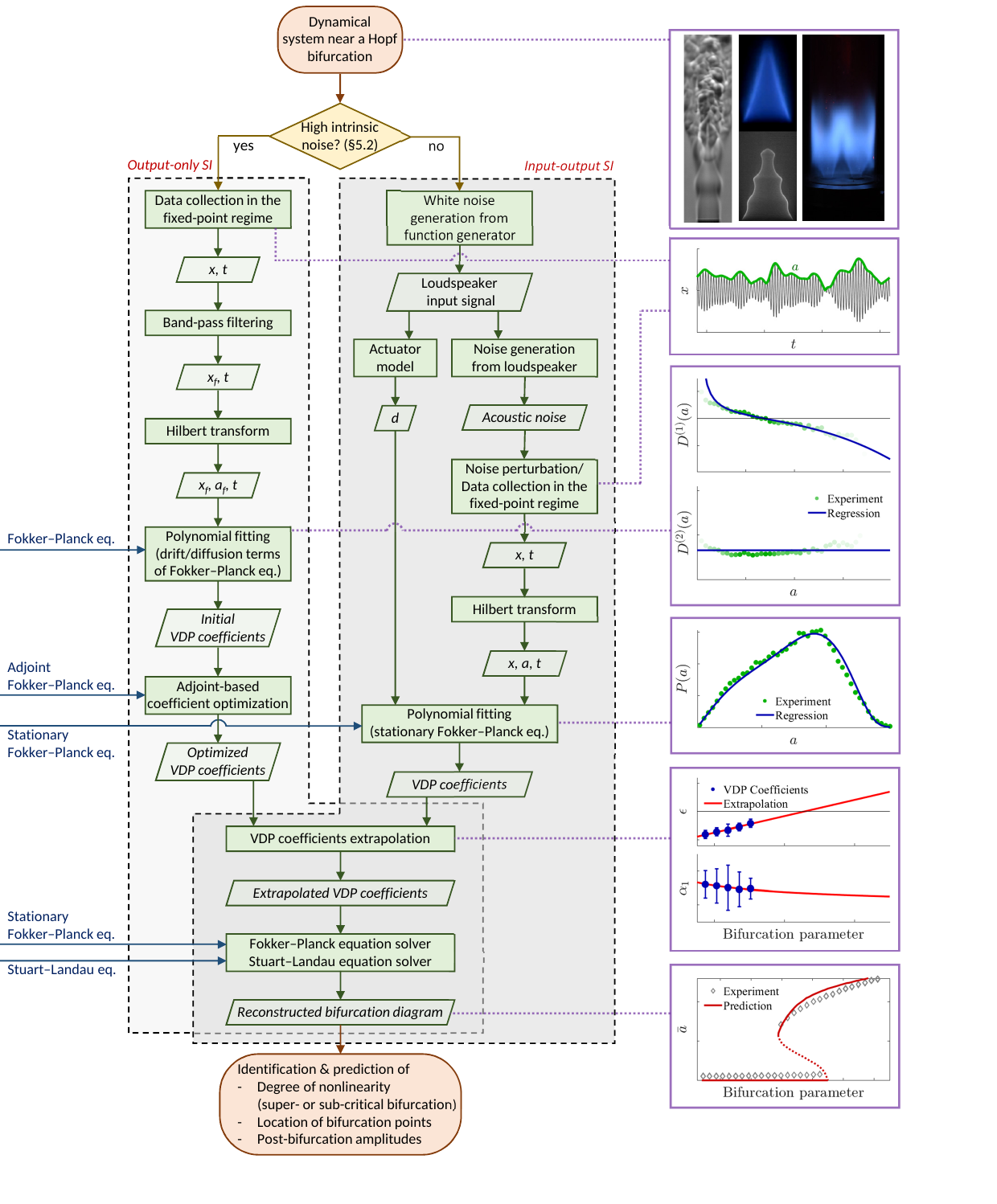}%
    \caption[]{For caption see next page.}
\end{figure}
\clearpage
\begin{figure}
    \captionsetup{labelformat=adja-page}
    \ContinuedFloat
    \caption[Block diagram showing the SI algorithm. The rectangles denote mathematical and experimental processes, and the parallelograms denote inputs and outputs. The blocks in blue show mathematical models and their treatments, and the blocks in green show experimental data and their measurements/treatments. For full caption see p.87.]{Block diagram showing the SI algorithm. The rectangles denote mathematical and experimental processes, and the parallelograms denote inputs and outputs. The blocks in blue show mathematical models and their treatments, and the blocks in green show experimental data and their measurements/treatments. In our SI framework, a high-order stochastic van der Pol (VDP) equation is used as the system model, and it is transformed into the amplitude and phase equations using the method of variation of parameters. When there is no noise acting on the system, the Stuart--Landau equation is derived from the amplitude equation via time averaging, under the assumption that the growth rate is small compared with the oscillation frequency, which is generally true near a Hopf point. For the case of non-zero noise, the Fokker--Planck equation can be derived from the amplitude and phase equations via stochastic averaging, which is also conducted by assuming that the growth rate is small (i.e. close to a Hopf bifurcation). The derived equations are used for two different versions of the SI framework: (i) input-output SI, which is applied for systems with a low level of intrinsic noise, and (ii) output-only SI, which is applied for systems with a high level of intrinsic noise. In input-output SI, external noise whose amplitude is known from the actuator model is fed into the system. The system coefficients (i.e. the VDP coefficients) are obtained by fitting the experimental data to the probability density function of the amplitude fluctuations, computed with the stationary Fokker--Planck equation. In output-only SI, the system's intrinsic noise acts as the noise source, and the VDP coefficients are found by fitting the band-pass filtered data to the drift and diffusion terms of the Fokker--Planck equation. The obtained coefficients are then optimized with the adjoint-based algorithm. In both versions of the SI framework, the VDP coefficients are extrapolated from the fixed-point regime to the limit-cycle regime. With the extrapolated coefficients, we reconstruct the (noise-free) bifurcation diagram via the Stuart--Landau equation using data from only the fixed-point regime. Alternatively, if the noise amplitudes near the Hopf bifurcation are known, we can reconstruct the bifurcation diagram via the (stationary) Fokker--Planck equation using the same VDP coefficients and data, thus taking into account the effect of noise. Either way, from the reconstructed bifurcation diagram, we identify and predict the key features of the Hopf bifurcation using only pre-bifurcation data.}\label{fig:block}
\end{figure}

\section{Minimum noise level for system identification using the noise-induced dynamics}

Let us consider an oscillatory system whose dynamics is governed by the following stochastic van der Pol (VDP) equation with cubic nonlinearity:
\begin{equation}\label{goveq}
    \Ddot{x}-(\epsilon+\alpha_1x^2)\Dot{x}+\omega^2  x=\sqrt{2d}\eta,\\
\end{equation}
where $x$ is the state variable, $\omega$ is the angular frequency, $\eta$ is a unit white Gaussian noise term, $d$ is the noise amplitude, $\epsilon$ and $\alpha_1$ are the linear and cubic VDP coefficients that govern the system, and the overdots denote differentiation with respect to time. 

By analyzing the signal $x$, we determine the VDP coefficients in equation \ref{goveq} using the SI framework illustrated in figure \ref{fig:block}. In \S\ref{chap:low}--\ref{chap:gas}, these coefficients were found by fitting experimental data to the stationary Fokker--Planck equation (if $d$ is known \textit{a priori}: input-output SI) or to the drift and diffusion terms of the Fokker--Planck equation (output-only SI). However, because both of these SI approaches use the system's noise-induced dynamics, $d$ should be of a sufficiently high amplitude that a measurable and informative response is induced in the system by noise.

\begin{figure}
    \centering
    \includegraphics[width=\textwidth,trim={0 0 0 0},clip]{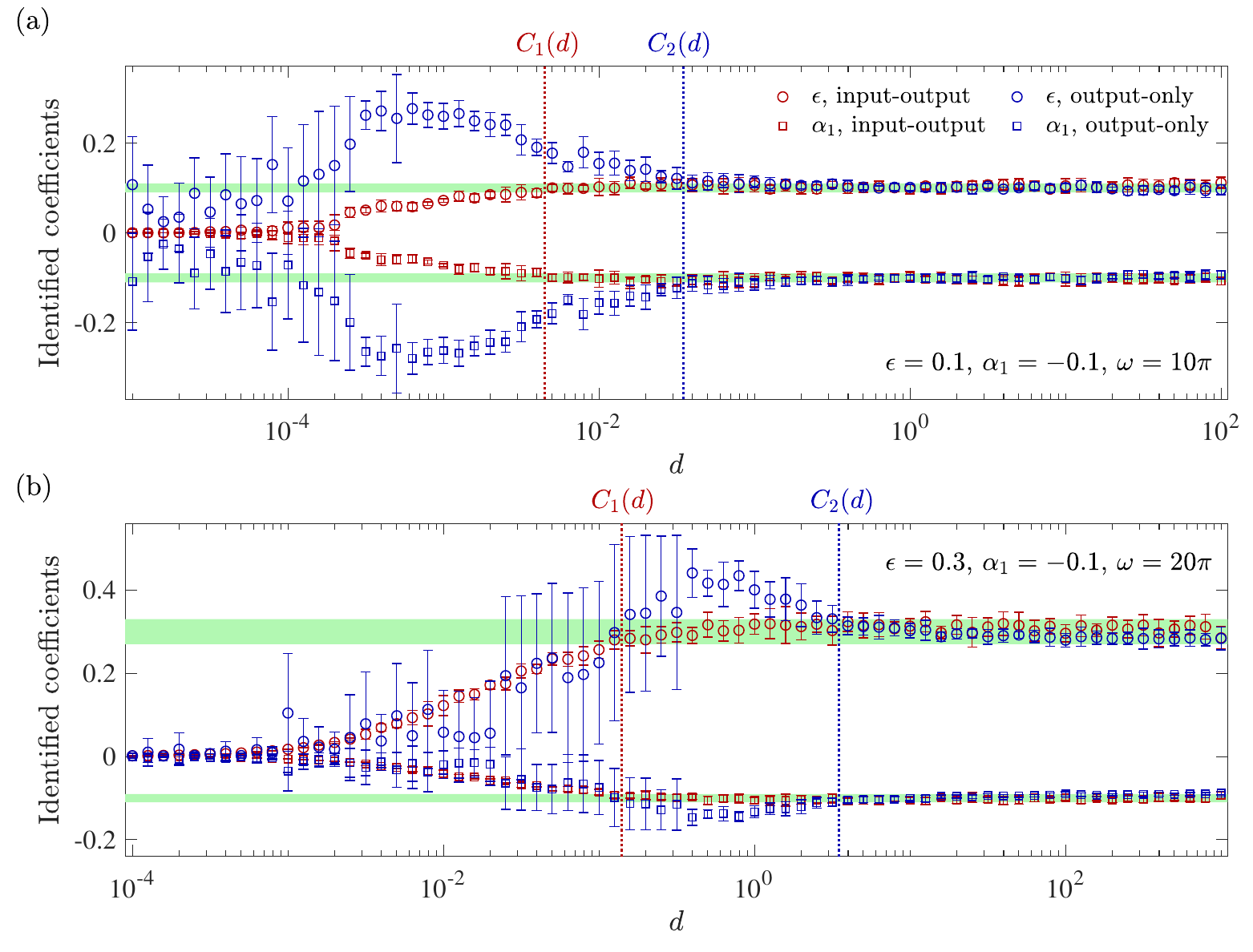}%
    \caption{Identified VDP coefficients from SI under varying $d$ for a numerically simulated system (equation \ref{goveq}). Red markers show the VDP coefficients identified from input-output SI (i.e. when $d$ is known \textit{a priori}), and blue markers show the VDP coefficients identified from output-only SI. Circular and square makers show the identified $\epsilon$ and $\alpha_1$ coefficients, respectively. Whiskers show the standard deviation calculated from 10 numerical repetitions, and the green shaded areas denote $\pm10\%$ interval from the true values (i.e. input VDP coefficients). $C_1(d)$ and $C_2(d)$ denote the minimum $d$ that yields reliable SI results from the input-output and output-only SI, respectively. Input parameters are set to (a) $\epsilon=0.1$, $\alpha_1=-0.1$, $\omega=10\pi$ and (b) $\epsilon=0.3$, $\alpha_1=-0.1$, $\omega=20\pi$. The horizontal axis is on a logarithmic scale.}\label{fig:effnoise1_1}
\end{figure}

To determine this critical amplitude of noise, we numerically generate time-series data ($x$) by solving the stochastic cubic VDP model of equation \ref{goveq} using a 4th-order Runge-Kutta (RK4) algorithm in the time span $0<t<5000$, with a time step of $dt=0.001$ ($t$ is in arbitrary units). Figure \ref{fig:effnoise1_1}a shows the SI results for this VDP model when its parameters are set to $\epsilon=0.1$, $\alpha_1=-0.1$, $\omega=10\pi$. It can be seen that the identified VDP coefficients ($\epsilon$ and $\alpha_1$) are reliable only when $d$ is sufficiently high. From this observation, we define two thresholds: $C_1$ and $C_2$. $C_1$ is the minimum amplitude of noise required for reliable\footnote{In this chapter, the word `reliable' is defined as when every identified coefficient shows less than 10\% discrepancy from its true (input) value.} input-output SI. Therefore, when using input-output SI, one must perturb the system with noise stronger than $C_1$ in order to obtain accurate results. The second threshold, $C_2$, is the minimum amplitude of noise required for reliable output-only SI. In other words, the intrinsic noise of the system should have an amplitude greater than $C_2$ for accurate identification of the VDP coefficients from output-only SI. 

\begin{figure}
    \centering
    \includegraphics[width=\textwidth,trim={0 0 0 0},clip]{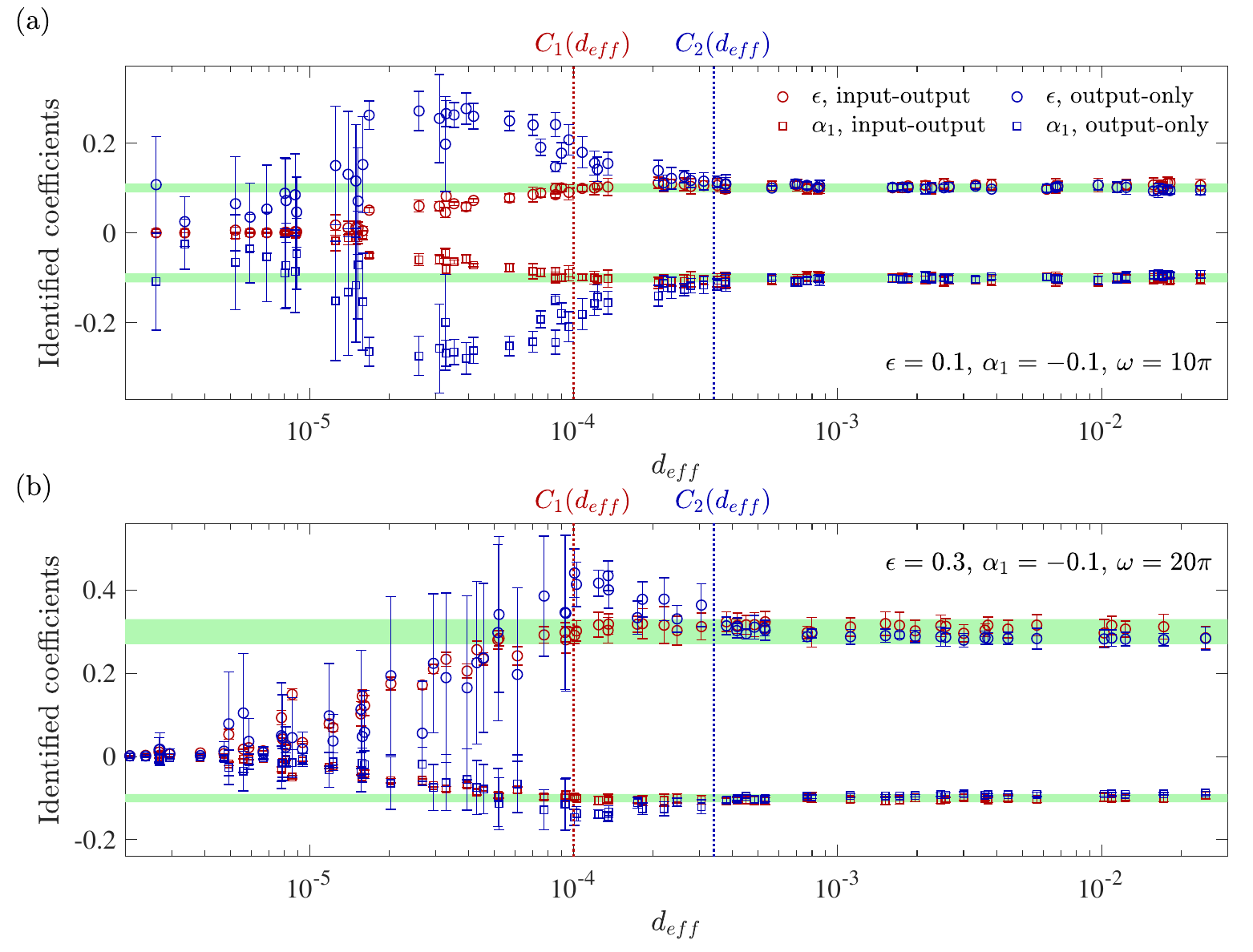}%
    \caption{Identified VDP coefficients from SI under varying $d_{eff}$ for a numerically simulated system (equation \ref{goveq}). Red markers show the VDP coefficients identified from input-output SI, and blue markers show the VDP coefficients identified from output-only SI. Circular and square makers show the identified $\epsilon$ and $\alpha_1$ coefficients, respectively. Whiskers show the standard deviation calculated from 10 numerical repetitions, and the green shaded areas denote $\pm10\%$ interval from the true values (i.e. input VDP coefficients). $C_1(d_{eff})$ and $C_2(d_{eff})$ denote the minimum $d_{eff}$ required for reliable SI results from input-output and output-only SI, respectively. Input parameters are set to (a) $\epsilon=0.1$, $\alpha_1=-0.1$, $\omega=10\pi$ and (b) $\epsilon=0.3$, $\alpha_1=-0.1$, $\omega=20\pi$. The horizontal axis is on a logarithmic scale.}\label{fig:effnoise1_2}
\end{figure}

In figure \ref{fig:effnoise1_1}a, $C_1$ and $C_2$ are found by using the noise amplitude $d$ as an independent variable ($C_1(d)=4.5\times10^{-3}$, $C_2(d)=3.5\times10^{-2}$). However, it should be noted that $C_1(d)$ and $C_2(d)$ are not invariant criteria, as they depend on the parameters characterizing the system (i.e. they are system specific). For example, if $\epsilon$ and $\omega$ are changed to $0.3$ and $20\pi$, respectively, $C_1(d)$ and $C_2(d)$ change to $1.4\times10^{-1}$ and $3.5\times10^{0}$, respectively (see figure \ref{fig:effnoise1_1}b). This means that the minimum values of $d$ required to induce a sufficient response in the system for reliable input-output and output-only SI have increased by factors of 30 and 100, respectively. In practical applications, because $C_1(d)$ and $C_2(d)$ are specific to each system, it is difficult to determine whether the amplitude of noise ($d$) acting on a given system is above or below these critical thresholds. Therefore, it is important to express $C_1$ and $C_2$ using an alternative measure of noise, one that is independent of the system parameters. 

\begin{figure}
    \centering
    \includegraphics[width=0.7\textwidth,trim={0 0 0 0},clip]{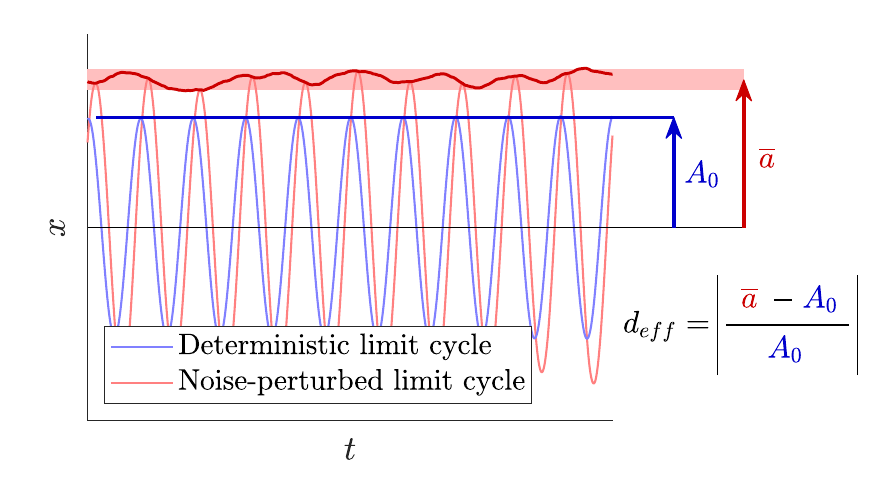}%
    \caption{Illustration for the calculation of the effective noise level ($d_{eff}$). $A_0$ is the deterministic limit-cycle amplitude, and $\overline{a}$ is the time-averaged amplitude of the noise-perturbed limit cycle.}\label{fig:deff}
\end{figure}

For this reason, we quantify the critical noise amplitude by calculating how strongly the noise can kick a system away from its original dynamical state. In particular, we define the effective noise amplitude ($d_{eff}$) by evaluating the contribution of stochastic effects to the limit-cycle amplitude (see figure \ref{fig:deff} for illustration):
\begin{equation}\label{deff}
    d_{eff} = \abs{\frac{\overline{a} - A_0}{A_0}},
\end{equation}
where $\overline{a}$ is the time-averaged amplitude of the noise-perturbed limit cycle, which can be obtained from the Hilbert transform of the output signal $x$. $A_0$ is the deterministic limit-cycle amplitude, which can be obtained from the joint probability density function of [$x$,$\Dot{x}$], or from the exponential decay rate of the autocorrelation function \citep{NOIRAY2013152}. Alternatively, if the VDP coefficients are known, $A_0$ can be found by solving the Stuart--Landau equation (see \S\ref{sec:extrapol}).

\begin{figure}
    \centering
    \includegraphics[width=\textwidth,trim={0 0 0 0},clip]{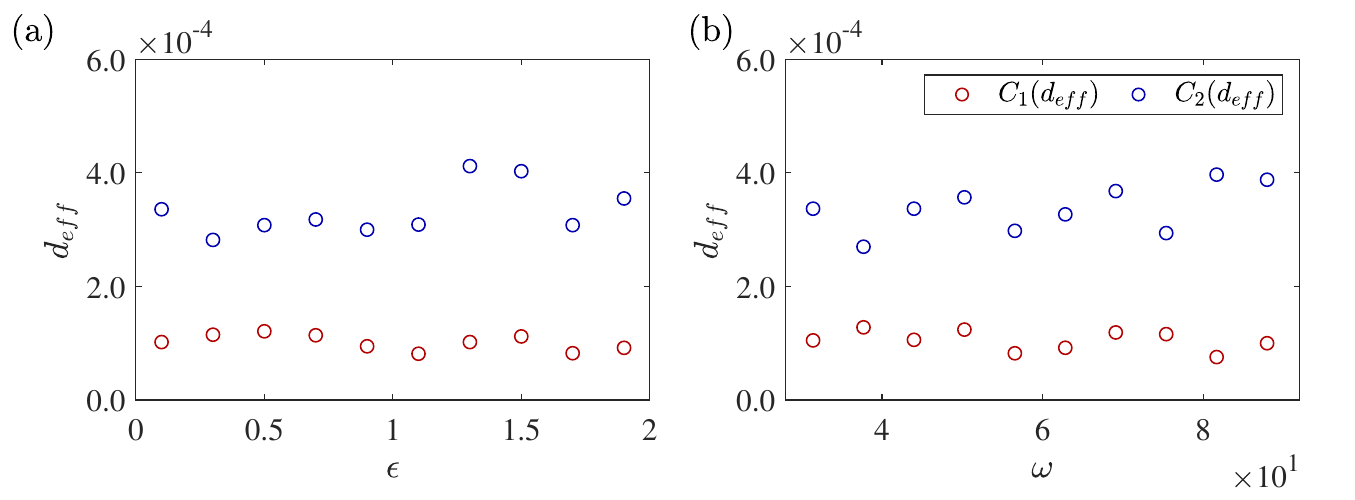}%
    \caption{$C_1(d_{eff})$ and $C_2(d_{eff})$ values for varying (a) $\epsilon$ and (b) $\omega$. }\label{fig:effnoise2}
\end{figure}

Figure \ref{fig:effnoise1_2}a shows the same SI results as figure \ref{fig:effnoise1_1}a, but with $d_{eff}$ on the horizontal axis. The minimum noise amplitudes defined with $d_{eff}$ are $C_1(d_{eff})=1.0\times10^{-4}$ and $C_2(d_{eff})=3.3\times10^{-4}$. These thresholds---$C_1(d_{eff})$ and $C_2(d_{eff})$---are more useful than $C_1(d)$ and $C_2(d)$, because they are independent of the system parameters. For example, in figure \ref{fig:effnoise1_2}b where the linear VDP coefficient and the oscillation frequency are changed into $0.3$ and $20\pi$, respectively, $C_1(d_{eff})$ and $C_2(d_{eff})$ remain unchanged. The independence of $C_1(d_{eff})$ and $C_2(d_{eff})$ from the system parameters is further demonstrated in figure \ref{fig:effnoise2}. Regardless of the strength or frequency of the oscillation, if the noise fed into the system can change the limit-cycle amplitude by more than $0.010\%$\footnote{Obtained by averaging $C_1(d_{eff})$ values shown in figure \ref{fig:effnoise2}.} ($C_1(d_{eff})$), then input-output SI can be reliably performed. Similarly, if the intrinsic noise of the system can change the limit-cycle amplitude by more than $0.033\%$\footnote{Obtained by averaging $C_2(d_{eff})$ values shown in figure \ref{fig:effnoise2}.} ($C_2(d_{eff})$), then output-only SI can be reliably performed.

\begin{figure}
    \centering
    \includegraphics[width=0.85\textwidth,trim={0 0 0 0},clip]{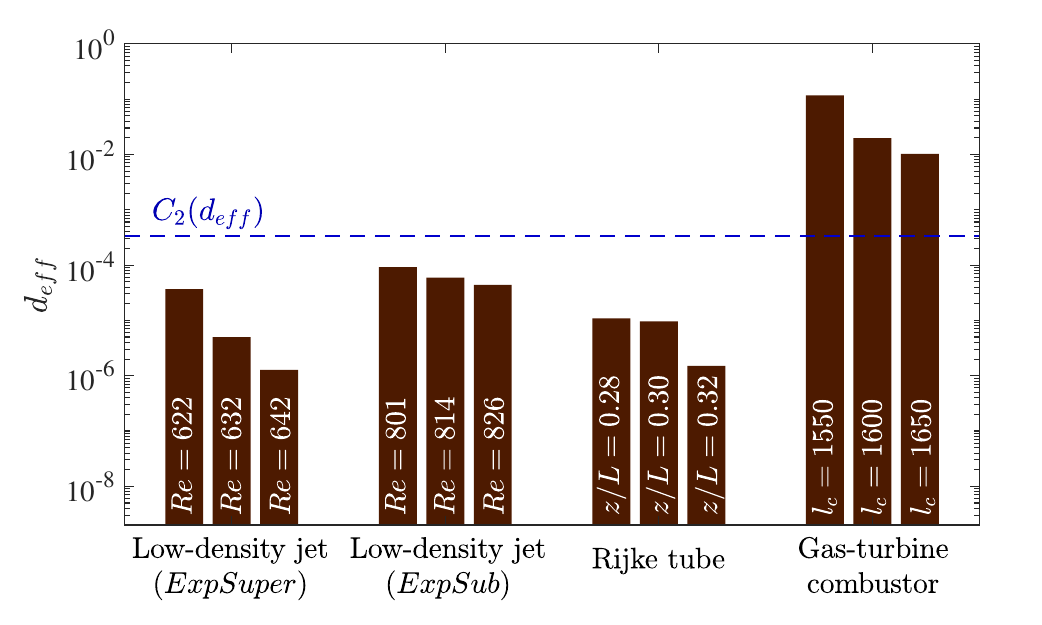}%
    \caption{Effective noise level ($d_{eff}$) corresponding to the intrinsic noise in three experimental systems: low-density jet(\S\ref{chap:low}), Rijke tube (\S\ref{chap:Rij}) and gas turbine combustor (\S\ref{chap:gas}). The horizontal dashed line denotes the $C_2(d_{eff})$ threshold for reliable output-only SI.}\label{fig:crit}
\end{figure}

Now, using the measure of $d_{eff}$ and the $C_2(d_{eff})$ criterion, we verify whether our choice of SI strategy---input-output or output-only---was appropriate in the previous chapters. Figure \ref{fig:crit} shows $d_{eff}$ values for the three experimental systems---the low-density jet(\S\ref{chap:low}), the Rijke tube (\S\ref{chap:Rij}) and the gas turbine combustor (\S\ref{chap:gas}). In the low-density jet and the Rijke tube, $d_{eff}$ is lower than $C_2(d_{eff})$ identified above. This implies that the intrinsic noise in these two laminar systems is insufficient for output-only SI to work reliably. Therefore, it was indeed appropriate to use input-output SI, where additional noise is externally applied. By contrast, in the gas turbine combustor (\S\ref{chap:gas}), $d_{eff}$ is at least 30 times higher than $C_2(d_{eff})$. This means that the intrinsic noise of the combustor is strong enough for output-only SI to work reliably, as shown in \S\ref{chap:gas}.

\section{Conclusions}

In this chapter, we defined an effective noise amplitude, $d_{eff}$, by quantifying the change to the limit-cycle amplitude induced by noise. Using this measure, we found the minimum noise level required for reliable SI. It is shown that input-output SI and output-only SI can be successfully performed if the noise acting on the system changes the limit-cycle amplitude by more than $0.010\%$ and $0.033\%$, respectively.

In practical applications of input-output SI, $C_1(d_{eff})$ can serve as a reference threshold for how strongly the system should be perturbed with external noise. Similarly, $C_2(d_{eff})$ can aid the choice between input-output and output-only SI. In particular, if a system's intrinsic noise is higher than $C_2(d_{eff})$, output-only SI can be reliably performed. However, if a system's intrinsic noise is lower than $C_2(d_{eff})$, then an additional injection of noise is required, so input-output SI should be chosen.

A limitation in determining the SI strategy using $C_1(d_{eff})$ and $C_2(d_{eff})$ is that $d_{eff}$ can only be computed when the deterministic limit-cycle amplitude is real and positive. Therefore, in order to calculate $d_{eff}$, the system must be tested, at least once, in the limit-cycle regime. In future studies, these minimum noise levels ($C_1$, $C_2$) can be defined using measures for which the limit-cycle amplitude is not available (e.g. in the fixed-point regime), as it can enable the determination of the SI strategy even before the birth of any limit-cycle oscillations.

\chapter{Fokker--Planck equation for two coupled oscillators} \label{chap:coupled}

\section{Introduction}

In previous chapters, we assumed that a single mode of oscillation dominates the system dynamics (equations \ref{fracdp}, \ref{eq:4vdp} and \ref{fracdp5}a). In some systems, however, multiple modes can coexist. For example, secondary or harmonic modes are frequently encountered in thermoacoustic systems, as evidenced in the experiments of \S\ref{chap:Rij}. Such modes are sometimes strong enough to violate the single-mode approximation (see figure \ref{fig:4_5}), requiring the data to be band-pass filtered. Band-pass filtering, however, may also remove important information about the system itself (see \S\ref{future:band} for a detailed discussion). This is because secondary modes often contain valuable information about the system dynamics, and the growth of such modes can sometimes lead to mode switching or secondary bifurcations \citep{moeck2012nonlinear,acharya2018, guan2019control}. It is therefore important to be able to capture the dynamics of such modes through an appropriate modeling framework. In addition, the Fokker--Planck equation, which is used for system identification (SI) from the noise-induced dynamics, should be derived for each mode, so as to identify and predict the coupled dynamics.

In pioneering work, \citet{culick1992} modeled two longitudinal modes in a combustion chamber, where the frequency of the second mode was twice that of the first mode, with a set of coupled stochastic differential equations containing parametric and additive noise terms. To reduce the mathematical complexity in deriving the Fokker--Planck equation for the first mode, the authors neglected all the stochastic terms in the oscillator equation for the second mode. Hence, it was assumed that the stochastic effect on the second mode is only delivered to it by the first mode via coupling terms. The analytical solution for the probability density function (PDF) of the amplitude of the first mode was shown to match the numerical results well. However, in this approach, the Fokker--Planck solution could be found for only the first mode, making it impractical for SI of the secondary modes.


In an investigation of annular thermoacoustic modes in gas-turbine combustors, \citet{noiray2013dynamic} derived an individual Fokker--Planck equation for each oscillator. They modeled two azimuthal wave components at the same frequency with two coupled stochastic oscillator equations. Assuming that the noise is additive, the authors derived the drift and diffusion terms of the Fokker--Planck equation for each oscillator. Subsequently, output-only SI was performed using those derivations. \citet{Pau2017} went on to improve this method by applying the adjoint-based optimization algorithm proposed by \citet{boujo2017robust} for a single oscillator. The results were shown to reproduce the dynamics of numerically simulated standing and spinning modes in a gas turbine engine. However, because the Fokker--Planck equations in these studies \citep{noiray2013dynamic, Pau2017} were derived under the assumption that both oscillators have the same frequency, this framework does not allow for the coexistence of two modes at different frequencies.

To model multi-frequency thermoacoustic oscillations, \citet{bonciolini2019modelling} proposed a model consisting of three stochastic oscillators, each of which has a different frequency but is subjected to the same additive noise term. The pressure signal was then modeled by superpositioning the three oscillators, which represent three different acoustic modes. It was shown that this model could reproduce the results from experiments, specifically in terms of the power spectral density, PDF, and phase portrait. Also, \citet{gopalakrishnan2015} studied the effect of noise on a prototypical thermoacoustic system by modeling its dynamics with a set of stochastic oscillator equations. \citet{thomas2018} used a similar approach to investigate amplitude death in two coupled prototypical thermoacoustic systems; the authors used coupled stochastic oscillator equations to model different acoustic modes in two Rijke tubes. These studies \citep{gopalakrishnan2015, thomas2018, bonciolini2019modelling} were successful in modeling the effect of noise on experimental or numerical thermoacoustic systems exhibiting multiple modes of oscillation. However, a Fokker--Planck equation that can analytically describe the noise-induced dynamics of such multi-mode systems has yet to be derived or studied.

In this chapter, we model the dynamics of a system featuring two modes of oscillation using two coupled oscillators. Specifically, we consider two stochastic van der Pol (VDP) oscillators \citep{anishchenko2007,lakshmanan2011,guan2019QP} coupled in three different ways: distance coupling, velocity coupling, and nonlinear coupling. These coupling terms represent different types of interaction between the two modes. Using a fast-slow variables assumption and the stochastic averaging technique, we derive the Fokker--Planck equations for the PDF of the oscillation amplitude, which correspond to the two coupled VDP oscillators. Finally, we numerically validate the derived Fokker--Planck equation and investigate the effect of three factors: (i) noise amplitude, (ii) coupling strength, and (iii) coupling type.

\section{Distance coupling} 

\subsection{Derivation of the Fokker--Planck equation} \label{der_discpl}

We first consider two distance-coupled VDP oscillators, which could be used to model two different modes in an oscillatory system:
\begin{subequations} \label{dist_cpl}
\begin{align}
    \ddot{x_1}-(\epsilon_1 + \alpha_{11}{x_1}^2)\dot{x_1} +& {\omega_1}^2 x_1 + k (x_1 - x_2) = \sqrt{2 d_1} \eta_1, \\
    \ddot{x_2}-(\epsilon_2 + \alpha_{12}{x_2}^2)\dot{x_2} +& {\omega_2}^2 x_2 + k (x_2 - x_1) = \sqrt{2 d_2} \eta_2,
\end{align}
\end{subequations}
where [$x_1$, $x_2$] are the state variables, [$\omega_1$, $\omega_2$] are the angular frequencies, [$\epsilon_1$, $\epsilon_2$] and [$\alpha_{11}$, $\alpha_{12}$] are the VDP coefficients that characterize the system, $k$ is the coupling strength, and the overdots denote differentiation with respect to time. [$\eta_1$, $\eta_2$] are unit white Gaussian noise terms and [$d_1$, $d_2$] are the noise amplitudes applied to the first and second oscillators, respectively. In this chapter, the maximum order of nonlinearity is set to third order ($\alpha_1$). This is done for simplicity, and one may add higher-order nonlinear terms to equation \ref{dist_cpl} (see \S\ref{num} for a transformation of the higher-order terms to the Fokker--Planck representation).

It should be noted that different noise terms are introduced for the two oscillators, as per \citet{culick1992}. Because both oscillators belong to the same system, it is intuitive to assume that they sense the same noise source ($d_1 = d_2$, $\eta_1 = \eta_2$). This assumption of a common noise source is standard, having previously been used by \citet{gopalakrishnan2015}, \citet{bonciolini2017output}, \citet{thomas2018} and \citet{bonciolini2019modelling}. In practice, however, it might be necessary to apply band-pass filters with different bandwidths to the two oscillators. For example, a narrower filter should be used for the second mode if it is significantly weaker than the primary mode. Moreover, the existence of neighboring modes can affect the bandwidth as well. When different band-pass filters are applied to two oscillators, the properties of the noise applied to them will also differ. For these reasons, we use individual noise terms in the two equations. Nevertheless, it is worth mentioning that the choice of the noise terms does not affect the derivation below, so long as both terms are additive and white. In other words, even if $d_1 = d_2$ and $\eta_1 = \eta_2$, we will still arrive at the same Fokker--Planck equation (equation \ref{cpl_sta}).

In equation \ref{dist_cpl}, $x_1$ and $x_2$ can be transformed into amplitude ($a$) and phase ($\phi$) representation using the following relationship (see \S\ref{num} for justification):
\begin{subequations} \label{rel}
\begin{align}
    x_1 &= a_1 \cos{(\omega_1 t + \phi_1)}, \\
    x_2 &= a_2 \cos{(\omega_2 t + \phi_2)}, \\
    \dot{x_1} &= - a_1 \omega_1 \sin{(\omega_1 t + \phi_1)}, \\
    \dot{x_2} &= - a_2 \omega_2 \sin{(\omega_2 t + \phi_2)}.
\end{align}
\end{subequations}

By combining equations \ref{dist_cpl} and \ref{rel}, we obtain the following equations: 
\begin{subequations} \label{eq5_4}
\begin{align}
    \dot{a_1} \cos{\Phi_1} - a_1 \dot{\phi_1} \sin{\Phi_1} &= 0, \\
    \dot{a_2} \cos{\Phi_2} - a_2 \dot{\phi_2} \sin{\Phi_2} &= 0, \\
    \dot{a_1} \sin{\Phi_1} + a_1 \dot{\phi_1} \cos{\Phi_1} &= \epsilon_1 a_1 \sin{\Phi_1} + \alpha_{11} {a_1}^3 \sin{\Phi_1} \cos^2{\Phi_1} \notag \\
    &\quad+ \frac{k a_1}{\omega_1} \cos{\Phi_1} - \frac{k a_2}{\omega_1} \cos{\Phi_2} - \frac{\sqrt{2 d_1}}{\omega_1} \eta_1,\\
    \dot{a_2} \sin{\Phi_2} + a_2 \dot{\phi_2} \cos{\Phi_2} &= \epsilon_2 a_2 \sin{\Phi_2} + \alpha_{12} {a_2}^3 \sin{\Phi_2} \cos^2{\Phi_2} \notag \\
    &\quad- \frac{k a_1}{\omega_2} \cos{\Phi_1} + \frac{k a_2}{\omega_2} \cos{\Phi_2} - \frac{\sqrt{2 d_2}}{\omega_2} \eta_2,
\end{align}
\end{subequations}
where $\Phi_1=\omega_1 t + \phi_1$ and $\Phi_2=\omega_2 t + \phi_2$. We can now solve equation \ref{eq5_4} and apply trigonometric identities, yielding the following set of ordinary differential equations:
\begin{subequations} \label{eq5_5}
\begin{align}
    \dot{a_1} &= \frac{\epsilon_1 a_1}{2} + \frac{\alpha_{11} {a_1}^3}{8} - \frac{\epsilon_1 a_1}{2} \cos{2\Phi_1} - \frac{\alpha_{11} {a_1}^3}{8} \cos{4\Phi_1} + \frac{k a_1}{2 \omega_1} \sin{2\Phi_1} \notag \\ 
    &\quad- \frac{k a_2}{2\omega_1} \sin{(\Phi_1+\Phi_2)} - \frac{k a_2}{2\omega_1} \sin{(\Phi_1-\Phi_2)} - \frac{\sqrt{2d_1}}{\omega_1}(\sin{\Phi_1}) \eta_1,\\
    \dot{a_2} &= \frac{\epsilon_2 a_2}{2} + \frac{\alpha_{12} {a_2}^3}{8} - \frac{\epsilon_2 a_2}{2} \cos{2\Phi_2} - \frac{\alpha_{12} {a_2}^3}{8} \cos{4\Phi_2} + \frac{k a_2}{2 \omega_2} \sin{2\Phi_2} \notag \\ 
    &\quad- \frac{k a_1}{2\omega_2} \sin{(\Phi_2+\Phi_1)} - \frac{k a_1}{2\omega_2} \sin{(\Phi_2-\Phi_1)} - \frac{\sqrt{2d_2}}{\omega_2}(\sin{\Phi_2}) \eta_2,\\
    \dot{\phi_1} &= \frac{k}{2\omega_1} + \frac{\epsilon_1}{2}\sin{2\Phi_1} + \frac{\alpha_{11} {a_1}^2}{4}\sin{2\Phi_1} + \frac{\alpha_{11} {a_1}^2}{8}\sin{4\Phi_1} + \frac{k}{2\omega_1}\cos{2\Phi_1} \notag \\
    &\quad- \frac{k a_2}{2\omega_1 a_1}\cos{(\Phi_1+\Phi_2)} - \frac{k a_2}{2\omega_1 a_1}\cos{(\Phi_1-\Phi_2)} - \frac{\sqrt{2d_1}}{\omega_1 a_1} (\cos{\Phi_1}) \eta_1, \\
    \dot{\phi_2} &= \frac{k}{2\omega_2} + \frac{\epsilon_2}{2}\sin{2\Phi_2} + \frac{\alpha_{12} {a_2}^2}{4}\sin{2\Phi_2} + \frac{\alpha_{12} {a_2}^2}{8}\sin{4\Phi_2} + \frac{k}{2\omega_2}\cos{2\Phi_2} \notag \\
    &\quad- \frac{k a_1}{2\omega_2 a_2}\cos{(\Phi_2+\Phi_1)} - \frac{k a_1}{2\omega_2 a_2}\cos{(\Phi_2-\Phi_1)} - \frac{\sqrt{2d_2}}{\omega_2 a_2} (\cos{\Phi_2}) \eta_2. 
\end{align}
\end{subequations}

The last terms in equation \ref{eq5_5} are the stochastic components, whereas the other terms are the deterministic components. By applying stochastic averaging \citep{stratonovich1963, stratonovich1967, roberts1986stochastic}, we obtain the drift and diffusion terms of the amplitude ($a$):
\begin{subequations} \label{eq5_6}
\begin{align}
    \mathbf{m}(a_1) &= \frac{\epsilon_1 a_1}{2} + \frac{\alpha_{11} {a_1}^3}{8} + \frac{d_1}{2 \omega_1 a_1}, \\
    \mathbf{m}(a_2) &= \frac{\epsilon_2 a_2}{2} + \frac{\alpha_{12} {a_2}^3}{8} + \frac{d_2}{2 \omega_2 a_2}, \\
    \mathbf{\sigma}^2(a_1) &= \frac{d_1}{{\omega_1}^2}, \\
    \mathbf{\sigma}^2(a_2) &= \frac{d_2}{{\omega_2}^2},
\end{align}
\end{subequations}
where $\mathbf{m}$ and $\mathbf{\sigma}$ represent the drift and diffusion terms, respectively. Therefore, the PDFs for $a_1$ and $a_2$ can be found via the standard Fokker--Planck equations:
\begin{subequations} \label{eq5_7}
\begin{align}
    \pdv{}{t}P(a_1,t)&=-\pdv{}{a_1}\Big[\boldsymbol{m}(a_1)P(a_1,t)\Big]+\pdv[2]{}{a_1}\Big[\frac{\boldsymbol{\sigma}^2(a_1)}{2}P(a_1,t)\Big], \\
    \pdv{}{t}P(a_2,t)&=-\pdv{}{a_2}\Big[\boldsymbol{m}(a_2)P(a_2,t)\Big]+\pdv[2]{}{a_2}\Big[\frac{\boldsymbol{\sigma}^2(a_2)}{2}P(a_2,t)\Big],
\end{align}
\end{subequations}
where $P(a_1,t)$ and $P(a_2,t)$ are the PDFs of $a_1$ and $a_2$ at time $t$, respectively. When the PDF does not change with time, the stationary solution of equation \ref{eq5_7} can be obtained by integration:
\begin{subequations} \label{cpl_sta}
\begin{align}
    P(a_1) = C_1 a_1 \exp\Big[\frac{\epsilon_1 {\omega_1}^2}{2d_1} {a_1}^2 + \frac{\alpha_{11} {\omega_1}^2}{16d_1} {a_1}^4 \Big], \\
    P(a_2) = C_2 a_2 \exp\Big[\frac{\epsilon_2 {\omega_2}^2}{2d_2} {a_2}^2 + \frac{\alpha_{12} {\omega_2}^2}{16d_2} {a_2}^4 \Big],
\end{align}
\end{subequations}
where $P(a_1)$ and $P(a_2)$ are the stationary probability functions of $a_1$ and $a_2$, respectively, and $C_1$ and $C_2$ are normalization constants. It can be seen that both $P(a_1)$ and $P(a_2)$ are independent of the coupling coefficient $k$, which is an important result and will be examined below.

\subsection{Numerical analysis and validation} \label{num_ana_dis}

We perform numerical simulations to analyze the effect of the coupling strength ($k$) and the noise amplitude ($d_1$, $d_2$), and to validate the Fokker--Planck equation derived in \S\ref{der_discpl}. We set the angular frequency of the first mode to $\omega_1=2\pi$, and that of the second mode to $\omega_2 = 3.1\omega_1$. These choices, respectively, represent the primary mode and the third harmonic mode with a small frequency detuning, which we observed in \S\ref{chap:Rij}. 

Near a Hopf bifurcation, the dynamical states of a system with two modes of oscillation can be classified into three types, which are shown in table \ref{num_cases1}\footnote{Although the coefficients could be set to any values, we show one representative example for each case ($\abs{\epsilon_1} = 2 \abs{\epsilon_2}$), which allows for a clear comparison of the dynamics of two oscillators.}. The first is when both oscillators are in the fixed-point regime, prior to the Hopf point. The second is when one oscillator has crossed the Hopf
point, but the other is still in the fixed-point regime. In this case, only a single oscillator
is exhibiting limit-cycle oscillations. The third case is when both oscillators have crossed
the Hopf point and are in the limit-cycle regime.

\renewcommand{\arraystretch}{1.5}

\begin{table}[h]
\centering
    \begin{tabular}{|c|c|c|}
    \hline
         &  Oscillator 1 ($x_1$) & Oscillator 2 ($x_2$)\\
    \hline
    Case 1 & Fixed-Point ($\epsilon_1 = -0.2$) & Fixed-Point ($\epsilon_2 = -0.1$) \\
    \hline
    Case 2 & Limit-Cycle ($\epsilon_1 = +0.2$) & Fixed-Point ($\epsilon_2 = -0.1$) \\
    \hline
    Case 3 & Limit-Cycle ($\epsilon_1 = +0.2$) & Limit-Cycle ($\epsilon_2 = +0.1$) \\
    \hline
    \end{tabular}
    \caption{Three test cases for the numerical analysis and validation.}
    \label{num_cases1}
\end{table}

For each case, we conduct simulations with three coupling strengths and two noise amplitudes, which are shown in table \ref{num_cases2}. In the simulation, equation \ref{dist_cpl} is numerically solved with a 4th-order Runge-Kutta (RK4) algorithm in the time span $0<t<1000$, with a time step of $dt=0.001$ ($t$ in arbitrary units). We ignore the initial transient ($0<t<150$), so as to focus on the stationary dynamics, and extract the instantaneous amplitude ($a$) with the Hilbert transform. We then compare the results from the numerical simulations with the analytical solution given by the stationary Fokker--Planck equation (equation \ref{cpl_sta}).

\begin{table}[h]
\centering
    \begin{tabular}{|c|c|c|}
        \hline
         & Low Noise & High Noise \\
        \hline
        Uncoupled & $k=0$, $d=0.1$ & $k=0$, $d=0.5$ \\
        \hline
        Weakly Coupled & $k=0.08$, $d=0.1$ & $k=0.08$, $d=0.5$ \\
        \hline
        Strongly Coupled & $k=0.25$, $d=0.1$ & $k=0.25$, $d=0.5$ \\
        \hline
    \end{tabular}
    \caption{Combinations of the coupling strengths and noise amplitudes for the numerical analysis and validation. $d_1 = d_2 = d$ is assumed in all cases.}
    \label{num_cases2}
\end{table}

\renewcommand{\arraystretch}{1}

Next, in order to investigate the effect of the coupling strength and noise amplitude in more detail, we run simulations with fixed $d$ and varying $k$, and vice versa. The colormaps of $P(a_1)$ and $P(a_2)$ obtained from simulations are then compared with the analytical solution obtained from the stationary Fokker--Planck equation (equation \ref{cpl_sta}).

\begin{figure}
    \centering
    \includegraphics[clip,trim=1mm 1mm 1mm 1mm,width=\textwidth]{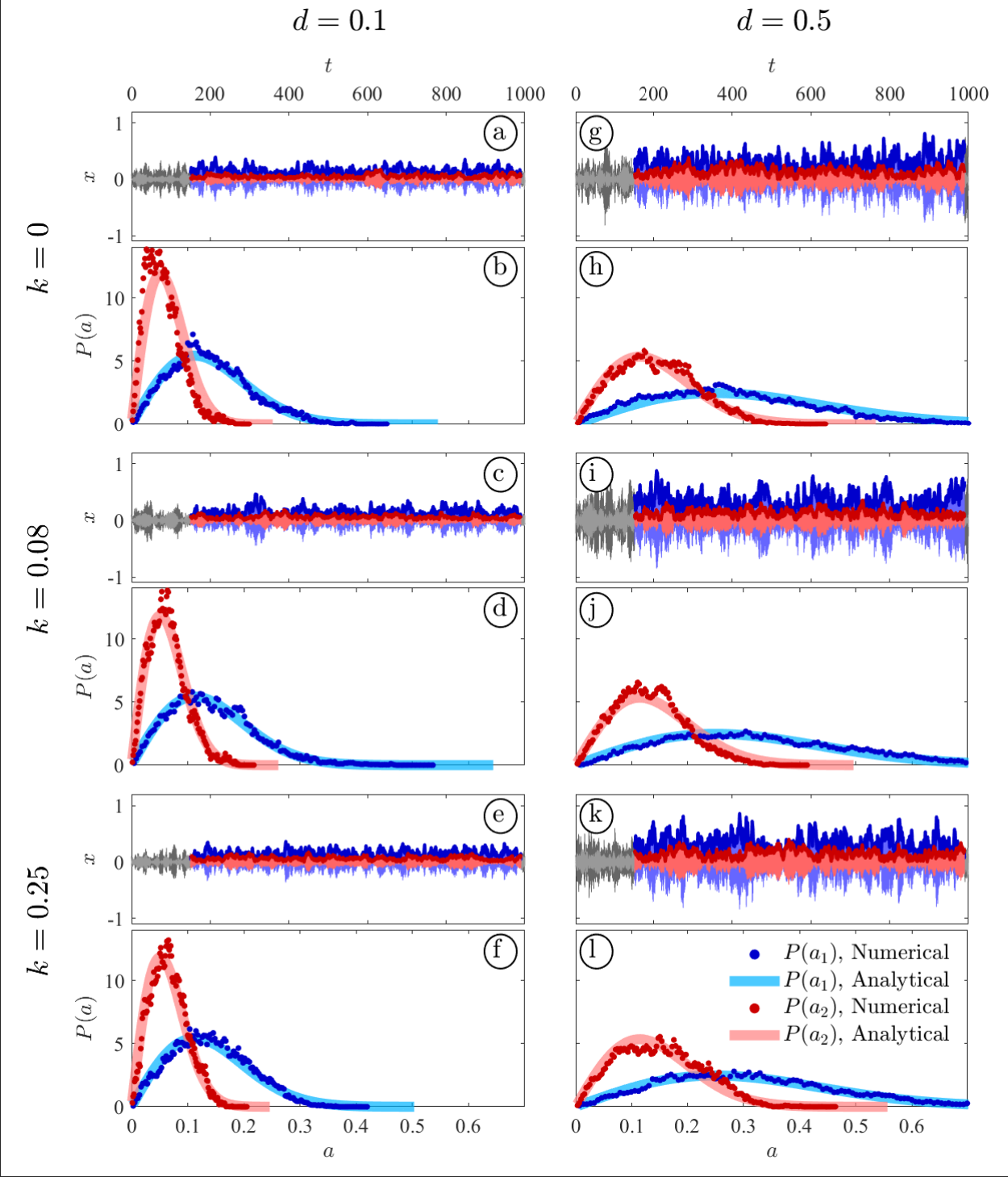}
    \caption{Distance coupling, case 1: fixed-point/fixed-point coupling. $\epsilon_1=-0.2$, $\epsilon_2=-0.1$, $\alpha_{11}=\alpha_{12}=-0.2$, $\omega_1=2\pi$,  $\omega_2=6.2\pi$, $d_1= d_2= d$. (a,g,c,i,e,k) time traces and (b,h,d,j,f,l) PDFs of the amplitude for (a,b,g,h) uncoupled, (c,d,i,j) weakly coupled, and (e,f,k.l) strongly coupled oscillators under (a-f) low noise and (g-l) high noise. Blue and red markers denote the first ($x_1$) and the second ($x_2$) oscillators, respectively. Scatter dots and thick lines show the numerical and analytical results, respectively.  Grey lines show the transient interval ($t<150$), which is not used in the analysis.}
\label{fig:d_fpfp}
\end{figure}

\begin{figure}
    \centering
    \includegraphics[clip,trim=1mm 1mm 1mm 1mm,width=0.85\textwidth]{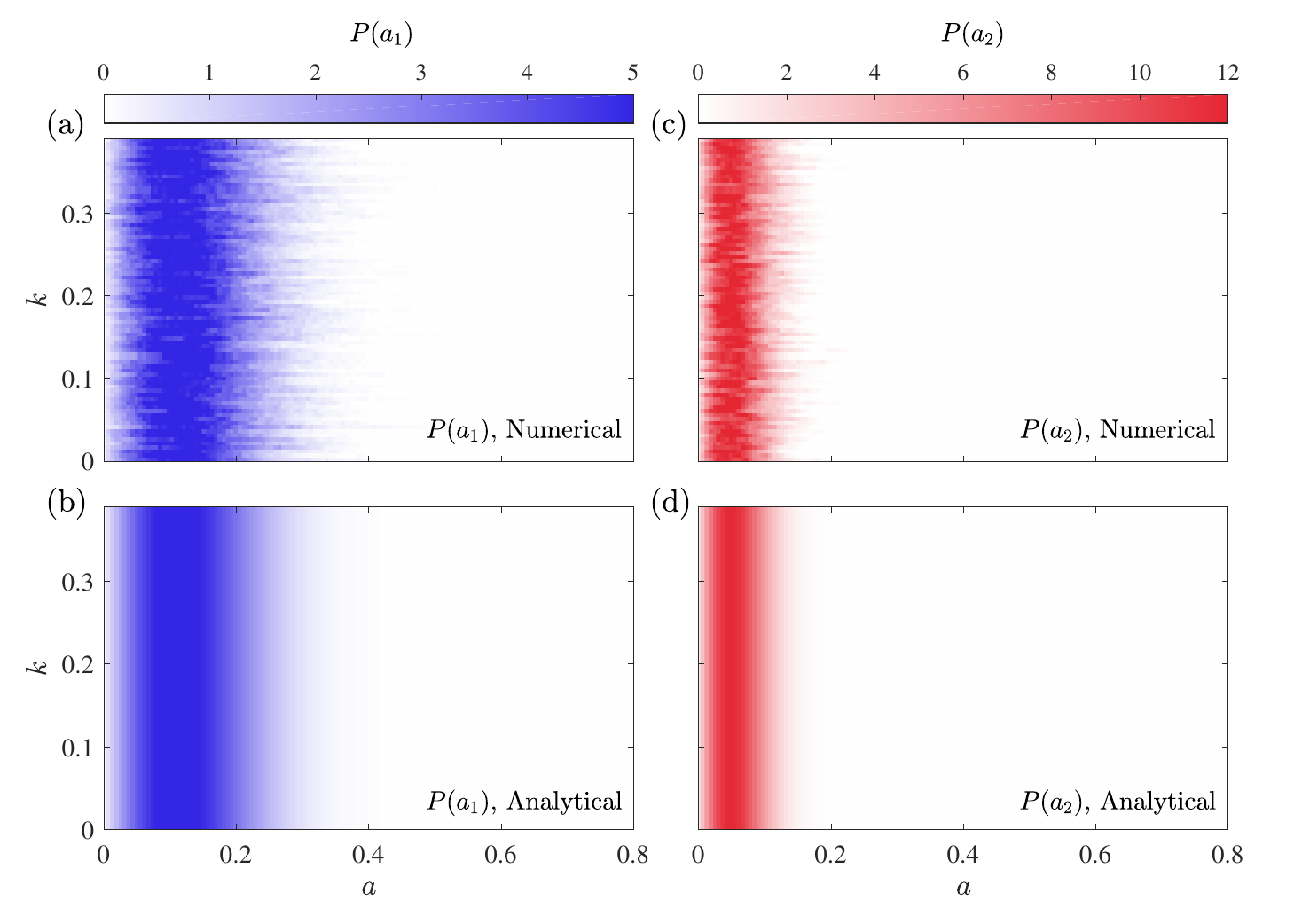}
    \caption{Distance coupling, case 1: fixed-point/fixed-point coupling. Effect of varying $k$ on (a,b) $P(a_1)$ and (c,d) $P(a_2)$, obtained from (a,c) numerical simulations and (b,d) equation \ref{cpl_sta}. $d_1=d_2=0.1$ and all other parameters are equal to figure \ref{fig:d_fpfp}.}
\label{fig:d_fpfp2}
\end{figure}

\begin{figure}
    \centering
    \includegraphics[clip,trim=1mm 1mm 1mm 1mm,width=0.85\textwidth]{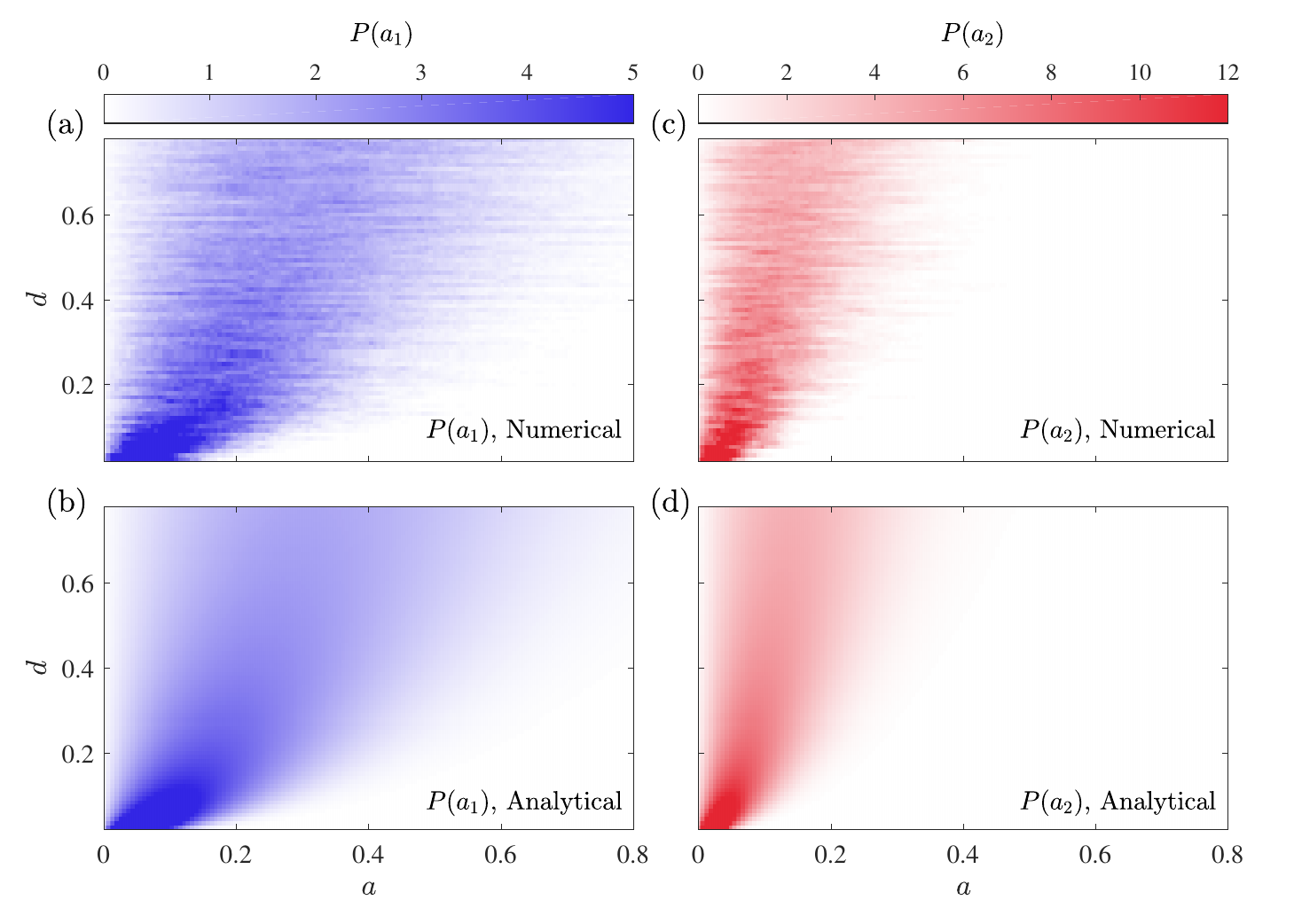}
    \caption{Distance coupling, case 1: fixed-point/fixed-point coupling. Effect of varying $d$ on (a,b) $P(a_1)$ and (c,d) $P(a_2)$, obtained from (a,c) numerical simulations and (b,d) equation \ref{cpl_sta}. $d_1=d_2=d$, $k=0.1$ and all other parameters are equal to figure \ref{fig:d_fpfp}.}
\label{fig:d_fpfp3}
\end{figure}

\begin{figure}
    \centering
    \includegraphics[clip,trim=1mm 1mm 1mm 1mm,width=\textwidth]{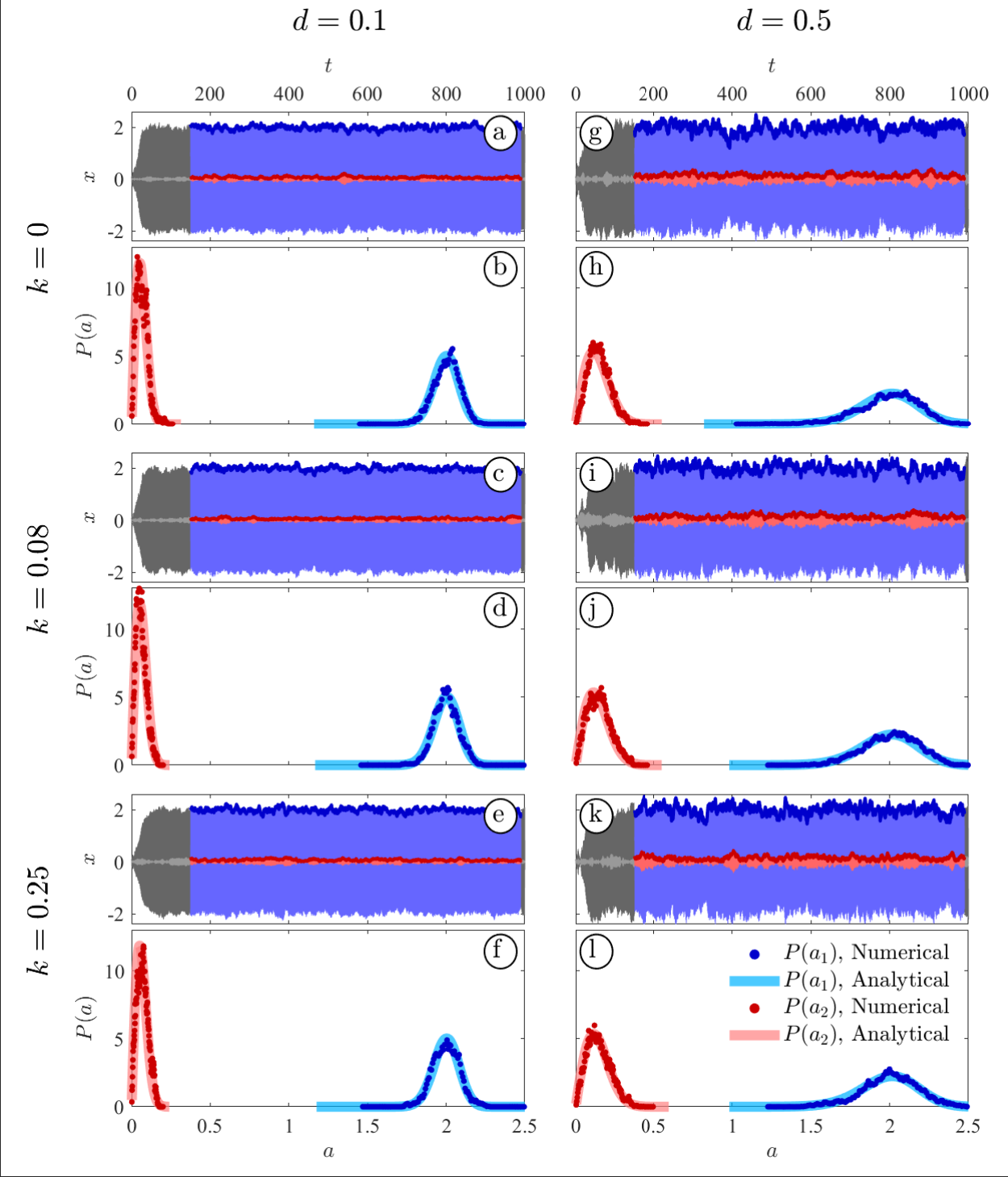}
    \caption{Distance coupling, case 2: limit-cycle/fixed-point coupling. $\epsilon_1=0.2$, $\epsilon_2=-0.1$, $\alpha_{11}=\alpha_{12}=-0.2$, $\omega_1=2\pi$,  $\omega_2=6.2\pi$, $d_1= d_2= d$. (a,g,c,i,e,k) time traces and (b,h,d,j,f,l) PDFs of the amplitude for (a,b,g,h) uncoupled, (c,d,i,j) weakly coupled, and (e,f,k.l) strongly coupled oscillators under (a-f) low noise and (g-l) high noise. Blue and red markers denote the first ($x_1$) and the second ($x_2$) oscillators, respectively. Scatter dots and thick lines show the numerical and analytical results, respectively.  Grey lines show the transient interval ($t<150$), which is not used in the analysis.}
\label{fig:d_lcfp}
\end{figure}

\begin{figure}
    \centering
    \includegraphics[clip,trim=1mm 1mm 1mm 1mm,width=0.85\textwidth]{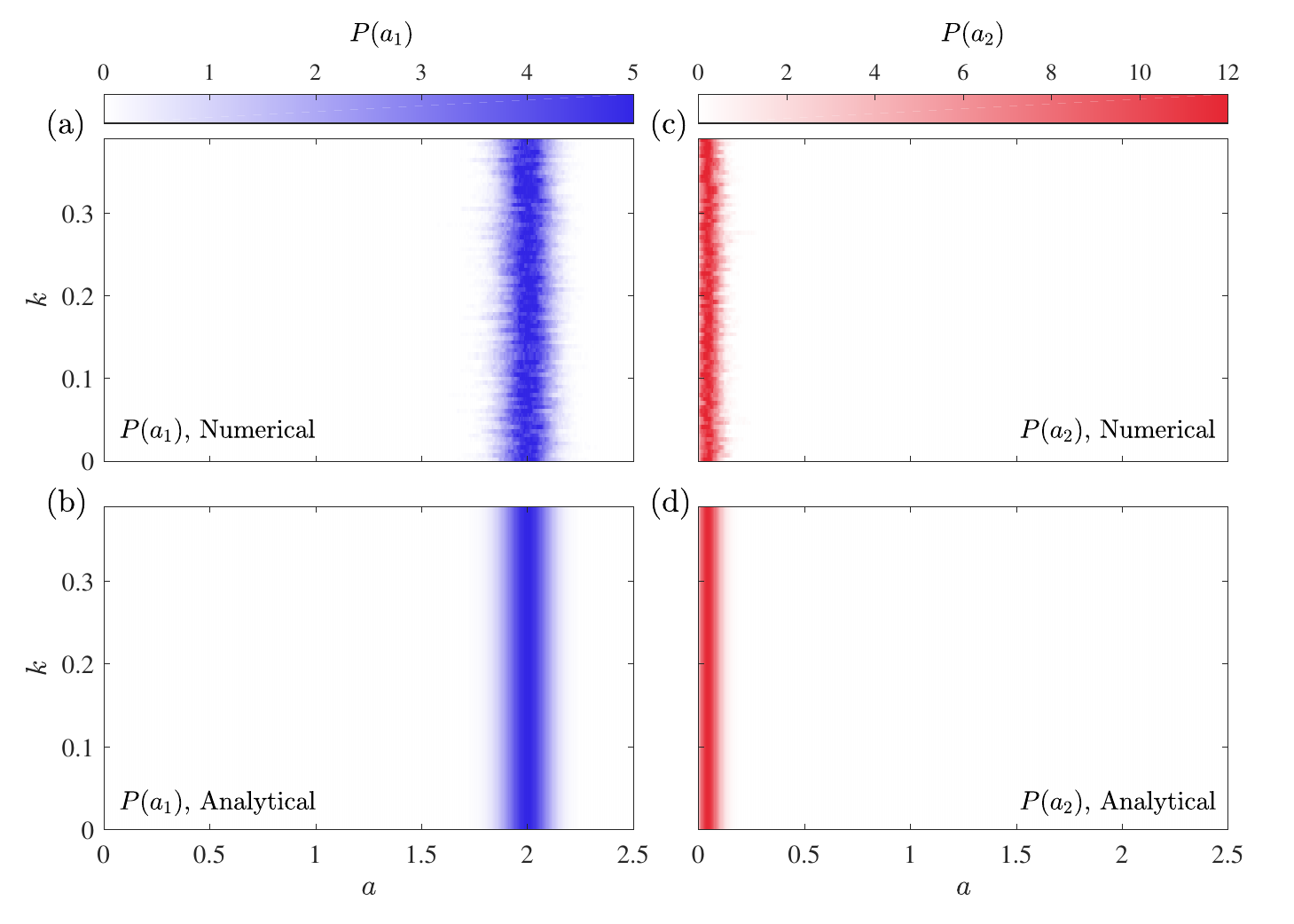}
    \caption{Distance coupling, case 2: limit-cycle/fixed-point coupling. Effect of varying $k$ on (a,b) $P(a_1)$ and (c,d) $P(a_2)$, obtained from (a,c) numerical simulations and (b,d) equation \ref{cpl_sta}. $d_1=d_2=0.1$ and all other parameters are equal to figure \ref{fig:d_lcfp}.}
\label{fig:d_lcfp2}
\end{figure}

\begin{figure}
    \centering
    \includegraphics[clip,trim=1mm 1mm 1mm 1mm,width=0.85\textwidth]{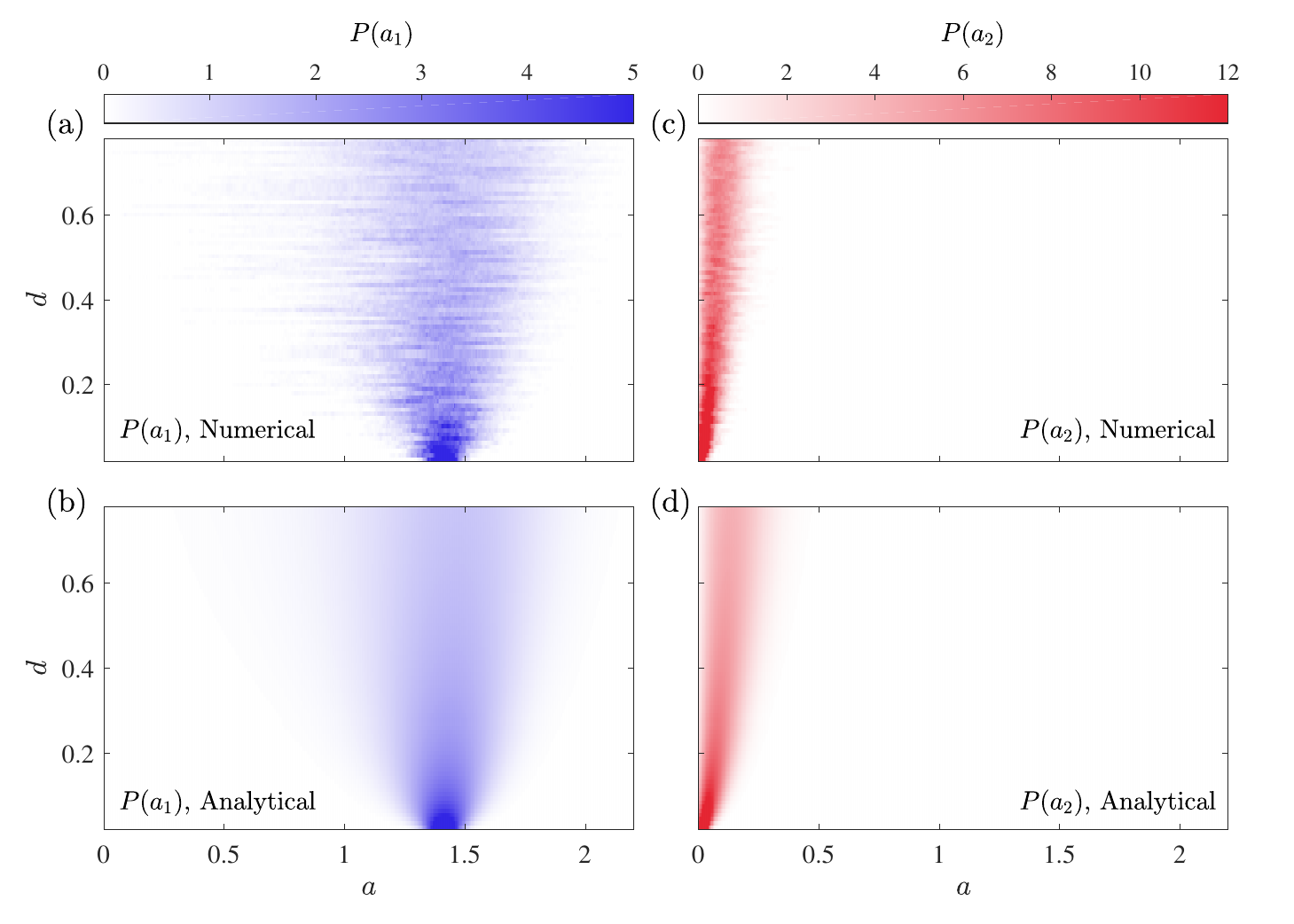}
    \caption{Distance coupling, case 2: limit-cycle/fixed-point coupling. Effect of varying $d$ on (a,b) $P(a_1)$ and (c,d) $P(a_2)$, obtained from (a,c) numerical simulations and (b,d) equation \ref{cpl_sta}. $d_1=d_2=d$, $k=0.1$ and all other parameters are equal to figure \ref{fig:d_lcfp}.}
\label{fig:d_lcfp3}
\end{figure}

\begin{figure}
    \centering
    \includegraphics[clip,trim=1mm 1mm 1mm 1mm,width=\textwidth]{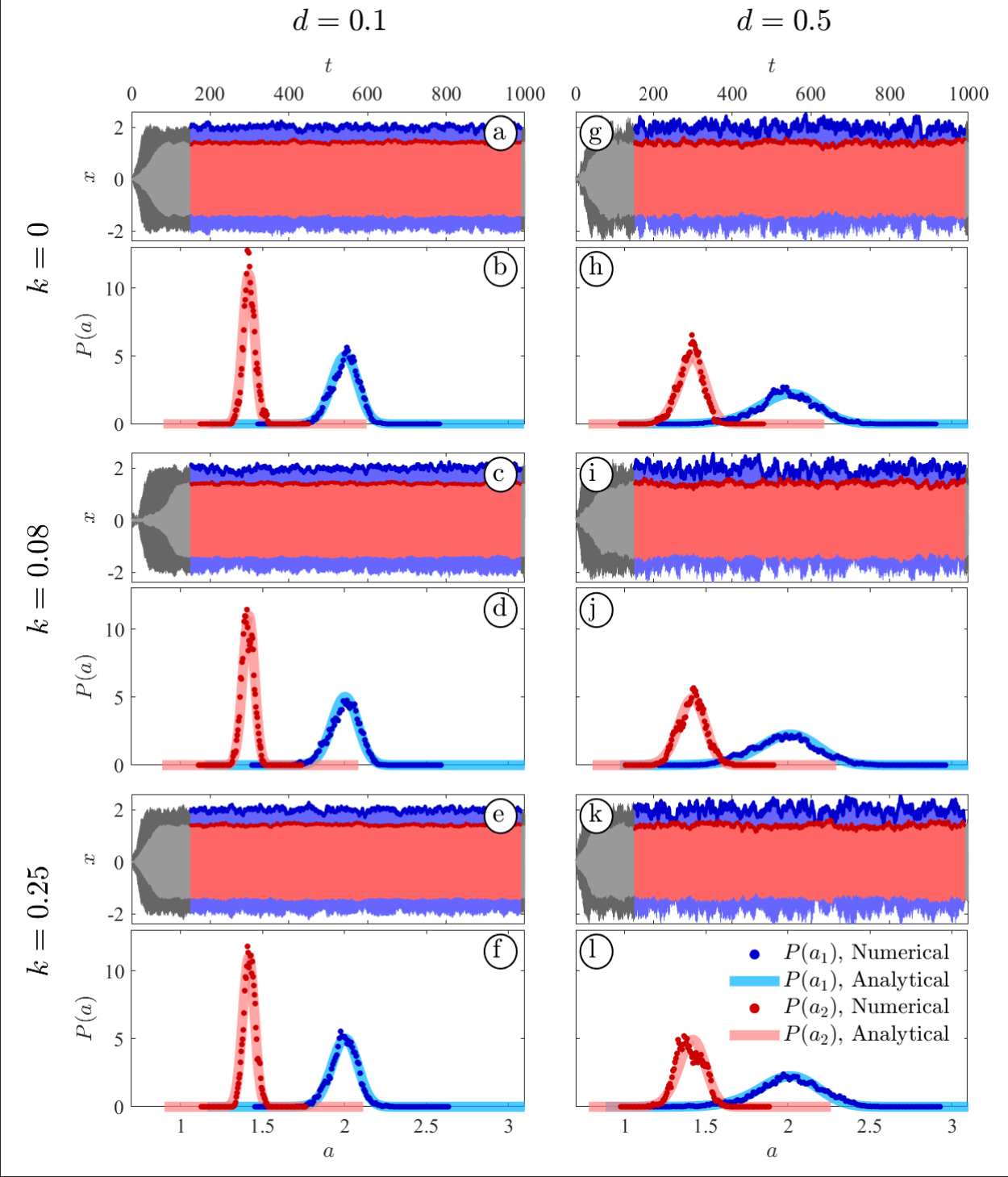}
    \caption{Distance coupling, case 3: limit-cycle/limit-cycle coupling. $\epsilon_1=0.2$, $\epsilon_2=0.1$, $\alpha_{11}=\alpha_{12}=-0.2$, $\omega_1=2\pi$,  $\omega_2=6.2\pi$, $d_1= d_2= d$. (a,g,c,i,e,k) time traces and (b,h,d,j,f,l) PDFs of the amplitude for (a,b,g,h) uncoupled, (c,d,i,j) weakly coupled, and (e,f,k.l) strongly coupled oscillators under (a-f) low noise and (g-l) high noise. Blue and red markers denote the first ($x_1$) and the second ($x_2$) oscillators, respectively. Scatter dots and thick lines show the numerical and analytical results, respectively. Grey lines show the transient interval ($t<150$), which is not used in the analysis.}
\label{fig:d_lclc}
\end{figure}

\begin{figure}
    \centering
    \includegraphics[clip,trim=1mm 1mm 1mm 1mm,width=0.85\textwidth]{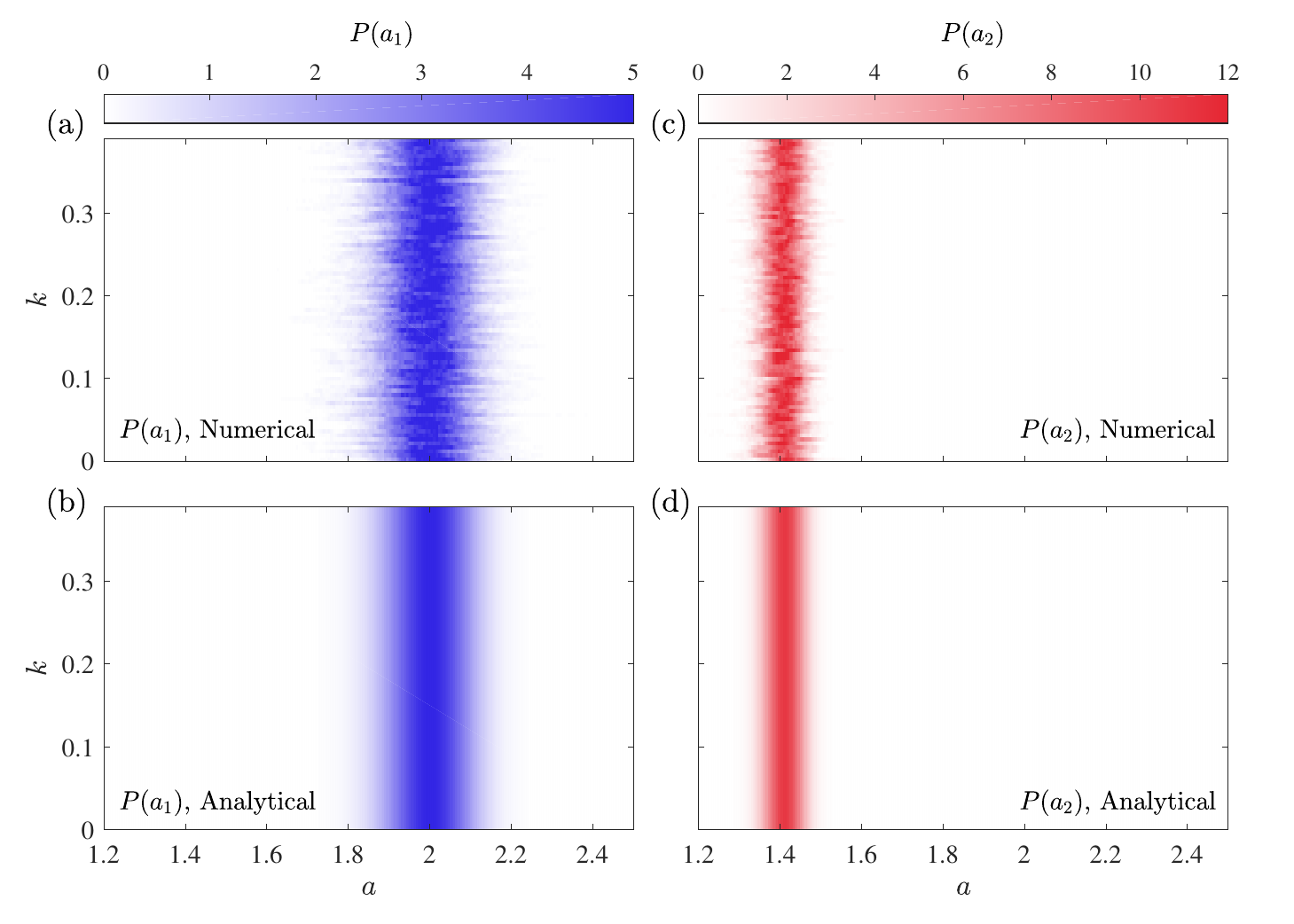}
    \caption{Distance coupling, case 3: limit-cycle/limit-cycle coupling. Effect of varying $k$ on (a,b) $P(a_1)$ and (c,d) $P(a_2)$, obtained from (a,c) numerical simulations and (b,d) equation \ref{cpl_sta}. $d_1=d_2=0.1$ and all other parameters are equal to figure \ref{fig:d_lclc}.}
\label{fig:d_lclc2}
\end{figure}

\begin{figure}
    \centering
    \includegraphics[clip,trim=1mm 1mm 1mm 1mm,width=0.85\textwidth]{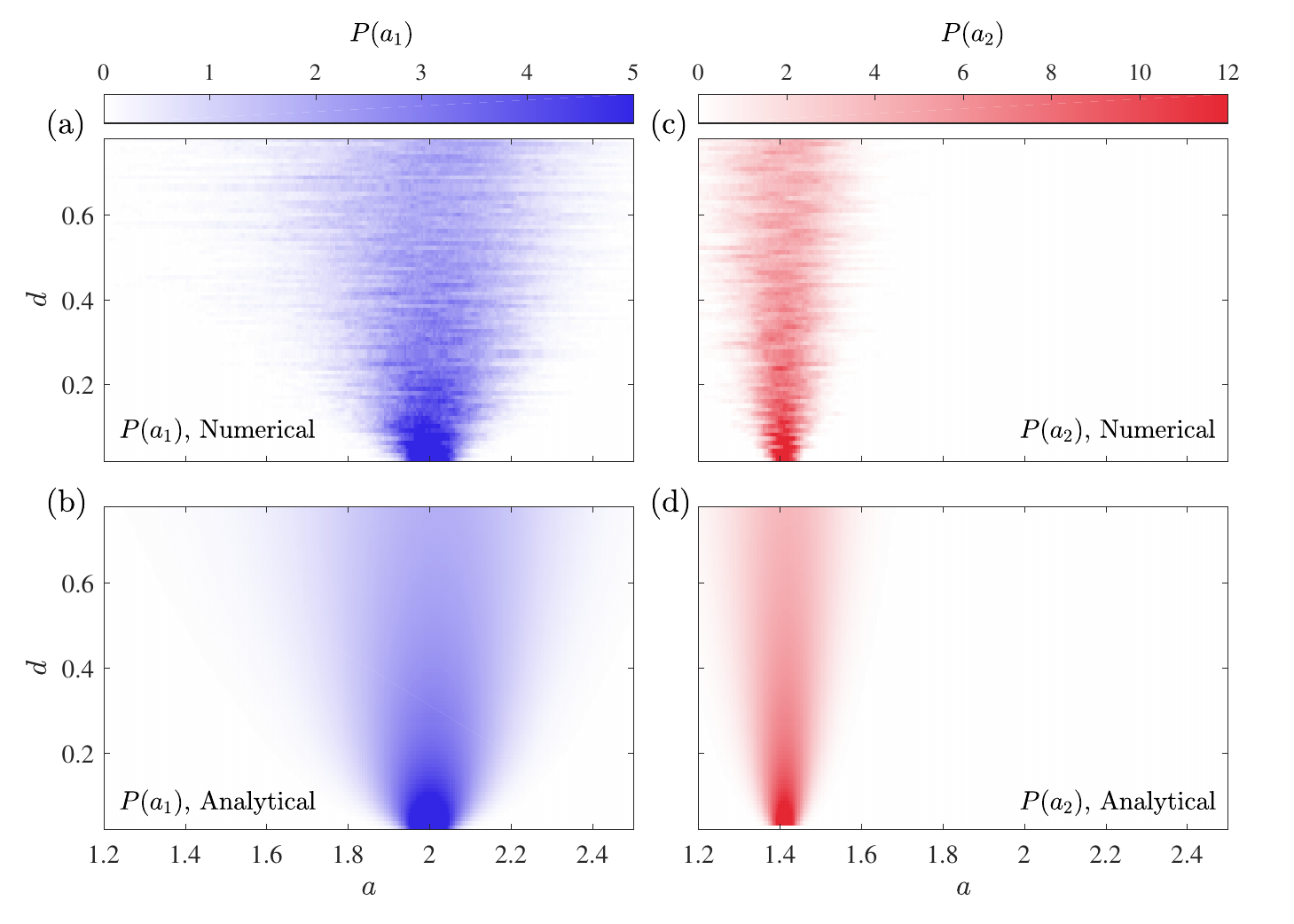}
    \caption{Distance coupling, case 3: limit-cycle/limit-cycle coupling. Effect of varying $d$ on (a,b) $P(a_1)$ and (c,d) $P(a_2)$, obtained from (a,c) numerical simulations and (b,d) equation \ref{cpl_sta}. $d_1=d_2=d$, $k=0.1$ and all other parameters are equal to figure \ref{fig:d_lclc}.}
\label{fig:d_lclc3}
\end{figure}

The time traces and the numerical/analytical PDFs for cases 1, 2 and 3 are shown in figures \ref{fig:d_fpfp}, \ref{fig:d_lcfp} and \ref{fig:d_lclc}, respectively. In all three cases, the analytical Fokker--Planck solutions agree well with the numerical simulations, regardless of the coupling strength ($k$) and the noise amplitude ($d$). 

The effect of varying $k$ is further shown in figures \ref{fig:d_fpfp2}, \ref{fig:d_lcfp2} and \ref{fig:d_lclc2}. It can be seen from all three cases that the coupling strength ($k$) does not affect the PDF of the fluctuation amplitude ($P(a)$). This observation is consistent with the fact that $k$ does not appear in the Fokker--Planck equation (equation \ref{cpl_sta}). Because the effect of $k$ is not reflected in $P(a)$, the distance-coupled VDP oscillators cannot be used as a system model if one intends to identify the coupling strength between the two modes via the Fokker-Planck equation. 

In addition, it is found that higher noise amplitudes ($d$) result in a higher mean and variation of $P(a)$, as shown in figures \ref{fig:d_fpfp3}, \ref{fig:d_lcfp3} and \ref{fig:d_lclc3}. This shows that the effect of noise on two distance-coupled VDP oscillators is well modeled with their corresponding Fokker--Planck equations.

\section{Velocity coupling}

\subsection{Derivation of the Fokker--Planck equation}

We consider two VDP oscillators coupled via the velocity terms:
\begin{subequations} \label{vel_cpl}
\begin{align}
    \ddot{x_1}-(\epsilon_1 + \alpha_{11}{x_1}^2)\dot{x_1} +& {\omega_1}^2 x_1 + k (\dot{x_1} - \dot{x_2}) = \sqrt{2 d_1} \eta_1, \\
    \ddot{x_2}-(\epsilon_2 + \alpha_{12}{x_2}^2)\dot{x_2} +& {\omega_2}^2 x_2 + k (\dot{x_2} - \dot{x_1}) = \sqrt{2 d_2} \eta_2,
\end{align}
\end{subequations}
where [$x_1$, $x_2$] are the state variables, [$\omega_1$, $\omega_2$] are the angular frequencies, [$\epsilon_1$, $\epsilon_2$] and [$\alpha_{11}$, $\alpha_{12}$] are the VDP coefficients, $k$ is the coupling strength, [$\eta_1$, $\eta_2$] are unit white Gaussian noise terms, and [$d_1$, $d_2$] are their amplitudes. The transformations in equation \ref{rel} can be used to express equation \ref{vel_cpl} in terms of the amplitude ($a$) and phase ($\phi$):
\begin{subequations} \label{eq5_9}
\begin{align}
    \dot{a_1} \cos{\Phi_1} - a_1 \dot{\phi_1} \sin{\Phi_1} &= 0, \\
    \dot{a_2} \cos{\Phi_2} - a_2 \dot{\phi_2} \sin{\Phi_2} &= 0, \\
    \dot{a_1} \sin{\Phi_1} + a_1 \dot{\phi_1} \cos{\Phi_1} &= \epsilon_1 a_1 \sin{\Phi_1} + \alpha_{11} {a_1}^3 \sin{\Phi_1} \cos^2{\Phi_1} \notag \\
    &\quad- k a_1 \sin{\Phi_1} + \frac{\omega_2}{\omega_1} k a_2 \sin{\Phi_2} - \frac{\sqrt{2 d_1}}{\omega_1} \eta_1,\\
    \dot{a_2} \sin{\Phi_2} + a_2 \dot{\phi_2} \cos{\Phi_2} &= \epsilon_2 a_2 \sin{\Phi_2} + \alpha_{12} {a_2}^3 \sin{\Phi_2} \cos^2{\Phi_2} \notag \\
    &\quad- k a_2 \sin{\Phi_2} + \frac{\omega_1}{\omega_2} k a_1 \sin{\Phi_1} - \frac{\sqrt{2 d_2}}{\omega_2} \eta_2,
\end{align}
\end{subequations}
where $\Phi_1=\omega_1 t + \phi_1$ and $\Phi_2=\omega_2 t + \phi_2$. Equation \ref{eq5_9} can be transformed into the following explicit ordinary differential equations for the amplitude and phase:
\begin{subequations} \label{eq5_10}
\begin{align}
    \dot{a_1} &= \frac{\epsilon_1 a_1}{2} + \frac{\alpha_{11} {a_1}^3}{8} - \frac{\epsilon_1 a_1}{2} \cos{2\Phi_1} - \frac{\alpha_{11} {a_1}^3}{8} \cos{4\Phi_1} - \frac{k a_1}{2} +\frac{k a_1}{2} \cos{2 \Phi_1} \notag \\ 
    &\quad- \frac{k \omega_2 a_2}{2 \omega_1} \cos{(\Phi_2 + \Phi_1)} + \frac{k \omega_2 a_2}{2 \omega_1} \cos{(\Phi_2 - \Phi_1)} - \frac{\sqrt{2d_1}}{\omega_1}(\sin{\Phi_1}) \eta_1,\\
    \dot{a_2} &= \frac{\epsilon_2 a_2}{2} + \frac{\alpha_{12} {a_2}^3}{8} - \frac{\epsilon_2 a_2}{2} \cos{2\Phi_2} - \frac{\alpha_{12} {a_2}^3}{8} \cos{4\Phi_2} - \frac{k a_2}{2} +\frac{k a_2}{2} \cos{2 \Phi_2} \notag \\ 
    &\quad- \frac{k \omega_1 a_1}{2 \omega_2} \cos{(\Phi_1 + \Phi_2)} + \frac{k \omega_1 a_1}{2 \omega_2} \cos{(\Phi_1 - \Phi_2)} - \frac{\sqrt{2d_2}}{\omega_2}(\sin{\Phi_2}) \eta_2,\\
    \dot{\phi_1} &= \frac{\epsilon_1}{2}\sin{2\Phi_1} + \frac{\alpha_{11} {a_1}^2}{4}\sin{2\Phi_1} + \frac{\alpha_{11} {a_1}^2}{8}\sin{4\Phi_1} - \frac{k}{2}\sin{2\Phi_1} \notag \\
    &\quad+ \frac{k \omega_2 a_2}{2 \omega_1 a_1} \sin{(\Phi_2 + \Phi_1)} + \frac{k \omega_2 a_2}{2 \omega_1 a_1} \sin{(\Phi_2 - \Phi_1)} - \frac{\sqrt{2d_1}}{\omega_1 a_1} (\cos{\Phi_1}) \eta_1, \\
    \dot{\phi_2} &= \frac{\epsilon_2}{2}\sin{2\Phi_2} + \frac{\alpha_{12} {a_2}^2}{4}\sin{2\Phi_2} + \frac{\alpha_{12} {a_2}^2}{8}\sin{4\Phi_2} - \frac{k}{2}\sin{2\Phi_2} \notag \\
    &\quad+ \frac{k \omega_1 a_1}{2 \omega_2 a_2} \sin{(\Phi_1 + \Phi_2)} + \frac{k \omega_1 a_1}{2 \omega_2 a_2} \sin{(\Phi_1 - \Phi_2)} - \frac{\sqrt{2d_2}}{\omega_2 a_2} (\cos{\Phi_2}) \eta_2.
\end{align}
\end{subequations}

The last terms in equation \ref{eq5_10} are the stochastic components, and the other terms are the deterministic components. The drift and diffusion terms of the amplitude ($a$) can be found via stochastic averaging \citep{stratonovich1963, stratonovich1967, roberts1986stochastic}:
\begin{subequations} 
\begin{align}
    \mathbf{m}(a_1) &= \frac{\epsilon_1 a_1}{2} + \frac{\alpha_{11} {a_1}^3}{8} - \frac{k a_1}{2} + \frac{d_1}{2 \omega_1 a_1}, \\
    \mathbf{m}(a_2) &= \frac{\epsilon_2 a_2}{2} + \frac{\alpha_{12} {a_2}^3}{8} - \frac{k a_2}{2} + \frac{d_2}{2 \omega_2 a_2}, \\
    \mathbf{\sigma}^2(a_1) &= \frac{d_1}{{\omega_1}^2}, \\
    \mathbf{\sigma}^2(a_2) &= \frac{d_2}{{\omega_2}^2},
\end{align}
\end{subequations}
where $\mathbf{m}$ and $\mathbf{\sigma}$ represent the drift and diffusion terms, respectively, which form part of the Fokker--Planck equation (equation \ref{eq5_7}). The corresponding stationary solution is given by:
\begin{subequations} \label{cpl_sta_vel}
\begin{align}
    P(a_1) = C_1 a_1 \exp\Big[\frac{(\epsilon_1 - k){\omega_1}^2}{2d_1} {a_1}^2 + \frac{\alpha_{11} {\omega_1}^2}{16d_1} {a_1}^4 \Big], \\
    P(a_2) = C_2 a_2 \exp\Big[\frac{(\epsilon_2 - k){\omega_2}^2}{2d_2} {a_2}^2 + \frac{\alpha_{12} {\omega_2}^2}{16d_2} {a_2}^4 \Big],
\end{align}
\end{subequations}
where $P(a_1)$ and $P(a_2)$ are the stationary probability functions of $a_1$ and $a_2$, respectively, and $C_1$ and $C_2$ are normalization constants. Unlike in the distance-coupling case (equation \ref{cpl_sta}), the coupling strength $k$ appears in the Fokker--Planck equation of the velocity-coupled oscillators. 

\subsection{Numerical analysis and validation} \label{num_ana_vel}
To numerically validate equations \ref{vel_cpl} and \ref{cpl_sta_vel}, we consider the same three cases considered earlier in \S\ref{num_ana_dis} for distance coupling (see table \ref{num_cases1}). For all three cases, six combinations of the coupling strengths and noise amplitudes are tested (see table \ref{num_cases2}). In addition, the effects of the coupling strength and noise amplitude are individually analyzed. In these analyses, numerical simulations are conducted with varying $k$ and fixed $d$, and vice versa. The results of the simulations are compared with the analytical solution obtained from the Fokker--Planck equation (equation \ref{cpl_sta_vel}).

\begin{figure}
    \centering
    \includegraphics[clip,trim=1mm 1mm 1mm 1mm,width=\textwidth]{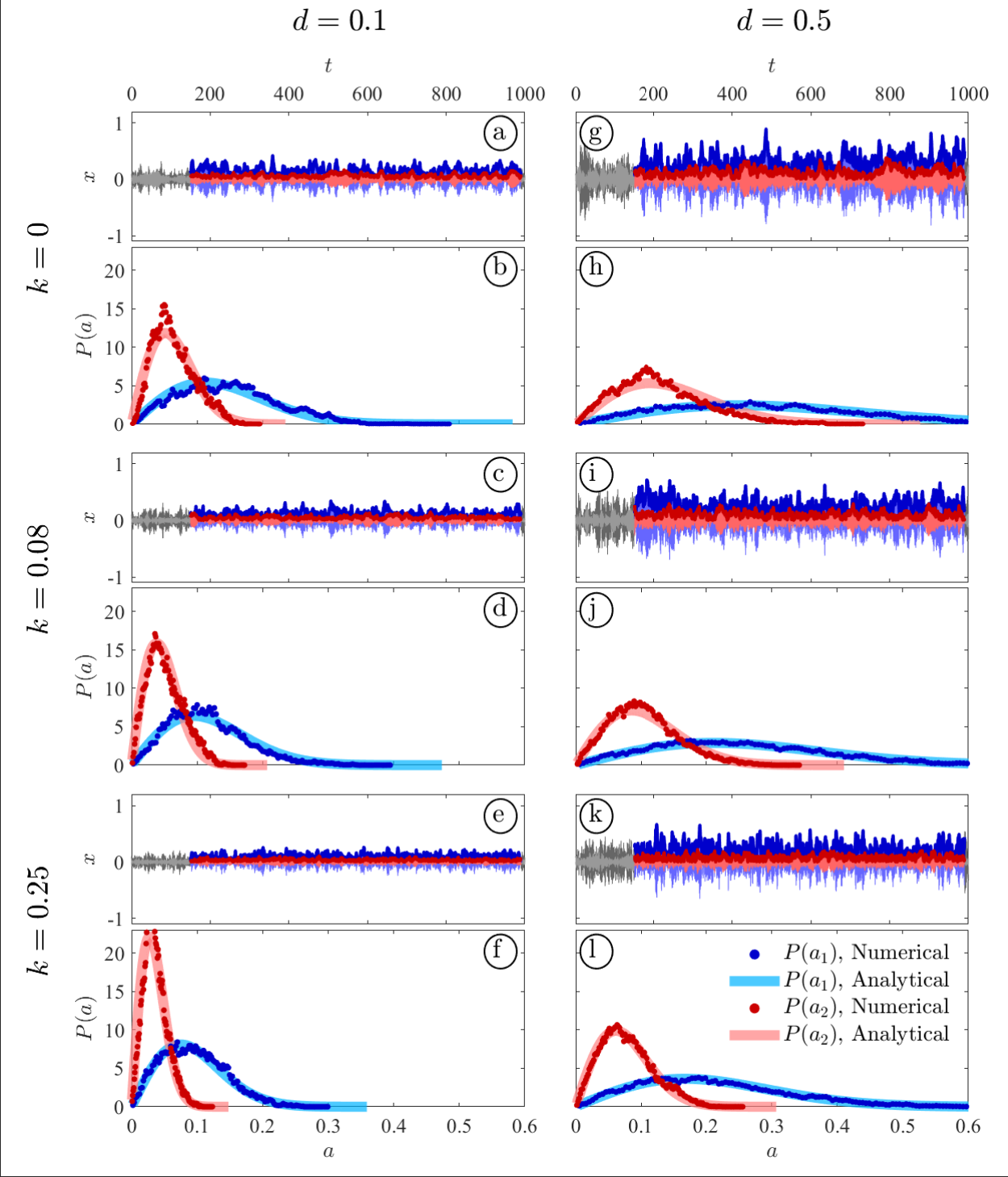}
    \caption{Velocity coupling, case 1: fixed-point/fixed-point coupling. $\epsilon_1=-0.2$, $\epsilon_2=-0.1$, $\alpha_{11}=\alpha_{12}=-0.2$, $\omega_1=2\pi$,  $\omega_2=6.2\pi$, $d_1= d_2= d$. (a,g,c,i,e,k) time traces and (b,h,d,j,f,l) PDFs of the amplitude for (a,b,g,h) uncoupled, (c,d,i,j) weakly coupled, and (e,f,k.l) strongly coupled oscillators under (a-f) low noise and (g-l) high noise. Blue and red markers denote the first ($x_1$) and the second ($x_2$) oscillators, respectively. Scatter dots and thick lines show the numerical and analytical results, respectively.  Grey lines show the transient interval ($t<150$), which is not used in the analysis.}
\label{fig:v_fpfp}
\end{figure}

\begin{figure}
    \centering
    \includegraphics[clip,trim=1mm 1mm 1mm 1mm,width=0.85\textwidth]{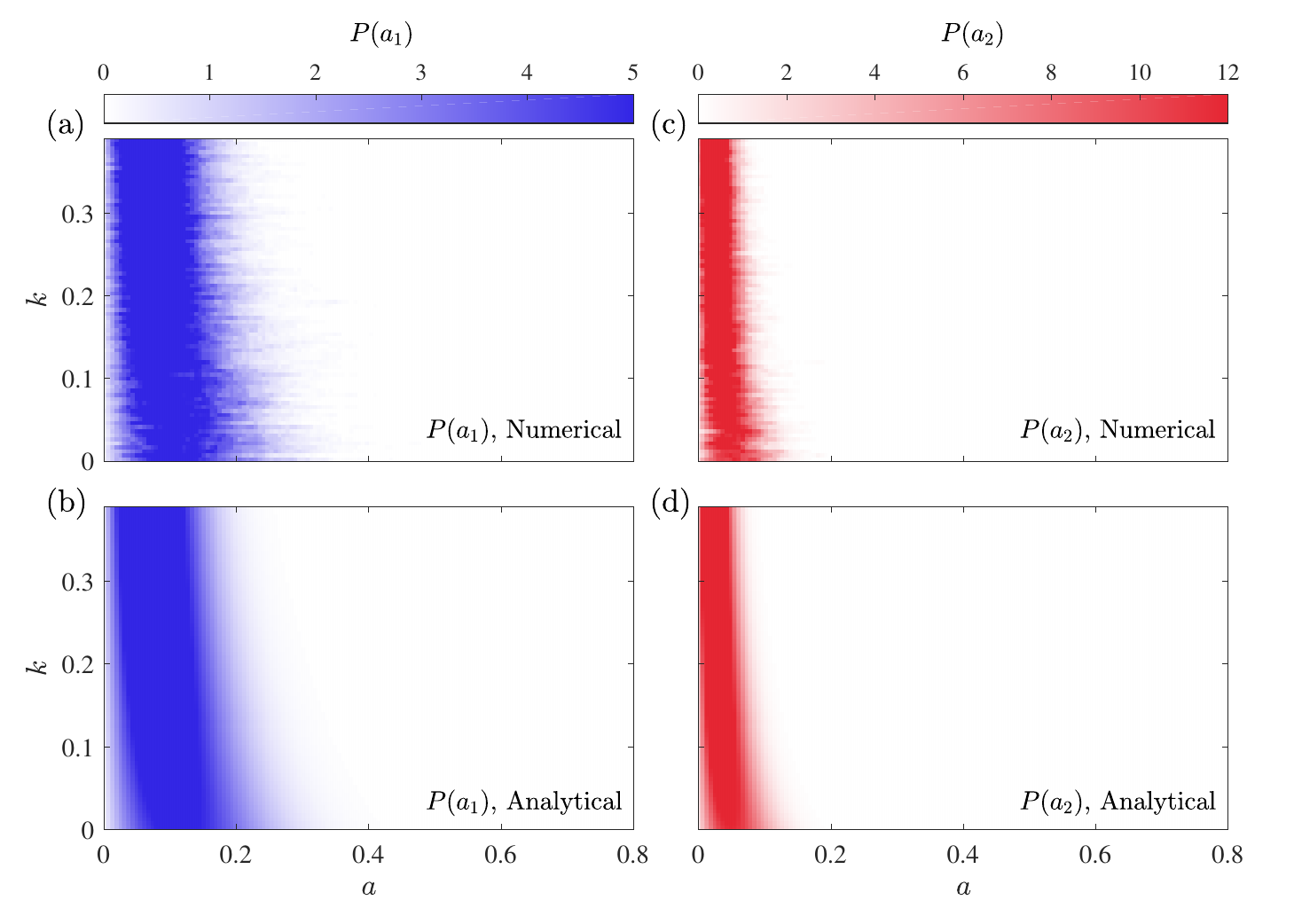}
    \caption{Velocity coupling, case 1: fixed-point/fixed-point coupling. Effect of varying $k$ on (a,b) $P(a_1)$ and (c,d) $P(a_2)$, obtained from (a,c) numerical simulations and (b,d) equation \ref{cpl_sta_vel}. $d_1=d_2=0.1$ and all other parameters are equal to figure \ref{fig:v_fpfp}.}
\label{fig:v_fpfp2}
\end{figure}

\begin{figure}
    \centering
    \includegraphics[clip,trim=1mm 1mm 1mm 1mm,width=0.85\textwidth]{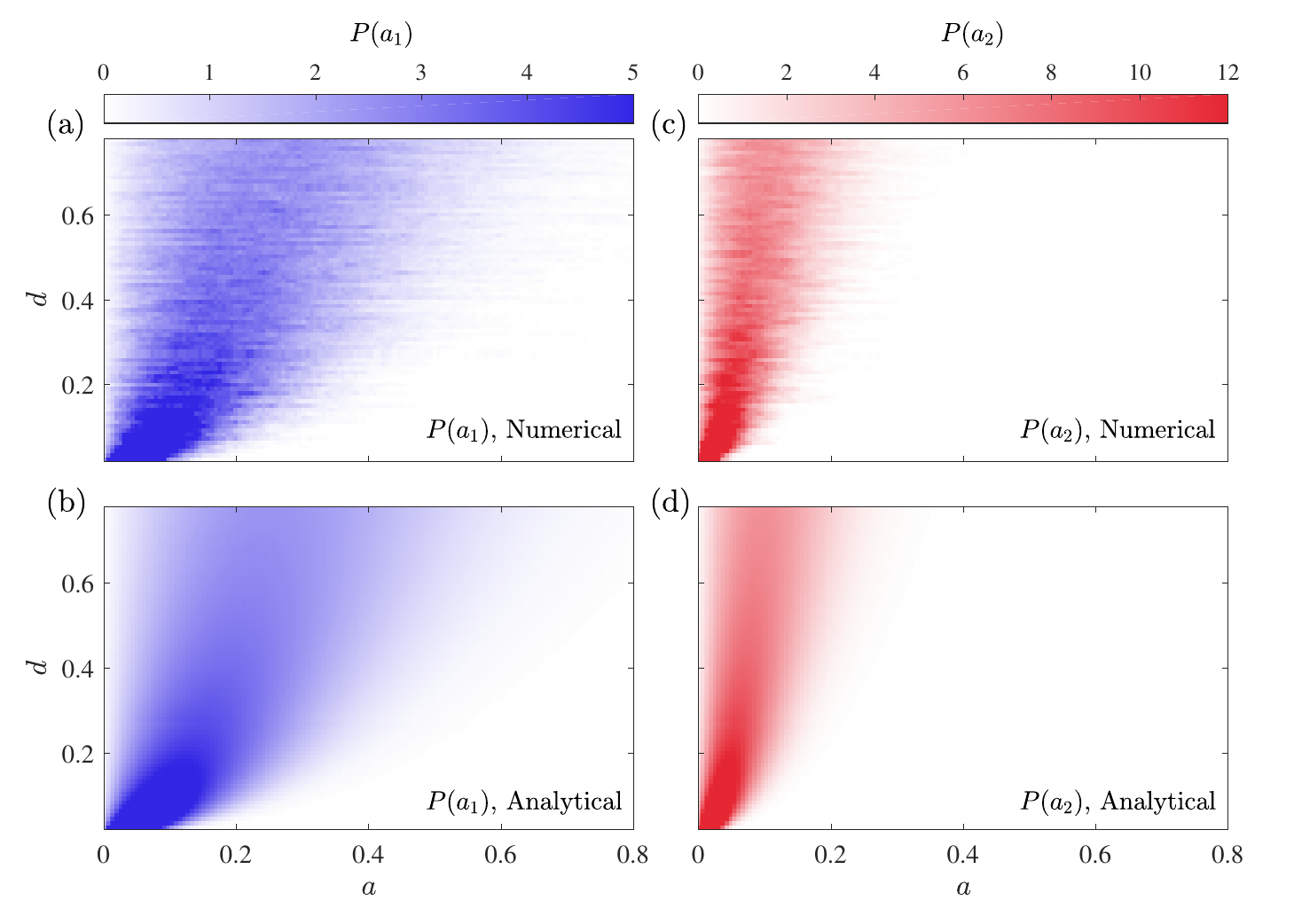}
    \caption{Velocity coupling, case 1: fixed-point/fixed-point coupling. Effect of varying $d$ on (a,b) $P(a_1)$ and (c,d) $P(a_2)$, obtained from (a,c) numerical simulations and (b,d) equation \ref{cpl_sta_vel}. $d_1=d_2=d$, $k=0.1$ and all other parameters are equal to figure \ref{fig:v_fpfp}.}
\label{fig:v_fpfp3}
\end{figure}

\begin{figure}
    \centering
    \includegraphics[clip,trim=1mm 1mm 1mm 1mm,width=\textwidth]{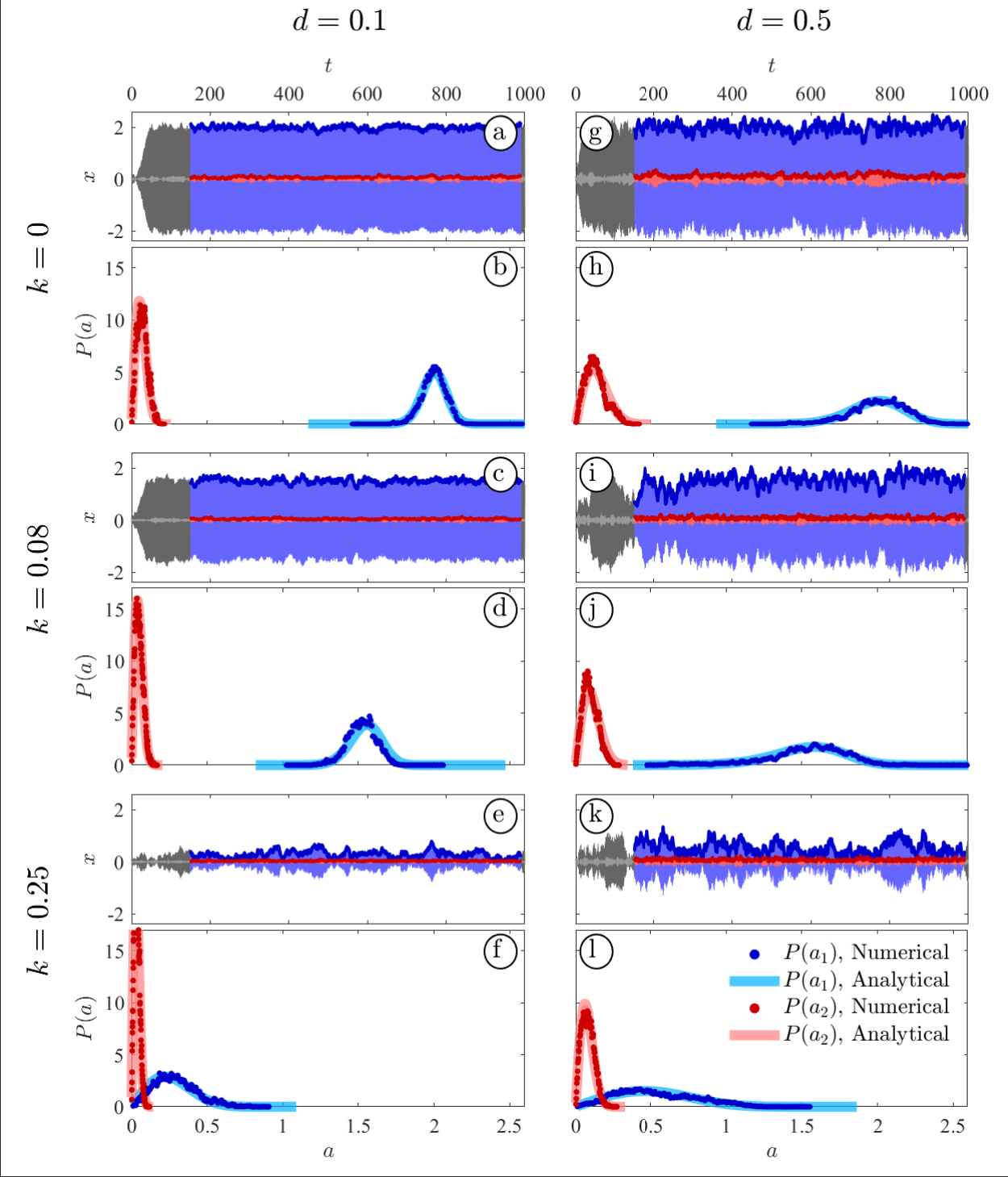}
    \caption{Velocity coupling, case 2: limit-cycle/fixed-point coupling. $\epsilon_1=0.2$, $\epsilon_2=-0.1$, $\alpha_{11}=\alpha_{12}=-0.2$, $\omega_1=2\pi$,  $\omega_2=6.2\pi$, $d_1= d_2= d$. (a,g,c,i,e,k) time traces and (b,h,d,j,f,l) PDFs of the amplitude for (a,b,g,h) uncoupled, (c,d,i,j) weakly coupled, and (e,f,k.l) strongly coupled oscillators under (a-f) low noise and (g-l) high noise. Blue and red markers denote the first ($x_1$) and the second ($x_2$) oscillators, respectively. Scatter dots and thick lines show the numerical and analytical results, respectively.  Grey lines show the transient interval ($t<150$), which is not used in the analysis.}
\label{fig:v_lcfp}
\end{figure}

\begin{figure}
    \centering
    \includegraphics[clip,trim=1mm 1mm 1mm 1mm,width=0.85\textwidth]{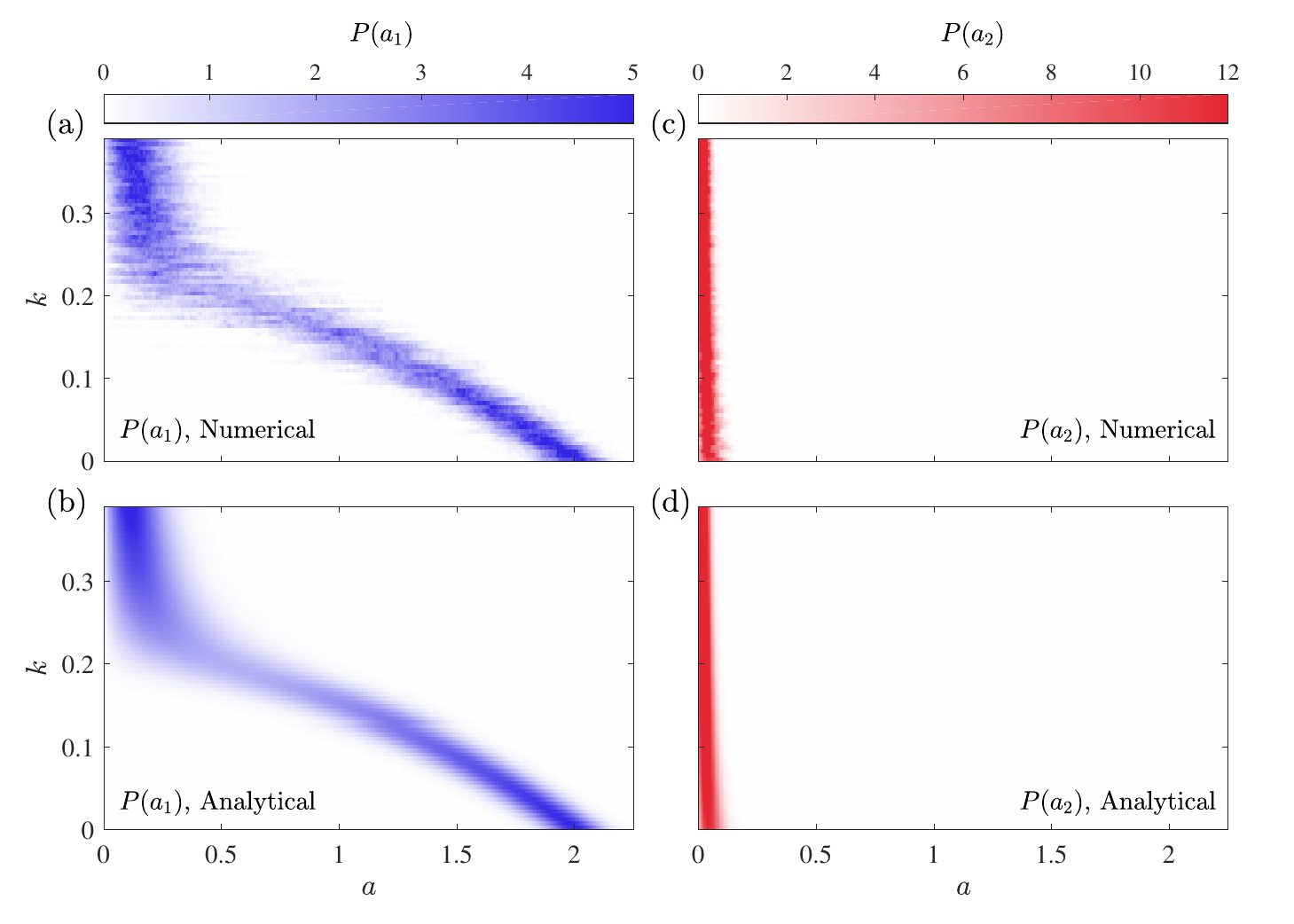}
    \caption{Velocity coupling, case 2: limit-cycle/fixed-point coupling. Effect of varying $k$ on (a,b) $P(a_1)$ and (c,d) $P(a_2)$, obtained from (a,c) numerical simulations and (b,d) equation \ref{cpl_sta_vel}. $d_1=d_2=0.1$ and all other parameters are equal to figure \ref{fig:v_lcfp}.}
\label{fig:v_lcfp2}
\end{figure}

\begin{figure}
    \centering
    \includegraphics[clip,trim=1mm 1mm 1mm 1mm,width=0.85\textwidth]{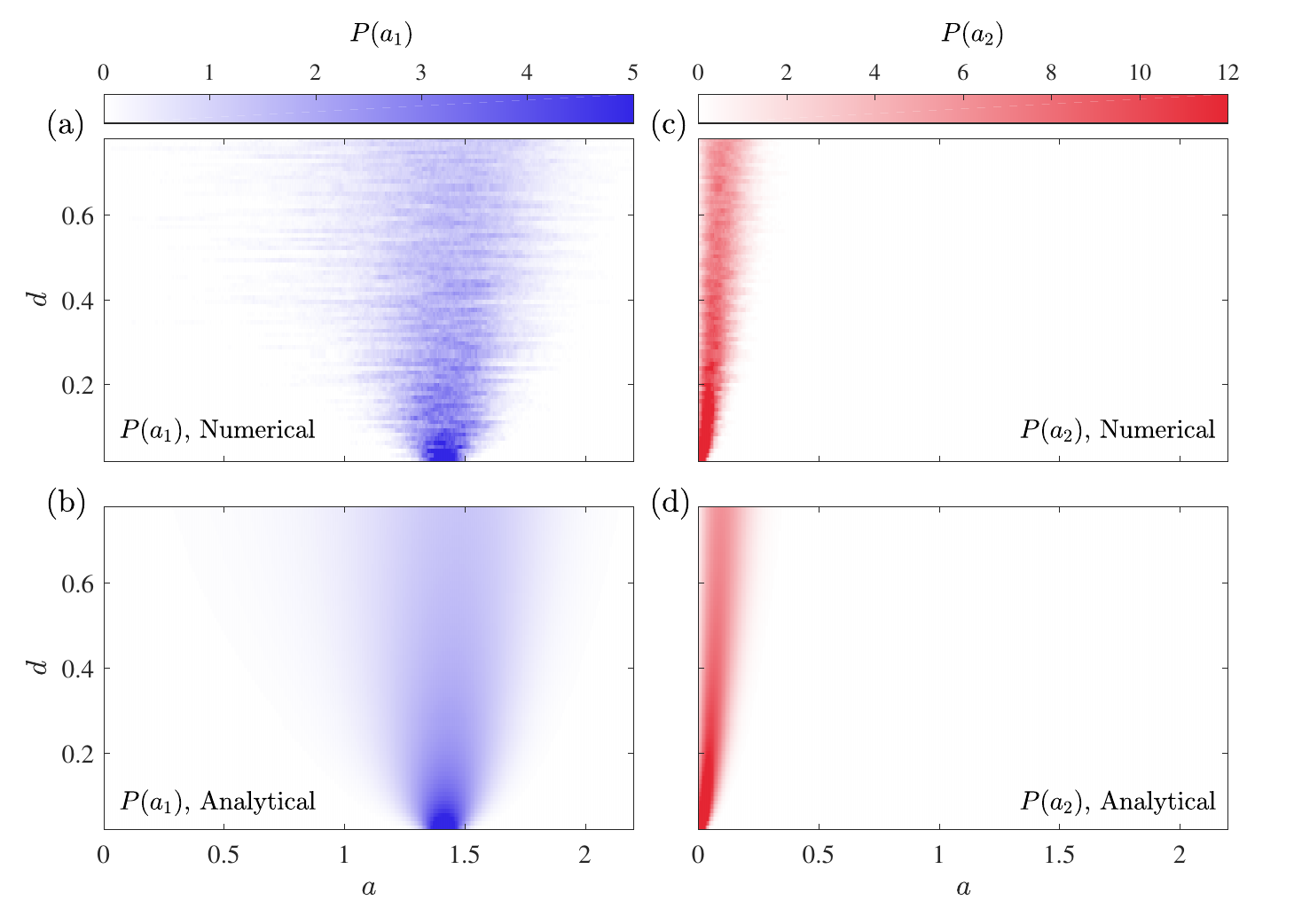}
    \caption{Velocity coupling, case 2: limit-cycle/fixed-point coupling. Effect of varying $d$ on (a,b) $P(a_1)$ and (c,d) $P(a_2)$, obtained from (a,c) numerical simulations and (b,d) equation \ref{cpl_sta_vel}. $d_1=d_2=d$, $k=0.1$ and all other parameters are equal to figure \ref{fig:v_lcfp}.}
\label{fig:v_lcfp3}
\end{figure}

\begin{figure}
    \centering
    \includegraphics[clip,trim=1mm 1mm 1mm 1mm,width=\textwidth]{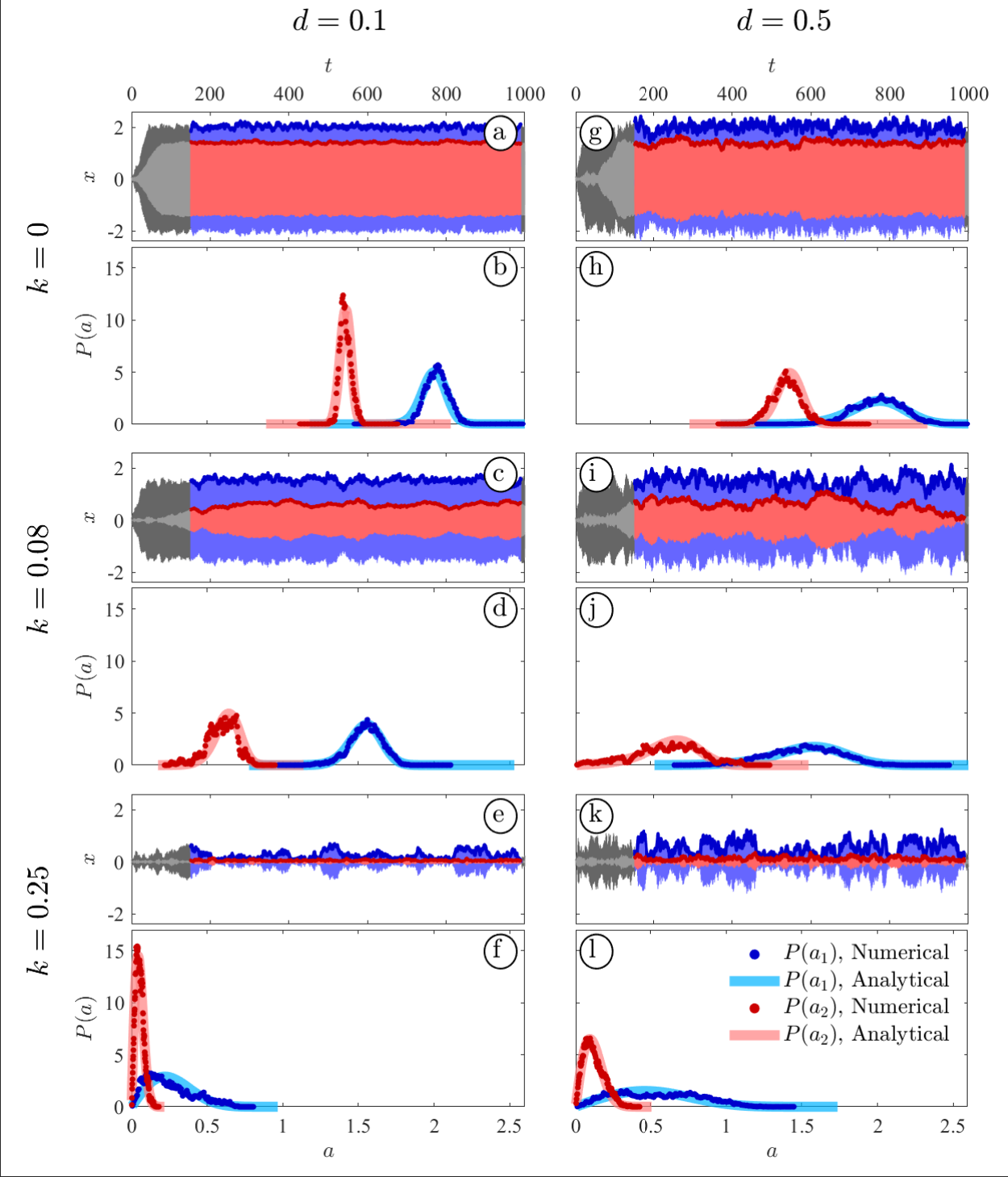}
    \caption{Velocity coupling, case 3: limit-cycle/limit-cycle coupling. $\epsilon_1=0.2$, $\epsilon_2=0.1$, $\alpha_{11}=\alpha_{12}=-0.2$, $\omega_1=2\pi$,  $\omega_2=6.2\pi$, $d_1= d_2= d$. (a,g,c,i,e,k) time traces and (b,h,d,j,f,l) PDFs of the amplitude for (a,b,g,h) uncoupled, (c,d,i,j) weakly coupled, and (e,f,k.l) strongly coupled oscillators under (a-f) low noise and (g-l) high noise. Blue and red markers denote the first ($x_1$) and the second ($x_2$) oscillators, respectively. Scatter dots and thick lines show the numerical and analytical results, respectively.  Grey lines show the transient interval ($t<150$), which is not used in the analysis.}
\label{fig:v_lclc}
\end{figure}

\begin{figure}
    \centering
    \includegraphics[clip,trim=1mm 1mm 1mm 1mm,width=0.85\textwidth]{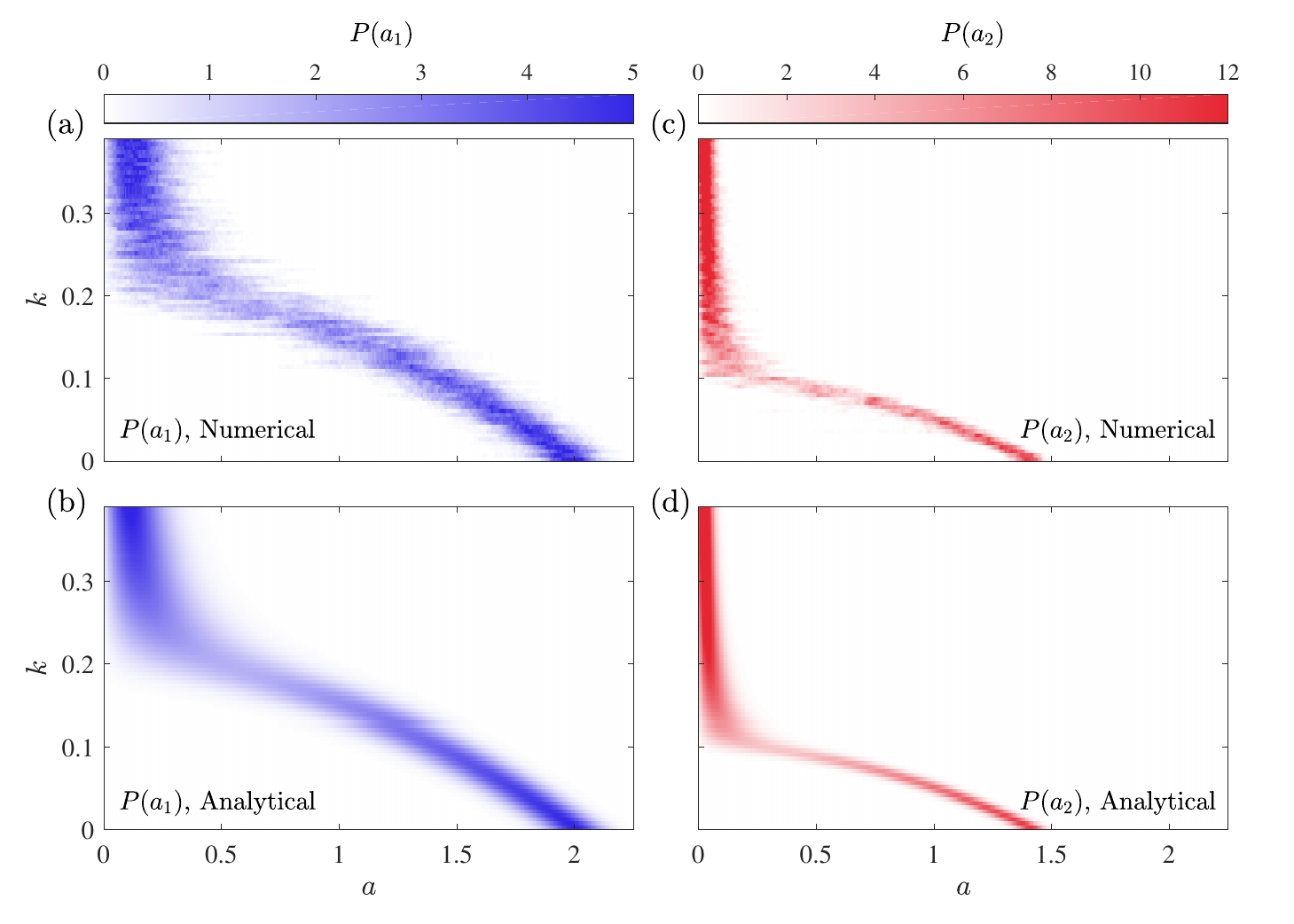}
    \caption{Velocity coupling, case 3: limit-cycle/limit-cycle coupling. Effect of varying $k$ on (a,b) $P(a_1)$ and (c,d) $P(a_2)$, obtained from (a,c) numerical simulations and (b,d) equation \ref{cpl_sta_vel}. $d_1=d_2=0.1$ and all other parameters are equal to figure \ref{fig:v_lclc}.}
\label{fig:v_lclc2}
\end{figure}

\begin{figure}
    \centering
    \includegraphics[clip,trim=1mm 1mm 1mm 1mm,width=0.85\textwidth]{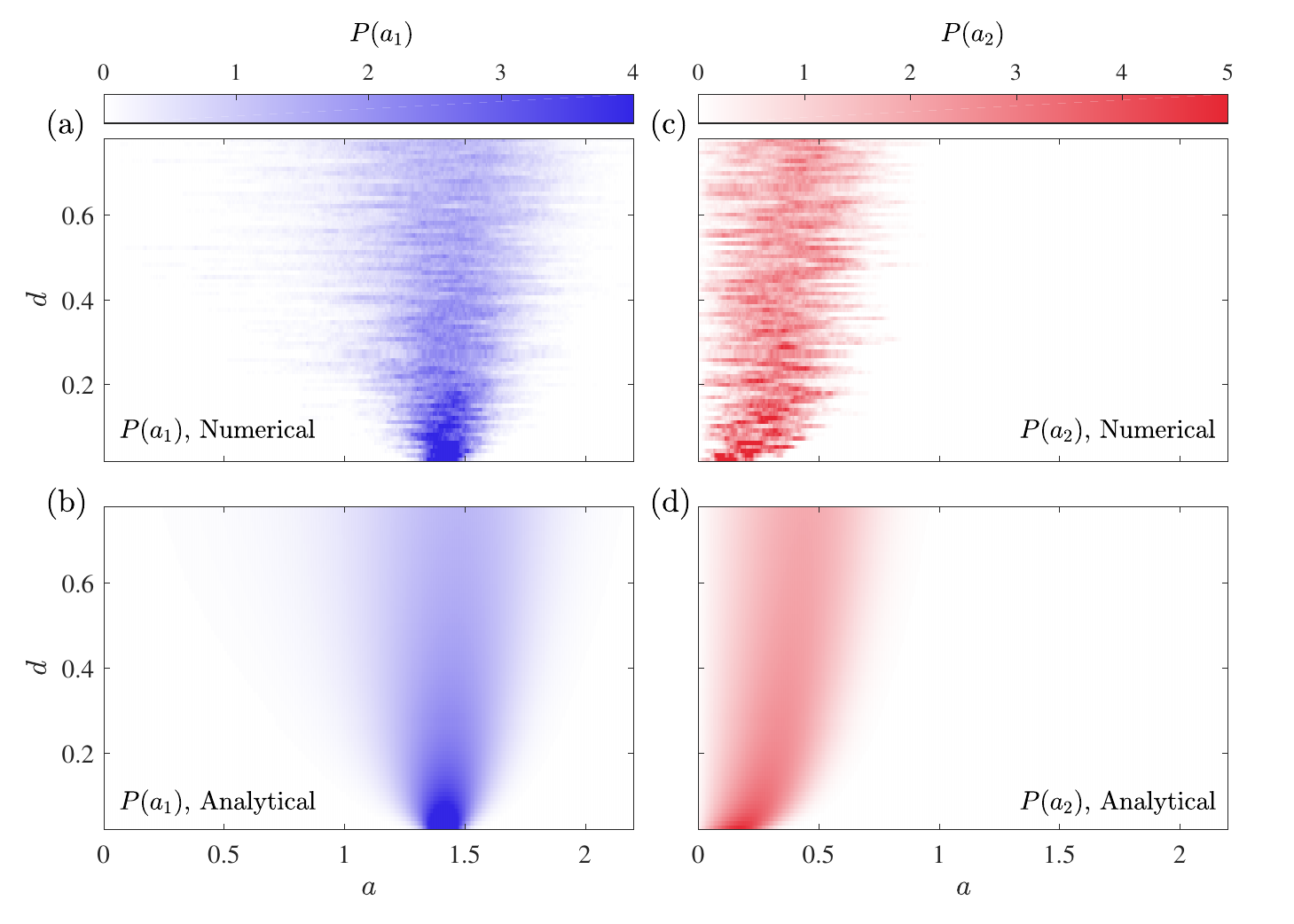}
    \caption{Velocity coupling, case 3: limit-cycle/limit-cycle coupling. Effect of varying $d$ on (a,b) $P(a_1)$ and (c,d) $P(a_2)$, obtained from (a,c) numerical simulations and (b,d) equation \ref{cpl_sta_vel}. $d_1=d_2=d$, $k=0.1$ and all other parameters are equal to figure \ref{fig:v_lclc}.}
\label{fig:v_lclc3}
\end{figure}

The time traces and the numerical/analytical PDFs for cases 1, 2 and 3 are shown in figures \ref{fig:v_fpfp}, \ref{fig:v_lcfp} and \ref{fig:v_lclc}, respectively. It can be seen that the analytical Fokker--Planck equation captures the dynamics of the numerical oscillators well. As with the distance-coupled oscillators, cases with a higher noise amplitude ($d$) show a greater mean and variation in $P(a)$ (see figures \ref{fig:v_fpfp3}, \ref{fig:v_lcfp3} and \ref{fig:v_lclc3}). 

However, contrary to the distance-coupled case, a significant effect of the coupling strength ($k$) is found. In particular, a higher $k$ results in a lower mean of the PDF of the amplitude ($P(a)$). In other words, strengthening the coupling between the two oscillators tends to suppress their oscillation (see figures \ref{fig:v_fpfp2}, \ref{fig:v_lcfp2} and \ref{fig:v_lclc2}). 

A mathematical explanation for the suppression effect of $k$ can be found from the stationary Fokker--Planck equation. In equation \ref{cpl_sta_vel}, $k$ is subtracted from $\epsilon_1$ and $\epsilon_2$, making the overall linear growth rates of the first and second oscillators ($\epsilon_1-k$) and ($\epsilon_2-k$), respectively. Therefore, positive $k$ acts as a damping parameter, and varying $k$ serves as a bifurcation parameter when $\epsilon_1$ and $\epsilon_2$ are fixed. Consequently, limit-cycle oscillations can occur only when the linear VDP coefficient ($\epsilon_1$, $\epsilon_2$) is larger than $k$ (see figure \ref{fig:v_lclc2}). 

The results of this section show that the Fokker--Planck equations corresponding to the velocity-coupled VDP equations are capable of describing not only the effect of noise but also the effect of the coupling between two oscillators. Specifically, suppression due to the coupling between two modes can be modeled successfully.

\section{Nonlinear coupling}

\subsection{Derivation of the Fokker--Planck equation}
In this section, we consider the nonlinear coupling term ($k_1 {x_1}^2 \dot{x_1} + k_2 {x_2}^2 \dot{x_2}$), which \citet{bonciolini2017output} used to model a multi-mode combustor.

\begin{subequations} \label{nl_cpl}
\begin{align}
    \ddot{x_1}-(\epsilon_1 + \alpha_{11}{x_1}^2)\dot{x_1} +& {\omega_1}^2 x_1 + k_1 {x_1}^2 \dot{x_1} + k_2 {x_2}^2 \dot{x_2} = \sqrt{2 d_1} \eta_1, \\
    \ddot{x_2}-(\epsilon_2 + \alpha_{12}{x_2}^2)\dot{x_2} +& {\omega_1}^2 x_1 + k_1 {x_1}^2 \dot{x_1} + k_2 {x_2}^2 \dot{x_2} = \sqrt{2 d_2} \eta_2,
\end{align}
\end{subequations}
where [$x_1$, $x_2$] are the state variables, [$\omega_1$, $\omega_2$] are the angular frequencies, [$\epsilon_1$, $\epsilon_2$] and [$\alpha_{11}$, $\alpha_{12}$] are the VDP coefficients, [$k_1$, $k_2$] are the coupling coefficients, [$\eta_1$, $\eta_2$] are unit white Gaussian noise terms, and [$d_1$, $d_2$] are their amplitudes. Equation \ref{nl_cpl} can be expressed in terms of the amplitude ($a$) and phase ($\phi$) using the transformations in equation \ref{rel}:
\begin{subequations} \label{eq5_14}
\begin{align}
    \dot{a_1} \cos{\Phi_1} - a_1 \dot{\phi_1} \sin{\Phi_1} &= 0, \\
    \dot{a_2} \cos{\Phi_2} - a_2 \dot{\phi_2} \sin{\Phi_2} &= 0, \\
    \dot{a_1} \sin{\Phi_1} + a_1 \dot{\phi_1} \cos{\Phi_1} &= \epsilon_1 a_1 \sin{\Phi_1} + \alpha_{11} {a_1}^3 \sin{\Phi_1} \cos^2{\Phi_1} - k_1 {a_1}^3 \sin{\Phi_1} \cos^2{\Phi_1} \notag \\
    &\quad- \frac{k_2 \omega_2}{\omega_1} {a_2}^3 \sin{\Phi_2} \cos^2{\Phi_2} - \frac{\sqrt{2d_1}}{\omega_1} \eta_1, \\ 
    \dot{a_2} \sin{\Phi_2} + a_2 \dot{\phi_2} \cos{\Phi_2} &= \epsilon_2 a_2 \sin{\Phi_2} + \alpha_{12} {a_2}^3 \sin{\Phi_2} \cos^2{\Phi_2} - k_2 {a_2}^3 \sin{\Phi_2} \cos^2{\Phi_2} \notag \\
    &\quad- \frac{k_1 \omega_1}{\omega_2} {a_1}^3 \sin{\Phi_1} \cos^2{\Phi_1} - \frac{\sqrt{2d_2}}{\omega_2} \eta_2,
\end{align}
\end{subequations}
where $\Phi_1=\omega_1 t + \phi_1$ and $\Phi_2=\omega_2 t + \phi_2$. The ordinary differential equations for $a$ and $\phi$ are given as:
\begin{subequations} \label{eq5_15}
\begin{align}
    \dot{a_1} &= \frac{\epsilon_1 a_1}{2} + \frac{\alpha_{11} {a_1}^3}{8} - \frac{k_1 {a_1}^3}{8} - \frac{\epsilon_1 a_1}{2} \cos{2\Phi_1} - \frac{\alpha_{11} {a_1}^3}{8} \cos{4\Phi_1} + \frac{k_1 {a_1}^3}{8} \cos{4\Phi_1} \notag \\
    &\quad+ \frac{k_2 \omega_2  {a_2}^3}{8 \omega_1} \cos{(\Phi_1 + \Phi_2)} - \frac{k_2 \omega_2  {a_2}^3}{8 \omega_1} \cos{(\Phi_1 - \Phi_2)} + \frac{k_2 \omega_2  {a_2}^3}{8 \omega_1} \cos{(\Phi_1 + 3\Phi_2)} \notag \\
    &\quad- \frac{k_2 \omega_2  {a_2}^3}{8 \omega_1} \cos{(\Phi_1 - 3\Phi_2)} - \frac{\sqrt{2d_1}}{\omega_1}(\sin{\Phi_1}) \eta_1,\\
    \dot{a_2} &= \frac{\epsilon_2 a_2}{2} + \frac{\alpha_{12} {a_2}^3}{8} - \frac{k_2 {a_2}^3}{8} - \frac{\epsilon_2 a_2}{2} \cos{2\Phi_2} - \frac{\alpha_{12} {a_2}^3}{8} \cos{4\Phi_2} + \frac{k_2 {a_2}^3}{8} \cos{4\Phi_2} \notag \\
    &\quad+ \frac{k_1 \omega_1  {a_1}^3}{8 \omega_2} \cos{(\Phi_2 + \Phi_1)} - \frac{k_1 \omega_1  {a_1}^3}{8 \omega_2} \cos{(\Phi_2 - \Phi_1)} + \frac{k_1 \omega_1  {a_1}^3}{8 \omega_2} \cos{(\Phi_2 + 3\Phi_1)} \notag \\
    &\quad- \frac{k_1 \omega_1  {a_1}^3}{8 \omega_2} \cos{(\Phi_2 - 3\Phi_1)} - \frac{\sqrt{2d_2}}{\omega_2}(\sin{\Phi_2}) \eta_2,\\
    \dot{\phi_1} &= \frac{\epsilon_1}{2}\sin{2\Phi_1} + \frac{\alpha_{11} {a_1}^2}{4}\sin{2\Phi_1} + \frac{\alpha_{11} {a_1}^2}{8}\sin{4\Phi_1} - \frac{k_1 {a_1}^2}{4}\sin{2\Phi_1} - \frac{k_1 {a_1}^2}{8}\sin{4\Phi_1}  \notag \\
    &\quad- \frac{k_2 \omega_2 {a_2}^3}{8 \omega_1 a_1} \sin{(\Phi_2 + \Phi_1)} - \frac{k_2 \omega_2 {a_2}^3}{8 \omega_1 a_1} \sin{(\Phi_2 - \Phi_1)} - \frac{k_2 \omega_2 {a_2}^3}{8 \omega_1 a_1} \sin{(3\Phi_2 + \Phi_1)} \notag \\
    &\quad- \frac{k_2 \omega_2 {a_2}^3}{8 \omega_1 a_1} \sin{(3\Phi_2 - \Phi_1)} - \frac{\sqrt{2d_1}}{\omega_1 a_1} (\cos{\Phi_1}) \eta_1, \\
    \dot{\phi_2} &= \frac{\epsilon_2}{2}\sin{2\Phi_2} + \frac{\alpha_{12} {a_2}^2}{4}\sin{2\Phi_2} + \frac{\alpha_{12} {a_2}^2}{8}\sin{4\Phi_2} - \frac{k_2 {a_2}^2}{4}\sin{2\Phi_2} - \frac{k_2 {a_2}^2}{8}\sin{4\Phi_2}  \notag \\
    &\quad- \frac{k_1 \omega_1 {a_1}^3}{8 \omega_2 a_2} \sin{(\Phi_1 + \Phi_2)} - \frac{k_1 \omega_1 {a_1}^3}{8 \omega_2 a_2} \sin{(\Phi_1 - \Phi_2)} - \frac{k_1 \omega_1 {a_1}^3}{8 \omega_2 a_2} \sin{(3\Phi_1 + \Phi_2)} \notag \\
    &\quad- \frac{k_1 \omega_1 {a_1}^3}{8 \omega_2 a_2} \sin{(3\Phi_1 - \Phi_2)} - \frac{\sqrt{2d_2}}{\omega_2 a_2} (\cos{\Phi_2}) \eta_2.
\end{align}
\end{subequations}

The last terms in equation \ref{eq5_15} are the stochastic components, and the other terms are the deterministic components. The drift ($\mathbf{m}$) and diffusion ($\mathbf{\sigma}$) terms of the amplitude ($a$) can be found via stochastic averaging \citep{stratonovich1963, stratonovich1967, roberts1986stochastic}:
\begin{subequations} 
\begin{align}
    \mathbf{m}(a_1) &= \frac{\epsilon_1 a_1}{2} + \frac{\alpha_{11} {a_1}^3}{8} - \frac{k_1 {a_1}^3}{8} + \frac{d_1}{2 \omega_1 a_1}, \\
    \mathbf{m}(a_2) &= \frac{\epsilon_2 a_2}{2} + \frac{\alpha_{12} {a_2}^3}{8} - \frac{k_2 {a_2}^3}{8} + \frac{d_2}{2 \omega_2 a_2}, \\
    \mathbf{\sigma}^2(a_1) &= \frac{d_1}{{\omega_1}^2}, \\
    \mathbf{\sigma}^2(a_2) &= \frac{d_2}{{\omega_2}^2}.
\end{align}
\end{subequations}

For the stationary case, solutions of the Fokker--Planck equation (equation \ref{eq5_7}) are given as:
\begin{subequations} \label{cpl_sta_nl}
\begin{align}
    P(a_1) = C_1 a_1 \exp\Big[\frac{\epsilon_1 {\omega_1}^2}{2d_1} {a_1}^2 + \frac{(\alpha_{11} - k_1) {\omega_1}^2}{16d_1} {a_1}^4 \Big], \\
    P(a_2) = C_2 a_2 \exp\Big[\frac{\epsilon_2 {\omega_2}^2}{2d_2} {a_2}^2 + \frac{(\alpha_{12} - k_2) {\omega_2}^2}{16d_2} {a_2}^4 \Big],
\end{align}
\end{subequations}
where $P(a_1)$ and $P(a_2)$ are the stationary probability functions of $a_1$ and $a_2$, respectively, and $C_1$ and $C_2$ are normalization constants. It is worth noting that unlike the distance-coupled case (equation \ref{cpl_sta}) but like the velocity-coupled case (equation \ref{cpl_sta_vel}), here the coupling strength ($k$) appears in the expressions for $P(a_1)$ and $P(a_2)$.

\subsection{Numerical analysis and validation}

Again, we validate the stationary Fokker--Planck solution (equation \ref{cpl_sta_nl}) by comparing it with the results from numerical simulations. We consider the same three cases considered earlier for distance coupling (\S\ref{num_ana_dis}) and velocity coupling (\S\ref{num_ana_vel}): see table \ref{num_cases1}. In all three cases, we perform simulations with three coupling strengths and two noise amplitudes (see table \ref{num_cases2}). To solve equation \ref{nl_cpl}, we use a 4th-order Runge-Kutta (RK4) algorithm in the time interval $0<t<1000$, with a time step of $dt=0.001$ ($t$ in arbitrary units). The amplitude ($a$) is then obtained via the Hilbert transform, and compared with the analytical Fokker--Planck solution (equation \ref{cpl_sta_nl}). Finally, the effects of the coupling strength and noise amplitude are individually analyzed by running simulations with varying $k$ and $d$, respectively, with all other parameters fixed. The results obtained from these simulations are compared with the analytical solution obtained from equation \ref{cpl_sta_nl}.

\begin{figure}
    \centering
    \includegraphics[clip,trim=1mm 1mm 1mm 1mm,width=\textwidth]{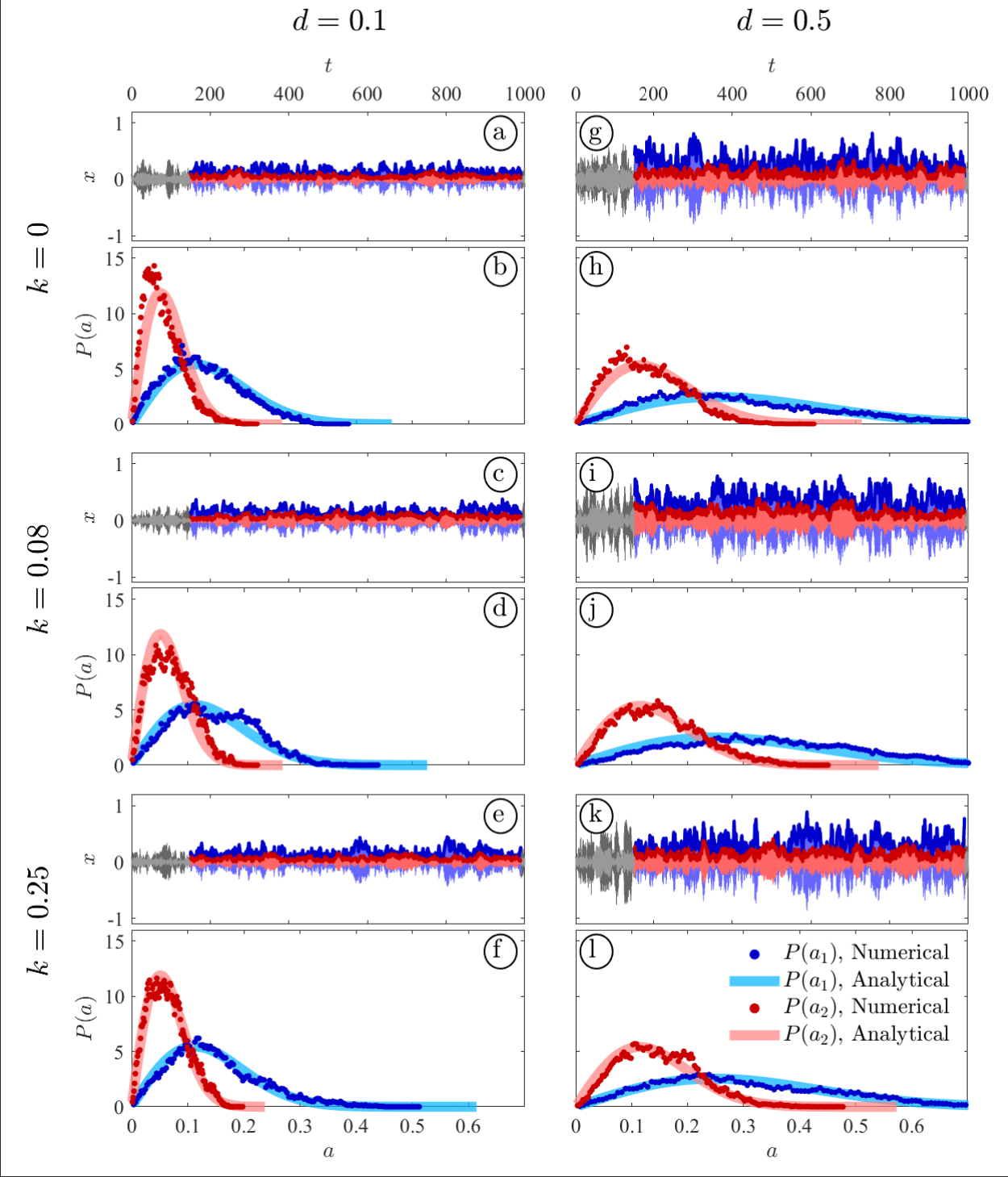}
    \caption{Nonlinear coupling, case 1: fixed-point/fixed-point coupling. $\epsilon_1=-0.2$, $\epsilon_2=-0.1$, $\alpha_{11}=\alpha_{12}=-0.2$, $\omega_1=2\pi$, $\omega_2=6.2\pi$, $k_1 = k_2 = k$, $d_1= d_2= d$. (a,g,c,i,e,k) time traces and (b,h,d,j,f,l) PDFs of the amplitude for (a,b,g,h) uncoupled, (c,d,i,j) weakly coupled, and (e,f,k.l) strongly coupled oscillators under (a-f) low noise and (g-l) high noise. Blue and red markers denote the first ($x_1$) and the second ($x_2$) oscillators, respectively. Scatter dots and thick lines show the numerical and analytical results, respectively.  Grey lines show the transient interval ($t<150$), which is not used in the analysis.}
\label{fig:nl_fpfp}
\end{figure}

\begin{figure}
    \centering
    \includegraphics[clip,trim=1mm 1mm 1mm 1mm,width=0.85\textwidth]{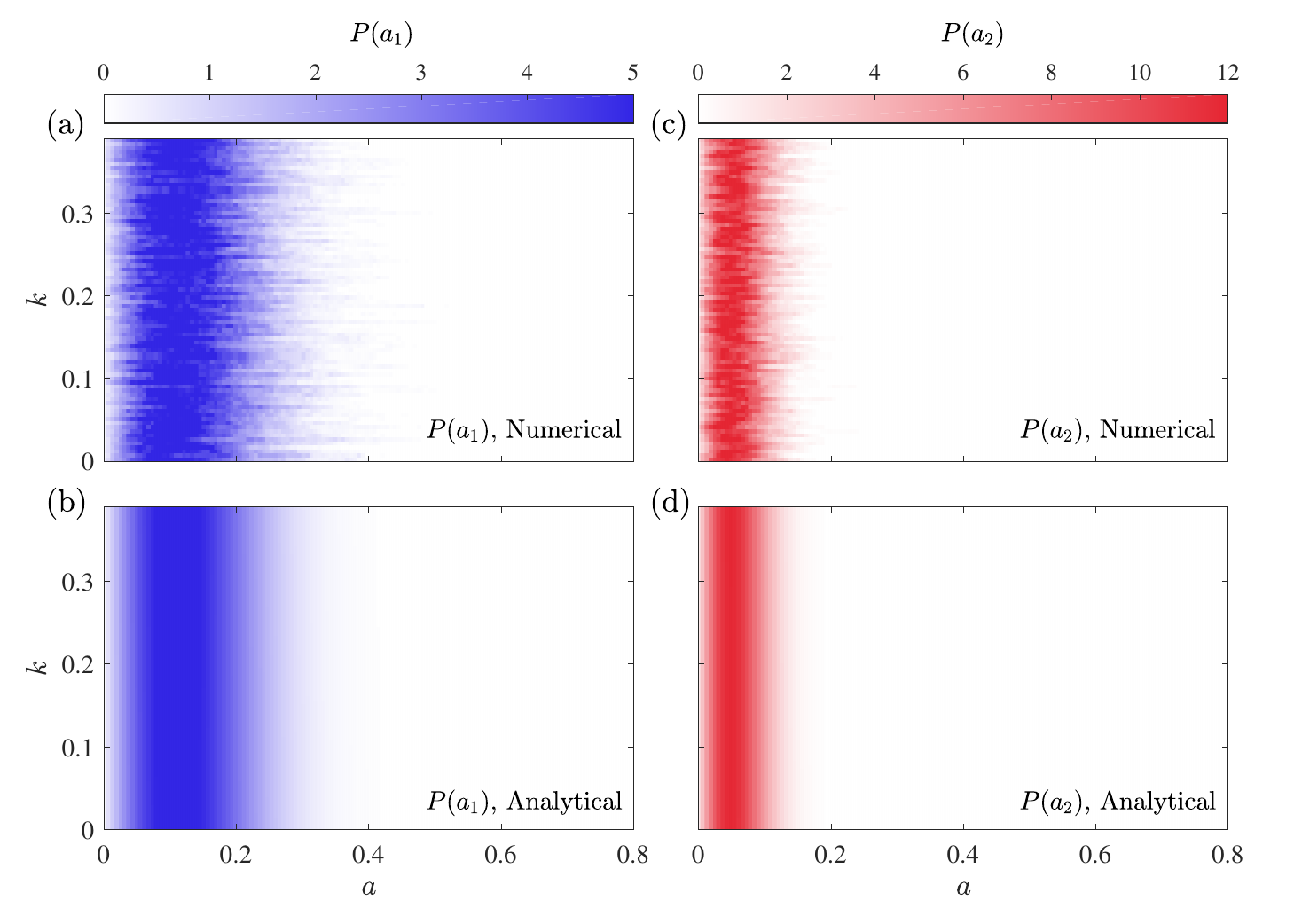}
    \caption{Nonlinear coupling, case 1: fixed-point/fixed-point coupling. Effect of varying $k$ on (a,b) $P(a_1)$ and (c,d) $P(a_2)$, obtained from (a,c) numerical simulations and (b,d) equation \ref{cpl_sta_nl}. $d_1=d_2=0.1$ and all other parameters are equal to figure \ref{fig:nl_fpfp}.}
\label{fig:nl_fpfp2}
\end{figure}

\begin{figure}
    \centering
    \includegraphics[clip,trim=1mm 1mm 1mm 1mm,width=0.85\textwidth]{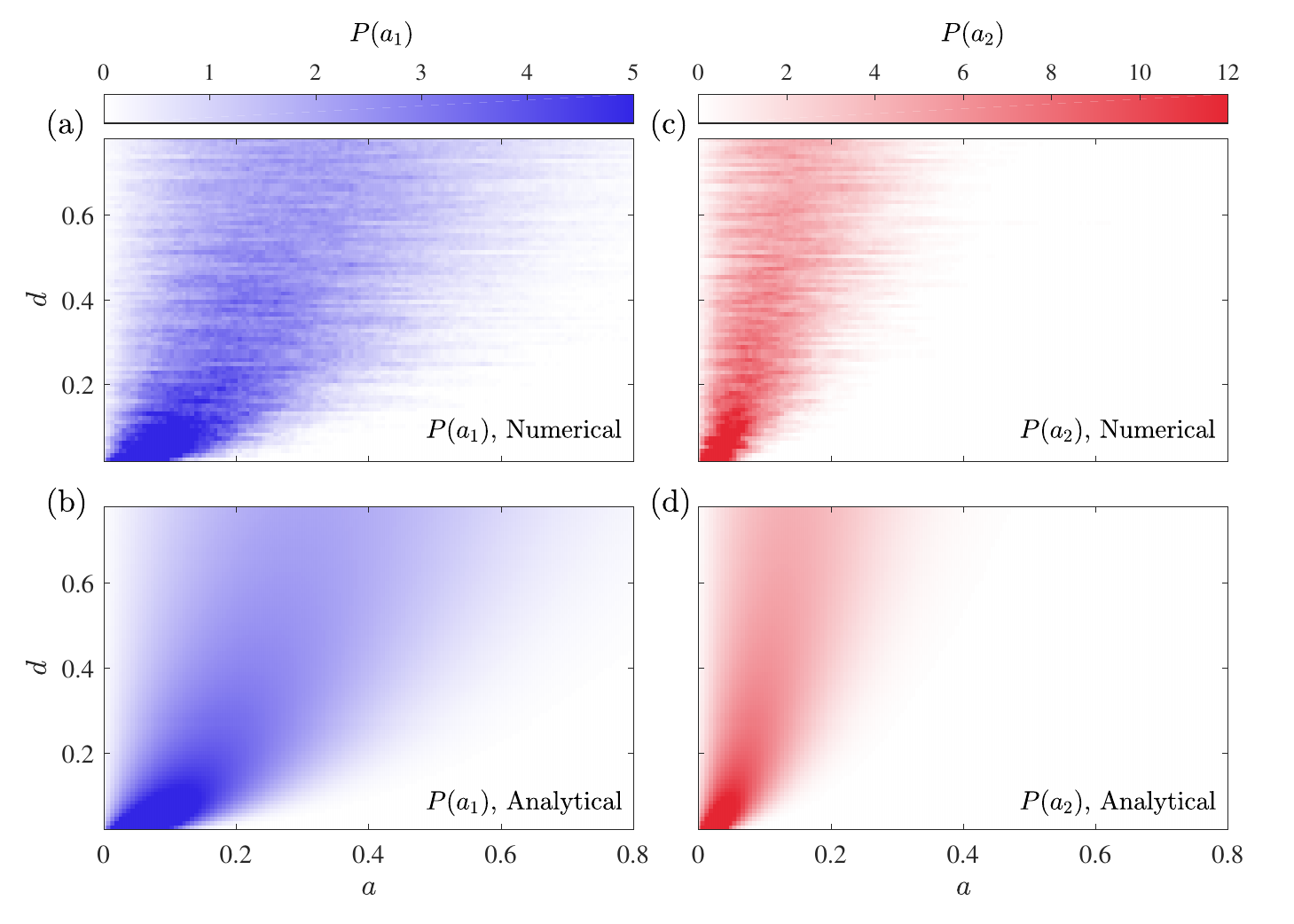}
    \caption{Nonlinear coupling, case 1: fixed-point/fixed-point coupling. Effect of varying $d$ on (a,b) $P(a_1)$ and (c,d) $P(a_2)$, obtained from (a,c) numerical simulations and (b,d) equation \ref{cpl_sta_nl}. $d_1=d_2=d$, $k=0.1$ and all other parameters are equal to figure \ref{fig:nl_fpfp}.}
\label{fig:nl_fpfp3}
\end{figure}

\begin{figure}
    \centering
    \includegraphics[clip,trim=1mm 1mm 1mm 1mm,width=\textwidth]{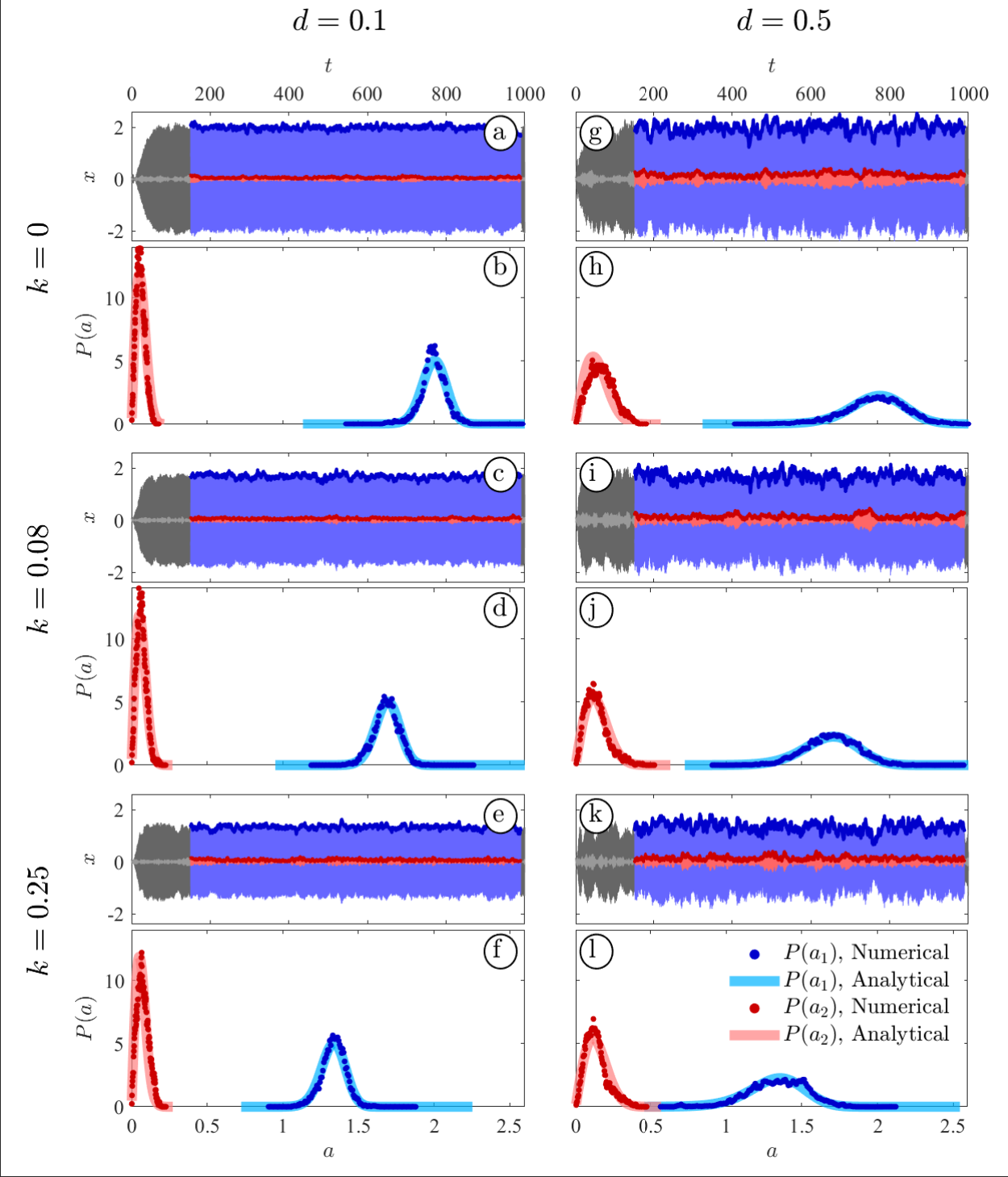}
    \caption{Nonlinear coupling, case 2: limit-cycle/fixed-point coupling. $\epsilon_1=0.2$, $\epsilon_2=-0.1$, $\alpha_{11}=\alpha_{12}=-0.2$, $\omega_1=2\pi$, $\omega_2=6.2\pi$, $k_1 = k_2 = k$, $d_1= d_2= d$. (a,g,c,i,e,k) time traces and (b,h,d,j,f,l) PDFs of the amplitude for (a,b,g,h) uncoupled, (c,d,i,j) weakly coupled, and (e,f,k.l) strongly coupled oscillators under (a-f) low noise and (g-l) high noise. Blue and red markers denote the first ($x_1$) and the second ($x_2$) oscillators, respectively. Scatter dots and thick lines show the numerical and analytical results, respectively.  Grey lines show the transient interval ($t<150$), which is not used in the analysis.}
\label{fig:nl_lcfp}
\end{figure}

\begin{figure}
    \centering
    \includegraphics[clip,trim=1mm 1mm 1mm 1mm,width=0.85\textwidth]{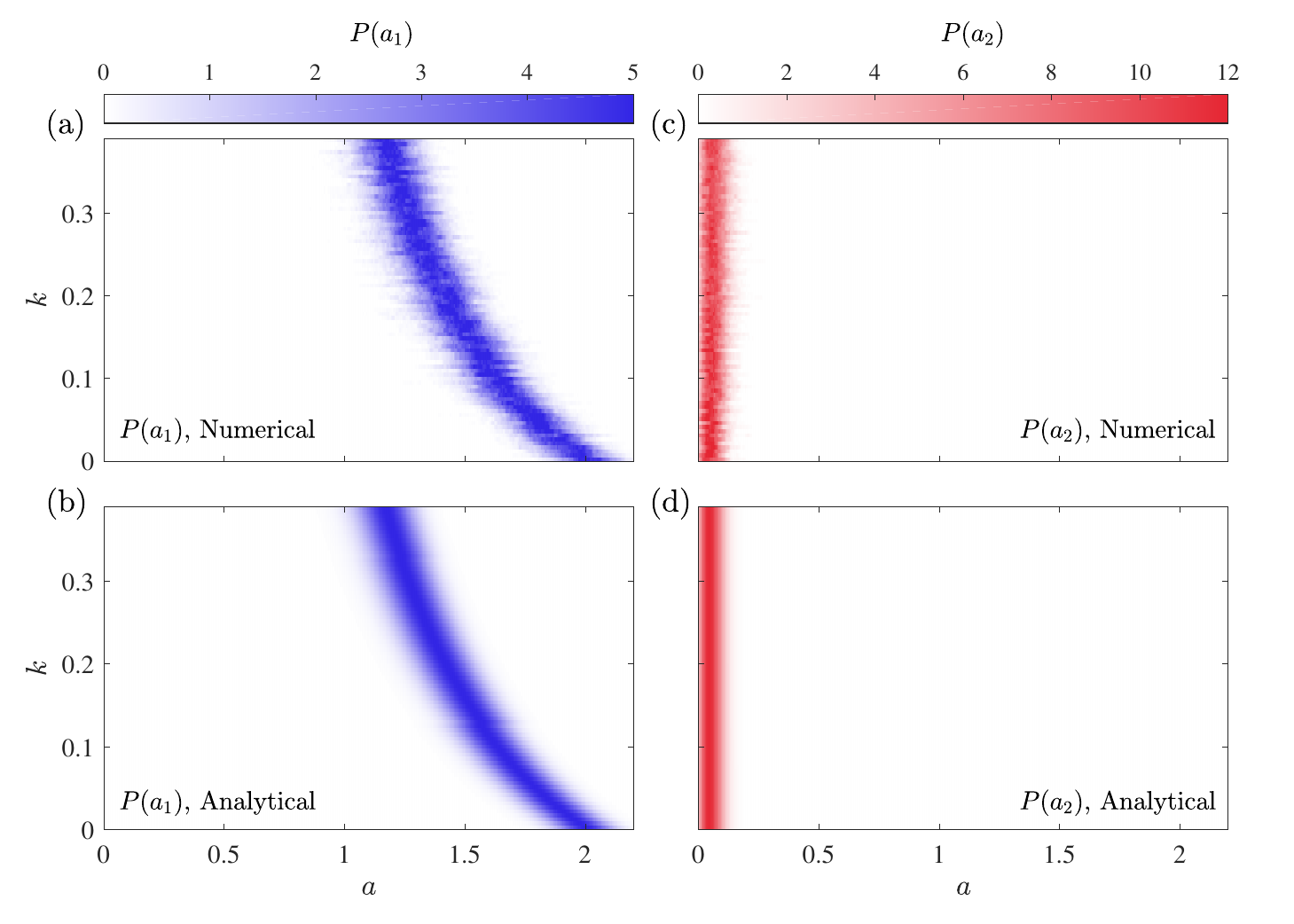}
    \caption{Nonlinear coupling, case 2: limit-cycle/fixed-point coupling. Effect of varying $k$ on (a,b) $P(a_1)$ and (c,d) $P(a_2)$, obtained from (a,c) numerical simulations and (b,d) equation \ref{cpl_sta_nl}. $d_1=d_2=0.1$ and all other parameters are equal to figure \ref{fig:nl_lcfp}.}
\label{fig:nl_lcfp2}
\end{figure}

\begin{figure}
    \centering
    \includegraphics[clip,trim=1mm 1mm 1mm 1mm,width=0.85\textwidth]{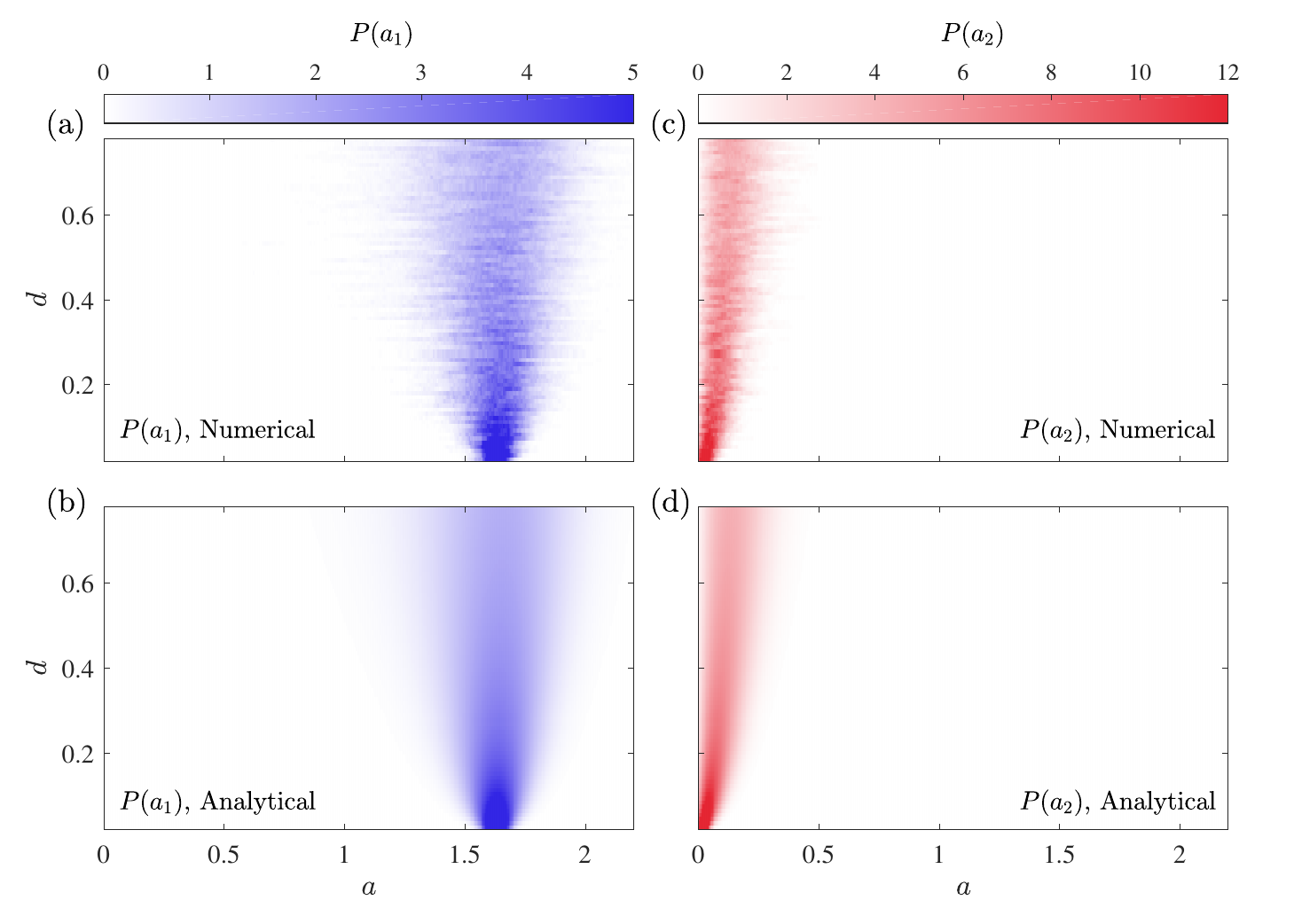}
    \caption{Nonlinear coupling, case 2: limit-cycle/fixed-point coupling. Effect of varying $d$ on (a,b) $P(a_1)$ and (c,d) $P(a_2)$, obtained from (a,c) numerical simulations and (b,d) equation \ref{cpl_sta_nl}. $d_1=d_2=d$, $k=0.1$ and all other parameters are equal to figure \ref{fig:nl_lcfp}.}
\label{fig:nl_lcfp3}
\end{figure}

\begin{figure}
    \centering
    \includegraphics[clip,trim=1mm 1mm 1mm 1mm,width=\textwidth]{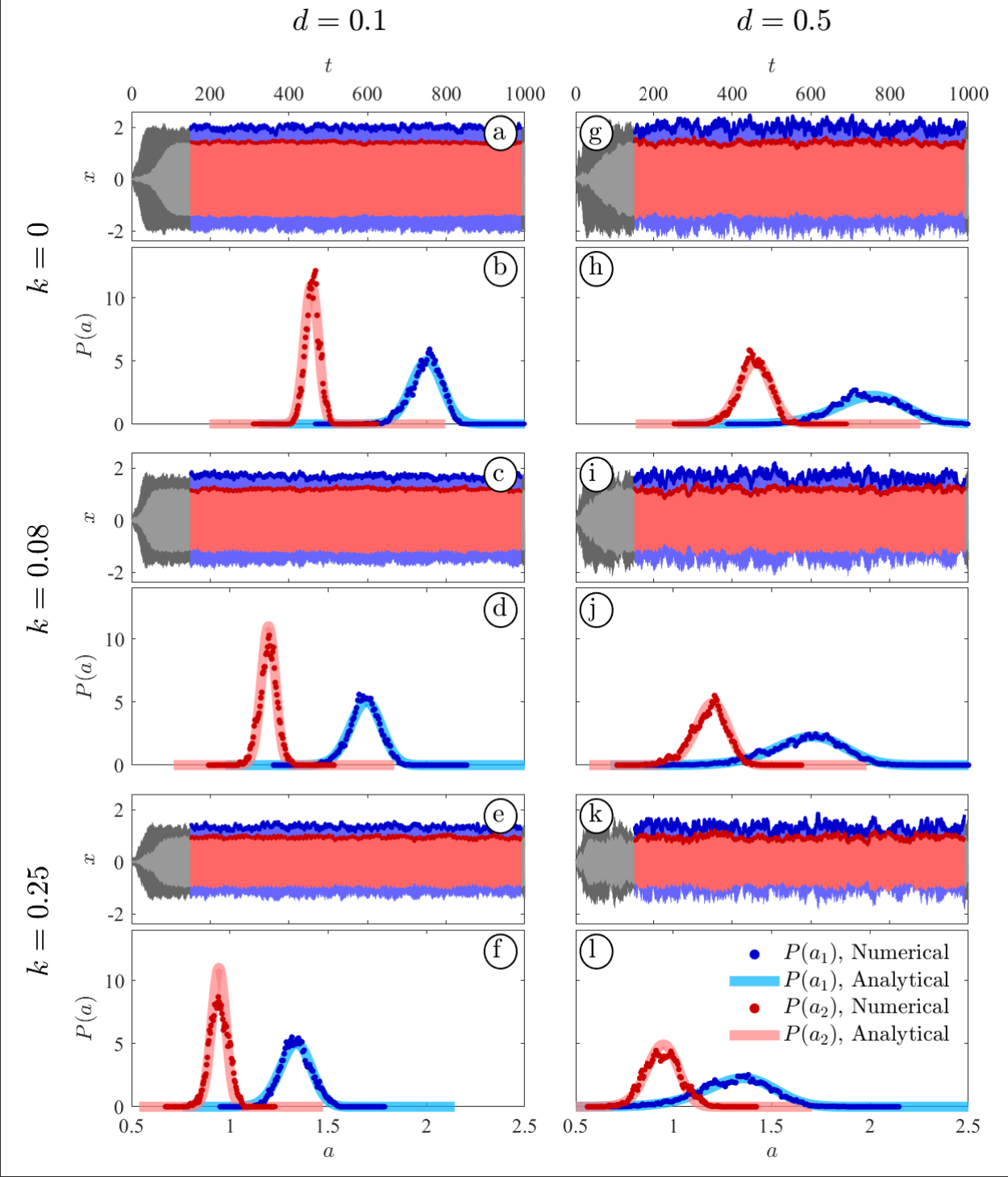}
    \caption{Nonlinear coupling, case 3: limit-cycle/limit-cycle coupling. $\epsilon_1=0.2$, $\epsilon_2=0.1$, $\alpha_{11}=\alpha_{12}=-0.2$, $\omega_1=2\pi$, $\omega_2=6.2\pi$, $k_1 = k_2 = k$, $d_1= d_2= d$. (a,g,c,i,e,k) time traces and (b,h,d,j,f,l) PDFs of the amplitude for (a,b,g,h) uncoupled, (c,d,i,j) weakly coupled, and (e,f,k.l) strongly coupled oscillators under (a-f) low noise and (g-l) high noise. Blue and red markers denote the first ($x_1$) and the second ($x_2$) oscillators, respectively. Scatter dots and thick lines show the numerical and analytical results, respectively.  Grey lines show the transient interval ($t<150$), which is not used in the analysis.}
\label{fig:nl_lclc}
\end{figure}

\begin{figure}
    \centering
    \includegraphics[clip,trim=1mm 1mm 1mm 1mm,width=0.85\textwidth]{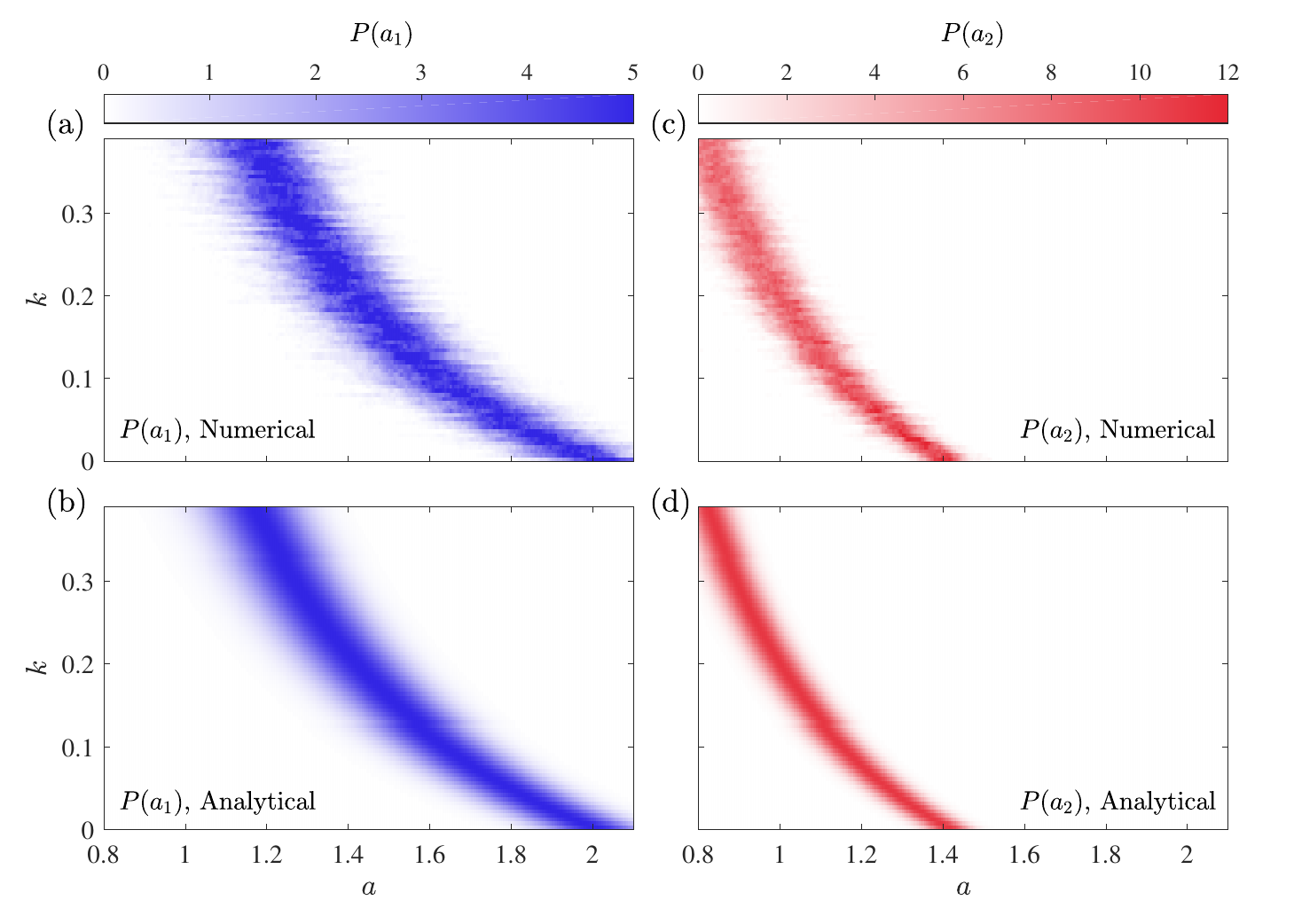}
    \caption{Nonlinear coupling, case 3: limit-cycle/limit-cycle coupling. Effect of varying $k$ on (a,b) $P(a_1)$ and (c,d) $P(a_2)$, obtained from (a,c) numerical simulations and (b,d) equation \ref{cpl_sta_nl}. $d_1=d_2=0.1$ and all other parameters are equal to figure \ref{fig:nl_lclc}.}
\label{fig:nl_lclc2}
\end{figure}

\begin{figure}
    \centering
    \includegraphics[clip,trim=1mm 1mm 1mm 1mm,width=0.85\textwidth]{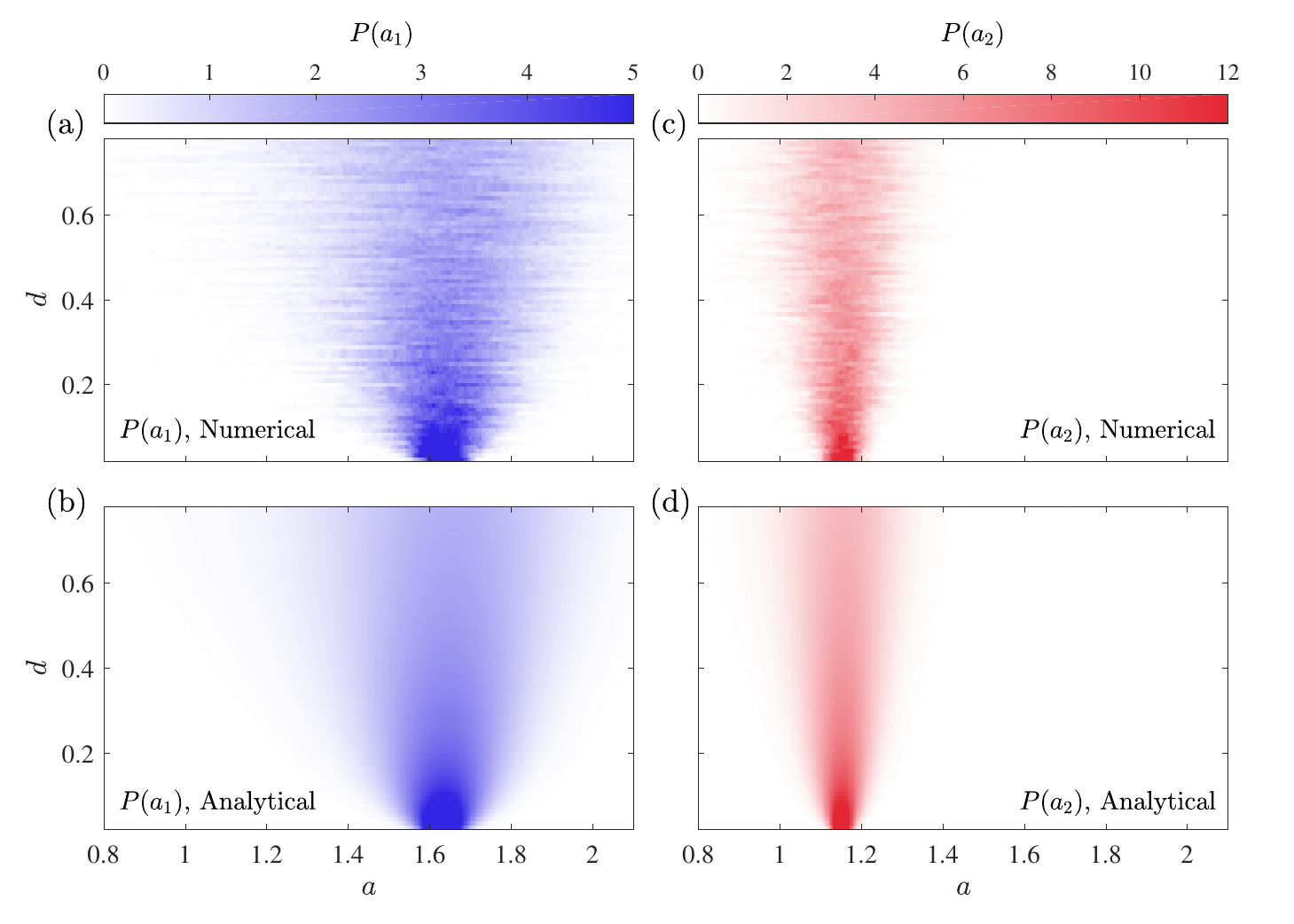}
    \caption{Nonlinear coupling, case 3: limit-cycle/limit-cycle coupling. Effect of varying $d$ on (a,b) $P(a_1)$ and (c,d) $P(a_2)$, obtained from (a,c) numerical simulations and (b,d) equation \ref{cpl_sta_nl}. $d_1=d_2=d$, $k=0.1$ and all other parameters are equal to figure \ref{fig:nl_lclc}.}
\label{fig:nl_lclc3}
\end{figure}

The time traces and the numerical/analytical PDFs for cases 1, 2 and 3 are shown in figures \ref{fig:nl_fpfp}, \ref{fig:nl_lcfp} and \ref{fig:nl_lclc}, respectively. Regardless of the dynamical states of the two oscillators, the analytical Fokker--Planck solutions are in excellent agreement with the numerical simulations. The effect of the coupling strength ($k_1$, $k_2$) on two oscillators is shown in figures \ref{fig:nl_fpfp2}, \ref{fig:nl_lcfp2} and \ref{fig:nl_lclc2}. The significant effect of noise is found, especially when one oscillator is in the limit-cycle regime (see figures \ref{fig:nl_lcfp2} and \ref{fig:nl_lclc2}). Specifically, when the two oscillators are strongly coupled, the oscillation amplitudes are lower than those of the uncoupled case. Mathematically, this is because the coupling coefficients ($k_1$, $k_2$) are subtracted from the stabilizing nonlinear parameters ($\alpha_1$, $\alpha_2$), as shown in equation \ref{cpl_sta_nl}. As a result, stronger nonlinear coupling results in stronger damping, leading to stronger suppression of the oscillation amplitude. 

\section{Conclusions}
In this chapter, we have modeled the dynamics of two interacting modes using two coupled VDP oscillators. Three different types of coupling---distance coupling, velocity coupling, and nonlinear coupling---are considered in the analysis, and the governing equations are numerically solved. In the distance-coupled model, the fluctuation amplitude is independent of the coupling constant $k$. This implies that the energy transfer between two modes is purely in the form of frequency and phase variation. In the velocity-coupled model and the nonlinear coupled model, however, stronger coupling leads to a smaller mean and deviation of the fluctuation amplitude. In particular, in the velocity-coupled model, $k$ is found to act as a bifurcation parameter. The results of the numerical simulations imply that the type of coupling can be inferred by analyzing the probability density function of the fluctuation amplitude, provided that the coupling strength is known. 

Furthermore, we showed that the analytical solutions obtained from the Fokker--Planck equation, which are derived from the coupled VDP model, are in good agreement with the numerical results. In particular, the effects of the coupling type, the coupling strength, and the noise amplitude are successfully reproduced with the analytical solutions. Therefore, it can be concluded that the Fokker--Planck equations derived above can successfully capture the dynamics of a dynamical system with two dominant interacting modes.

In future work, the derived equations can be used for SI of such systems. For example, one can apply band-pass filters around the primary and secondary oscillation frequencies to isolate these two modes. The coupling strength and the VDP parameters characterizing these modes can then be found by fitting the data to the stationary (\S\ref{chap:low}, \S\ref{chap:Rij}) or transitional (\S\ref{chap:gas}) Fokker--Planck equations. Subsequently, one can analyze the relationship between the obtained coefficients and the bifurcation parameter, and extrapolate the former to predict the bifurcation point and the post-bifurcation dynamics (\S\ref{sec:extrapol}). In this way, the growth of a secondary mode hidden behind the primary mode can be identified, and potential mode switching or secondary bifurcation can be predicted before it occurs.

\chapter{Fokker--Planck equation for the thermoacoustic oscillations of a Rijke tube} \label{chap:ntau}


\section{Introduction}
The Rijke tube, which consists of an acoustic resonator and a heat source, is known to exhibit a variety of dynamical states and bifurcations. As such, it is generally considered an ideal platform for studying thermoacoustic oscillations \citep{balasubramanian2008, subramanian2013, gopalakrishnan2015, Gopalakrishnan2016, rigas2016experimental, guan2016forcedsync, guan2017openloop, guan2018suppression, guan2018strange, guan2019open, guan2019forcedand, guan2019QP, guan2019control, yin2019asymmetric}. In \S\ref{chap:Rij}, we modeled the dynamics of an externally forced Rijke tube using a self-excited oscillator model (i.e. the van der Pol oscillator) \citep{NOIRAY2013152}. In this chapter, we model the same system using the momentum and energy equations, along with a white noise term. Then, we derive the corresponding Fokker--Planck equation, which we use to investigate the noise-induced dynamics of a Rijke tube. Finally, we validate the derived Fokker--Planck equation with numerical simulations.

\section{Rijke tube model and derivation of the Fokker--Planck equation} \label{der_ntau}

\begin{figure}
    \centering
    \includegraphics[clip,width=0.55\textwidth]{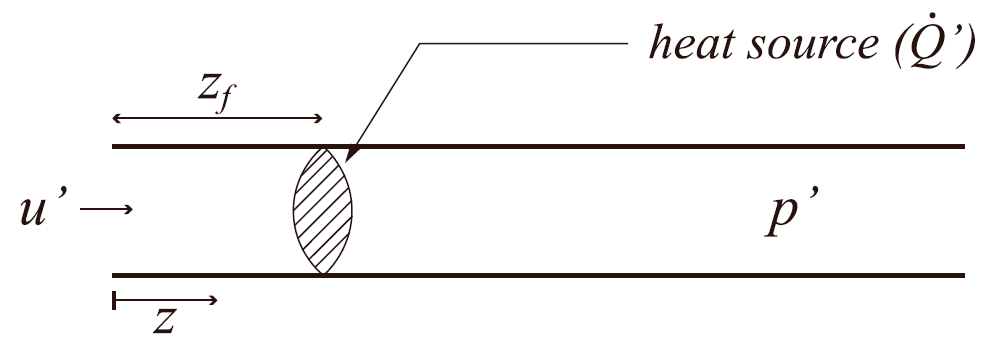}
    \caption{Schematic of the horizontal Rijke tube}
\label{fig:6Rij}
\end{figure}

We consider a horizontal Rijke tube \citep{matveev2003, balasubramanian2008, subramanian2013, Gopalakrishnan2016}, where a rightward mean flow is produced in an open-open duct by a blower. We assume that the Mach number of the mean flow is small, so that the duct acoustic field can be considered linear. The acoustic field is also assumed to be one-dimensional. The heat source is assumed to be compact, compared with the acoustic length-scales. Unlike a vertical Rijke tube, this setup allows us to justifiably neglect the effect of the mean flow gradient on the duct acoustics caused by natural convection \citep{balasubramanian2008}. When white Gaussian noise of amplitude $d$ is introduced into the system, the linearized momentum and energy equations, respectively, can be written as \citep{balasubramanian2008, subramanian2013, Gopalakrishnan2016, Gupta2017}: 
\begin{subequations} \label{gov_Rij}
\begin{align}
    \gamma M \pdv{u'}{t} + \pdv{p'}{z} &= 0, \\
    \pdv{p'}{t} + \gamma M \pdv{u'}{z} + \epsilon p' &= (\dot{Q'} + \sqrt{2d} \eta) \delta(z-z_f),
\end{align}
\end{subequations}
where $u'$ and $p'$ are the normalized acoustic velocity and pressure fluctuations, $z$ is the normalized distance from the left end of the tube in the axial direction, $t$ is time, $\epsilon$ is the acoustic damping rate, $z_f$ is the normalized flame location, $\gamma$ is the specific heat ratio of the air and $M$ is the Mach number of the mean flow. $\delta$ is the Dirac delta function and $\eta$ is a unit white Gaussian noise term. $\dot{Q'}$ denotes the heat release rate of the heat source, which can be represented using a modified version of King's law \citep{balasubramanian2008, subramanian2013, Gopalakrishnan2016, Gupta2017}:
\begin{equation}
    \dot{Q'}=k_Q\Big(\sqrt{\abs{\frac{1}{3}+u'(t-\tau)}}-\sqrt{\frac{1}{3}}\:\Big)
\end{equation}
where $k_Q$ is the heater power coefficient determined by the thermal properties of the heat source. The effect of acoustic perturbations arises in the heat release rate fluctuation after a time-delay $\tau$. Equation \ref{gov_Rij} can be reduced to ordinary differential equations using the Galerkin technique \citep{zinn1971, lores1973}. In particular, $u'$ and $p'$ can be expressed as a superposition of expansion functions:
\begin{subequations} \label{Galer}
\begin{align}
    u' &= \sum_{j=1}^{j_{max}} x_j \cos{(j \pi z)}, \\
    p' &= -\sum_{j=1}^{j_{max}} \frac{\gamma M}{j \pi} \dot{x_j} \sin{(j \pi z)},
\end{align}
\end{subequations}
where $j_{max}$ is the number of Galerkin modes and $x_j$ is a state variable of the $j^{th}$ Galerkin mode. Combining equations \ref{gov_Rij} and \ref{Galer}, we obtain:
\begin{subequations} \label{gov}
\begin{align}
    \Ddot{x_j} + {\omega_j}^2 x_j + \epsilon_j \dot{x_j} &= -k_Q \frac{2 j \pi}{\gamma M} \sin{(j \pi z_f)} \Big(\sqrt{\abs{\frac{1}{3}+u_f'(t-\tau)}}-\sqrt{\frac{1}{3}}\:\Big) \notag \\
    &\quad - \frac{2 j \pi}{\gamma M} \sin{(j \pi z_f)} \sqrt{2d} \eta,
\end{align}
\end{subequations}
where $\omega_j$ is the non-dimensional angular frequency of the $j^{th}$ duct mode, which is equal to $j\pi$ for an open-open duct \citep{subramanian2013}. $\epsilon_j$ is the damping term for the $j^{th}$ mode, which can be represented as $\epsilon_j= \epsilon_a + \epsilon_b \sqrt{j}$ \citep{gopalakrishnan2015, Gupta2017}. The heater power term in equation (\ref{gov}a), which contains a square root, can be approximated with a Maclaurin series expansion:
\begin{equation} \label{macl}
\begin{split}
    k_Q \frac{2 j \pi}{\gamma M} &\sin{(j \pi z_f)} \Big(\sqrt{\abs{\frac{1}{3}+u_f'(t-\tau)}} - \sqrt{\frac{1}{3}}\:\Big)\\
    &= \frac{2\sqrt{3} k_Q j \pi}{3 \gamma M} \sin{(j \pi z_f)} \big(\sqrt{\abs{1 + 3\cos{(j \pi z_f) x_j (t-\tau)}}} -1 \big) \\
    &\approx \frac{2\sqrt{3} k_Q j \pi}{3 \gamma M} \sin{(j \pi z_f)} \Big( \frac{3}{2} \cos{(j \pi z_f) x_j (t-\tau)} - \frac{9}{8} \cos^2{(j \pi z_f)} {x_j}^2 (t-\tau) \\
    &\hspace{18em} + \frac{27}{16} \cos^3{(j \pi z_f)} {x_j}^3 (t-\tau) \Big) \\
    &= \frac{\sqrt{3} k_Q j \pi}{2 \gamma M} \sin{(2j\pi z_f)} \Big( x_j (t-\tau) - \frac{3}{4} \cos{(j \pi z_f)} {x_j}^2 (t-\tau)  \\
    &\hspace{18em} + \frac{9}{8} \cos^2{(j \pi z_f)} {x_j}^3 (t-\tau) \Big). \\    
\end{split}
\end{equation}

To simplify the governing equation, we introduce the following parameters:
\begin{subequations}
\begin{align}
    c_{j1} &= \frac{\sqrt{3} k_Q j \pi}{2 \gamma M} \sin{(2 j \pi z_f)}, \\
    c_{j2} &= -\frac{3\sqrt{3} k_Q j \pi}{8 \gamma M} \sin{(2 j \pi z_f)} \cos{(j \pi z_f)}, \\
    c_{j3} &= \frac{9\sqrt{3} k_Q j \pi}{16 \gamma M} \sin{(2 j \pi z_f)} \cos^2{(j \pi z_f)}, \\
    c_{j4} &= -\frac{2 j \pi}{\gamma M} \sin{(\pi z_f)}.
\end{align}
\end{subequations}

Then, equation \ref{gov} can be written as:
\begin{equation} \label{gov_simp}
\begin{split}
    \Ddot{x_j} + {\omega_j}^2 x_j + \epsilon_j \dot{x_j} + c_{j1} x_j (t-\tau) + c_{j2} {x_j}^2(t-\tau) + c_{j3} & {x_j}^3(t-\tau) \\
    &- c_{j4}\sqrt{2d}\eta_j = 0.
\end{split}
\end{equation}

For small $\tau$, $x(t-\tau)$ can be approximated as:
\begin{equation} \label{tau_approx}
    x(t-\tau) \approx x(t) - \tau \dot{x}(t).
\end{equation}

Combining equations \ref{gov_simp} and \ref{tau_approx}, we obtain the following linearized equation for the $j^{th}$ Galerkin mode:
\begin{equation} \label{lin_gov}
    \begin{split}
        \Ddot{x_j} + \epsilon_j \dot{x_j} &- c_{j1}\tau\dot{x_j} + {\omega_j}^2 \dot{x} + c_{j1} x_j + c_{j2} {x_j}^2 - 2 c_{j2} \tau x_j \dot{x_j} + c_{j2} \tau^2 \dot{x_j}^2 \\
        &+ c_{j3} {x_j}^3 - 3c_{j3}\tau {x_j}^2 \dot{x_j} + 3c_{j3}\tau^2 x_j \dot{x_j}^2 - c_{j3} \tau^3 \dot{x_j}^3 - c_{j4}\sqrt{2d}\eta_j =0.
    \end{split}
\end{equation}

Next, we derive the Fokker--Planck equation representing the probability density function (PDF) of the amplitude of the pressure fluctuation inside the Rijke tube. To obtain an explicit form of the stationary Fokker--Planck equation, we derive the Fokker--Planck equation for the $j^{th}$ mode. 

The state variable for the $j^{th}$ mode can be transformed into amplitude and phase terms using the following relationship (see \S\ref{num} for justification):
\begin{subequations} \label{amp_ph}
\begin{align}
    x_j &= a_j \cos{(\omega_j t + \phi_j)}, \\
    \dot{x_j} &= -a_j \omega_j \sin{(\omega_j t + \phi_j)},
\end{align}
\end{subequations}
where $a_j$ and $\phi_j$ are the instantaneous amplitude and phase of $x_j$. From equations \ref{lin_gov} and \ref{amp_ph}, the following equations are obtained:
\begin{subequations} \label{eqss}
\begin{align}
    \dot{a_j} \cos{\Phi_j} &- a_j \dot{\phi_j} \sin{\Phi_j} = 0, \\
    \dot{a_j} \sin{\Phi_j} &+ a_j \dot{\phi_j} \cos{\Phi_j} \notag \\
    &= -\epsilon_j a_j \sin{\Phi_j} + c_{j1} \tau a_j \sin{\Phi_j} + \frac{c_{j1}}{\omega_j} a_j \cos{\Phi_j} + \frac{c_{j2}}{\omega_j} {a_j}^2 \cos^2{\Phi_j} \notag \\
    & \quad + 2 c_{j2} \tau {a_j}^2 \sin{\Phi_j} \cos{\Phi_j} + c_{j2} \tau^2 {a_j}^2 \omega_j \sin^2{\Phi_j} + \frac{c_{j3}}{\omega_j} {a_j}^3 \cos^3{\Phi_j} \\
    & \quad + 3 c_{j3} \tau {a_j}^3 \sin{\Phi_j} \cos^2{\Phi_j} + 3 c_{j3} \tau^2 {a_j}^3 \omega_j \sin^2{\Phi_j} \cos{\Phi_j} \notag \\
    & \quad + c_{j3} \tau^3 {a_j}^3 {\omega_j}^2 \sin^3{\Phi_j} + \frac{c_{j4}\sqrt{2d}}{\omega_j} \eta, \notag
\end{align}
\end{subequations}
where $\Phi_j=\omega_j t + \phi_j$. The explicit ordinary differential equations for $a_j$ and $\phi_j$ can be obtained from equation \ref{eqss}:
\begin{subequations} \label{ode_aphi}
\begin{align}
    \dot{a_j} &= -\frac{\epsilon_j a_j}{2} + \frac{c_{j1} \tau a_j}{2} + \frac{3 c_{j3} \tau {a_j}^3}{8} + \frac{3c_{j3} \tau^3 {\omega_j}^2 {a_j}^3}{8} + Q_a(a,\Phi) + \frac{c_{j4} \sqrt{2d}}{\omega_j}(\sin{\Phi}) \eta, \\
    \dot{\phi_j} &= \frac{c_{j1}}{2 \omega_j} + \frac{3 c_{j3} {a_j}^2}{8 \omega_j} + \frac{3 c_{j3} \tau^2 {a_j}^2 \omega_j}{8} + Q_{\phi}(a,\Phi) + \frac{c_{j4} \sqrt{2d}}{\omega_j a_j}(\cos{\Phi}) \eta,
\end{align}
\end{subequations}
where $Q_a(a,\Phi)$ and $Q_{\phi}(a,\Phi)$ are the sum of first-order sine and cosine terms that become zero after time-averaging. By stochastically averaging equation \ref{ode_aphi} \citep{stratonovich1963, stratonovich1967, roberts1986stochastic}, we can compute the drift and diffusion terms:
\begin{subequations}
\begin{align}
    \mathbf{m}(a_j) &= -\frac{\epsilon_j a_j}{2} + \frac{c_{j1} \tau a_j}{2} + \frac{3 c_{j3} \tau {a_j}^3}{8} + \frac{3c_{j3} \tau^3 {\omega_j}^2 {a_j}^3}{8} + \frac{{c_{j4}}^2 d}{2 {\omega_j}^2 a_j}, \\
    \mathbf{\sigma}(a_j) &= \frac{{c_{j4}}^2 d}{{\omega_j}^2},
\end{align}
\end{subequations}
where $\mathbf{m}(a_j)$ and $\mathbf{\sigma}(a_j)$ are the drift and diffusion components, respectively. The standard form of the Fokker--Planck equation is:
\begin{equation} \label{trans_pdf}
    \pdv{}{t}P(a_j,t)=-\pdv{}{a_j}\Big[\boldsymbol{m}(a_j)P(a_j,t)\Big]+\pdv[2]{}{a_j}\Big[\frac{\boldsymbol{\sigma}^2(a_j)}{2}P(a_j,t)\Big], 
\end{equation}
where $P(a_j,t)$ is the transitional PDF of $a_j$ at time $t$. When stationary, equation \ref{trans_pdf} can be integrated, yielding:
\begin{equation} \label{Gal_stat}
    P(a_j) = C a_j \exp \Big[ (-\epsilon + c_{j1}\tau) \frac{{\omega_j}^2 {a_j}^2}{2d c_{j4}^2} + ({\omega_j}^2 + \tau^2) \frac{3 c_{j3} \tau {a_j}^4}{16d c_{j4}^2} \Big],
\end{equation}
where $P(a_j)$ is the stationary PDF of $a_j$. At the pressure antinode of the first duct mode ($z=0.5$), the PDF of the pressure fluctuation amplitude for the $j^{th}$ Galerkin mode is:
\begin{equation}
    P(a_{pj}) = \frac{1}{\gamma M} P(a_j),
\end{equation}
where $P(a_{pj})$ is the stationary PDF of the pressure fluctuation amplitude for the $j^{th}$ mode.  

The overall pressure fluctuation amplitude of the Rijke tube ($P(a_p)$) is given by the combination of $P(a_{p1})$, $P(a_{p2})$, $\cdots$. However, the exact analytical function that combines $P(a_{p1})$, $P(a_{p2})$, $\cdots$ is unknown. Alternatively, we can calculate the maximum and minimum PDF profiles of $P(a_p)$. Specifically, we focus on the fact that---for the independent random variables $X$, $Y$ and their PDFs $P(X)$ and $P(Y)$---$P(X+Y)$ is given as the convolution between $P(X)$ and $P(Y)$ \citep{hogg2005}. Therefore, the maximum PDF of $P(a_p)$ is:
\begin{equation}
    P(a_{p,max}) = P(a_{p1}) * P(a_{p2}) * \cdots *P(a_{pn}) \qquad (P(a_{pj}\neq0)),
\end{equation}
where $P(a_{p,max})$ is the maximum PDF of the pressure fluctuation amplitude and $P(a_{pn})$ is the PDF of the pressure fluctuation amplitude for the $n^{th}$ mode. More specifically, the maximum and minimum PDF that can appear at amplitude $A$ by combining $P(a_{p1})$ and $P(a_{p2})$ are given as:
\begin{subequations} \label{convo}
\begin{align}
    P(a_{p,max}=A) &= \sum_{k=-\infty}^{\infty} P(a_{p1}=A) P(a_{p2}=A-k), \\
    P(a_{p,min}=A) &= \sum_{k=-\infty}^{\infty} P(a_{p1}=A) P(a_{p2}=k-A),
\end{align}
\end{subequations}
where $P(a_{p1})\neq0$ and $P(a_{p2})\neq0$. The convolution in equation \ref{convo} can be repeated to find the appropriate maximum and minimum PDF profiles.

\section{Numerical analysis and validation}
\begin{figure}
    \centering
    \includegraphics[clip,width=\textwidth]{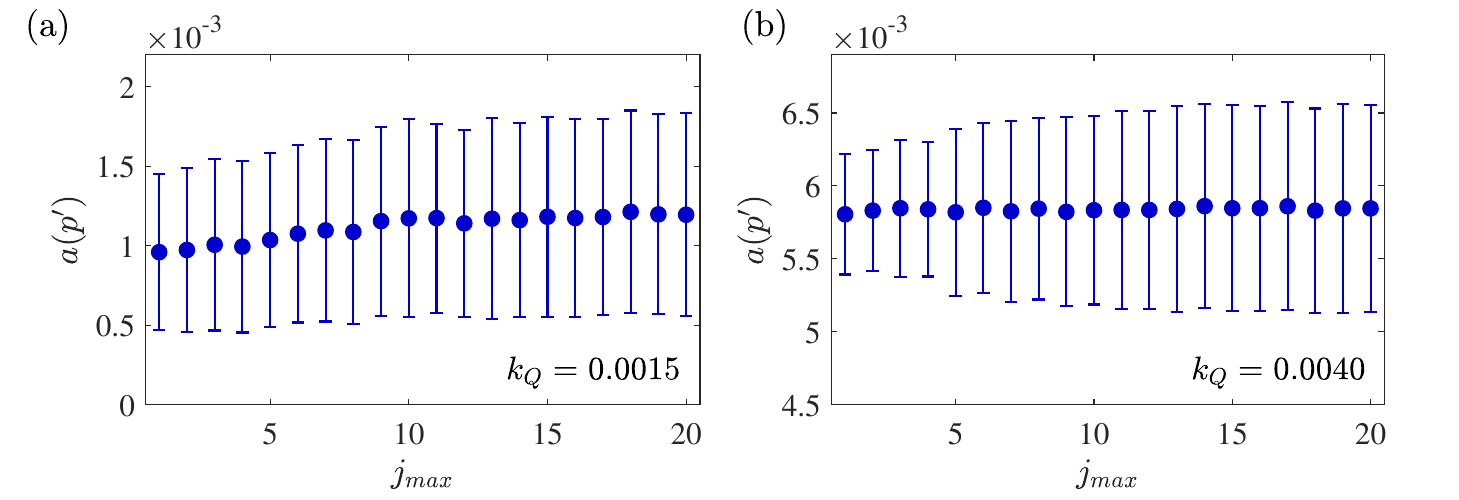}
    \caption{Convergence of the pressure fluctuation amplitude ($a(p')$) in (a) the fixed-point regime and (b) the limit-cycle regime. The round markers denote the mean and the whiskers show the standard deviation. $j_{max}$ is the number of Galerkin modes used for the simulation.}
\label{fig:converg}
\end{figure}

To validate the Fokker--Planck equation derived in \S\ref{der_ntau}, we run numerical simulations and compare the results with the analytical solution. For the numerical simulations, we solve equation \ref{gov} using a 4th-order Runge-Kutta (RK4) algorithm in the time span $0<t<1050$ with a time step of $dt=0.001$ ($t$ in arbitrary units). 20 Galerkin modes are superpositioned (equation \ref{Galer}) for convergence (see figure \ref{fig:converg}). 

\begin{figure}
    \centering
    \includegraphics[clip,width=0.65\textwidth]{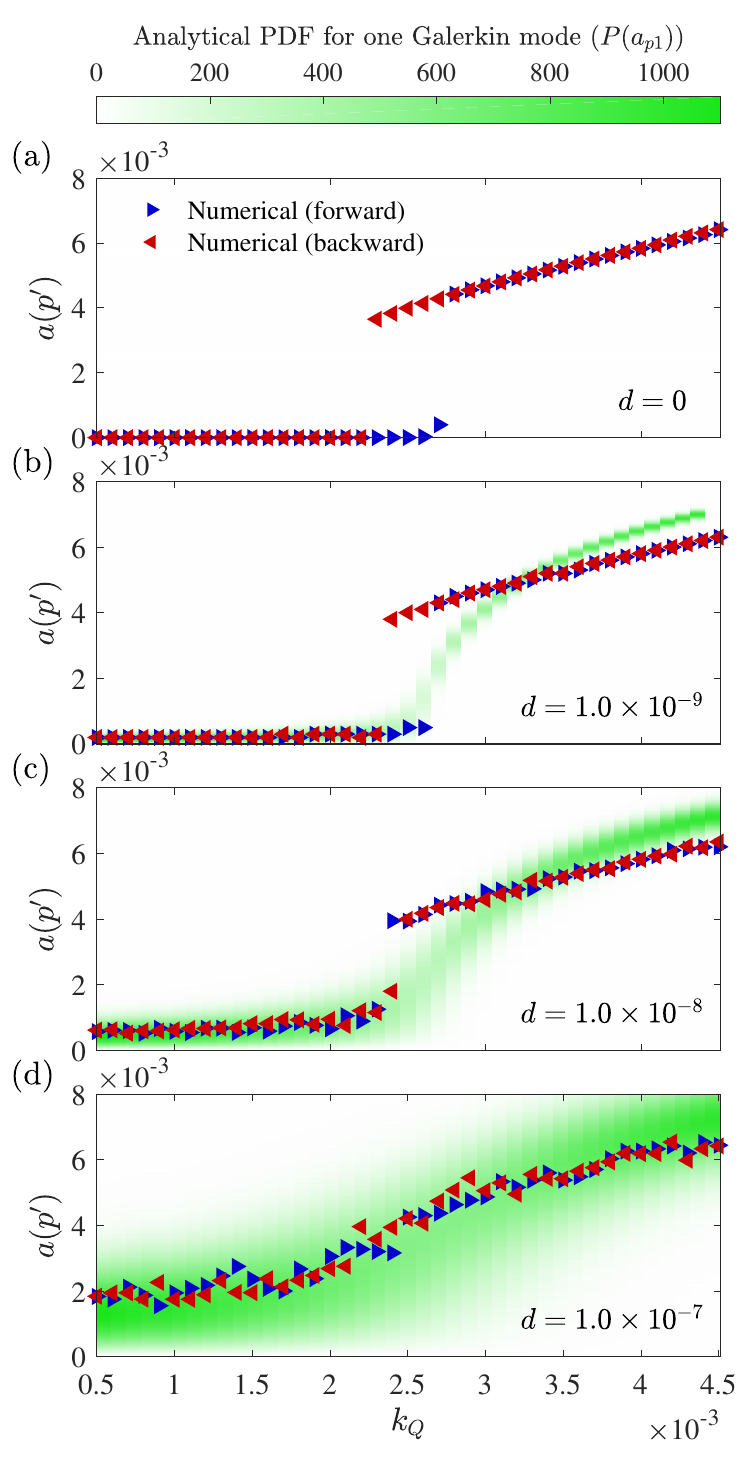}
    \caption{Subcritical bifurcation diagram of the Rijke tube with varying $k_Q$. Blue and red markers show the peak amplitude in the numerically obtained PDF, calculated with 20 Galerkin modes. Green bands show the analytical PDF, which is calculated with the Fokker--Planck equation of the first Galerkin mode.}
\label{fig:d_bif}
\end{figure}

As the heater power ($k_Q$) increases, the system bifurcates from a fixed point to limit-cycle oscillations via a subcritical Hopf bifurcation, (see figure \ref{fig:d_bif}). As the noise amplitude increases, the hysteretic region shrinks, consistent with the results of \citet{gopalakrishnan2015}. We compare these numerical results with the analytical Fokker--Planck solution in equation \ref{Gal_stat}. We can see from figure \ref{fig:d_bif} that the analytical result with only the first Galerkin mode matches the numerical result obtained with 20 modes. This means that the first Galerkin mode largely determines the dynamics of this Rijke tube, as shown in figure \ref{fig:converg}. 

\begin{figure}
    \centering
    \includegraphics[clip,trim=1mm 1mm 1mm 1mm,width=\textwidth]{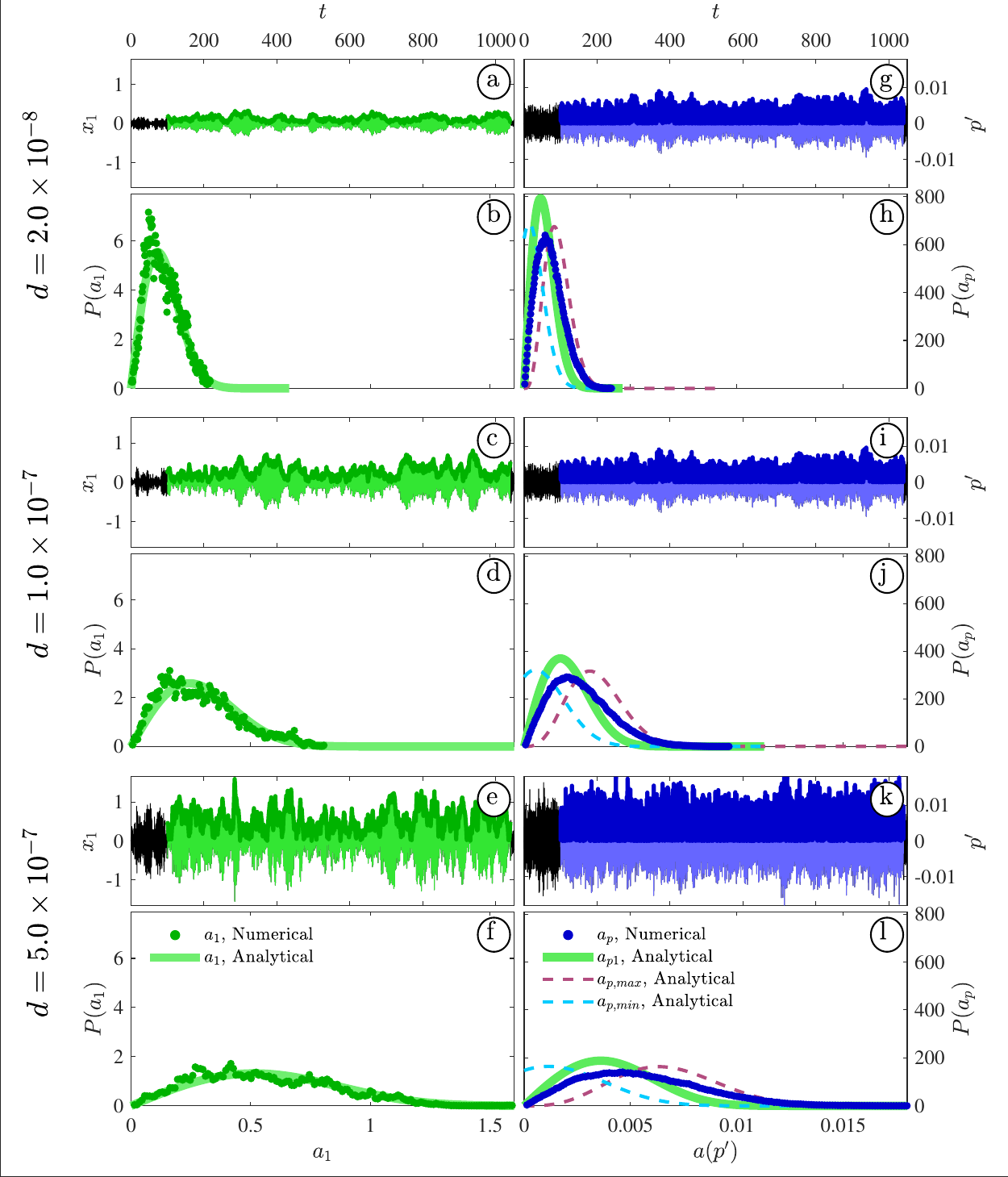}
    \caption{Comparison between the numerical and the analytical results in the fixed-point regime, far from the bifurcation ($k_Q=0.0015$). (a,c,e) Time traces for numerically obtained $x_1$ and (g,i,k) $p'$. (b,d,f) Numerical (green scatter dots) and analytical (green bands) PDFs of the state variable of the first Galerkin mode. (h,j,l) Numerically obtained $P(a_p)$ (blue scatter dots), analytical $P(a_{p1})$ obtained from the first Galerkin mode (green bands). Red and blue dashed lines show the maximum and the minimum PDF profiles that are analytically obtained by combining up to the fifth Galerkin mode. $z=0.5$, $z_f=0.25$, $\gamma=1.4$, $M=0.005$, $\tau=0.16$, $\epsilon_a=0.1$, $\epsilon_a=0.06$.}
\label{fig:d_fp}
\end{figure}

\begin{figure}
    \centering
    \includegraphics[clip,trim=1mm 1mm 1mm 1mm,width=\textwidth]{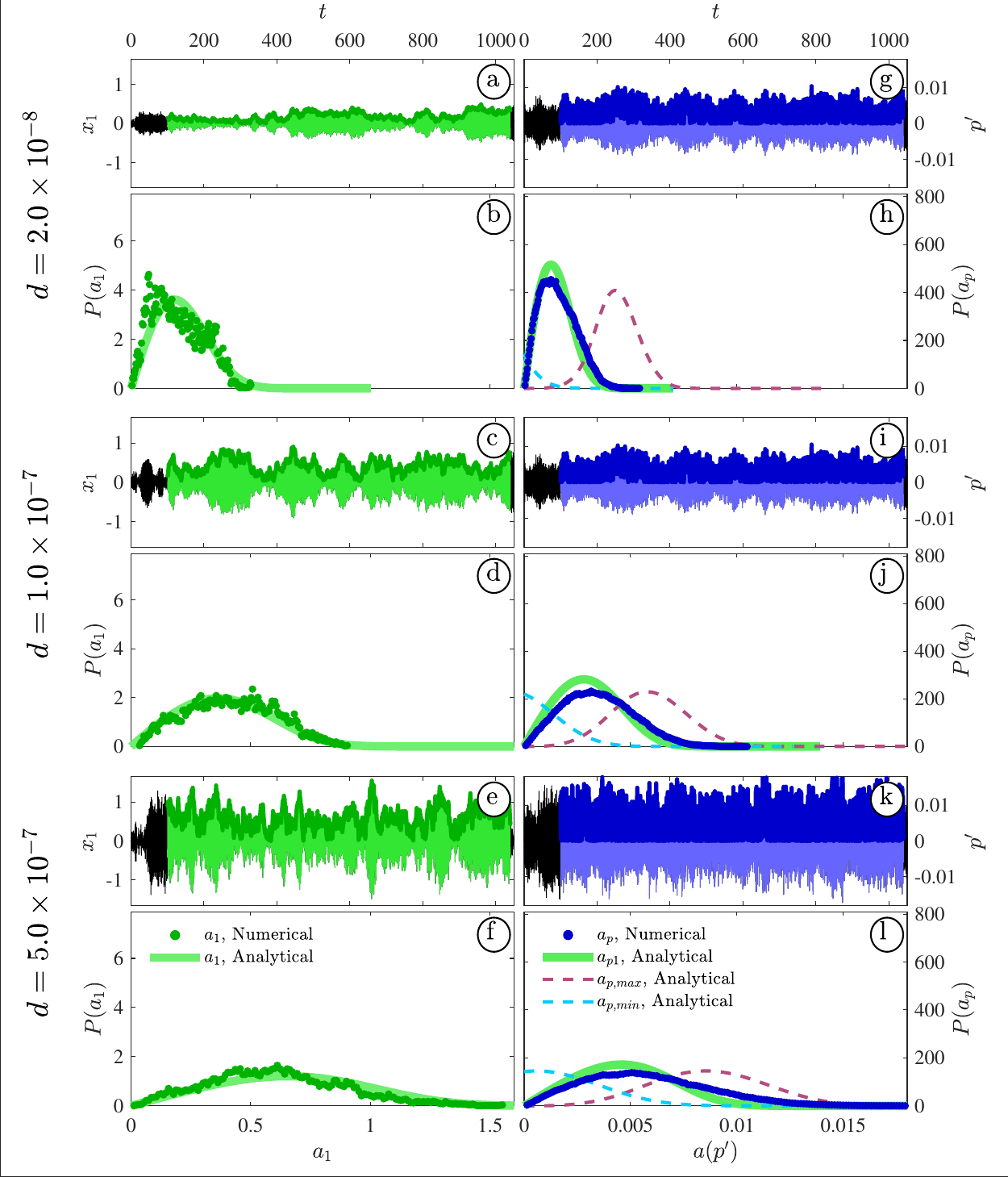}
    \caption{Comparison between the numerical and the analytical results in the fixed-point regime, close to the bifurcation ($k_Q=0.0022$). (a,c,e) Time traces for numerically obtained $x_1$ and (g,i,k) $p'$. (b,d,f) Numerical (green scatter dots) and analytical (green bands) PDFs of the state variable of the first Galerkin mode. (h,j,l) Numerically obtained $P(a_p)$ (blue scatter dots), analytical $P(a_{p1})$ obtained from the first Galerkin mode (green bands). Red and blue dashed lines show the maximum and the minimum PDF profiles that are analytically obtained by combining up to the fifth Galerkin mode. $z=0.5$, $z_f=0.25$, $\gamma=1.4$, $M=0.005$, $\tau=0.16$, $\epsilon_a=0.1$, $\epsilon_a=0.06$.}
\label{fig:d_fp2}
\end{figure}

\begin{figure}
    \centering
    \includegraphics[clip,trim=1mm 1mm 1mm 1mm,width=\textwidth]{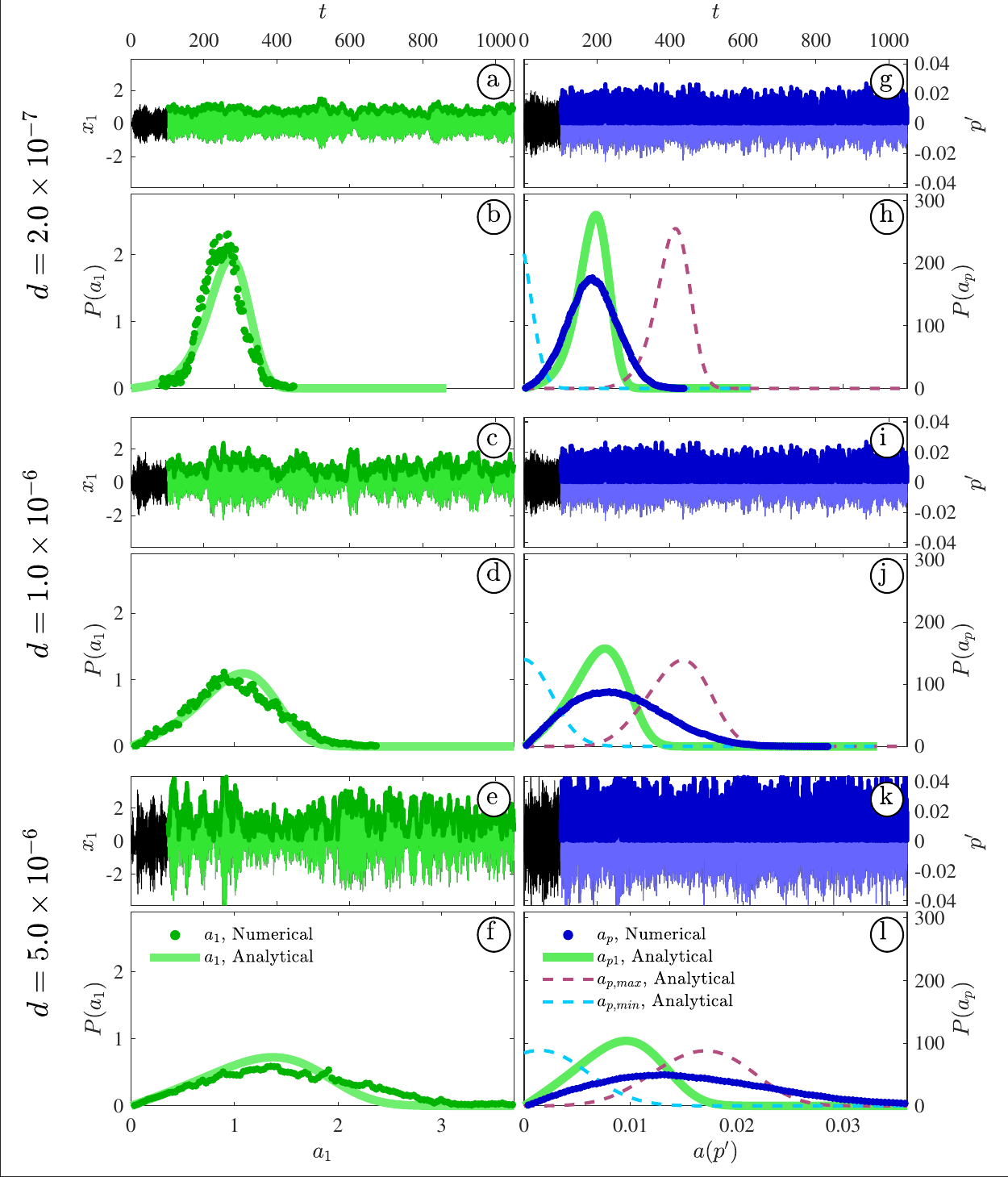}
    \caption{Comparison between the numerical and the analytical results in the limit-cycle regime ($k_Q=0.0040$). (a,c,e) Time traces for numerically obtained $x_1$ and (g,i,k) $p'$. (b,d,f) Numerical (green scatter dots) and analytical (green bands) PDFs of the state variable of the first Galerkin mode. (h,j,l) Numerically obtained $P(a_p)$ (blue scatter dots), analytical $P(a_{p1})$ obtained from the first Galerkin mode (green bands). Red and blue dashed lines show the maximum and the minimum PDF profiles that are analytically obtained by combining up to the fifth Galerkin mode. $z=0.5$, $z_f=0.25$, $\gamma=1.4$, $M=0.005$, $\tau=0.16$, $\epsilon_a=0.1$, $\epsilon_a=0.06$.}
\label{fig:d_lc}
\end{figure}

In further analysis, we choose three $k_Q$ values: the first in the fixed-point regime far before the bifurcation ($k_Q=1.5 \times 10^{-3}$), the second in the fixed-point regime close to the bifurcation ($k_Q=2.2 \times 10^{-3}$), and the third in the limit-cycle regime ($k_Q=4.0 \times 10^{-3}$). First, we validate the Fokker--Planck equation for $x_j$ (equation \ref{Gal_stat}) with one Galerkin mode. In figures \ref{fig:d_fp}(a-f), \ref{fig:d_fp2}(a-f) and \ref{fig:d_lc}(a-f), we find that the numerical and analytical $P(a_1)$ agree well with each other, confirming that the approximations made during the derivation (e.g. Maclaurin expansion in equation \ref{macl}, linearization in equation \ref{tau_approx}) are sufficiently valid.

Second, we analyze the PDF of the pressure fluctuation amplitude computed from the numerical simulation with 20 Galerkin modes ($P(a_p)$), and compare it with the analytical PDF derived for only one Galerkin mode ($P(a_{p1})$). Furthermore, to approximate the actual profile of $P(a_p)$, we include up to the fifth Galerkin mode ($P(a_{p1})$ - $P(a_{p5})$), extracting the maximum and the minimum PDF profiles (see equation \ref{convo}). The results are shown in figures \ref{fig:d_fp}(g-l), \ref{fig:d_fp2}(g-l) and \ref{fig:d_lc}(g-l).

Far before the bifurcation ($k_Q=1.5 \times 10^{-3}$, figure \ref{fig:d_fp}(g-l)), the maximum and minimum PDF profiles are close to each other. This implies that the maximum fluctuation amplitudes of non-primary modes are low, compared with those in the limit-cycle regime (figure \ref{fig:d_lc}(g-l)). Consequently, the analytical PDF with only the first Galerkin mode ($P(a_{p1})$) exhibits a similar trend to the numerical $P(a_p)$. However, because of the effect of the other modes, discrepancies arise between the two PDFs. Closer to the bifurcation point ($k_Q=2.2 \times 10^{-3}$, figure \ref{fig:d_fp2}(g-l)), the discrepancies between $P(a_{p1})$ and $P(a_p)$ are smaller. In other words, the system dynamics are better represented with only the first mode. This means that the noise-induced dynamics of the primary mode dominates the system dynamics when the system is close to the bifurcation point. Finally, when the system is in the limit-cycle regime ($k_Q=4.0 \times 10^{-3}$, figure \ref{fig:d_lc}(g-l)), the effect of non-primary modes becomes stronger, and the discrepancies between $P(a_{p1})$ and $P(a_p)$ are larger than those in the fixed-point regime. Even so, regardless of $k_Q$, the numerical $P(a_p)$ is located between the maximum and minimum PDF profiles, as shown in figures \ref{fig:d_fp}(h,j,l), \ref{fig:d_fp2}(h,j,l) and \ref{fig:d_lc}(h,j,l).


\section{Conclusions}

In this chapter, we have numerically analyzed a prototypical thermoacoustic system, a Rijke tube, undergoing a subcritical Hopf bifurcation as the heater power is varied. We modeled the dynamics of the Rijke tube using the momentum and energy equations. We expanded these governing equations with the Galerkin technique, and derived the corresponding Fokker--Planck equation. 

From the numerical analysis, we found that the Fokker--Planck equation for only the first Galerkin mode can represent reasonably well the full dynamics of a noise-perturbed Rijke tube, which is numerically reproduced with 20 Galerkin modes. Physically, this means that the pressure oscillation in the duct is mainly governed by the fundamental acoustic mode. Notably, we found that this single-mode approximation works best when the system is in the fixed-point regime, close to the bifurcation. From a practical perspective, this observation implies that a single-mode approximation can be used for system identification of a Rijke tube, and is most accurate in this particular region. In other words, when the system approaches the bifurcation point, one can find the deterministic parameters, such as the damping rate and the time delay, by fitting the experimental data to the Fokker--Planck equation of only the fundamental mode. The full dynamics of a Rijke tube can then be approximated from the governing equation for the first mode (equation \ref{gov}). A proper extrapolation of the extracted coefficients will lead to the prediction of the bifurcation point and the post-bifurcation dynamics.

Furthermore, by combining the PDFs of the different acoustic modes, the analytical functions showing the maximum and the minimum PDF profiles can be found. The obtained functions provide information on the actual PDF, without the need to run computationally expensive numerical simulations.

\chapter{Conclusions} \label{chap:concl}

\section{Summary}
In this thesis, we present a system identification (SI) framework that exploits the noise-induced dynamics in the fixed-point regime of a Hopf bifurcation. We model the system using a self-excited oscillator, a high-order van der Pol (VDP) equation, perturbed by an additive white Gaussian noise term. Assuming that the amplitude of the oscillations varies much more slowly than the oscillations themselves (i.e. assuming weak nonlinearity, or close to the Hopf point), we apply stochastic averaging, yielding the Fokker--Planck equation that describes the response of the system to noise. 

Two different versions of our SI framework---input-output and output-only---are presented. In the input-output version, white Gaussian noise is fed into the system by an external actuator, and the stochastic response of the system is measured. An actuator model that converts the actuator (loudspeaker) input voltage into the noise amplitude ($d$) is developed for this. The VDP coefficients are then identified by fitting the experimental data to the stationary Fokker--Planck equation. In the output-only version, the system's intrinsic noise is used as the noise source, and only the output signal is measured. An adjoint-based optimization algorithm is applied to accurately extract the VDP coefficients from the output signal. In both versions of the SI framework, we extrapolate the identified VDP coefficients to predict the locations and types of the Hopf bifurcation, and the post-bifurcation behavior. 

We demonstrated the SI framework on three experimental systems: a laminar hydrodynamic system (a low-density jet), a laminar thermoacoustic system (a flame-driven Rijke tube), and a turbulent thermoacoustic system (a gas turbine combustor). In \S\ref{chap:low}, we applied input-output SI to a low-density jet. By analyzing the system's noise-induced dynamics in the pre-bifurcation (fixed-point) regime, we were able to determine the order of nonlinearity and, hence, whether the Hopf bifurcation was supercritical or subcritical. We were also able to predict the location of the Hopf bifurcation and the post-bifurcation dynamics of the resultant limit cycle. We then applied the SI framework to a laminar thermoacoustic system (\S\ref{chap:Rij}), where we identified and predicted a supercritical Hopf bifurcation in a flame-driven Rijke tube. We showed that it is possible to capture the dynamics of the main oscillatory mode, even when it coexists with other modes. In \S\ref{chap:gas}, we applied output-only SI to a turbulent gas turbine combustor and showed that the proposed framework could work on practical systems. Furthermore, in \S\ref{chap:io_oo}, we determined the minimum level of noise required for reliable SI.

To prepare for future work, we derived and tested additional system models in \S\ref{chap:coupled} and \S\ref{chap:ntau}. In \S\ref{chap:coupled}, we modeled the dynamics of a system with two modes of oscillation using two coupled stochastic oscillators. The effects of coupling type, coupling strength, and noise amplitude were parametrically studied using the corresponding Fokker--Planck equations. In \S\ref{chap:ntau}, we modeled the dynamics of a Rijke tube using the relevant momentum and energy equations. We then derived the Fokker--Planck equation corresponding to these governing equations, and used it to study the effect of noise. These additional system models and their corresponding Fokker--Planck equations provide alternative approaches to SI. Specifically, if a system cannot be adequately modeled with a single self-excited oscillator, these system models may be considered as the governing equation, enabling SI to be performed using their corresponding Fokker--Planck equations.

The SI framework presented here has three significant advantages. First, it has been experimentally proven to be a robust forecasting tool for an impending bifurcation, without the need to set ad-hoc instability thresholds -- in contrast to most existing early-warning indicators (see \S\ref{sec:hopf_pr}). Second, to predict the location of the bifurcation point and the post-bifurcation dynamics, the presented SI framework uses data only from the fixed-point regime where the oscillation amplitude is small, rather than from the limit-cycle regime where the oscillation amplitude is potentially high enough to be dangerous. Lastly, the order and the signs of the nonlinear terms in the high-order VDP oscillator model can be identified, revealing the type of Hopf bifurcation--whether it is supercritical or subcritical. 

In this thesis, the proposed SI framework is applied to three experimental systems to demonstrate its applicability and versatility. It is worth noting, though, that the framework is applicable to other nonlinear dynamical systems as well, so long as the system obeys the normal-form equation for a Hopf bifurcation (i.e. the Stuart--Landau equation; equation~\ref{norm_Hopf}). Examples of such systems include chemical reactions \citep{kuramoto2003chemical}, optical lasers \citep{ludge2012nonlinear}, open shear flows \citep{Provansal1987}, and flow-induced vibrations \citep{liu2017}, among many others found in nature and engineering.

The SI framework presented here has three notable limitations. First, the location of the Hopf point and the post-bifurcation dynamics can only be predicted when the system is sufficiently close to the Hopf point (see \S\ref{future:extrap} for more discussions). Second, because the framework makes use of data sampled at a single location, the framework can only be applied to systems with a simple spatial structure. For systems with complex spatial dynamics, one may need to switch to other SI strategies that can process larger data matrices (e.g. SINDy algorithm of \citealp{Brunton2016}). Third, it should be noted that our system model is phenomenological, which is based on the normal-form equation of a Hopf bifurcation (i.e. the Stuart--Landau equation). Although this model is universally valid for all nonlinear dynamical systems near a Hopf bifurcation, it is not derived from first principles in most cases.\footnote{A counterexample to this is plane Poiseuille flow, for which the Stuart--Landau equation was derived from hydrodynamic equations by \citet{stuart1960non}.} Therefore, the SI framework cannot explicitly identify the physical mechanisms governing the system dynamics. 



\section{Future work} \label{future}

The SI methodology and applications presented in this thesis provide a solid foundation for future studies. Below, several questions arising from this study are discussed.

\subsection{Band-pass filtering} \label{future:band}

In the proposed input-output and output-only SI framework, a system is modeled with a self-excited oscillator equation featuring a single oscillation frequency. In other words, we assume that a single mode of oscillation dominates the system dynamics. However, if the primary mode is weak, or if other active modes coexist in the system, one might have to consider applying a band-pass filter. This is because a band-pass filter around the main oscillation frequency removes the effect of other modes, enabling the data to be more consistent with the single-mode approximation. However, applying such a filter removes a significant amount of information about the system itself, and violates the Markovian properties of noise. In other words, the effect of noise on the system is confined within the bandwidth of the filter. Considering that the Fokker--Planck equation is derived under the white noise assumption (i.e. noise being delta-correlated), it is likely that a band-pass filter will give rise to inconsistencies with the Fokker--Planck equation. Therefore, these two aspects of band-pass filtering should be considered when performing SI using the noise-induced dynamics.

In this thesis, a band-pass filter was not applied for the low-density jet (\S\ref{chap:low}) and the Rijke tube (\S\ref{chap:Rij}), but one was applied for the gas turbine combustor (\S\ref{chap:gas}). This is because the first two systems are laminar with weak background noise, which implies that the main oscillatory mode is dominant even in the fixed-point regime. By contrast, in the third system, which is turbulent, the background noise is strong and the primary mode is comparatively weak, making band-pass filtering essential. Accordingly, we applied adjoint-based optimization to the turbulent gas-turbine combustor in order to mitigate the adverse effect of band-pass filtering. However, a comprehensive parametric analysis of the effect of band-pass filtering on the SI framework has not been carried out yet. In future work, one may apply a progressively narrower bandwidth around the main oscillation frequency and find an optimal filter that yields the most reliable predictions.

\subsection{Noise properties} \label{future:noise}

In future studies, different types of noise can be used in the system model. In the present SI framework, it is assumed that the noise is additive. Such an assumption allowed for a simplification of the mathematical treatment (e.g. derivation of the Fokker--Planck equation), and was justified by the following grounds: (i) in the input-output version, the extrinsic loudspeaker-generated noise fed into the quiescent system was independent of the system state or its parameters. (ii) in the output-only version, turbulence was used as the noise source, and previous studies have shown that it is valid to assume that additive noise is dominant in such a turbulent environment \citep{NOIRAY2013152, noiray2017method}.

However, in practical systems, noise originates from several different sources, and can therefore be not only additive but also parametric or multiplicative. For example, \citet{clavin1994} suggested that pressure oscillations in a combustion chamber could arise as a result of fluctuations in the linear growth rate in the Stuart--Landau equation. On that basis, \citet{clavin1994} derived a Fokker--Planck equation from the stochastic differential equation containing a multiplicative noise term. However, SI using such a multiplicatively forced model remains to be explored. 



\subsection{System model}

In this thesis, model-based SI was conducted using the VDP oscillator, the Stuart--Landau equation, and the corresponding Fokker--Planck equation. These equations can successfully reproduce and predict the system dynamics, but can only permit a single dominant mode of oscillation. Future studies may consider relaxing this limitation. For example, systems with two modes of oscillation can be modeled with two coupled stochastic oscillators and their corresponding Fokker--Planck equation, as shown in \S\ref{chap:coupled}. SI using this system model, however, has yet to be performed but should be feasible. Specifically, if two peaks coexist in the power spectrum of a system, band-pass filters can be applied to isolate the two modes. Then, the deterministic parameters characterizing these two modes can be extracted by fitting the experimental probability density function of the oscillation amplitude to the Fokker--Planck equations derived in \S\ref{chap:coupled}.

In addition, it could be interesting to adopt data-driven computational tools to identify the system model. For example, \citet{Brunton2016} recognized that most physical systems can be represented with a system model containing just a few terms. With the aid of sparsity-promoting tools and machine learning, those authors showed that these terms could be identified without assumptions on the form of the governing equations. However, prediction of the bifurcation point and the post-bifurcation dynamics using such a data-driven approach has yet to be comprehensively explored. Given that sufficient data is available, combining such computational techniques with the bifurcation prediction methods presented in this thesis could potentially enable the identification and prediction of various types of bifurcations.

\subsection{Extrapolation-based prediction} \label{future:extrap}

To predict the Hopf point and the post-bifurcation dynamics, we extrapolate the VDP coefficients to the limit-cycle regime (figures \ref{fig:coeff}, \ref{fig:4_7} and \ref{fig:fig5_4}). This extrapolation-based SI has two notable limitations.

First, the Hopf point can be found only when the system is sufficiently close to the bifurcation. This is because the Fokker--Planck equation used for SI is derived under the assumption that the growth rate of the system is small ($|\epsilon| \approx 0$). Therefore, this SI framework is valid only near the Hopf point, and the prediction can only be initiated when a sufficient number of consecutive increases in $\epsilon$ have occurred (see \S\ref{sec:5_4}). Consequently, five, seven and five data points before the Hopf point are used for SI in chapters \S\ref{chap:low}, \S\ref{chap:Rij} and \S\ref{chap:gas}, respectively (see figures \ref{fig:coeff}, \ref{fig:4_7} and \ref{fig:fig5_4}). Further mathematical and experimental analysis can be performed in order to expand the region where the SI framework is valid, so as to enable prediction of the Hopf bifurcation further in advance.

Second, for prediction of the post-bifurcation dynamics, the nonlinear coefficients ($\alpha_1$, $\alpha_2$, $\cdots$) are extrapolated using power-law regression. This is done because the empirical relationship between the nonlinear coefficients and the bifurcation parameter obeyed a power-law trend. Although this trend was found to be universal---appearing in the low-density jet (figure \ref{fig:coeff}), Rijke tube (figure \ref{fig:4_7}), and gas turbine combustor (figure \ref{fig:fig5_4})---it is purely empirical, with no physical justification. It could be interesting to derive an analytical relationship between the nonlinear coefficients and the bifurcation parameter, as that could provide valuable physical insight into the system.

\newpage
\addcontentsline{toc}{chapter}{References}
\bibliographystyle{agsmmodify}
\bibliography{reference} 

\begin{thebibliography}{}

\bibitem[Abel \emph{et~al.}, 2009]{abel2009}
Abel, M., Ahnert, K., and Bergweiler, S. (2009).
\newblock Synchronization of sound sources.
\newblock {\em Phys. Rev. Lett.}, 103(11):114301.

\bibitem[Acharya \emph{et~al.}, 2018]{acharya2018}
Acharya, V.~S., Bothien, M.~R., and Lieuwen, T. (2018).
\newblock Non-linear dynamics of thermoacoustic eigen-mode interactions.
\newblock {\em Combust. Flame}, 194:309--321.

\bibitem[Albert \emph{et~al.}, 1999]{albert1999}
Albert, R., Jeong, H., and Barab{\'a}si, A.-L. (1999).
\newblock Internet: Diameter of the world-wide web.
\newblock {\em Nature}, 401(6749):130.

\bibitem[Anishchenko \emph{et~al.}, 2007]{anishchenko2007}
Anishchenko, V., Nikolaev, S., and Kurths, J. (2007).
\newblock Peculiarities of synchronization of a resonant limit cycle on a
  two-dimensional torus.
\newblock {\em Phys. Rev. E}, 76(4):046216.

\bibitem[{\AA}str{\"o}m and Eykhoff, 1971]{Astrom1971}
{\AA}str{\"o}m, K. and Eykhoff, P. (1971).
\newblock System identification--a survey.
\newblock {\em Automatica}, 7(2):123 -- 162.

\bibitem[Balanov \emph{et~al.}, 2008]{balanov2008}
Balanov, A., Janson, N., Postnov, D., and Sosnovtseva, O. (2008).
\newblock {\em Synchronization: from simple to complex}.
\newblock Springer Science \& Business Media.

\bibitem[Balasubramanian and Sujith, 2008]{balasubramanian2008}
Balasubramanian, K. and Sujith, R.~I. (2008).
\newblock Thermoacoustic instability in a {R}ijke tube: Non-normality and
  nonlinearity.
\newblock {\em Phys. Fluids}, 20(4):044103.

\bibitem[Balusamy \emph{et~al.}, 2017]{balusamy2017extracting}
Balusamy, S., Li, L. K.~B., Han, Z., and Hochgreb, S. (2017).
\newblock Extracting flame describing functions in the presence of self-excited
  thermoacoustic oscillations.
\newblock {\em P. Combust. Inst.}, 36(3):3851--3861.

\bibitem[Balusamy \emph{et~al.}, 2015]{balusamy2015nonlinear}
Balusamy, S., Li, L. K.~B., Han, Z., Juniper, M.~P., and Hochgreb, S. (2015).
\newblock Nonlinear dynamics of a self-excited thermoacoustic system subjected
  to acoustic forcing.
\newblock {\em P. Combust. Inst.}, 35(3):3229--3236.

\bibitem[Bandt and Pompe, 2002]{bandt2002}
Bandt, C. and Pompe, B. (2002).
\newblock Permutation entropy: a natural complexity measure for time series.
\newblock {\em Phys. Rev. Lett.}, 88(17):174102.

\bibitem[Bassingthwaighte \emph{et~al.}, 1994]{bassingthwaighte1994}
Bassingthwaighte, J.~B., Liebovitch, L.~S., and West, B.~J. (1994).
\newblock {\em Fractal Physiology}.
\newblock Oxford University Press, USA.

\bibitem[Beck and van Straten, 1983]{Beck1983}
Beck, M. and van Straten, G. (1983).
\newblock {\em Uncertainty and forecasting of water quality}.
\newblock Springer-Verlag, Berlin ; New York.

\bibitem[Bekey and Beneken, 1978]{BEKEY1978}
Bekey, G.~A. and Beneken, J.~E. (1978).
\newblock Identification of biological systems: a survey.
\newblock {\em Automatica}, 14(1):41 -- 47.

\bibitem[Benzi, 2010]{Benzi2010}
Benzi, R. (2010).
\newblock Stochastic resonance: from climate to biology.
\newblock {\em Nonlinear Proc. Geoph.}, 17(5):431.

\bibitem[Benzi \emph{et~al.}, 1982]{Benzi1982}
Benzi, R., Parisi, G., Sutera, A., and Vulpiani, A. (1982).
\newblock Stochastic resonance in climatic change.
\newblock {\em Tellus}, 34(1):10--16.

\bibitem[Benzi \emph{et~al.}, 1983]{Benzi1983}
Benzi, R., Parisi, G., Sutera, A., and Vulpiani, A. (1983).
\newblock A theory of stochastic resonance in climatic change.
\newblock {\em SIAM J. appl. math.}, 43(3):565--578.

\bibitem[Benzi \emph{et~al.}, 1981]{Benzi1981}
Benzi, R., Sutera, A., and Vulpiani, A. (1981).
\newblock The mechanism of stochastic resonance.
\newblock {\em J. Phys. A}, 14(11):L453.

\bibitem[Bhaban \emph{et~al.}, 2016]{Bhaban2016}
Bhaban, S., Talukdar, S., and Salapaka, M. (2016).
\newblock Noise induced transport at microscale enabled by optical fields.
\newblock In {\em P. Amer. Contr. Conf.}, pages 5823--5829.

\bibitem[Blevins and Scanlan, 1977]{blevins1977}
Blevins, R. and Scanlan, R. (1977).
\newblock {\em Flow-induced vibration}.
\newblock American Society of Mechanical Engineers Digital Collection.

\bibitem[Boccaletti \emph{et~al.}, 2006]{boccaletti2006}
Boccaletti, S., Latora, V., Moreno, Y., Chavez, M., and Hwang, D.-U. (2006).
\newblock Complex networks: Structure and dynamics.
\newblock {\em Phys. Rep.}, 424(4-5):175--308.

\bibitem[Bonciolini, 2019]{bonciolini2019modelling}
Bonciolini, G. (2019).
\newblock {\em Modelling of thermoacoustic dynamics in gas turbines
  applications}.
\newblock PhD thesis, ETH Zurich.

\bibitem[Bonciolini \emph{et~al.}, 2017]{bonciolini2017output}
Bonciolini, G., Boujo, E., and Noiray, N. (2017).
\newblock Output-only parameter identification of a colored-noise-driven
  {V}an-der-{P}ol oscillator: Thermoacoustic instabilities as an example.
\newblock {\em Phys. Rev. E}, 95(6):062217.

\bibitem[Bonciolini \emph{et~al.}, 2018]{Bonciolini172078}
Bonciolini, G., Ebi, D., Boujo, E., and Noiray, N. (2018).
\newblock Experiments and modelling of rate-dependent transition delay in a
  stochastic subcritical bifurcation.
\newblock {\em R. Soc. Open Sci.}, 5(3):172078.

\bibitem[B{\"o}ttcher \emph{et~al.}, 2006]{bottcher2006}
B{\"o}ttcher, F., Peinke, J., Kleinhans, D., Friedrich, R., Lind, P.~G., and
  Haase, M. (2006).
\newblock Reconstruction of complex dynamical systems affected by strong
  measurement noise.
\newblock {\em Phys. Rev. Lett.}, 97(9):090603.

\bibitem[Boujo \emph{et~al.}, 2020]{Boujo2020}
Boujo, E., Bourquard, C., Xiong, Y., and Noiray, N. (2020).
\newblock Processing time-series of randomly forced self-oscillators: The
  example of beer bottle whistling.
\newblock {\em J. Sound Vib.}, 464:114981.

\bibitem[Boujo and Noiray, 2017]{boujo2017robust}
Boujo, E. and Noiray, N. (2017).
\newblock Robust identification of harmonic oscillator parameters using the
  adjoint {F}okker--{P}lanck equation.
\newblock {\em Proc. R. Soc. A}, 473(2200):20160894.

\bibitem[Brunton \emph{et~al.}, 2016]{Brunton2016}
Brunton, S.~L., Proctor, J.~L., and Kutz, J.~N. (2016).
\newblock Discovering governing equations from data by sparse identification of
  nonlinear dynamical systems.
\newblock {\em Proc. Natl. Acad. Sci. USA}, 113(15):3932--3937.

\bibitem[Burnley and Culick, 2000]{burnley2000}
Burnley, V.~S. and Culick, F.~E. (2000).
\newblock Influence of random excitations on acoustic instabilities in
  combustion chambers.
\newblock {\em AIAA J.}, 38(8):1403--1410.

\bibitem[Candel, 2002]{Candel2002}
Candel, S. (2002).
\newblock Combustion dynamics and control: {Progress} and challenges.
\newblock {\em P. Combust. Inst.}, 29(1):1--28.

\bibitem[Carpenter \emph{et~al.}, 2011]{carpenter2011}
Carpenter, S.~R., Cole, J.~J., Pace, M.~L., Batt, R., Brock, W.~A., Cline, T.,
  Coloso, J., Hodgson, J.~R., Kitchell, J.~F., Seekell, D.~A., \emph{et~al.}
  (2011).
\newblock Early warnings of regime shifts: a whole-ecosystem experiment.
\newblock {\em Science}, 332(6033):1079--1082.

\bibitem[Charakopoulos \emph{et~al.}, 2014]{charakopoulos2014}
Charakopoulos, A.~K., Karakasidis, T.~E., Papanicolaou, P.~N., and Liakopoulos,
  A. (2014).
\newblock The application of complex network time series analysis in turbulent
  heated jets.
\newblock {\em Chaos}, 24(2):024408.

\bibitem[Chisholm and Filotas, 2009]{chisholm2009}
Chisholm, R.~A. and Filotas, E. (2009).
\newblock Critical slowing down as an indicator of transitions in two-species
  models.
\newblock {\em J. Theor. Biol.}, 257(1):142--149.

\bibitem[Chomaz, 2005]{chomaz2005}
Chomaz, J.-M. (2005).
\newblock Global instabilities in spatially developing flows: non-normality and
  nonlinearity.
\newblock {\em Annu. Rev. Fluid Mech.}, 37:357--392.

\bibitem[Clavin \emph{et~al.}, 1994]{clavin1994}
Clavin, P., Kim, J.~S., and Williams, F.~A. (1994).
\newblock Turbulence-induced noise effects on high-frequency combustion
  instabilities.
\newblock {\em Combust. Sci. Technol.}, 96(1-3):61--84.

\bibitem[Cordier \emph{et~al.}, 2005]{cordier2005}
Cordier, S., Pareschi, L., and Toscani, G. (2005).
\newblock On a kinetic model for a simple market economy.
\newblock {\em J. Stat. Phys.}, 120(1-2):253--277.

\bibitem[Cuenod and Sage, 1968]{Cuenod1968}
Cuenod, M. and Sage, A. (1968).
\newblock Comparison of some methods used for process identification.
\newblock {\em Automatica}, 4(4):235 -- 269.

\bibitem[Culick, 2006]{culick2006}
Culick, F. E.~C. (2006).
\newblock Unsteady motions in combustion chambers for propulsion systems.
\newblock AGARDograph RTO-AG-AVT-039, {NATO} Research and Technology
  Organization.

\bibitem[Culick \emph{et~al.}, 1992]{culick1992}
Culick, F. E.~C., Paparizos, L., Sterling, J., and Burnley, V. (1992).
\newblock Combustion noise and combustion instabilities in propulsion systems.
\newblock AGARD N93-10666 01-71, {NATO} Research and Technology Organization.

\bibitem[Dai \emph{et~al.}, 2012]{dai2012}
Dai, L., Vorselen, D., Korolev, K.~S., and Gore, J. (2012).
\newblock Generic indicators for loss of resilience before a tipping point
  leading to population collapse.
\newblock {\em Science}, 336(6085):1175--1177.

\bibitem[Dakos \emph{et~al.}, 2008]{dakos2008}
Dakos, V., Scheffer, M., van Nes, E.~H., Brovkin, V., Petoukhov, V., and Held,
  H. (2008).
\newblock Slowing down as an early warning signal for abrupt climate change.
\newblock {\em Proc. Natl. Acad. Sci. USA}, 105(38):14308--14312.

\bibitem[Dakos \emph{et~al.}, 2013]{dakos2013}
Dakos, V., van Nes, E.~H., and Scheffer, M. (2013).
\newblock Flickering as an early warning signal.
\newblock {\em Theor. ecol.}, 6(3):309--317.

\bibitem[Dijkstra \emph{et~al.}, 2014]{dijkstra_2014}
Dijkstra, H.~A., Wubs, F.~W., Cliffe, A.~K., Doedel, E., Dragomirescu, I.~F.,
  Eckhardt, B., Gelfgat, A.~Y., Hazel, A.~L., Lucarini, V., Salinger, A.~G.,
  Phipps, E.~T., Sanchez-Umbria, J., Schuttelaars, H., Tuckerman, L.~S., and
  Thiele, U. (2014).
\newblock Numerical bifurcation methods and their application to fluid
  dynamics: analysis beyond simulation.
\newblock {\em Commun. Comput. Phys.}, 15:1 –-- 45.

\bibitem[Divshali \emph{et~al.}, 2009]{Divshali2009}
Divshali, P.~H., Hosseinian, S.~H., Nasr, E., and Vahidi, B. (2009).
\newblock Reliable prediction of {H}opf bifurcation in power systems.
\newblock {\em Electr. Eng.}, 91(2):61--68.

\bibitem[Doering, 1986]{Doering1986}
Doering, C.~R. (1986).
\newblock Stability and dynamics of a noise-induced stationary state.
\newblock {\em Phys. Rev. A}, 34:2564--2567.

\bibitem[Dong and Long, 2018]{dong2018}
Dong, K. and Long, L. (2018).
\newblock Complexity-entropy causality plane based on return intervals: A
  useful approach to quantify the aeroengine gas path parameters.
\newblock {\em Math. Probl. Eng.}, 2018.

\bibitem[Dowling and Mahmoudi, 2015]{dowling2015combustion}
Dowling, A.~P. and Mahmoudi, Y. (2015).
\newblock Combustion noise.
\newblock {\em P. Combust. Inst.}, 35(1):65--100.

\bibitem[Drazin and Reid, 2004]{drazin2004hydrodynamic}
Drazin, P.~G. and Reid, W.~H. (2004).
\newblock {\em Hydrodynamic stability}.
\newblock Cambridge university press.

\bibitem[Dudley, 1983]{DUDLEY1983}
Dudley, D. (1983).
\newblock Parametric identification of transient electromagnetic systems.
\newblock {\em Wave Motion}, 5(4):369 -- 384.

\bibitem[Dusek \emph{et~al.}, 1994]{Dusek1994}
Dusek, J., Le~Gal, P., and Fraune, P. (1994).
\newblock A numerical and theoretical study of the first {H}opf bifurcation in
  a cylinder wake.
\newblock {\em J. Fluid Mech.}, 264:59--80.

\bibitem[Einstein, 1905]{einstein1905}
Einstein, A. (1905).
\newblock On the movement of small particles suspended in stationary liquids
  required by the molecular-kinetic theory of heat.
\newblock {\em Ann. Phys.}, 17:549--560.

\bibitem[Etikyala and Sujith, 2017]{Etikyala2017}
Etikyala, S. and Sujith, R.~I. (2017).
\newblock Change of criticality in a prototypical thermoacoustic system.
\newblock {\em Chaos}, 27(2):023106.

\bibitem[Eykhoff, 1968]{Eykhoff1968}
Eykhoff, P. (1968).
\newblock Process parameter and state estimation.
\newblock {\em Automatica}, 4(4):205 -- 233.

\bibitem[Falconer, 2004]{falconer2004}
Falconer, K. (2004).
\newblock {\em Fractal geometry: mathematical foundations and applications}.
\newblock John Wiley \& Sons.

\bibitem[Fard \emph{et~al.}, 2004]{Fard2004}
Fard, M.~A., Ishihara, T., and Inooka, H. (2004).
\newblock Identification of the head-neck complex in response to trunk
  horizontal vibration.
\newblock {\em Biol. Cybern.}, 90(6):418--426.

\bibitem[Fauve and Heslot, 1983]{Fauve1983}
Fauve, S. and Heslot, F. (1983).
\newblock Stochastic resonance in a bistable system.
\newblock {\em Phys. Lett. A}, 97(1):5 -- 7.

\bibitem[Fokker, 1914]{fokker1914}
Fokker, A.~D. (1914).
\newblock Die mittlere energie rotierender elektrischer dipole im
  strahlungsfeld.
\newblock {\em Ann. Phys.}, 348(43):810--820.

\bibitem[Fraser and Swinney, 1986]{fraser1986}
Fraser, A.~M. and Swinney, H.~L. (1986).
\newblock Independent coordinates for strange attractors from mutual
  information.
\newblock {\em Phys. Rev. A}, 33:1134--1140.

\bibitem[Friedrich and Peinke, 1997]{friedrich1997}
Friedrich, R. and Peinke, J. (1997).
\newblock Statistical properties of a turbulent cascade.
\newblock {\em Physica D}, 102(1-2):147--155.

\bibitem[Friedrich \emph{et~al.}, 2000]{friedrich2000}
Friedrich, R., Siegert, S., Peinke, J., Siefert, M., Lindemann, M., Raethjen,
  J., Deuschl, G., Pfister, G., \emph{et~al.} (2000).
\newblock Extracting model equations from experimental data.
\newblock {\em Phys. Lett. A}, 271(3):217--222.

\bibitem[Fu and Li, 2013]{Fu2013}
Fu, L. and Li, P. (2013).
\newblock The research survey of system identification method.
\newblock {\em 2013 5th International Conference on Intelligent Human-Machine
  Systems and Cybernetics}, 2:397--401.

\bibitem[Fussmann \emph{et~al.}, 2000]{fussmann2000}
Fussmann, G.~F., Ellner, S.~P., Shertzer, K.~W., and Hairston~Jr, N.~G. (2000).
\newblock Crossing the {H}opf bifurcation in a live predator-prey system.
\newblock {\em Science}, 290(5495):1358--1360.

\bibitem[Gammaitoni \emph{et~al.}, 1998]{Gammaitoni1998}
Gammaitoni, L., H{\"a}nggi, P., Jung, P., and Marchesoni, F. (1998).
\newblock Stochastic resonance.
\newblock {\em Rev. Mod. Phys.}, 70:223--287.

\bibitem[Gang \emph{et~al.}, 1993]{gang1993}
Gang, H., Ditzinger, T., Ning, C.-Z., and Haken, H. (1993).
\newblock Stochastic resonance without external periodic force.
\newblock {\em Phys. Rev. Lett.}, 71(6):807.

\bibitem[Ganopolski and Rahmstorf, 2002]{Ganopolski2002}
Ganopolski, A. and Rahmstorf, S. (2002).
\newblock Abrupt glacial climate changes due to stochastic resonance.
\newblock {\em Phys. Rev. Lett.}, 88(3):123903.

\bibitem[Gao and Ma, 2009]{gao2009}
Gao, Q. and Ma, J. (2009).
\newblock Chaos and {H}opf bifurcation of a finance system.
\newblock {\em Nonlinear Dyn.}, 58(1-2):209.

\bibitem[Ghadami \emph{et~al.}, 2018]{ghadami2018}
Ghadami, A., Cesnik, C. E.~S., and Epureanu, B.~I. (2018).
\newblock Model-less forecasting of {H}opf bifurcations in fluid-structural
  systems.
\newblock {\em J. Fluid Struct.}, 76:1--13.

\bibitem[Ghadami and Epureanu, 2016]{ghadami2016}
Ghadami, A. and Epureanu, B.~I. (2016).
\newblock Bifurcation forecasting for large dimensional oscillatory systems:
  forecasting flutter using gust responses.
\newblock {\em J. Comput. Nonlin. Dyn.}, 11(6):061009.

\bibitem[Ghadami and Epureanu, 2017]{ghadami2017}
Ghadami, A. and Epureanu, B.~I. (2017).
\newblock Forecasting the post-bifurcation dynamics of large-dimensional
  slow-oscillatory systems using critical slowing down and center space
  reduction.
\newblock {\em Nonlinear Dyn.}, 88(1):415--431.

\bibitem[Girvan and Newman, 2002]{girvan2002}
Girvan, M. and Newman, M. E.~J. (2002).
\newblock Community structure in social and biological networks.
\newblock {\em Proc. Natl. Acad. Sci. U.S.A}, 99(12):7821--7826.

\bibitem[Gitterman, 2013]{gitterman2013}
Gitterman, M. (2013).
\newblock {\em The noisy oscillator: random mass, frequency, damping}.
\newblock World Scientific, 2nd edition.

\bibitem[Gopalakrishnan \emph{et~al.}, 2016a]{Gopalakrishnan2016b}
Gopalakrishnan, E.~A., Sharma, Y., John, T., Dutta, P.~S., and Sujith, R.~I.
  (2016a).
\newblock Early warning signals for critical transitions in a thermoacoustic
  system.
\newblock {\em Scientific Reports}, 6(1).

\bibitem[Gopalakrishnan and Sujith, 2015]{gopalakrishnan2015}
Gopalakrishnan, E.~A. and Sujith, R.~I. (2015).
\newblock Effect of external noise on the hysteresis characteristics of a
  thermoacoustic system.
\newblock {\em J. Fluid Mech.}, 776:334--353.

\bibitem[Gopalakrishnan \emph{et~al.}, 2016b]{Gopalakrishnan2016}
Gopalakrishnan, E.~A., Tony, J., Sreelekha, E., and Sujith, R.~I. (2016b).
\newblock Stochastic bifurcations in a prototypical thermoacoustic system.
\newblock {\em Phys. Rev. E}, 94:022203.

\bibitem[Gotoda \emph{et~al.}, 2012]{gotoda2012}
Gotoda, H., Amano, M., Miyano, T., Ikawa, T., Maki, K., and Tachibana, S.
  (2012).
\newblock Characterization of complexities in combustion instability in a lean
  premixed gas-turbine model combustor.
\newblock {\em Chaos}, 22(4):043128.

\bibitem[Granger and Newbold, 1986]{Granger1986}
Granger, C. and Newbold, P. (1986).
\newblock {\em Forecasting Economic Time Series}.
\newblock Elsevier, 2nd edition.

\bibitem[Guan \emph{et~al.}, 2016]{guan2016forcedsync}
Guan, Y., Gupta, V., Kashinath, K., and Li, L. K.~B. (2016).
\newblock Forced synchronization of thermoacoustic oscillations in a ducted
  flame.
\newblock {\em Bull. Am. Phys. Soc.}, 61.

\bibitem[Guan \emph{et~al.}, 2017]{guan2017openloop}
Guan, Y., Gupta, V., Kashinath, K., and Li, L. K.~B. (2017).
\newblock Open-loop control of quasiperiodic thermoacoustic oscillations.
\newblock {\em Bull. Am. Phys. Soc.}, 62.

\bibitem[Guan \emph{et~al.}, 2018a]{guan2018suppression}
Guan, Y., Gupta, V., Kashinath, K., and Li, L. K.~B. (2018a).
\newblock Suppression of chaotic thermoacoustic oscillations by external
  acoustic forcing.
\newblock {\em Bull. Am. Phys. Soc.}, 63.

\bibitem[Guan \emph{et~al.}, 2019a]{guan2019open}
Guan, Y., Gupta, V., Kashinath, K., and Li, L. K.~B. (2019a).
\newblock Open-loop control of periodic thermoacoustic oscillations:
  experiments and low-order modelling in a synchronization framework.
\newblock {\em P. Combust. Inst.}, 37(4):5315--5323.

\bibitem[Guan \emph{et~al.}, 2019b]{guan2019forcedand}
Guan, Y., Gupta, V., and Li, L. K.~B. (2019b).
\newblock Forced and mutual synchronization of periodic and aperiodic
  oscillations in a self-excited thermoacoustic system.
\newblock {\em Bull. Am. Phys. Soc.}, 64.

\bibitem[Guan \emph{et~al.}, 2020]{guan2020intermittency}
Guan, Y., Gupta, V., and Li, L. K.~B. (2020).
\newblock Intermittency route to self-excited chaotic thermoacoustic
  oscillations.
\newblock {\em J. Fluid Mech.}, 894:R3.

\bibitem[Guan \emph{et~al.}, 2019c]{guan2019QP}
Guan, Y., Gupta, V., Wan, M., and Li, L. K.~B. (2019c).
\newblock Forced synchronization of quasiperiodic oscillations in a
  thermoacoustic system.
\newblock {\em J. Fluid Mech.}, 879:390--421.

\bibitem[Guan \emph{et~al.}, 2019d]{guan2019control}
Guan, Y., He, W., Murugesan, M., Li, Q., Liu, P., and Li, L. K.~B. (2019d).
\newblock Control of self-excited thermoacoustic oscillations using transient
  forcing, hysteresis and mode switching.
\newblock {\em Combust. Flame}, 202:262--275.

\bibitem[Guan \emph{et~al.}, 2019e]{guan2019chaos}
Guan, Y., Li, L. K.~B., Ahn, B., and Kim, K.~T. (2019e).
\newblock Chaos, synchronization, and desynchronization in a liquid-fueled
  diffusion-flame combustor with an intrinsic hydrodynamic mode.
\newblock {\em Chaos}, 29(5):053124.

\bibitem[Guan \emph{et~al.}, 2018b]{guan2018nonlinear}
Guan, Y., Liu, P., Jin, B., Gupta, V., and Li, L. K.~B. (2018b).
\newblock Nonlinear time-series analysis of thermoacoustic oscillations in a
  solid rocket motor.
\newblock {\em Exp. Therm. Fluid Sci.}, 98:217--226.

\bibitem[Guan \emph{et~al.}, 2018c]{guan2018strange}
Guan, Y., Murugesan, M., and Li, L. K.~B. (2018c).
\newblock Strange nonchaotic and chaotic attractors in a self-excited
  thermoacoustic oscillator subjected to external periodic forcing.
\newblock {\em Chaos}, 28(9):093109.

\bibitem[Gupta \emph{et~al.}, 2017]{Gupta2017}
Gupta, V., Saurabh, A., Paschereit, C.~O., and Kabiraj, L. (2017).
\newblock Numerical results on noise-induced dynamics in the subthreshold
  regime for thermoacoustic systems.
\newblock {\em J. Sound Vib.}, 390:55--66.

\bibitem[Guttal and Jayaprakash, 2008]{guttal2008}
Guttal, V. and Jayaprakash, C. (2008).
\newblock Changing skewness: an early warning signal of regime shifts in
  ecosystems.
\newblock {\em Ecol. lett.}, 11(5):450--460.

\bibitem[Guttal \emph{et~al.}, 2013]{guttal2013}
Guttal, V., Jayaprakash, C., and Tabbaa, O.~P. (2013).
\newblock Robustness of early warning signals of regime shifts in time-delayed
  ecological models.
\newblock {\em Theor. ecol.}, 6(3):271--283.

\bibitem[Hachijo \emph{et~al.}, 2019]{hachijo2019}
Hachijo, T., Masuda, S., Kurosaka, T., and Gotoda, H. (2019).
\newblock Early detection of thermoacoustic combustion oscillations using a
  methodology combining statistical complexity and machine learning.
\newblock {\em Chaos}, 29(10):103123.

\bibitem[Hallberg and Strykowski, 2006]{hallberg2006universality}
Hallberg, M.~P. and Strykowski, P.~J. (2006).
\newblock On the universality of global modes in low-density axisymmetric jets.
\newblock {\em J. Fluid Mech.}, 569:493--507.

\bibitem[Harrje and Reardon, 1972]{harrje1972}
Harrje, D.~T. and Reardon, F.~H. (1972).
\newblock Liquid propellant rocket combustion instability.
\newblock Technical Report NASA-SP-194, NASA.

\bibitem[Harte, 2001]{harte2001}
Harte, D. (2001).
\newblock {\em Multifractals: theory and applications}.
\newblock Chapman and Hall/CRC.

\bibitem[Hashimoto \emph{et~al.}, 2019]{hashimoto2019spatiotemporal}
Hashimoto, T., Shibuya, H., Gotoda, H., Ohmichi, Y., and Matsuyama, S. (2019).
\newblock Spatiotemporal dynamics and early detection of thermoacoustic
  combustion instability in a model rocket combustor.
\newblock {\em Phys. Rev. E}, 99(3):032208.

\bibitem[He \emph{et~al.}, 2019a]{he2019stability}
He, W., Guan, Y., Theofilis, V., and Li, L. K.~B. (2019a).
\newblock Stability of low-reynolds-number separated flow around an airfoil
  near a wavy ground.
\newblock {\em AIAA J.}, 57(1):29--34.

\bibitem[He \emph{et~al.}, 2018a]{he2018non}
He, W., P{\'e}rez, J.~M., Yu, P., and Li, L. K.~B. (2018a).
\newblock Non-modal stability analysis of low-re separated flow around a naca
  4415 airfoil in ground effect.
\newblock {\em Bull. Am. Phys. Soc.}, 63.

\bibitem[He \emph{et~al.}, 2019b]{he2019non}
He, W., P{\'e}rez, J.~M., Yu, P., and Li, L. K.~B. (2019b).
\newblock Non-modal stability analysis of low-re separated flow around a naca
  4415 airfoil in ground effect.
\newblock {\em Aerosp. Sci. Technol.}, 92:269--279.

\bibitem[He \emph{et~al.}, 2017]{he2017ground}
He, W., Yu, P., and Li, L. K.~B. (2017).
\newblock Ground effects on the stability of separated flow around an airfoil
  at low reynolds numbers.
\newblock {\em Bull. Am. Phys. Soc.}, 62.

\bibitem[He \emph{et~al.}, 2018b]{he2018ground}
He, W., Yu, P., and Li, L. K.~B. (2018b).
\newblock Ground effects on the stability of separated flow around a naca 4415
  airfoil at low reynolds numbers.
\newblock {\em Aerosp. Sci. Technol.}, 72:63--76.

\bibitem[Hempstead and Lax, 1967]{hempstead1967}
Hempstead, R.~D. and Lax, M. (1967).
\newblock Classical noise. vi. noise in self-sustained oscillators near
  threshold.
\newblock {\em Phys. Rev.}, 161(2):350.

\bibitem[Henry and Judge, 2019]{henry2019}
Henry, M. and Judge, G. (2019).
\newblock Permutation entropy and information recovery in nonlinear dynamic
  economic time series.
\newblock {\em Econometrics}, 7(1):10.

\bibitem[Hikihara \emph{et~al.}, 1997]{hikihara1997}
Hikihara, T., Fujinami, T., and Moon, F.~C. (1997).
\newblock Bifurcation and multifractal vibration in dynamics of a high-tc
  superconducting levitation system.
\newblock {\em Phys. Lett. A}, 231(3-4):217--223.

\bibitem[Hines \emph{et~al.}, 2011]{hines2011}
Hines, P., Cotilla-Sanchez, E., and Blumsack, S. (2011).
\newblock Topological models and critical slowing down: Two approaches to power
  system blackout risk analysis.
\newblock In {\em 44th Hawaii International Conference on System Sciences},
  pages 1--10. IEEE.

\bibitem[Hogg \emph{et~al.}, 2005]{hogg2005}
Hogg, R.~V., Mc{K}ean, J., and Craig, A.~T. (2005).
\newblock {\em Introduction to mathematical statistics}.
\newblock Pearson Education.

\bibitem[Horsthemke, 1984]{Horsthemke1984}
Horsthemke, W.~W. (1984).
\newblock {\em Noise-induced transitions: theory and applications in physics,
  chemistry, and biology}.
\newblock Springer-Verlag, Berlin.

\bibitem[Huerre and Monkewitz, 1990]{huerre1990local}
Huerre, P. and Monkewitz, P.~A. (1990).
\newblock Local and global instabilities in spatially developing flows.
\newblock {\em Annu. Rev. Fluid Mech.}, 22(1):473--537.

\bibitem[Huygens, 1986]{huygens1986}
Huygens, C. (1986).
\newblock {\em The pendulum clock}.
\newblock The Iowa State University Press.

\bibitem[Jackson, 1987]{jackson_1987}
Jackson, C.~P. (1987).
\newblock A finite-element study of the onset of vortex shedding in flow past
  variously shaped bodies.
\newblock {\em J. Fluid Mech.}, 182:23–--45.

\bibitem[Jaensch \emph{et~al.}, 2014]{jaensch2014grey}
Jaensch, S., Emmert, T., Silva, C.~F., and Polifke, W. (2014).
\newblock A grey-box identification approach for thermoacoustic network models.
\newblock In {\em ASME Turbo Expo}. ASME Digital Collection.

\bibitem[Jafari \emph{et~al.}, 2003]{jafari2003}
Jafari, G.~R., Fazeli, S.~M., Ghasemi, F., Allaei, S. M.~V., Tabar, M. R.~R.,
  Kavei, G., \emph{et~al.} (2003).
\newblock Stochastic analysis and regeneration of rough surfaces.
\newblock {\em Phys. Rev. Lett.}, 91(22):226101.

\bibitem[Jegadeesan and Sujith, 2013]{jegadeesan2013}
Jegadeesan, V. and Sujith, R.~I. (2013).
\newblock Experimental investigation of noise induced triggering in
  thermoacoustic systems.
\newblock {\em P. Combust. Inst.}, 34(2):3175 -- 3183.

\bibitem[Jegal \emph{et~al.}, 2019]{jegal2019mutual}
Jegal, H., Moon, K., Gu, J., Li, L. K.~B., and Kim, K.~T. (2019).
\newblock Mutual synchronization of two lean-premixed gas turbine combustors:
  Phase locking and amplitude death.
\newblock {\em Combust. Flame}, 206:424--437.

\bibitem[Jenkins, 2013]{jenkins2013}
Jenkins, A. (2013).
\newblock Self-oscillation.
\newblock {\em Phys. Rep.}, 525(2):167--222.

\bibitem[Jerome and Ayyagari, 2014]{jerome2014}
Jerome, M.~M. and Ayyagari, R. (2014).
\newblock A brief survey of stochastic resonance and its application to
  control.
\newblock {\em IFAC Proc. Vol.}, 47(1):313--320.

\bibitem[Johnson \emph{et~al.}, 2000]{johnson2000}
Johnson, C.~E., Neumeier, Y., Lieuwen, T., and Zinn, B.~T. (2000).
\newblock Experimental determination of the stability margin of a combustor
  using exhaust flow and fuel injection rate modulations.
\newblock {\em P. Combust. Inst.}, 28(1):757--763.

\bibitem[Juniper \emph{et~al.}, 2009]{juniper2009forcing}
Juniper, M.~P., Li, L. K.~B., and Nichols, J.~W. (2009).
\newblock Forcing of self-excited round jet diffusion flames.
\newblock {\em P. Combust. Inst.}, 32(1):1191--1198.

\bibitem[Juniper and Sujith, 2018]{juniper2018}
Juniper, M.~P. and Sujith, R.~I. (2018).
\newblock Sensitivity and nonlinearity of thermoacoustic oscillations.
\newblock {\em Annu. Rev. Fluid Mech.}, 50:661--689.

\bibitem[Kabiraj \emph{et~al.}, 2015]{Kabiraj2015}
Kabiraj, L., Steinert, R., Saurabh, A., and Paschereit, C.~O. (2015).
\newblock Coherence resonance in a thermoacoustic system.
\newblock {\em Phys. Rev. E}, 92:042909.

\bibitem[Kabiraj \emph{et~al.}, 2020]{Kabiraj2020}
Kabiraj, L., Vishnoi, N., and Saurabh, A. (2020).
\newblock {\em A review on noise-induced dynamics of thermoacoustic systems},
  pages 265--281.
\newblock Springer.

\bibitem[Kalaba and Spingarn, 1982]{Kalaba1982}
Kalaba, R. and Spingarn, K. (1982).
\newblock {\em Applications of System Identification}, pages 195--222.
\newblock Springer US.

\bibitem[Karaaslanl{\i}, 2012]{karaaslanli2012}
Karaaslanl{\i}, C.~{\c{C}}. (2012).
\newblock {\em Bifurcation analysis and its applications}.
\newblock INTECH Open Access Publisher.

\bibitem[Kashinath \emph{et~al.}, 2018]{kashinath2018forced}
Kashinath, K., Li, L. K.~B., and Juniper, M.~P. (2018).
\newblock Forced synchronization of periodic and aperiodic thermoacoustic
  oscillations: lock-in, bifurcations and open-loop control.
\newblock {\em J. Fluid Mech.}, 838:690--714.

\bibitem[Kim and Moin, 1985]{kim1985application}
Kim, J. and Moin, P. (1985).
\newblock Application of a fractional-step method to incompressible
  {Navier--Stokes} equations.
\newblock {\em J. Comput. Phys.}, 59(2):308--323.

\bibitem[Kobayashi \emph{et~al.}, 2019]{kobayashi2019}
Kobayashi, T., Murayama, S., Hachijo, T., and Gotoda, H. (2019).
\newblock Early detection of thermoacoustic combustion instability using a
  methodology combining complex networks and machine learning.
\newblock {\em Phys. Rev. Appl.}, 11(6):064034.

\bibitem[Kopell and Howard, 1973]{Kopell1973}
Kopell, N. and Howard, L.~N. (1973).
\newblock Horizontal bands in the {B}elousov reaction.
\newblock {\em Science}, 180(4091):1171--1173.

\bibitem[Kramer and Ross, 1985]{kramer1985}
Kramer, J. and Ross, J. (1985).
\newblock Stabilization of unstable states, relaxation, and critical slowing
  down in a bistable system.
\newblock {\em J. Chem. Phys.}, 83(12):6234--6241.

\bibitem[Krediet \emph{et~al.}, 2012]{krediet2012}
Krediet, H.~J., Beck, C.~H., Krebs, W., Schimek, S., Paschereit, C.~O., and
  Kok, J. B.~W. (2012).
\newblock Identification of the flame describing function of a premixed swirl
  flame from {LES}.
\newblock {\em Combust. Sci. Technol.}, 184(7-8):888--900.

\bibitem[Kuramoto, 2003]{kuramoto2003chemical}
Kuramoto, Y. (2003).
\newblock {\em Chemical oscillations, waves, and turbulence}.
\newblock Courier Corporation.

\bibitem[Kyle and Sreenivasan, 1993]{kyle_sreenivasan_1993}
Kyle, D.~M. and Sreenivasan, K.~R. (1993).
\newblock The instability and breakdown of a round variable-density jet.
\newblock {\em J. Fluid Mech.}, 249:619–--664.

\bibitem[Lade, 2009]{lade2009}
Lade, S.~J. (2009).
\newblock Finite sampling interval effects in {K}ramers--{M}oyal analysis.
\newblock {\em Phys. Lett. A}, 373(41):3705--3709.

\bibitem[Lagarias \emph{et~al.}, 1998]{lagarias1998}
Lagarias, J.~C., Reeds, J.~A., Wright, M.~H., and Wright, P.~E. (1998).
\newblock Convergence properties of the {N}elder--{M}ead simplex method in low
  dimensions.
\newblock {\em SIAM J. optimiz.}, 9(1):112--147.

\bibitem[Lakshmanan and Senthilkumar, 2011]{lakshmanan2011}
Lakshmanan, M. and Senthilkumar, D.~V. (2011).
\newblock {\em Dynamics of nonlinear time-delay systems}.
\newblock Springer Science \& Business Media.

\bibitem[Lamberti \emph{et~al.}, 2004]{lamberti2004}
Lamberti, P.~W., Martin, M.~T., Plastino, A., and Rosso, O.~A. (2004).
\newblock Intensive entropic non-triviality measure.
\newblock {\em Physica A}, 334(1-2):119--131.

\bibitem[Landa \emph{et~al.}, 2000]{Landa2000}
Landa, P.~S., Zaikin, A.~A., Ushakov, V.~G., and Kurths, J. (2000).
\newblock Influence of additive noise on transitions in nonlinear systems.
\newblock {\em Phys. Rev. E}, 61:4809--4820.

\bibitem[Landau, 1944]{landau1944problem}
Landau, L.~D. (1944).
\newblock On the problem of turbulence.
\newblock {\em Dokl. Akad. Nauk SSSR}, 44(8):339--349.

\bibitem[Lang \emph{et~al.}, 2010]{Lang2010}
Lang, X., Lu, Q., and Kurths, J. (2010).
\newblock Phase synchronization in noise-driven bursting neurons.
\newblock {\em Phys. Rev. E}, 82:021909.

\bibitem[Langevin, 1908]{langevin1908}
Langevin, P. (1908).
\newblock Sur la th{\'e}orie du mouvement brownien.
\newblock {\em C. R. Acad. Sci. Paris}, 146:530--533.

\bibitem[Lax \emph{et~al.}, 2006]{lax2006}
Lax, M., Cai, W., and Xu, M. (2006).
\newblock {\em Random processes in physics and finance}.
\newblock Oxford University Press.

\bibitem[Lee \emph{et~al.}, 2016]{lee2016nonlinear}
Lee, C.~Y., Li, L. K.~B., Juniper, M.~P., and Cant, R.~S. (2016).
\newblock Nonlinear hydrodynamic and thermoacoustic oscillations of a
  bluff-body stabilised turbulent premixed flame.
\newblock {\em Combust. Theor. Model.}, 20(1):131--153.

\bibitem[Lee \emph{et~al.}, 2019a]{lee2019systemt}
Lee, M., Guan, Y., Gupta, V., and Li, L. K.~B. (2019a).
\newblock System identification of a thermoacoustic system using its
  noise-induced dynamics.
\newblock {\em The 16th International Conference on Flow Dynamics}.

\bibitem[Lee \emph{et~al.}, 2020]{lee_tbr}
Lee, M., Guan, Y., Gupta, V., and Li, L. K.~B. (2020).
\newblock Input-output system identification of a thermoacoustic oscillator
  near a {H}opf bifurcation using only fixed-point data.
\newblock {\em Phys. Rev. E}, 101(1):013102.

\bibitem[Lee \emph{et~al.}, 2019b]{lee2019exploiting}
Lee, M., Zhu, Y., Guan, Y., Li, L. K.~B., and Gupta, V. (2019b).
\newblock Exploiting noise-induced dynamics for system identification near a
  hopf bifurcation.
\newblock {\em Bull. Am. Phys. Soc.}, 64.

\bibitem[Lee \emph{et~al.}, 2018]{lee2018system}
Lee, M., Zhu, Y., Li, L. K.~B., and Gupta, V. (2018).
\newblock System identification of a low-density jet via its noise-induced
  dynamics.
\newblock {\em Bull. Am. Phys. Soc.}, 63.

\bibitem[Lee \emph{et~al.}, 2019c]{lee_2019}
Lee, M., Zhu, Y., Li, L. K.~B., and Gupta, V. (2019c).
\newblock System identification of a low-density jet via its noise-induced
  dynamics.
\newblock {\em J. Fluid Mech.}, 862:200--215.

\bibitem[Lee \emph{et~al.}, 2019d]{LEE20195137}
Lee, T., Park, J., Han, D., and Kim, K.~T. (2019d).
\newblock The dynamics of multiple interacting swirl-stabilized flames in a
  lean-premixed gas turbine combustor.
\newblock {\em P. Combust. Inst.}, 37(4):5137 -- 5145.

\bibitem[Leonard and Reichl, 1994]{leonard1994}
Leonard, D.~S. and Reichl, L.~E. (1994).
\newblock Stochastic resonance in a chemical reaction.
\newblock {\em Phys. Rev. E}, 49(2):1734.

\bibitem[Li and Juniper, 2013a]{Li13a}
Li, L. K.~B. and Juniper, M.~P. (2013a).
\newblock Lock-in and quasiperiodicity in a forced hydrodynamically
  self-excited jet.
\newblock {\em J. Fluid Mech.}, 726:624--655.

\bibitem[Li and Juniper, 2013b]{Li13}
Li, L. K.~B. and Juniper, M.~P. (2013b).
\newblock Lock-in and quasiperiodicity in hydrodynamically self-excited flames:
  Experiments and modelling.
\newblock {\em P. Combust. Inst.}, 34(1):947--954.

\bibitem[Li and Juniper, 2013c]{Li13b}
Li, L. K.~B. and Juniper, M.~P. (2013c).
\newblock Phase trapping and slipping in a forced hydrodynamically self-excited
  jet.
\newblock {\em J. Fluid Mech.}, 735.

\bibitem[Li \emph{et~al.}, 2007]{li2007}
Li, X., Ouyang, G., and Richards, D.~A. (2007).
\newblock Predictability analysis of absence seizures with permutation entropy.
\newblock {\em Epilepsy Res.}, 77(1):70--74.

\bibitem[Li \emph{et~al.}, 2019]{li2019coherence}
Li, X., Zhao, D., and Shi, B. (2019).
\newblock Coherence resonance and stochastic bifurcation behaviors of
  simplified standing-wave thermoacoustic systems.
\newblock {\em J. Acoust. Soc. Am.}, 145(2):692--702.

\bibitem[Lieuwen, 2002]{lieuwen2002}
Lieuwen, T. (2002).
\newblock Experimental investigation of limit-cycle oscillations in an unstable
  gas turbine combustor.
\newblock {\em J. Propul. Power}, 18(1):61--67.

\bibitem[Lieuwen, 2005]{Lieuwen2005a}
Lieuwen, T. (2005).
\newblock Online combustor stability margin assessment using dynamic pressure
  data.
\newblock {\em J. Eng. Gas Turbine Power}, 127(3):478--482.

\bibitem[Lieuwen, 2012]{lieuwen2012unsteady}
Lieuwen, T. (2012).
\newblock {\em Unsteady Combustor Physics}.
\newblock Cambridge University Press, Cambridge.

\bibitem[Lieuwen and Banaszuk, 2005]{lieuwen2005b}
Lieuwen, T. and Banaszuk, A. (2005).
\newblock Background noise effects on combustor stability.
\newblock {\em J. Propul. Power}, 21(1):25--31.

\bibitem[Lieuwen and Yang, 2005]{Lieuwen2005}
Lieuwen, T. and Yang, V. (2005).
\newblock {\em Combustion instabilities in gas turbine engines: operational
  experience, fundamental mechanisms and modeling}.
\newblock AIAA, Reston.

\bibitem[Lim and Epureanu, 2011]{lim2011}
Lim, J. and Epureanu, B.~I. (2011).
\newblock Forecasting a class of bifurcations: Theory and experiment.
\newblock {\em Phys. Rev. E}, 83(1):016203.

\bibitem[Lin and Chen, 2013]{lin2013}
Lin, J. and Chen, Q. (2013).
\newblock Fault diagnosis of rolling bearings based on multifractal detrended
  fluctuation analysis and {M}ahalanobis distance criterion.
\newblock {\em Mech. Syst. Signal Process}, 38(2):515--533.

\bibitem[Liu \emph{et~al.}, 2011]{liu2011}
Liu, W., Fu, C., and Chen, B. (2011).
\newblock Hopf bifurcation for a predator--prey biological economic system with
  holling type ii functional response.
\newblock {\em J. Franklin I.}, 348(6):1114--1127.

\bibitem[Liu \emph{et~al.}, 2017]{liu2017}
Liu, W.~B., Dai, H.~L., and Wang, L. (2017).
\newblock Suppressing wind-induced oscillations of prismatic structures by
  dynamic vibration absorbers.
\newblock {\em Int. J. of Struct. Stab. Dyn.}, 17(06):1750056.

\bibitem[Ljung, 1999]{Ljung1999}
Ljung, L. (1999).
\newblock {\em System identification : theory for the user}.
\newblock Prentice-Hall information and system sciences series. Prentice Hall,
  Upper Saddle River, N.J., 2nd edition.

\bibitem[Longtin, 1997]{longtin1997}
Longtin, A. (1997).
\newblock Autonomous stochastic resonance in bursting neurons.
\newblock {\em Phys. Rev. E}, 55(1):868.

\bibitem[Lores and Zinn, 1973]{lores1973}
Lores, M.~E. and Zinn, B.~T. (1973).
\newblock Nonlinear longitudinal combustion instability in rocket motors.
\newblock {\em Combust. Sci. Technol.}, 7(6):245--256.

\bibitem[Los, 2006]{LOS2006}
Los, C.~A. (2006).
\newblock System identification in noisy data environments: An application to
  six asian stock markets.
\newblock {\em J. Bank. Finance}, 30(7):1997 -- 2024.

\bibitem[L{\"o}tstedt and Ferm, 2006]{lotstedt2006}
L{\"o}tstedt, P. and Ferm, L. (2006).
\newblock Dimensional reduction of the {F}okker--{P}lanck equation for
  stochastic chemical reactions.
\newblock {\em Multiscale Model. Sim.}, 5(2):593--614.

\bibitem[L{\"u}dge and Schuster, 2012]{ludge2012nonlinear}
L{\"u}dge, K. and Schuster, H. (2012).
\newblock {\em Nonlinear laser dynamics: from quantum dots to cryptography},
  volume~5.
\newblock John Wiley \& Sons.

\bibitem[Ma, 1976]{ma1976}
Ma, S.-K. (1976).
\newblock {\em Modern theory of critical phenomena}.
\newblock Routledge.

\bibitem[Marsden and McCracken, 1976]{marsden1976}
Marsden, J.~E. and McCracken, M. (1976).
\newblock {\em The {H}opf Bifurcation and Its Applications}.
\newblock Springer-Verlag, New York.

\bibitem[Mathis \emph{et~al.}, 1984]{mathis1984}
Mathis, C., Provansal, M., and Boyer, L. (1984).
\newblock The {B}{\'e}nard-von {K}{\'a}rm{\'a}n instability: an experimental
  study near the threshold.
\newblock {\em J. Phys. Paris Lett.}, 45(10):483--491.

\bibitem[Matveev, 2003]{matveev2003}
Matveev, K.~I. (2003).
\newblock Energy consideration of the nonlinear effects in a {R}ijke tube.
\newblock {\em J. Fluid Struct.}, 18(6):783--794.

\bibitem[McDonnell and Abbott, 2009]{mcdonnell2009}
McDonnell, M.~D. and Abbott, D. (2009).
\newblock What is stochastic resonance? definitions, misconceptions, debates,
  and its relevance to biology.
\newblock {\em PLoS comput. biol.}, 5(5):e1000348.

\bibitem[McNamara \emph{et~al.}, 1988]{mcnamara1988}
McNamara, B., Wiesenfeld, K., and Roy, R. (1988).
\newblock Observation of stochastic resonance in a ring laser.
\newblock {\em Phys. Rev. Lett.}, 60(25):2626.

\bibitem[Mendel, 2013]{mendel2013}
Mendel, J.~M. (2013).
\newblock {\em Optimal Seismic Deconvolution: An Estimation-Based Approach}.
\newblock Elsevier Science.

\bibitem[Merk \emph{et~al.}, 2019]{merk2019}
Merk, M., Gaudron, R., Silva, C., Gatti, M., Mirat, C., Schuller, T., and
  Polifke, W. (2019).
\newblock Prediction of combustion noise of an enclosed flame by simultaneous
  identification of noise source and flame dynamics.
\newblock {\em P. Combust. Inst.}, 37(4):5263--5270.

\bibitem[Mevel \emph{et~al.}, 2006]{MEVEL2006531}
Mevel, L., Benveniste, A., Basseville, M., Goursat, M., Peeters, B., Van~der
  Auweraer, H., and Vecchio, A. (2006).
\newblock Input/output versus output-only data processing for structural
  identification -- {Application} to in-flight data analysis.
\newblock {\em J. Sound Vib.}, 295(3):531--552.

\bibitem[Moeck and Paschereit, 2012]{moeck2012nonlinear}
Moeck, J.~P. and Paschereit, C.~O. (2012).
\newblock Nonlinear interactions of multiple linearly unstable thermoacoustic
  modes.
\newblock {\em Int. J. Spray Combust.}, 4(1):1--27.

\bibitem[Monkewitz \emph{et~al.}, 1990]{monk1990self}
Monkewitz, P.~A., Bechert, D.~W., Barsikow, B., and Lehmann, B. (1990).
\newblock Self-excited oscillations and mixing in a heated round jet.
\newblock {\em J. Fluid Mech.}, 213:611--639.

\bibitem[Moon \emph{et~al.}, 2020]{moon2020mutual}
Moon, K., Guan, Y., Li, L. K.~B., and Kim, K.~T. (2020).
\newblock Mutual synchronization of two flame-driven thermoacoustic
  oscillators: Dissipative and time-delayed coupling effects.
\newblock {\em Chaos}, 30(2):023110.

\bibitem[Morelli and Klein, 2005]{morelli2005}
Morelli, E.~A. and Klein, V. (2005).
\newblock Application of system identification to aircraft at {NASA} {L}angley
  research center.
\newblock {\em J. Aircraft}, 42(1):12--25.

\bibitem[Murayama \emph{et~al.}, 2018]{murayama2018characterization}
Murayama, S., Kinugawa, H., Tokuda, I., and Gotoda, H. (2018).
\newblock Characterization and detection of thermoacoustic combustion
  oscillations based on statistical complexity and complex-network theory.
\newblock {\em Phys. Rev. E}, 97(2):022223.

\bibitem[Murthy \emph{et~al.}, 2019]{murthy2019analysis}
Murthy, S.~R., Sayadi, T., Le~Chenadec, V., Schmid, P.~J., and Bodony, D.~J.
  (2019).
\newblock Analysis of degenerate mechanisms triggering finite-amplitude
  thermo-acoustic oscillations in annular combustors.
\newblock {\em J. Fluid Mech.}, 881:384--419.

\bibitem[Murugesan and Sujith, 2016]{murugesan2016}
Murugesan, M. and Sujith, R.~I. (2016).
\newblock Detecting the onset of an impending thermoacoustic instability using
  complex networks.
\newblock {\em J. Propul. Power}, 32(1):707--712.

\bibitem[Murugesan \emph{et~al.}, 2016]{murugesan2016recurrence}
Murugesan, M., Zhu, Y., and Li, L. K.~B. (2016).
\newblock A recurrence network approach to analyzing forced synchronization in
  hydrodynamic systems.
\newblock {\em Bull. Am. Phys. Soc.}, 61.

\bibitem[Murugesan \emph{et~al.}, 2017]{murugesan2017intermittency}
Murugesan, M., Zhu, Y., and Li, L. K.~B. (2017).
\newblock An intermittency route to global instability in low-density jets.
\newblock {\em Bull. Am. Phys. Soc.}, 62.

\bibitem[Murugesan \emph{et~al.}, 2019]{murugesan2019complex}
Murugesan, M., Zhu, Y., and Li, L. K.~B. (2019).
\newblock Complex network analysis of forced synchronization in a
  hydrodynamically self-excited jet.
\newblock {\em Int. J. Heat Fluid Fl.}, 76:14--25.

\bibitem[Nair and Sujith, 2014]{nair2014b}
Nair, V. and Sujith, R.~I. (2014).
\newblock Multifractality in combustion noise: predicting an impending
  combustion instability.
\newblock {\em J. Fluid Mech.}, 747:635--655.

\bibitem[Nair \emph{et~al.}, 2013]{nair2013}
Nair, V., Thampi, G., Karuppusamy, S., Gopalan, S., and Sujith, R.~I. (2013).
\newblock Loss of chaos in combustion noise as a precursor of impending
  combustion instability.
\newblock {\em Int. J. Spray Combust.}, 5(4):273--290.

\bibitem[Nair \emph{et~al.}, 2014]{nair2014a}
Nair, V., Thampi, G., and Sujith, R.~I. (2014).
\newblock Intermittency route to thermoacoustic instability in turbulent
  combustors.
\newblock {\em J. Fluid Mech.}, 756:470--487.

\bibitem[Najafian, 2007a]{NAJAFIAN2007a}
Najafian, G. (2007a).
\newblock Application of system identification techniques in efficient
  modelling of offshore structural response. part i: Model development.
\newblock {\em Appl. Ocean Res.}, 29(1):1 -- 16.

\bibitem[Najafian, 2007b]{NAJAFIAN2007b}
Najafian, G. (2007b).
\newblock Application of system identification techniques in efficient
  modelling of offshore structural response. part ii: Model validation.
\newblock {\em Appl. Ocean Res.}, 29(1):17 -- 36.

\bibitem[Nayfeh, 1981]{nayfeh1981introduction}
Nayfeh, A.~H. (1981).
\newblock {\em Introduction to Perturbation Techniques}.
\newblock John Wiley, New York.

\bibitem[Nayfeh and Mook, 1979]{nayfeh1979nonlinear}
Nayfeh, A.~H. and Mook, D.~T. (1979).
\newblock {\em Nonlinear Oscillations}.
\newblock John Wiley, New York.

\bibitem[Neiman \emph{et~al.}, 1997]{Neiman1997}
Neiman, A., Saparin, P.~I., and Stone, L. (1997).
\newblock Coherence resonance at noisy precursors of bifurcations in nonlinear
  dynamical systems.
\newblock {\em Phys. Rev. E}, 56:270--273.

\bibitem[Newman, 2003]{newman2003}
Newman, M. E.~J. (2003).
\newblock The structure and function of complex networks.
\newblock {\em SIAM Rev.}, 45(2):167--256.

\bibitem[Noiray, 2017]{noiray2017linear}
Noiray, N. (2017).
\newblock Linear growth rate estimation from dynamics and statistics of
  acoustic signal envelope in turbulent combustors.
\newblock {\em J. Eng. Gas Turb. Power}, 139(4):041503.

\bibitem[Noiray and Denisov, 2017]{noiray2017method}
Noiray, N. and Denisov, A. (2017).
\newblock A method to identify thermoacoustic growth rates in combustion
  chambers from dynamic pressure time series.
\newblock {\em P. Combust. Inst.}, 36(3):3843--3850.

\bibitem[Noiray and Schuermans, 2013a]{NOIRAY2013152}
Noiray, N. and Schuermans, B. (2013a).
\newblock Deterministic quantities characterizing noise driven {H}opf
  bifurcations in gas turbine combustors.
\newblock {\em Int. J. Nonlin. Mech.}, 50:152 -- 163.

\bibitem[Noiray and Schuermans, 2013b]{noiray2013dynamic}
Noiray, N. and Schuermans, B. (2013b).
\newblock On the dynamic nature of azimuthal thermoacoustic modes in annular
  gas turbine combustion chambers.
\newblock {\em Proc. R. Soc. A}, 469(2151):20120535.

\bibitem[Nurujjaman \emph{et~al.}, 2008]{Nurujjaman2008}
Nurujjaman, M., Sekar~Iyengar, A.~N., and Parmananda, P. (2008).
\newblock Noise-invoked resonances near a homoclinic bifurcation in the glow
  discharge plasma.
\newblock {\em Phys. Rev. E}, 78:026406.

\bibitem[Orchini \emph{et~al.}, 2016]{orchini2016}
Orchini, A., Rigas, G., and Juniper, M.~P. (2016).
\newblock Weakly nonlinear analysis of thermoacoustic bifurcations in the
  {R}ijke tube.
\newblock {\em J. Fluid Mech.}, 805:523--550.

\bibitem[Pagani and Aiello, 2013]{pagani2013}
Pagani, G.~A. and Aiello, M. (2013).
\newblock The power grid as a complex network: a survey.
\newblock {\em Physica A}, 392(11):2688--2700.

\bibitem[Paladin and Vulpiani, 1987]{paladin1987}
Paladin, G. and Vulpiani, A. (1987).
\newblock Anomalous scaling laws in multifractal objects.
\newblock {\em Phys. Rep.}, 156(4):147--225.

\bibitem[Paparizos and Culick, 1989]{paparizos1989}
Paparizos, L.~G. and Culick, F. E.~C. (1989).
\newblock The two-mode approximation to nonlinear acoustics in combustion
  chambers i. exact solution for second order acoustics.
\newblock {\em Combust. Sci. Technol.}, 65(1-3):39--65.

\bibitem[Parrondo \emph{et~al.}, 1996]{Parrondo1996}
Parrondo, J. M.~R., van~den Broeck, C., Buceta, J., and de~la Rubia, F.~J.
  (1996).
\newblock Noise-induced spatial patterns.
\newblock {\em Physica A}, 224(1):153--161.

\bibitem[Parzen, 1999]{parzen1999}
Parzen, E. (1999).
\newblock {\em Stochastic processes}, volume~24.
\newblock SIAM, Philadelphia.

\bibitem[Pau, 2017]{Pau2017}
Pau, M.~K. (2017).
\newblock Model-based output-only identification of coupled thermoacoustic
  modes.
\newblock Master's thesis, ETH Zurich.

\bibitem[Peng \emph{et~al.}, 2019]{peng2019}
Peng, X., Small, M., Zhao, Y., and Moore, J.~M. (2019).
\newblock Detecting and predicting tipping points.
\newblock {\em Int. J. Bifurcat. Chaos}, 29(08):1930022.

\bibitem[Peslin \emph{et~al.}, 1975]{Peslin1975}
Peslin, R., Papon, J., Duviver, C., and Richalet, J. (1975).
\newblock Frequency response of the chest: modeling and parameter estimation.
\newblock {\em J. Appl. Physiol.}, 39 4:523--34.

\bibitem[Pikovsky \emph{et~al.}, 2001]{pikovsky2003}
Pikovsky, A., Rosenblum, M., and Kurths, J. (2001).
\newblock {\em Synchronization: a universal concept in nonlinear sciences},
  volume~12.
\newblock Cambridge university press.

\bibitem[Pikovsky and Kurths, 1997]{pikovsky1997coherence}
Pikovsky, A.~S. and Kurths, J. (1997).
\newblock Coherence resonance in a noise-driven excitable system.
\newblock {\em Phys. Rev. Lett.}, 78:775--778.

\bibitem[Planck, 1917]{planck1917}
Planck, M. (1917).
\newblock {\"U}ber einen satz der statistischen dynamik und seine erweiterung
  in der quantentheorie.
\newblock {\em Sitzber. Preu{\ss}. Akad. Wiss}, 24:324–--341.

\bibitem[Platt \emph{et~al.}, 1993]{platt1993}
Platt, N. S. E.~A., Spiegel, E.~A., and Tresser, C. (1993).
\newblock On-off intermittency: A mechanism for bursting.
\newblock {\em Phys. Rev. Lett.}, 70(3):279.

\bibitem[Poinsot, 2017]{Poinsot2017}
Poinsot, T. (2017).
\newblock Prediction and control of combustion instabilities in real engines.
\newblock {\em P. Combust. Inst.}, 36(1):1--28.

\bibitem[Polifke, 2010]{polifke2010system}
Polifke, W. (2010).
\newblock System identification for aero-and thermo-acoustic applications.
\newblock In {\em Advances in Aero-Acoustics and Thermo-Acoustics}. von Karman
  Institute for Fluid Dynamics, Brussels.

\bibitem[Pomeau and Manneville, 1980]{pomeau1980}
Pomeau, Y. and Manneville, P. (1980).
\newblock Intermittent transition to turbulence in dissipative dynamical
  systems.
\newblock {\em Commun. Math. Phys.}, 74(2):189--197.

\bibitem[Price and Valerio, 1990]{PRICE1990419}
Price, S. and Valerio, N. (1990).
\newblock A non-linear investigation of single-degree-of-freedom instability in
  cylinder arrays subject to cross-flow.
\newblock {\em J. Sound Vib.}, 137(3):419 -- 432.

\bibitem[Provansal \emph{et~al.}, 1987]{Provansal1987}
Provansal, M., Mathis, C., and Boyer, L. (1987).
\newblock {B}{\'e}rnard-von {K}{\'a}rm{\'a}n instability: transient and forced
  regimes.
\newblock {\em J. Fluid Mech.}, 182:1--22.

\bibitem[Putnam \emph{et~al.}, 1986]{putnam1986}
Putnam, A., Belles, F., and Kentfield, J. (1986).
\newblock Pulse combustion.
\newblock {\em Prog. Energy Combust. Sci.}, 12(1):43--79.

\bibitem[Raghu and Monkewitz, 1991]{Raghu1991}
Raghu, S. and Monkewitz, P.~A. (1991).
\newblock The bifurcation of a hot round jet to limit-cycle oscillations.
\newblock {\em Phys. Fluids A}, 3:501.

\bibitem[Rayleigh, 1878]{Rayleigh1878}
Rayleigh, L. (1878).
\newblock The explanation of certain acoustical phenomena.
\newblock {\em Nature}, 18:319--321.

\bibitem[Ren and Li, 2018a]{ren2018global}
Ren, D. D.~W. and Li, L. K.~B. (2018a).
\newblock Global helical modes in low-density jets.
\newblock {\em Bull. Am. Phys. Soc.}, 63.

\bibitem[Ren and Li, 2018b]{ren2018spatiotemporal}
Ren, D. D.~W. and Li, L. K.~B. (2018b).
\newblock Spatiotemporal intermittency of global helical modes in low-density
  jets.
\newblock {\em Bull. Am. Phys. Soc.}, 63.

\bibitem[Ribeiro \emph{et~al.}, 2012]{ribeiro2012}
Ribeiro, H.~V., Zunino, L., Lenzi, E.~K., Santoro, P.~A., and Mendes, R.~S.
  (2012).
\newblock Complexity-entropy causality plane as a complexity measure for
  two-dimensional patterns.
\newblock {\em PLoS one}, 7(8):e40689.

\bibitem[Rigas \emph{et~al.}, 2016]{rigas2016experimental}
Rigas, G., Jamieson, N.~P., Li, L. K.~B., and Juniper, M.~P. (2016).
\newblock Experimental sensitivity analysis and control of thermoacoustic
  systems.
\newblock {\em J. Fluid Mech.}, 787.

\bibitem[Risken, 1965]{risken1965}
Risken, H. (1965).
\newblock Distribution-and correlation-functions for a laser amplitude.
\newblock {\em Z. Physik}, 186(1):85--98.

\bibitem[Risken, 1984]{risken1984}
Risken, H. (1984).
\newblock {\em {F}okker--{P}lanck Equation}.
\newblock Springer, Berlin.

\bibitem[Roberts and Spanos, 1986]{roberts1986stochastic}
Roberts, J. and Spanos, P.~D. (1986).
\newblock Stochastic averaging: An approximate method of solving random
  vibration problems.
\newblock {\em Int. J. Nonlin. Mech.}, 21:111--134.

\bibitem[Robinson and Treitel, 2000]{robinson2000}
Robinson, E. and Treitel, S. (2000).
\newblock {\em Geophysical Signal Analysis}.
\newblock Society of Exploration Geophysicists.

\bibitem[Rosso \emph{et~al.}, 2007]{rosso2007}
Rosso, O.~A., Larrondo, H.~A., Martin, M.~T., Plastino, A., and Fuentes, M.~A.
  (2007).
\newblock Distinguishing noise from chaos.
\newblock {\em Phys. Rev. Lett.}, 99(15):154102.

\bibitem[Scheffer \emph{et~al.}, 2009]{scheffer2009}
Scheffer, M., Bascompte, J., Brock, W.~A., Brovkin, V., Carpenter, S.~R.,
  Dakos, V., Held, H., Van~Nes, E.~H., Rietkerk, M., and Sugihara, G. (2009).
\newblock Early-warning signals for critical transitions.
\newblock {\em Nature}, 461(7260):53--59.

\bibitem[Scheffer \emph{et~al.}, 2012]{scheffer2012}
Scheffer, M., Carpenter, S.~R., Lenton, T.~M., Bascompte, J., Brock, W., Dakos,
  V., Van~de Koppel, J., Van~de Leemput, I.~A., Levin, S.~A., Van~Nes, E.~H.,
  \emph{et~al.} (2012).
\newblock Anticipating critical transitions.
\newblock {\em Science}, 338(6105):344--348.

\bibitem[Schijve, 2009]{schijve2009fatigue}
Schijve, J. (2009).
\newblock Fatigue damage in aircraft structures, not wanted, but tolerated?
\newblock {\em Int. J. Fatigue}, 31(6):998--1011.

\bibitem[Schmidt and Lipson, 2009]{Schmidt2009}
Schmidt, M. and Lipson, H. (2009).
\newblock Distilling free-form natural laws from experimental data.
\newblock {\em Science}, 324(5923):81--85.

\bibitem[Schoen and Lee, 2017]{schoen2017}
Schoen, M.~P. and Lee, J.-C. (2017).
\newblock Application of system identification for modeling the dynamic
  behavior of axial flow compressor dynamics.
\newblock {\em Int. J. Rotating Mach.}, 2017.

\bibitem[Scholz \emph{et~al.}, 1987]{scholz1987}
Scholz, J.~P., Kelso, J. A.~S., and Sch{\"o}ner, G. (1987).
\newblock Nonequilibrium phase transitions in coordinated biological motion:
  critical slowing down and switching time.
\newblock {\em Phys. Lett. A}, 123(8):390--394.

\bibitem[Schuster and Just, 2006]{schuster2006}
Schuster, H.~G. and Just, W. (2006).
\newblock {\em Deterministic chaos: an introduction}.
\newblock John Wiley \& Sons.

\bibitem[Seywert, 2001]{seywert2001}
Seywert, C. N.~L. (2001).
\newblock {\em Combustion instabilities: issues in modeling and control}.
\newblock PhD thesis, California Institute of Technology.

\bibitem[Shimizu and Kawahara, 2018]{Shimizu2018}
Shimizu, M. and Kawahara, G. (2018).
\newblock Construction of low-dimensional system reproducing
  low-{R}eynolds-number turbulence by machine learning.
\newblock {\em Submitted to Phys. Rev. E}.

\bibitem[Siegert \emph{et~al.}, 1998]{siegert1998}
Siegert, S., Friedrich, R., and Peinke, J. (1998).
\newblock Analysis of data sets of stochastic systems.
\newblock {\em Phys. Lett. A}, 5(243):275--280.

\bibitem[Sikdar and Karmeshu, 1982]{sikdar1982}
Sikdar, P.~K. and Karmeshu, P.~K. (1982).
\newblock On population growth of cities in a region: a stochastic nonlinear
  model.
\newblock {\em Environ. Plan A}, 14(5):585--590.

\bibitem[Sipp and Lebedev, 2007]{Sipp2007}
Sipp, D. and Lebedev, A. (2007).
\newblock Global stability of base and mean flows: a general approach and its
  applications to cylinder and open cavity flows.
\newblock {\em J. Fluid Mech.}, 593:333--358.

\bibitem[Small, 2013]{small2013}
Small, M. (2013).
\newblock Complex networks from time series: Capturing dynamics.
\newblock In {\em 2013 IEEE International Symposium on Circuits and Systems
  (ISCAS2013)}, pages 2509--2512. IEEE.

\bibitem[S{\"o}derstr{\"o}m and Stoica, 1988]{soderstrom1988}
S{\"o}derstr{\"o}m, T. and Stoica, P. (1988).
\newblock {\em System identification}.
\newblock Prentice-Hall, Inc.

\bibitem[Sreenivasan, 1991]{Sreenivasan1991}
Sreenivasan, K.~R. (1991).
\newblock Fractals and multifractals in fluid turbulence.
\newblock {\em Annu. Rev. Fluid Mech.}, 23(1):539--604.

\bibitem[Sreenivasan and Meneveau, 1986]{Sreenivasan1986}
Sreenivasan, K.~R. and Meneveau, C. (1986).
\newblock The fractal facets of turbulence.
\newblock {\em J. Fluid Mech.}, 173:357--386.

\bibitem[Sreenivasan \emph{et~al.}, 1989]{Sreenivasan1989}
Sreenivasan, K.~R., Raghu, S., and Kyle, D. (1989).
\newblock Absolute instability in variable density round jets.
\newblock {\em Exp. Fluids}, 7(5):309--317.

\bibitem[Stratonovich, 1963]{stratonovich1963}
Stratonovich, R.~L. (1963).
\newblock {\em Topics in the Theory of Random Noise}.
\newblock Gordon and Breach.

\bibitem[Stratonovich, 1967]{stratonovich1967}
Stratonovich, R.~L. (1967).
\newblock {\em Topics in the Theory of Random Noise: General theory of random
  processes; Nonlinear transformations of signals and noise}.
\newblock Gordon and Breach.

\bibitem[Strogatz, 2000]{strogatz2000}
Strogatz, S.~H. (2000).
\newblock {\em Nonlinear dynamics and chaos: with applications to physics,
  biology, chemistry, and engineering}.
\newblock Westview Press, Colorado.

\bibitem[Strogatz, 2001]{strogatz2001}
Strogatz, S.~H. (2001).
\newblock Exploring complex networks.
\newblock {\em Nature}, 410(6825):268.

\bibitem[Stuart, 1960]{stuart1960non}
Stuart, J.~T. (1960).
\newblock On the non-linear mechanics of wave disturbances in stable and
  unstable parallel flows -- {Part} 1. the basic behaviour in plane
  {Poiseuille} flow.
\newblock {\em J. Fluid Mech.}, 9:353--370.

\bibitem[Subramanian \emph{et~al.}, 2013]{subramanian2013}
Subramanian, P., Sujith, R.~I., and Wahi, P. (2013).
\newblock Subcritical bifurcation and bistability in thermoacoustic systems.
\newblock {\em J. Fluid Mech.}, 715:210--238.

\bibitem[Sura and Barsugli, 2002]{sura2002}
Sura, P. and Barsugli, J. (2002).
\newblock A note on estimating drift and diffusion parameters from timeseries.
\newblock {\em Phys. Lett. A}, 305(5):304--311.

\bibitem[Takens, 1981]{takens1981detecting}
Takens, F. (1981).
\newblock {Detecting strange attractors in turbulence}.
\newblock {\em Lect. Notes Math.}, 898:366--381.

\bibitem[Thomas \emph{et~al.}, 2018]{thomas2018}
Thomas, N., Mondal, S., Pawar, S.~A., and Sujith, R.~I. (2018).
\newblock Effect of noise amplification during the transition to amplitude
  death in coupled thermoacoustic oscillators.
\newblock {\em Chaos}, 28(9):093116.

\bibitem[Thompson and Stewart, 2002]{Thompson2002}
Thompson, J. M.~T. and Stewart, H.~B. (2002).
\newblock {\em Nonlinear {Dynamics} and {Chaos}}.
\newblock John Wiley \& Sons, New York, NY, USA.

\bibitem[Thompson and Troian, 1997]{thompson1997general}
Thompson, P.~A. and Troian, S.~M. (1997).
\newblock A general boundary condition for liquid flow at solid surfaces.
\newblock {\em Nature}, 389(6649):360.

\bibitem[Thothadri and Moon, 2005]{thothadri2005nonlinear}
Thothadri, M. and Moon, F.~C. (2005).
\newblock Nonlinear system identification of systems with periodic limit-cycle
  response.
\newblock {\em Nonlinear Dynam.}, 39(1-2):63--77.

\bibitem[Tomim \emph{et~al.}, 2005]{tomim2005}
Tomim, M.~A., Lopes, B. I.~L., Leme, R.~C., Jovita, R., de~Souza, A. C.~Z.,
  de~Carvalho~Mendes, P.~P., and Lima, J. W.~M. (2005).
\newblock Modified {H}opf bifurcation index for power system stability
  assessment.
\newblock {\em IEE P. Gener. Transm. D.}, 152(6):906--912.

\bibitem[Tredicce \emph{et~al.}, 2004]{tredicce2004}
Tredicce, J.~R., Lippi, G.~L., Mandel, P., Charasse, B., Chevalier, A., and
  Picqu{\'e}, B. (2004).
\newblock Critical slowing down at a bifurcation.
\newblock {\em Am. J. Phys.}, 72(6):799--809.

\bibitem[Ushakov \emph{et~al.}, 2005]{ushakov2005coherence}
Ushakov, O.~V., W\"unsche, H.~J., Henneberger, F., Khovanov, I.~A.,
  Schimansky-Geier, L., and Zaks, M.~A. (2005).
\newblock Coherence resonance near a {H}opf bifurcation.
\newblock {\em Phys. Rev. Lett.}, 95:123903.

\bibitem[Van~Mourik \emph{et~al.}, 2006]{van2006}
Van~Mourik, A.~M., Daffertshofer, A., and Beek, P.~J. (2006).
\newblock Deterministic and stochastic features of rhythmic human movement.
\newblock {\em Biol. Cybern.}, 94(3):233--244.

\bibitem[Van~Nes and Scheffer, 2007]{vannes2007}
Van~Nes, E.~H. and Scheffer, M. (2007).
\newblock Slow recovery from perturbations as a generic indicator of a nearby
  catastrophic shift.
\newblock {\em Am. Nat.}, 169(6):738--747.

\bibitem[Vanag \emph{et~al.}, 2000]{vanag2000}
Vanag, V.~K., Zhabotinsky, A.~M., and Epstein, I.~R. (2000).
\newblock Pattern formation in the {B}elousov--{Z}habotinsky reaction with
  photochemical global feedback.
\newblock {\em J. Phys. Chem. A}, 104(49):11566--11577.

\bibitem[Venkatramani \emph{et~al.}, 2016]{venkatramani2016}
Venkatramani, J., Nair, V., Sujith, R.~I., Gupta, S., and Sarkar, S. (2016).
\newblock Precursors to flutter instability by an intermittency route: a model
  free approach.
\newblock {\em J. Fluid Struct.}, 61:376--391.

\bibitem[Venkatramani \emph{et~al.}, 2017]{venkatramani2017}
Venkatramani, J., Nair, V., Sujith, R.~I., Gupta, S., and Sarkar, S. (2017).
\newblock Multi-fractality in aeroelastic response as a precursor to flutter.
\newblock {\em J. Sound Vib.}, 386:390 -- 406.

\bibitem[Venkatramani \emph{et~al.}, 2018]{venkatramani2018}
Venkatramani, J., Sarkar, S., and Gupta, S. (2018).
\newblock Investigations on precursor measures for aeroelastic flutter.
\newblock {\em J. Sound Vib.}, 419:318--336.

\bibitem[Veraart \emph{et~al.}, 2012]{veraart2012}
Veraart, A.~J., Faassen, E.~J., Dakos, V., van Nes, E.~H., L{\"u}rling, M., and
  Scheffer, M. (2012).
\newblock Recovery rates reflect distance to a tipping point in a living
  system.
\newblock {\em Nature}, 481(7381):357.

\bibitem[Verhulst, 1990]{verhulst1990}
Verhulst, F. (1990).
\newblock {\em Nonlinear differential equations and dynamical systems}.
\newblock Springer.

\bibitem[Wang \emph{et~al.}, 2012]{wang2012}
Wang, R., Dearing, J.~A., Langdon, P.~G., Zhang, E., Yang, X., Dakos, V., and
  Scheffer, M. (2012).
\newblock Flickering gives early warning signals of a critical transition to a
  eutrophic lake state.
\newblock {\em Nature}, 492(7429):419.

\bibitem[Watson, 1960]{watson1960non}
Watson, J. (1960).
\newblock On the non-linear mechanics of wave disturbances in stable and
  unstable parallel flows part 2. the development of a solution for plane
  poiseuille flow and for plane couette flow.
\newblock {\em J. Fluid Mech.}, 9:371--389.

\bibitem[Wellens \emph{et~al.}, 2003]{wellens2003}
Wellens, T., Shatokhin, V., and Buchleitner, A. (2003).
\newblock Stochastic resonance.
\newblock {\em Rep. Prog. Phys}, 67(1):45.

\bibitem[Wiesenfeld, 1985]{Wiesenfeld1985}
Wiesenfeld, K. (1985).
\newblock Noisy precursors of nonlinear instabilities.
\newblock {\em J. Stat. Phys.}, 38(5):1071--1097.

\bibitem[Wissel, 1984]{wissel1984}
Wissel, C. (1984).
\newblock A universal law of the characteristic return time near thresholds.
\newblock {\em Oecologia}, 65(1):101--107.

\bibitem[Xu \emph{et~al.}, 2011]{xu2011stochastic}
Xu, Y., Gu, R., Zhang, H., Xu, W., and Duan, J. (2011).
\newblock Stochastic bifurcations in a bistable {D}uffing--van der {P}ol
  oscillator with colored noise.
\newblock {\em Phys. Rev. E}, 83(5):056215.

\bibitem[Yamapi \emph{et~al.}, 2012]{yamapi2012effective}
Yamapi, R., Filatrella, G., Aziz-Alaoui, M.~A., and Cerdeira, H.~A. (2012).
\newblock Effective {F}okker--{P}lanck equation for birhythmic modified van der
  {P}ol oscillator.
\newblock {\em Chaos}, 22(4):043114.

\bibitem[Yi and Gutmark, 2008]{yi2008}
Yi, T. and Gutmark, E.~J. (2008).
\newblock Online prediction of the onset of combustion instability based on the
  computation of damping ratios.
\newblock {\em J. Sound Vib.}, 310(1-2):442--447.

\bibitem[Yin \emph{et~al.}, 2019]{yin2019asymmetric}
Yin, B., Guan, Y., and Li, L. K.~B. (2019).
\newblock Asymmetric forcing of two coupled thermoacoustic oscillators.
\newblock {\em Bull. Am. Phys. Soc.}, 64.

\bibitem[Yoshida \emph{et~al.}, 2003]{yoshida2003}
Yoshida, T., Jones, L., Ellner, S., Fussmann, G., and Hairston~Jr, N. (2003).
\newblock Rapid evolution drives ecological dynamics in a predator--prey
  system.
\newblock {\em Nature}, 424(6946):303.

\bibitem[Yudianto and Xie, 2010]{yudianto2010comparison}
Yudianto, D. and Xie, Y. (2010).
\newblock A comparison of some numerical methods in solving 1-d steady-state
  advection dispersion reaction equation.
\newblock {\em Civ. Eng. Environ. Syst.}, 27(2):155--172.

\bibitem[Zakharova \emph{et~al.}, 2010]{Zakh2010}
Zakharova, A., Vadivasova, T., Anishchenko, V., Koseska, A., and Kurths, J.
  (2010).
\newblock Stochastic bifurcations and coherencelike resonance in a
  self-sustained bistable noisy oscillator.
\newblock {\em Phys. Rev. E}, 81(1):011106.

\bibitem[Zhu and Yu, 1987]{ZHU1987421}
Zhu, W. and Yu, J. (1987).
\newblock On the response of the van der {P}ol oscillator to white noise
  excitation.
\newblock {\em J. Sound Vib.}, 117(3):421 -- 431.

\bibitem[Zhu, 2017]{zhu2017mphil}
Zhu, Y. (2017).
\newblock Transition to global instability in low-density axisymmetric jets:
  Bistability, intermittency and coherence resonance.
\newblock Master's thesis, The Hong Kong University of Science and Technology.

\bibitem[Zhu \emph{et~al.}, 2016]{zhu2016subcritical}
Zhu, Y., Gupta, V., and Li, L. K.~B. (2016).
\newblock Subcritical hopf bifurcations in low-density jets.
\newblock {\em Bull. Am. Phys. Soc.}, 61.

\bibitem[Zhu \emph{et~al.}, 2017]{zhu2017onset}
Zhu, Y., Gupta, V., and Li, L. K.~B. (2017).
\newblock Onset of global instability in low-density jets.
\newblock {\em J. Fluid Mech.}, 828:R1.

\bibitem[Zhu \emph{et~al.}, 2019]{zhu2019}
Zhu, Y., Gupta, V., and Li, L. K.~B. (2019).
\newblock Coherence resonance in low-density jets.
\newblock {\em J. Fluid Mech.}, 881:R1.

\bibitem[Zhu \emph{et~al.}, 2018]{zhu2018noise}
Zhu, Y., Lee, M., Gupta, V., and Li, L. K.~B. (2018).
\newblock Noise-induced triggering in low-density jets.
\newblock {\em Bull. Am. Phys. Soc.}, 63.

\bibitem[Zinn and Lores, 1971]{zinn1971}
Zinn, B.~T. and Lores, M.~E. (1971).
\newblock Application of the galerkin method in the solution of non-linear
  axial combustion instability problems in liquid rockets.
\newblock {\em Combust. Sci. Technol.}, 4(1):269--278.

\end{thebibliography}

\newpage
\addcontentsline{toc}{chapter}{Publications from this thesis}
\null\skip0.2in
\begin{center}
{\bf \Large \underline{Publications arising from this thesis}}
\end{center}
\vspace{12mm}

\subsubsection*{Journal Papers}

\noindent 1. Lee, M., Kim, K. T., Gupta, V., and Li, L. K. B. (2020). System identification and early warning detection of thermoacoustic oscillations in a turbulent combustor using its noise-induced dynamics. Under review by the \textit{P. Combust. Inst.}

\noindent 2. Lee, M., Guan, Y., Gupta, V., and Li, L. K. B. (2020). Input-output system identification of a thermoacoustic oscillator near a Hopf bifurcation using only fixed-point data. \textit{Phys. Rev. E}, 101(1):013102.

\noindent 3. Lee, M., Li, L. K. B., and Song, W. (2019). Analysis of direct operating cost of wide-body passenger aircraft: A parametric study based on Hong Kong. \textit{Chinese J. Aeronaut.}, 32(5):1222--1243.

\noindent 4. Lee, M., Zhu, Y., Li, L. K. B., and Gupta, V. (2019). System identification of a low-density jet via its noise-induced dynamics. \textit{J. Fluid Mech.}, 862:200--215.

\subsubsection*{Conference Talks and Papers}

\noindent 1. Lee, M., Zhu, Y., Guan, Y., Li, L. K. B., and Gupta, V.  (2019). Exploiting noise-induced dynamics for system identification near a Hopf bifurcation. \textit{72$^{nd}$ Annual Meeting of the American Physical Society's Division of Fluid Dynamics}.

\noindent 2. Lee, M., Guan, Y., Gupta, V. and Li, L. K. B. (2019), System identification of a thermoacoustic system using its noise-induced dynamics. \textit{16$^{th}$ International Conference on Flow Dynamics}.

\noindent 3. Zhu, Y., Lee, M., and Li, L. K..B. (2018), Noise-induced triggering in low-density jets \textit{71$^{st}$ Annual Meeting of the American Physical Society's Division of Fluid Dynamics}.

\noindent 4. Lee, M., Zhu, Y., Li, L. K. B., and Gupta, V. (2018), System identification of a low-density jet via its noise-induced dynamics. \textit{71$^{st}$ Annual Meeting of the American Physical Society's Division of Fluid Dynamics}.

\noindent 5. Lee, M., Guan, Y., and Li, L. K. B. (2017), Nonlinear time-series analysis of self-excited thermoacoustic oscillations in a Rijke tube. \textit{2017 KSPE Fall Conference}.



\end{document}